\documentclass[sn-mathphys,Numbered]{sn-jnl}

\usepackage{graphicx}%
\usepackage{multirow}%
\usepackage{amsmath,amssymb,amsfonts}%
\usepackage{amsthm}%
\usepackage{mathrsfs}%
\usepackage[title]{appendix}%
\usepackage{xcolor}%
\usepackage{textcomp}%
\usepackage{manyfoot}%
\usepackage{booktabs}%
\usepackage{listings}%
\usepackage[ruled]{algorithm2e}

\theoremstyle{thmstyleone}%
\theoremstyle{thmstyletwo}%

\theoremstyle{thmstylethree}%

\usepackage{booktabs}
\newcommand{\bm}{\boldsymbol}
\newcommand{\bs}{\boldsymbol}

\makeatletter
\newcommand{\lrarrow}{\mathrel{\mathpalette\lrarrow@\relax}}
\newcommand{\lrarrow@}[2]{%
  \vcenter{\hbox{\ooalign{%
    $\m@th#1\mkern6mu\rightarrow$\cr
    \noalign{\vskip2pt}
    $\m@th#1\leftarrow\mkern6mu$\cr
  }}}%
}
\makeatother

\begin{document}

\title[Bayesian cure rate model]{Bayesian inference and cure rate
modeling  for event history data}


\author*[1]{\fnm{Panagiotis} \sur{Papastamoulis}}\email{papastamoulis@aueb.gr}

\author[2]{\fnm{Fotios S.} \sur{Milienos}}\email{milienos@panteion.gr}

\affil*[1]{\orgdiv{Department of Statistics}, \orgname{Athens University of Economics and Business}, \orgaddress{\street{76, Patission str.}, \city{Athens}, \postcode{10434}, \country{Greece}}}

\affil[2]{\orgdiv{Department of Sociology}, \orgname{Panteion University of Social and Political Sciences}, \orgaddress{\street{136, Syngrou Av.}, \city{Athens}, \postcode{17671}, \country{Greece}}}


\abstract{Estimating model parameters of a general family of cure models is always a challenging task mainly due to flatness and multimodality of the likelihood function. In this work, we propose a fully Bayesian approach in order to overcome these issues. Posterior inference is carried out by constructing a Metropolis-coupled Markov chain Monte Carlo (MCMC) sampler, which combines Gibbs sampling for the latent cure indicators and Metropolis-Hastings steps with Langevin diffusion dynamics for parameter updates. The main MCMC algorithm is embedded within a parallel tempering scheme by considering heated versions of the target posterior distribution. It is demonstrated that along the considered simulation study the proposed algorithm freely explores the multimodal posterior distribution and produces robust point estimates, while it outperforms maximum likelihood estimation via the Expectation-Maximization algorithm. A by-product of our Bayesian implementation is to control the False Discovery Rate  when classifying items as cured or not. Finally, the proposed method is illustrated in a real dataset which refers to recidivism for offenders released from prison; the event of interest is whether the offender was re-incarcerated after probation or not. }

\keywords{MCMC, Bayesian computation, censored data, cured fraction}


\pacs[MSC Classification]{62N02, 62F15, 62N01}

\maketitle

\section{Introduction} 
To deal with sources of heterogeneity, a critical issue in modeling time-to-event or event history data, it is not always an easy task; these sources are frequently unobservable (latent) and hence, difficult to manipulate. In this context, researchers typically follow a mixture model approach, especially when the population under study consists of distinct groups, which are considered as the main source of heterogeneity.  An emblematic problem of this nature is arising when a group of subjects do not face the event of interest, no matter how long they have been followed up, in contrast to the rest groups of the population; these subjects are mostly called as \textit{cured}, \textit{immune} or \textit{long-term survivors} \citep{Mall96, peng2021cure, pengtaylor14,amico2018cure}. It is reasonable to assume the existence of such subjects upon studying, for example, the time-to-divorce (not all couples get divorced; e.g., \citealp{rocha2017new}), recidivism  (not all offenders commit another crime; e.g., \citealp{schmidt1988}, Ch.~5), time-to-default (not all bank customers default; e.g., \citealp{pelaez2022probability}),  job mobility (not all employees change their jobs; e.g., \citealp{yamaguchi1992accelerated}), the residential mobility (not all people change residence; e.g., \citealp{yamaguchi2003accelerated}) or the duration of undergraduate studies (not all students graduate; e.g., \citealp{kalamatianou2003perpetual}), among others.

Obviously, the way we split the population into (mutually exclusive and exhaustive) groups, determines the model at hand. For example, if we assume that the population is only composed  of two groups, the cured (symb., $I=0$) and susceptibles (symb., $I=1$), then for the population survival function, $S_P(t)$, we have
\begin{align}\label{Sp00}
S_P(t)=\pi+(1-\pi) S(t|I=1),
\end{align} 
where $\pi\in[0,1]$ is the probability of being cured, known as \textit{cure rate}  (incidence), and $S(t|I=1)$ is the survival function of susceptibles (latency); due to the definition of cured subjects, we assume that   $S(t|I=0)=1$, for every $t$. In fact, the story of cure modeling started with model \eqref{Sp00}, the \textit{mixture cure model}, and the seminal works by Boag \cite{boag49} which rigorously treated the idea of the existence of cured patients, among those who get a cancer treatment. However, real-life applications make researchers consider different ways of splitting the population into groups, such the one deduced by a competing cause scenario (the roots of this approach can be found in \citealp{yako1994,yakovlev1993simple, hoang1996parametric}, although some earlier works of similar nature could also be met in \citealp{greenhouse1984competing, larson1985mixture}); specifically,  suppose that there is a number of competing causes, $M$, with each cause being able (at least, partially) to deliver the event of interest. Then, the population survival function is now given by
\begin{align}\label{SpM0}
S_P(t)=\sum_{m=0}^{L} S_m(t) P(M=m)=\pi+\sum_{m=1}^{L} S_m(t) P(M=m),
\end{align} 
where the cure rate is equal to the probability of having zero competing causes (i.e., $\pi=P(M=0)$), while $S_m(t)$ is the conditional survival function of subjects with $m$ competing causes (i.e., $S_m(t)=P(T>t|M=m)$, with $S_0(t)=P(T>t|M=0)=1$, for every $t$) and $L$ is the maximum number of $M$.  Hence, under this framework the population is now divided into $L+1$ groups, according to the number of competing causes which could be detected to each subject. Note that the limiting values (as $t\rightarrow \infty$) of the survival functions \eqref{Sp00} or \eqref{SpM0}, are equal to the cure rate,  given that $S(t|I=1)$ and $S_m(t)$ are proper survival functions (i.e., $\lim_{t\rightarrow \infty}=S(t|I=1)=S_m(t)=0$). The mixture model may be seen as a special case of the competing cause cure model \eqref{SpM0}, by assuming that either  the conditional survival function $S_m(t)$ is the same for each $m$, i.e., $S_m(t)=S(t)$,  or $M$ is a Bernoulli random variable, with success probability $1-\pi$. 

Although a great part of the literature refers to the study of the mixture cure model, also known as  \textit{split population model} \citep[Ch. 5]{schmidt1988}, or \textit{limited-failure population model} \citep{meeker1995accelerated}, the research interest, at least in the last two decades, is also heavily focused on the competing cause approach and its extensions. Typically, the time needed for the $j$th competing cause to being activated (\textit{promotion} or \textit{progression} time) is a random variable, $W_j$, and given $M$, $W_j, j=1,2, \ldots$ are assumed to be independent and identically distributed, and independent of $M$, with a common distribution function $F(t)=1-S(t)$. Then, if the first activation suffices for delivering the event of interest, i.e., $T = \min\{ W_0,\ldots,W_M \}$, with $P(W_0 = \infty) = 1$, model \eqref{SpM0} becomes
\begin{equation*}
S_{P}(t)=\pi+\sum_{m=1}^{L} S(t)^m P(M=m)=\varphi(S(t)),
\end{equation*}
where $\varphi(z)$ is the probability generating function of $M$ (e.g., \citealp{Tsodikov03}). Other approaches generalize the above classical competing cause scenario, by asking, for example, the activation of $r$ out of $M$ causes, for facing the event of interest, with $r\geq 1$ (e.g., \citealp{cooner2007flexible}; some generalizations could also be found in \citealp{Tsodikov03, tsod02,tsodikov2003semiparametric,balakrishnan2020class}). Obviously, the stochastic properties of $M$ play a critical role to the competing cause cure model, and one of the fundamental points for selecting a distribution for  $M$ (apart from being non-negative integer-valued random variable, with mass at zero) is whether this distribution is flexible enough, to deal with many different real-life settings. The case where $M$ follows a Poisson distribution, i.e., $S_{P}(t)=\exp\{-\vartheta F(t)\}$, with $\vartheta>0$ being the parameter of the Poisson distribution, known as \textit{bounded cumulative hazard} or \textit{promotion time} cure model, is one of the key references in the literature. Having said that, the research interest often focuses on proposing models which contain as special cases the two most studied cure models: the mixture cure model and the promotion time model. This is the case, for example, for the model derived by using the Box-Cox transformation on the population survival function, which is of the form  (\citealp{yin05b}) 
\begin{align*}
S_P(t)=\left(1-\gamma\vartheta F(t)\right)^{1/\gamma}, \vartheta>0, \gamma\in[0,1], 
\end{align*}
with  $\vartheta:\rightarrow\delta/(1+\delta \gamma)$, and $\delta>0$ (see also \citealp{taylor2007statistical,peng2012extended,diao2012general,pal2017expectation,wang2022two,pal2023parameter}). Note that both the promotion time ($\gamma \rightarrow 0$) and mixture cure model  ($\gamma=1$) are special cases of this family; $\vartheta$ was given by the above form due to the restriction $\gamma\vartheta \in [0,1]$. To mention that this model was estimated within a Bayesian framework in \cite{yin05b} and posterior inference was performed using Adaptive Rejection Metropolis sampling within the Gibbs sampler \citep{10.2307/2986138}. Of similar nature are the models derived by assuming a negative binomial distribution for the number of competing causes (e.g., \citealp{tournoud2008promotion,Castro09,d2018negative,leao2018negative,pal2021simplified}), or a Poisson distribution with random parameter (e.g., \citealp{zeng06}) or a generalized discrete Linnik distribution (\citealp{koutras2017flexible}). 

In \cite{milienos2022reparameterization}, motivated by the above family of cure models, the next model was introduced 
\begin{align}\label{mil22a}
S_P(t)=(1+\gamma c^{\gamma}  F(t)^\lambda)^{-\vartheta/\gamma},\gamma \in \Re,
\end{align}
or equivalently
\begin{align}\label{mil22}
S_P(t)=(1+\gamma \vartheta c^{\gamma \vartheta}  F(t)^\lambda)^{-1/\gamma},\gamma \in \Re,
\end{align}
with $\vartheta>0$, $\lambda>0$, and $c=e^{e^{-1}}$. The above model includes among its special cases the promotion time ($\gamma \rightarrow 0$, $\lambda=1$), the negative binomial ($\gamma>0$, $\lambda=1$) and the mixture cure model ($\gamma=-1$, $\lambda=1$; the binomial cure model for $\gamma<0$, $\lambda=1$). Moreover, the case of zero cure rate, i.e., when the population survival function is equal to $1-F(t)$, can also be covered, and it is not found at the boundary of the parameter
space, as it is usually happens to other classes of cure models; specifically, this case is described by the scenarios $(\gamma,\lambda,\vartheta)=(-e, 1, e)$, under \eqref{mil22a}, or $(\gamma,\lambda,\vartheta)=(-1, 1, e)$,  under \eqref{mil22}. The parameter estimation was carried out by the Expectation-Maximization (EM) algorithm (under to non-informative random right censoring),  wherein the cure indicator among the (right) censored subjects was being treated as the missing information. There are factors, such as, the selection of the initial values of the algorithm (and the corresponding selection for the numerical maximization techniques, involved into the M-step), the flatness and multimodality of the  likelihood function (especially, with respect to the parameter $\gamma$), which seem to affect point estimation accuracy; issues seem to exist also with the  standard errors, and the coverage probabilities of the confidence intervals, relying on the asymptotic normality of the estimators.

The contribution of this work is to propose a more accurate and robust estimation technique in the class of models introduced by \cite{milienos2022reparameterization}, following a fully Bayesian approach. Posterior inference is carried out using Markov chain Monte Carlo (MCMC) sampling. Our MCMC sampler combines standard Metropolis-Hastings moves for single-parameter updates as well as Gibbs sampling \citep{geman1984stochastic} for the latent cure indicators. Simultaneous updates of the parameter vector are also implemented using  Metropolis-Adjusted Langevin diffusion dynamics \citep{10.2307/3318418}. Nevertheless, this scheme will rarely cross low-probability regions; thus, escaping from a minor mode becomes practically unachievable.

We overcome this issue by embedding the main algorithm within a Metropolis-coupled MCMC \citep{altekar2004parallel} scheme. Heated versions of the posterior distribution are considered and multiple MCMC chains are ran in parallel. The heated chains boost the  tendency of the corresponding MCMC samples to deliberately explore the parameter space due to the fact that their target distribution are flatter compared to the cold chain (target posterior distribution). A Metropolis-Hastings step encourages these parallel chains to exchange states, thus, the cold chain has the potential to switch between modes. An extensive simulation study reveals that the proposed method freely explores the multimodal target posterior distribution and outperforms maximum likelihood estimation via the EM algorithm. 

Naturally, classification of subjects as cured or susceptibles is directly available from the MCMC output of the latent status indicators. Furthermore, we are also considering the problem of controlling the False Discovery Rate (FDR) within pre-specified tolerance limits, when the aim of the analysis is to produce a list of ``discoveries'', that is, cured individuals. For this purpose we borrow ideas from Bayesian FDR control procedures in statistical bioinformatics \citep{papastamoulis2018bayesian}.

The rest of the paper is organized as follows. Section \ref{sec:model} presents the main ingredients of our Bayesian approach, that is, the complete likelihood after augmenting the model with latent status indicators, as well as the prior assumptions. Section \ref{sec:mcmc} introduces the MCMC sampling schemes for performing posterior inference. Control of the False Discovery Rate is discussed in Section \ref{sec:fdr}. Illustrations of the proposed methodology are given in Section \ref{sec:app} using simulated datasets and comparing against the EM algorithm (Section \ref{sec:sims}); also, a real-life application on recidivism for offenders released from prison (parole/special sentence, work
release, or discharge) is presented in Section \ref{sec:recidivism}. Further details of the implementation and additional results on the simulation study are included in Supplementary Material.

\section{Bayesian model}\label{sec:model}
The estimation of the distribution of susceptibles and the recognition of the cured subjects, are conventionally  the most challenging tasks in cure modeling. The study of such populations is commonly characterized by the existence of censored data; this is because the cured subjects do not face the event of interest and therefore, no matter what censoring scheme (research design) has been followed up, their time-to-event are censored and their cure status is unknown (nevertheless, the cure status is being treated as partially known, in some approaches; e.g., \citealp{bernhardt2016flexible, laska1992nonparametric, safari2021product, safari2022nonparametric,safari2023latency,wu2014extension}).

In our approach, we assume that    the time-to-event is subject to non-informative random right censoring. Under the classical competing cause scenario and model \eqref{mil22}, we further assume that the promotion time $W_j$ follows a Weibull distribution; however, as can also be seen by the presentation of our estimation approach, this assumption can effortlessly be  relaxed. This is also the case about the role of covariates; although we assume that the set of covariates affects only  parameter $\vartheta$, through an exponential link function, the same or another set of covariates (under conditions), could also affect the rest parameters. Furthermore, it is known that there are some identifiability issues with the model under study  (see also \citealp{milienos2022reparameterization}). For example, under \eqref{mil22a}, we can find two different sets of parameters with exactly the same survival function, given that no covariates have been used in the model; also, when $\gamma=-e$, $\lambda=1$ and the distribution of the promotion time is a Weibull distribution, then a constant cannot be added to the exponential link function of $\vartheta$, in case of using covariates. Here, we use the parametrization \eqref{mil22} and further assuming the existence of a continuous covariate with non-negligible effect on $\vartheta$ through the exponential link function, the identifiability issues are solved.

\subsection{Observed likelihood and data augmentation}\label{sec:ll}
Denoting by $C_i$ and $T_i$ the censoring time and time-to-event of the $i$th subject, respectively, our observed data consist of $Y_i = \min\{T_i,C_i\}$, along with the censoring  indicator $\delta_i$, i.e., 
$$\delta_i = \begin{cases}1,& \mbox{if } Y_i \mbox{ is time-to-event } (T_i < C_i)\\ 0,& \mbox{if } Y_i \mbox{ is censoring time } (T_i > C_i)\end{cases}, i=1,2,\ldots.$$
The cure indicator, 
$$I_i = \begin{cases}1,&\mbox{if the $i$th subject is susceptible} \\ 0, & \mbox{if the $i$th subject is cured}\end{cases},i=1,2,\ldots,$$
is therefore a latent variable, among the censored subjects. Thus, although we know that every uncensored subject is susceptible (i.e., $i\in \Delta_1 \Rightarrow I_i=1$, where $\Delta_1 =\{i:\delta_i=1\}$), this is not the case for every censored subject, which may either be cured or not (i.e., $i\in \Delta_0 \Rightarrow I_i=1$ or $I_i=0$, where $\Delta_0 =\{i:\delta_i=0\}$).   

 Hence, from  $n$ independent pairs $(\delta_i,y_i)$, as well as a set of covariates ${\bm x}=({\bm x}_1,\ldots, {\bm x}_n)$, the likelihood function can be written as
\begin{align*}
L=L({\bm \theta}; {\bm y}, {\bm x}, {\bm \Delta})=\prod_{i=1}^{n}f_P(y_i;{\bm \theta},{\bm x}_i)^{\delta_i}S_P(y_i;{\bm \theta},{\bm x}_i)^{1-\delta_i},
\end{align*}
where  ${\bm y}=(y_1,\ldots,y_n)$, ${\bm \Delta}=(\delta_1,\ldots,\delta_n)$,  and $f_P(y_i;{\bm \theta},{\bm x}_i)$ denotes the population density function, i.e.,
$f_P(y;{\bm \theta},{\bm x})=-\partial S_P(y;{\bm \theta},{\bm x})/\partial y $,
with
$$S_P(y;{\bm \theta},{\bm x})=(1+\gamma\vartheta({\bm x}) c^{\gamma\vartheta({\bm x})} F(y;{\bm \theta})^\lambda)^{-1/\gamma}$$
while, $\bm\theta$ stands for the vector of model parameters.

The complete likelihood function (assuming that the cure indicator for the censored group is observed)  reads  
\begin{align*}
L_c({\bm \theta}; {\bm y}, {\bm x}, {\bm \Delta}, {\bm I})=&\prod_{i\in \Delta_1}(1-p_0({\bm x}_i;{\bm \theta}))\prod_{i\in \Delta_1}f_U(y_i;{\bm \theta},{\bm x}_i) \\
&\prod_{i\in \Delta_0}p_0({\bm x}_i;{\bm \theta})^{1-I_i} \prod_{i\in \Delta_0}\left\{(1-p_0({\bm x}_i;{\bm \theta}))S_U(y_i;{\bm \theta},{\bm x}_i)\right\}^{I_i},
\end{align*}
with ${\bm I}=(I_1,\ldots,I_n)$, and $p_0({\bm x}_i;{\bm \theta})=(1+\gamma\vartheta({\bm x}_i) c^{\gamma\vartheta({\bm x}_i)})^{-1/\gamma}$ being the cure rate, while 
\begin{align*}
S_U(y_i;{\bm \theta},{\bm x}_i)=\frac{S_P(y_i;{\bm \theta},{\bm x}_i)-p_0({\bm x}_i;{\bm \theta})}{1-p_0({\bm x}_i;{\bm \theta})}, f_U(y_i;{\bm \theta},{\bm x}_i)=\frac{f_P(y_i;{\bm \theta},{\bm x}_i)}{1-p_0({\bm x}_i;{\bm \theta})},
\end{align*}
are the (proper) survival and probability density functions of susceptibles, respectively.

\subsection{Prior assumptions}\label{sec:prior}
Let us assume that the promotion times are Weibull distributed (as we mentioned, this assumption could effortlessly be  relaxed), with cumulative distribution function
$$F(y_i;\alpha_1,\alpha_2)=1-\exp\{-(\alpha_1 y_i)^{\alpha_2}\}, \alpha_1>0, \alpha_2>0.$$
Furthermore, supposing that a set of $k$ covariates affects the parameter $\vartheta>0$ of model \eqref{mil22},  through the exponential function, we have 
$$\vartheta_{i} := \vartheta({\bm x}_i)=\exp\{{\bm \beta} {\bm x}_i'\}=\exp\{\beta_0+\beta_1x_{i1}+\ldots+\beta_kx_{ik}\},$$
where ${\bm \beta}=(\beta_0,\beta_1,\ldots,\beta_k)$, ${\bm x}_i=(1,x_{1i},\ldots,x_{ki})$, and $k\geq 1$. Thus, given the vector of parameters $\bm\theta=(\gamma, \lambda, \alpha_1, \alpha_2, {\bm \beta})$ and the (fixed) covariates $\bm x$, the joint density of the complete data $(\bm y, \bm I)$ is
\begin{align}
f(\bm y,\bm I|\bm\Delta, \bm\theta,\bm x) = & \prod_{i\in\Delta_1}(1-p_0(\bm x_i;\bm\theta))
\prod_{i\in\Delta_1}f_{U}(y_i;\bm\theta ,\bm x_i)\nonumber\\
&
\label{eq:completeL}
\prod_{i\in\Delta_0}p_0(\bm x_i;\bm\theta)^{1-I_i}
\prod_{i\in\Delta_0}\left\{(1-p_0(\bm x_i;\bm\theta))S_U(y_i;\bm\theta)\right\}^{I_i}.
\end{align}
All parameters are assumed a-priori independent, thus, the general form of the joint prior distribution factorizes as $\pi(\bs\theta)=\pi(\gamma)\pi(\lambda)\pi(\alpha_1)\pi(\alpha_2)\pi(\bs\beta)$. More specifically, each parameter is a-priori distributed as follows:
\begin{align}
\label{gamma_prior}
\gamma&\sim f_\gamma(a_\gamma,b_\gamma):=\frac{b_\gamma^{a_\gamma}}{2\Gamma(a_\gamma)}|\gamma|^{a_\gamma-1}\exp\{-b_\gamma|\gamma|\}\mathrm{I}_{\mathbb R^\star}(\gamma)\\
\nonumber
\alpha_1&\sim\mathcal{IG}(a_1,b_1)\\
\nonumber
\alpha_2&\sim\mathcal{IG}(a_2,b_2)\\
\nonumber
\lambda&\sim\mathcal{IG}(a_{\lambda},b_{\lambda})\\
\nonumber
\boldsymbol{\beta}=(\beta_0,\beta_1,\ldots,\beta_{k})^\top&\sim\mathcal{N}_p(\bm\mu,\Sigma),
\end{align}
where $a_\gamma > 0$, $b_\gamma > 0$, $a_1 > 0$, $a_2 > 0$, $b_1 > 0$, $b_2 > 0$, $a_\lambda > 0$, $b_\lambda > 0$, $\bm\mu\in\mathbb R^{k+1}$, $\Sigma\in \mathcal M_+$, with $\mathcal M_+$ denoting the space of $(k+1)\times(k+1)$ symmetric positive definite matrices ($G(\cdot, \cdot), \mathcal{IG}(\cdot, \cdot)$ and $\mathcal{N}_p(\cdot, \cdot)$, denote the Gamma (shape, rate), inverse Gamma and $p$th dimensional normal distribution, respectively). Finally, $f_{\gamma}$ corresponds to the  density of a random variable defined as $\gamma = ZU$
where $Z\sim\mathcal G(a_\gamma,b_\gamma)$ and $U \sim\mathcal Uniform_{\{-1,1\}}$, with $Z$ and $U$ independent. Note that \eqref{gamma_prior} reduces to a standard Laplace distribution when $a_\gamma=b_\gamma=1$ (a choice which will be considered in our simulations, among others).

The prior assumptions stated above are based on adopting  continuous distributions on the parameter space. Although this seems a reasonable approach, it is not always the case in cure modeling theory; for example, we refer the reader to \cite{yin05b} where a discrete prior distribution was assigned to $\gamma$, under the Box-Cox transformation cure model, in order to avoid numerical problems. In our applications we consider two distinct sets of hyperparameter values, corresponding to ``vague'' and ``regularized'' setups. The reader is referred to Section \ref{sec:mcmc_details} of supplementary material, for more details. Figure \ref{fig:dag} illustrates the previous hierarchical cure model as a directed acyclic graph, where circles are used to denote unobserved (partially, at least) random variables and squares represent fixed/observed quantities. 

\begin{figure}[t]
\centering
\includegraphics[scale = 0.45]{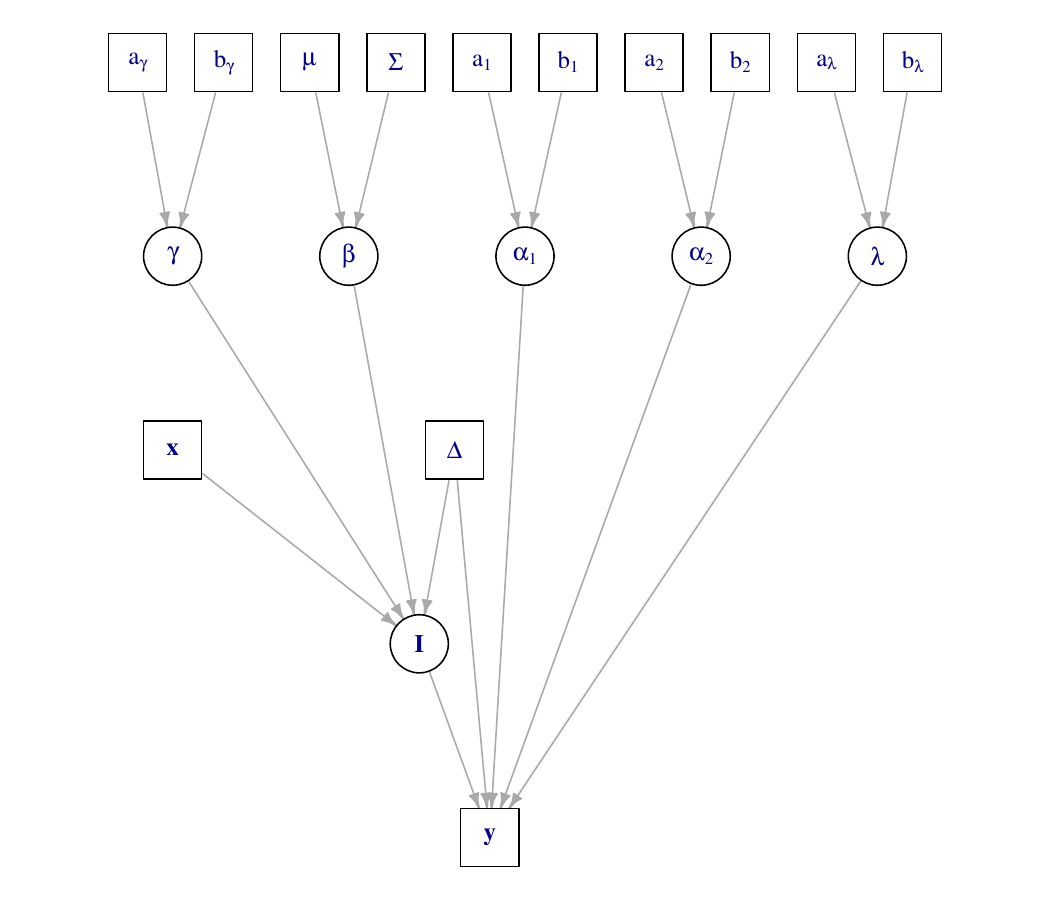}
\caption{Directed acyclic graph representation of the hierarchical Bayesian cure model.}
\label{fig:dag}
\end{figure}

\section{MCMC sampling schemes}\label{sec:mcmc}

The joint posterior distribution of latent indicator variables and model parameters, up to a normalizing constant, is written as

\begin{align}
\nonumber
\pi(\bs\theta,\bs I|\bs\Delta,\bs y,\bs x)
\propto& f(\boldsymbol y|\boldsymbol\theta,\boldsymbol I, \boldsymbol\Delta, \boldsymbol x)\pi(\boldsymbol I,\boldsymbol\theta)\\
\nonumber
\propto& f(\boldsymbol y|\boldsymbol\theta,\boldsymbol I, \boldsymbol\Delta, \boldsymbol x)\pi(\boldsymbol I|\boldsymbol\theta)\pi(\boldsymbol\theta)\\
\propto& f(\bs y, \bs I|\bs\Delta,\bs\theta, \bs x)\pi(\bs\theta)\\
\nonumber
\propto& 
\prod_{i\in\Delta_1}(1-p_0(\bs x_i;\bs\theta))\prod_{i\in\Delta_1}f_U(y_i,\bs x_i)\prod_{i\in\Delta_0}p_0(\bs x_i;\bs\theta)^{1-I_i}\\
&\prod_{i\in\Delta_0}\left\{(1-p_0(\bs x_i;\bs\theta))S_U(y_i,\bs x_i;\theta) \right\}^{I_i}\pi(\bs\theta).
\label{eq:joint_posterior}
\end{align}

In the following sections we will also consider the problem of simulating an MCMC sample which targets a power of the original posterior distribution. For a given constant $0\leqslant h \leqslant 1$, we define the corresponding heated posterior distribution $\pi(\bs\theta,\bs I|\bs\Delta,\bs y,\bs x)^h$, that is,
\begin{align}
\nonumber
\pi(\bs\theta,\bs I|\bs\Delta,\bs y,\bs x)^h\propto &f(\bs y, \bs I|\bs\Delta, \bs\theta, \bs x)^h\pi(\bs\theta)^h\\
\nonumber
\propto &
\prod_{i\in\Delta_1}(1-p_0(\bs x_i;\bs\theta))^h\prod_{i\in\Delta_1}f_U(y_i,\bs x_i)^h\prod_{i\in\Delta_0}p_0(\bs x_i;\bs\theta)^{h(1-I_i)}\\
&\prod_{i\in\Delta_0}\left\{(1-p_0(\bs x_i;\bs\theta))S_U(y_i,\bs x_i;\theta) \right\}^{hI_i}\pi(\bs\theta)^h.
\label{eq:joint_heated_posterior}
\end{align}
Note that 
\begin{align}
\nonumber
&f(\bs y, \bs I|\bs\Delta,\bs\theta, \bs x)^h\propto \dot f_h(\bs y, \bs I|\bs\Delta,\bs\theta, \bs x):=\\
&\prod_{i\in\Delta_1}(1-p_0(\bs x_i;\bs\theta))^h\prod_{i\in\Delta_1}f_U(y_i,\bs x_i)^h\prod_{i\in\Delta_0}p_0(\bs x_i;\bs\theta)^{h(1-I_i)}\prod_{i\in\Delta_0}\left\{(1-p_0(\bs x_i;\bs\theta))S_U(y_i,\bs x_i;\theta) \right\}^{hI_i}
\label{eq:f_h}
\end{align}
as well as
\begin{align*}
\pi(\bs\theta)^h\propto \dot\pi_h(\gamma) \dot\pi_h(\lambda) \dot\pi_h(\alpha_1) \dot\pi_h(\alpha_2)\dot\pi_h(\bs\beta) 
\end{align*}
where we have also defined
\begin{align*}
\dot\pi_h(\gamma)&:=|\gamma|^{\alpha_\gamma - 1} \exp\{-\beta_\gamma|\gamma|\}\mathrm{I}_{\mathbb R^\star}(\gamma)
\\
\dot\pi_h(\lambda)&:=\frac{1}{\lambda^{h(\alpha_\lambda+1)}}\exp\left\{-\frac{h\beta_\lambda}{\lambda}\right\}\mathrm{I}_{(0,\infty)}(\lambda)\\
\dot\pi_h(\alpha_1)&:=\frac{1}{\alpha_1^{h(a_1+1)}}\exp\left\{-\frac{hb_1}{\alpha_1}\right\}\mathrm{I}_{(0,\infty)}(\alpha_1)\\
\dot\pi_h(\alpha_2)&:=\frac{1}{\alpha_2^{h(a_2+1)}}\exp\left\{-\frac{hb_2}{\alpha_2}\right\}\mathrm{I}_{(0,\infty)}(\alpha_2)\\
\dot\pi_h(\bs\beta)&:=\exp\left\{
-\frac{h}{2}(\bs\beta-\bs\mu)^\top \Sigma^{-1}(\bs\beta-\bs\mu)
\right\}\mathrm{I}_{\mathbb{R}^p}(\bs \beta),
\end{align*}
with $\mathrm I_A(\cdot)$ denoting the indicator function of $A$.

\subsection{Metropolis-Hastings and MALA updates of parameters}\label{sec:metropolis_hastings}

The components of the parameter vector $\bs\theta = (\gamma,\lambda,\alpha_1,\alpha_2,\bs\beta)$ are updated using two schemes. First, a Metropolis-Hastings kernel updates each component separately. Second, a Metropolis-Adjusted Langevin diffusion updates simultaneously all of them. These moves generate candidate states by suitable proposal distributions, which are then accepted or rejected according to certain acceptance probabilities. 

More specifically, the first part of our MCMC sampler consists of standard Metropolis-Hastings updates. In brief, we have used a normal proposal distribution for $\gamma$, log-normal distributions for $\alpha_1$, $\alpha_2$ and $\lambda$ and a multivariate normal proposal for $\bs\beta$ (see Section \ref{sec:mh-details} of supplementary material for detailed descriptions of these moves).

The previous Metropolis-Hastings steps are based on random walk proposals per parameter and it is well known that may lead to slow-mixing and poor convergence of the MCMC sampler. In order to overcome this issue, we also use a proposal distribution which is based on the gradient information of the  (heated) posterior distribution, conditional on the current value of latent status indicators. The Metropolis Adjusted Langevin Algorithm (MALA) \citep{10.2307/3318418, https://doi.org/10.1111/1467-9868.00123, girolami2011riemann} is based on the following proposal mechanism 
\begin{equation}
\label{eq:proposal}
\tilde{\bs\theta} = \bs\theta + \tau\nabla\log \pi^h(\bs\theta|\bs\Delta,\bs y, \bs x, \bs I) + \sqrt{2\tau}\bs\varepsilon, 
\end{equation}
where $\bs\varepsilon\sim\mathcal N(\bs 0, \mathcal I_{p+4})$ and $\nabla\log \pi^h(\bs\theta|\bs\Delta,\bs y, \bs x, \bs I)$ denotes the gradient vector of the logarithm of the full conditional heated posterior distribution of $\bs\theta$ ($\mathcal I_{p+4}$ stands for the $(p+4)\times (p+4)$ identity matrix; see  Sections \ref{sec:mala-details} and \ref{ap2} of supplementary material for a detailed description of the implementation).

\subsection{Gibbs update for latent cured status indicators}\label{sec:gibbs}

Every recorded time-to-event ($i \in \Delta_1$) implies that the corresponding subject is susceptible ($I_i = 1$), thus, the full conditional distribution in this case is a degenerate distribution at $1$, i.e.,
\begin{align*}
\mathrm{P}(I_i = 1|\delta_i = 1) = 1.
\end{align*}
For the remaining (censored) times, according to Equation \eqref{eq:joint_posterior}, we have that
\begin{align*}
\mathrm{P}(I_i = 1|\delta_i = 0,\boldsymbol{\theta},\boldsymbol{y},\boldsymbol{x}) &\propto 
(1-p_0(\bs x_i;\bs\theta))S_U(y_i,\bs x_i;\theta) = S_P(y_i, x_i;\bs\theta)-p_0(x_i;\bs\theta)\\
\mathrm{P}(I_i = 0|\delta_i = 0,\boldsymbol{\theta},\boldsymbol{y},\boldsymbol{x}) &\propto p_0(x_i;\bs\theta).
\end{align*}
In summary, the full conditional distribution of latent status indicators is 
\begin{equation*}
I_i|(\delta_i,\boldsymbol{\theta},\boldsymbol{y},\boldsymbol{x})\sim\mathcal B(1,\rho_i), \quad \mbox{independent for } i=1,\ldots,n
\end{equation*}
where 
$$\rho_i = 
\begin{cases}
1, &\quad\mbox{ if } \delta_i = 1\\
\dfrac{S_P(y_i, x_i;\bs\theta)-p_0(x_i;\bs\theta)}{S_P(y_i,x_i;\bs\theta)}, &\quad\mbox{ if } \delta_i = 0
\end{cases}
$$
($\mathcal B(1,\rho_i)$ stands for the binomial distribution with parameters, 1 and $\rho_i$).

Generalizing this to the case where the target distribution is the heated posterior $\pi(\bs\theta,\bs I|\bs\Delta,\bs y,\bs x)^h$ in Equation \eqref{eq:joint_heated_posterior} for some $h\in[0,1]$, it readily follows that the relevant full conditional distribution becomes

\begin{equation}
I_i|(\delta_i,\boldsymbol{\theta},\boldsymbol{y},\boldsymbol{x})\sim\mathcal B(1,w_{hi}), \quad \mbox{independent for } i=1,\ldots,n
\label{eq:I_full_conditional}
\end{equation}
where $$w_{hi} = 
\begin{cases}
1, &\quad\mbox{ if } \delta_i = 1\\
\dfrac{\left\{S_P(y_i, x_i;\bs\theta)-p_0(x_i;\bs\theta)\right\}^{h}}{\left\{S_P(y_i, x_i;\bs\theta)-p_0(x_i;\bs\theta)\right\}^{h} + p_0(x_i;\bs\theta)^h}, &\quad\mbox{ if } \delta_i = 0
\end{cases}.
$$
Naturally, when $h = 1$ (which corresponds to the original posterior distribution) the expression for $w_{hi}$ collapses  to $\rho_i$ defined above.

Combining the single site Metropolis-Hastings updates, the MALA proposal and the Gibbs sampling step for the latent status gives rise to the Metropolis-Adjusted Langevin Within Gibbs MCMC sampling scheme summarized in Algorithm \ref{alg:mala}. Notice that in each iteration, a random selection is made between the  simple Metropolis-Hastings move (Step 1) and the MALA proposal (Step 2), with corresponding probabilities equal to $p_1$  and $1-p_1$, respectively. The value of $p_1 = 0.5$ is chosen as default in our applications.

\begin{algorithm*}[p]
\colorbox{gray!25}{\parbox{0.88\textwidth}{

\caption{MALA within Gibbs MCMC targeting $\pi(\bs\theta,\bs I|\bs\Delta,\bs y,\bs x)^h$, $0\leqslant h \leqslant 1$.}
\SetKwInOut{Input}{Input}
\SetKwInOut{Output}{Output}
\SetKwBlock{blocknotext}{~}{end}
\SetKwBlock{step}{Step}{~}
\SetKwBlock{stepo}{Step 0:  Initialization}{~}
\SetKwBlock{stepiii}{Step 3:  Gibbs sampling for $\bs I$}{~}
\SetKwBlock{stepi}{Step 1: Simple Metropolis-Hastings updates for $\bs \theta$}{~}
\SetKwBlock{stepii}{Step 2:  MALA proposal for $\bs \theta$}{~}
\SetKwBlock{stepiv}{Step 4:  While $t < M$}{~}
\SetKwFor{For}{for}{~}{endfor}
\SetKwFor{ForAll}{for~all}{~}{endfor}
\SetKwFor{ForEach}{for~each}{~}{endfor}

\Input{Data $\bs y, \bs x, \bs\Delta=(\bs\Delta_0,\bs\Delta_1)$\\ 
Prior hyper-parameters $\mu_\gamma,\sigma^2_\gamma, \alpha_\lambda,\beta_\lambda,a_1,b_1,a_2,b_2,\bm\mu,\Sigma$\\
Scale parameters of the proposal distributions $$\eta_h := (s^2_{g(h)}, s^2_{\lambda(h)},s^2_{\alpha_1(h)},s^2_{\alpha_2(h)}, s^2_{g(h)},\bs\nu_{(h)}, \tau_{(h)})$$ \\
(Inverse) temperature of the target distribution $0\leqslant h\leqslant 1$\\
Number of MCMC iterations $m$\\
Optional vector of starting values $(\bs I^{(0)}, \bs \theta^{(0)})$}
\Output{
MCMC sample $\{\bs\xi_h^{(t)} = (\bs I^{(t)}, \bs \theta^{(t)}), t = 1,\ldots,m\}$}
\stepo{
\SetAlgoVlined\If{Starting values not supplied}{
Obtain random starting values $\bs\xi_h^{(0)} = (\bs \theta^{(0)}, \bs I^{(0)})$}
}
\For{$t=1$ \KwTo $m$}{
With probability $p_1$ go to Step 1, otherwise go to Step 2.\\
\stepi{
\begin{enumerate}
\item[1.1] Generate $\tilde\gamma$ according to \eqref{eq:gamma_proposal} and set $\gamma^{(t)}=\tilde\gamma$ with probability $A_{1,h}(\gamma^{(t-1)},\tilde\gamma)$ in Equation \eqref{eq:gamma_prob}, otherwise set $\gamma^{(t)} = \gamma^{(t-1)}$.
\item[1.2] Generate $\tilde\lambda$ according to \eqref{eq:lambda_proposal} and set $\lambda^{(t)}=\tilde\lambda$ with probability $A_{2,h}(\lambda^{(t-1)},\tilde\lambda)$ in Equation \eqref{eq:lambda_prob}, otherwise set $\lambda^{(t)} = \lambda^{(t-1)}$.
\item[1.3] Generate $\tilde\alpha_1$ according to \eqref{eq:alpha1_proposal} and set $\alpha_1^{(t)}=\tilde\alpha_1$ with probability $A_{3,h}(\alpha_1^{(t-1)},\tilde\alpha_1)$ in Equation \eqref{eq:a1_prob}, otherwise set $\alpha_1^{(t)} = \alpha_1^{(t-1)}$.
\item[1.4] Generate $\tilde\alpha_2$ according to \eqref{eq:alpha2_proposal} and set $\alpha_2^{(t)}=\tilde\alpha_2$ with probability $A_{4,h}(\alpha_2^{(t-1)},\tilde\alpha_2)$ in Equation \eqref{eq:a2_prob}, otherwise set $\alpha_2^{(t)} = \alpha_2^{(t-1)}$.
\item[1.5] Generate $\tilde{\bs{\beta}}$ according to \eqref{eq:beta_proposal} and set $\bs\beta^{(t)}=\tilde{\bs{\beta}}$ with probability $A_{5,h}(\bs\beta^{(t-1)},\tilde{\bs{\beta}})$ in Equation \eqref{eq:beta_prob}, otherwise set $\bs\beta^{(t)} = \bs\beta^{(t-1)}$.
\item[1.6] Set $\bs\theta^{(t)} = \left(\gamma^{(t)},\lambda^{(t)},\alpha_1^{(t)},\alpha_2^{(t)},\bs\beta^{(t)}\right)$.
\end{enumerate}
}
\stepii{

Generate $\tilde{\bs{\theta}}$ according to \eqref{eq:proposal} and set $\bs\theta^{(t)}=\tilde{\bs{\theta}}$ with probability $A_{6,h}(\bs\theta^{(t-1)},\tilde{\bs{\theta}})$ in Equation \eqref{eq:map}, otherwise set $\bs\theta^{(t)} = \bs\theta^{(t-1)}$.

}
\stepiii{
Simulate $I^{(t)}_i$ from $I_i|\bs\Delta,\bs y, \bs x, \bs 
\theta^{(t)}$ in Equation \eqref{eq:I_full_conditional},  $i=1,\ldots,n$.

}
}
END     of algorithm
\label{alg:mala}
}}
\footnotesize{For Equations \eqref{eq:gamma_proposal}-\eqref{eq:map}, see supplementary material.}
\end{algorithm*}

\subsection{Metropolis-coupled MCMC algorithm}\label{sec:mc3}




In case of multimodal posterior distributions it is quite challenging to properly explore the posterior surface, since the simulated MCMC sample may be trapped in a local mode. In order to  improve mixing of the MCMC sampler, the Metropolis-coupled MCMC ($\mbox{MC}^{3}$) \citep{geyer1991markov, geyer1995annealing, altekar2004parallel} strategy is adopted. An $\mbox{MC}^{3}$ sampler runs $C$ chains with different posterior distributions $\pi_c(\xi),  c=1,\ldots,C$. The target posterior distribution corresponds to $c=1$, that is, $\pi_1(\xi) = \pi(\xi)$, while the rest of them correspond to  ``heated'' (flatter) versions of the original target. This is achieved by  considering $\pi_c(\xi) = \pi(\xi)^{h_c}$, where $h_1 = 1$ and $0<h_c < 1$ for $c = 2,\ldots,C$ represents the heat value of the chain. Only the chain that corresponds to the target posterior distribution is used for inference, however, after each iteration a proposal attempts to swap the states of two randomly chosen chains. 

Let us now denote by $\bs \xi_c := (\bs\theta_c,\bs I_c)$ the state of chain $c = 1,\ldots,C$, at a given iteration. A swap between chains $i$ and $j$ is  accepted with probability
\begin{equation}
\label{eq:mh_ar}
A(\xi_i,\xi_j)=
\min\left\{1,\frac{\pi\left(\bs\xi_j|\bs\Delta,\bs y,\bs x\right)^{h_i}\pi\left(\bs\xi_i|\bs\Delta,\bs y,\bs x\right)^{h_j}}{\pi\left(\bs\xi_i|\bs\Delta,\bs y,\bs x\right)^{h_i}\pi\left(\bs\xi_j|\bs\Delta,\bs y,\bs x\right)^{h_j}}\right\},
\end{equation}
where $\pi\left(\bs\xi_j|\bs\Delta,\bs y,\bs x\right)^{h_i}$ is defined in Equation \eqref{eq:joint_heated_posterior} for $h=h_i$, $\bs\theta = \bs\theta_j$ and $\bs I = \bs I_j$. The overall procedure is given at the form of a pseudocode in Algorithm \ref{alg:mc3}. 



\begin{algorithm*}[p]
\colorbox{gray!25}{\parbox{0.8\textwidth}{
\caption{MALA-within-Gibbs MC$^3$ targeting $\pi(\bs\theta,\bs I|\bs\Delta,\bs y,\bs x)$}
\SetKwInOut{Input}{Input}
\SetKwInOut{Output}{Output}
\SetKwBlock{blocknotext}{~}{end}
\SetKwBlock{step}{Step}{~}
\SetKwBlock{stepo}{Step 0:  Initialization}{~}
\SetKwBlock{stepi}{Step 1: perform a cycle of MALA within Gibbs per chain}{~}
\SetKwBlock{stepii}{Step 2: Chain swaping}{~}
\SetKwBlock{stepiii}{Step 3: Store target chain state}{~}
\SetKwFor{For}{for}{~}{endfor}
\SetKwFor{ForAll}{for~all}{~}{endfor}
\SetKwFor{ForEach}{for~each}{~}{endfor}

\Input{Data $\bs y, \bs x,\bs\Delta$\\ 
Number of parallel chains $C$\\
(Inverse) temperatures $1=h_1  > h_2 > \ldots>h_C\geqslant 0$\\
Number of MCMC cycles $N$\\
MCMC iterations per cycle $m_{1}$\\
Number of iterations that will be used for warm-up $m_{0}$\\
Values for the parameters of the proposal distributions $\eta_c$, $c=1,\ldots,C$.
}
\Output{MCMC sample $\{\bs\xi^{(t)}=(\bs I^{(t)}, \bs \theta^{(t)}), t = 1,\ldots,N\}$}

\stepo{
\For{chain $c=1$ \KwTo $C$}{
\begin{enumerate}
\item (Optional): Run Algorithm \ref{alg:mala} repeatedly for $m_0$ iterations and adjust $\eta_c$ until the acceptance rates of the proposed moves is within pre-specified limits.
\item Obtain a random starting value $\bs\xi_{c}^{(0)}$.
\end{enumerate}
}
}

\For{MCMC cycle $t=1$ \KwTo $N$}{
\stepi{
\For{chain $c=1$ \KwTo $C$}{
\begin{enumerate}
\item[1.1] Run Algorithm \ref{alg:mala} with $m = m_1$, $\eta_h = \eta_{h_c}$ and starting value $\bs\xi^{(t - 1)}_{(c)}$.
\item[1.2] Set $\bs\xi_c^{(t)}\leftarrow \left(\bs \theta^{(m_1)}, \bs I^{(m_1)}\right)$
\end{enumerate}
}
}
\stepii{
\begin{enumerate}
\item[2.1] Randomly choose $1\leqslant c\leqslant C - 1$ and set $c_1 = c$, $c_2 = c + 1$ 
\item[2.2] Generate $u\sim\mathcal U(0,1)$
\end{enumerate}

\SetAlgoVlined\If{$u < A(\bs\xi_{c_1}^{(t)},\bs\xi_{c_2}^{(t)})$ in Equation \eqref{eq:mh_ar}}{set $\bs\xi_{c_1}^{(t)}\lrarrow \bs\xi_{c_2}^{(t)}$}

}

\stepiii{
Set $\bs\xi^{(t)} = \bs \xi_1^{(t)}$
}
}
END of algorithm
\label{alg:mc3}
}}
\end{algorithm*}

\section{Controlling the False Discovery Rate}\label{sec:fdr}

Controlling the False Discovery Rate \citep{benjamini1995controlling,storey2003positive}  at a given tolerance $0<\alpha<1$  is of particular interest in cure modeling. In our context, a ``discovery'' corresponds to classifying a given censored individual as cured; hence, to effectively recognize someone as not-susceptible, for example, to a disease relapse, or committing again a crime, is of great importance in terms of decision making.   

Let $D_0$ denote the total number of censored items in our sample, i.e., the cardinality of set $\Delta_0$. For ease of notation, we will rearrange the observed times so that the first $D_0$ subjects correspond to censored ones, that is, $\delta_i = 0$ for $i = 1,\ldots,D_0$, and $\delta_i = 1$ for $i>D_0$.  Within our Bayesian approach (see also \citealp{muller2004optimal}), it follows that
\begin{equation*}
\hat{\mathbb{E}}(\mathrm{FDR}|\mathrm{data}) =  \sum_{j=1}^{D_0}\frac{\widehat{\mathrm{P}}(I_j = 1|\bs\Delta,\bs y,\bs x)d_j}{R},
\end{equation*}
where $d_j\in\{0,1\}$ and $R=\sum_{j=1}^{D_0}d_j$ denote the \textit{decision} for $j$th subject  and the inferred number of cured subjects (among the censored subjects), respectively. The decision  $d_j = 1$ means that $j$th subject is classified as being cured (a discovery),  while $d_j = 0$ means that $j$th subject  is classified as susceptible. In case where $R = 0$, we follow the typical FDR convention \citep{benjamini1995controlling} that $0/0=0$. 
 Notice that the estimated susceptible probabilities are directly available from our MCMC output as
\[
\widehat{\mathrm{P}}(I_j = 1|\bs\Delta,\bs y,\bs x) = \frac{1}{L_T-L_B}\sum_{t = L_B + 1}^{L_T}I_j^{(t)},\quad j =1,\ldots,D_0
\]
where $L_B$ denotes the length of the burn-in period, and $L_T$ the total length of the chain. 

Let now $q_1\geqslant\ldots\geqslant q_{D_0}$ denote the ordered values of $\widehat{\mathrm{P}}(I_j = 0|\bs\Delta,\bs y,\bs x)$, for $j = 1,\ldots,D_0$. Define $G_j:=\sum_{i=1}^j(1-q_i)/j$, $j=1,\ldots,D_0$. For any given tolerance level $0<\alpha<1$, consider the decision rule
\begin{equation}
d_j=\begin{cases}
1,&\quad 1\leqslant j \leqslant k_\alpha\\
0,&\quad k_\alpha+1\leqslant j\leqslant D_0
\end{cases},
\label{eq:fdr}
\end{equation}
where $k_\alpha = \max\{j=1,\ldots,D_0: G_j\leqslant\alpha\}$. It readily follows that $\hat{\mathbb{E}}(\mathrm{FDR}|\bs\Delta,\bs y,\bs x)\leqslant \alpha$ (see \citealp{papastamoulis2018bayesian}, Section 2.6).

\section{Results}\label{sec:app}

Section \ref{sec:sims} uses a large scale simulation study in order to evaluate the accuracy of the proposed method in terms of point estimation, assess the robustness to prior assumptions and also benchmark our results against the EM algorithm. The ability of controlling the FDR is also discussed. Finally, a real dataset application is presented in Section \ref{sec:recidivism}. Details of hyperparmeters of the prior distribution and the MCMC sampler are given in Section \ref{sec:mcmc_details} of supplementary material. Further simulation results and insights are given in Section \ref{sec:sim_more} of supplementary material.

\subsection{Simulation study}\label{sec:sims}

We generated a variety of synthetic datasets under a number of distinct settings; that is, we study datasets with low or high cure rates (from $0\%$ to $60\%$), low or high censoring proportions among susceptibles (from $10\%$ to $40\%$) and various sample sizes ($n=500, 2000, 5000$). As we have already mentioned, the promotion times follow a Weibull distribution; furthermore, a set of two covariates, one discrete (symb., $X_1$, following the discrete uniform distribution on $\{0,1,\ldots, K\}$, with $K=1$ or $K=5$) and one continuous (symb., $X_2$, following the uniform distribution on the interval $[0,1]$), affects the parameter $\vartheta$ through $\vartheta=\exp\{\beta_0+\beta_1 X_1+\beta_2 X_2\}$. In order to get the pre-specified censoring proportions among susceptibles, the censoring times are generated through an exponential distribution, with properly determined parameters. 

Therefore, the first step for the simulation of our datasets, is the generation of the covariate values (${\bm x}_i$) and then, the cured status (i.e., $I_i$) via a Bernoulli random variable with parameter  $P(I_i=0)=p_0({\bm x}_i;{\bm \theta})=(1+\gamma \exp\{{\bm \beta}{\bm x}_i^{'}\} c^{\gamma \exp\{{\bm \beta}{\bm x}_i^{'}\}})^{-1/\gamma}$.  If $I_i=0$, we correspond a censoring time to the $i$th subject (since it is cured), otherwise (i.e., $I_i=1$) the assigned time is the minimum between the censoring time and the time generated by the distribution of susceptibles (i.e., $S_U(t)$); Table \ref{t4b}  contains the parameter values and the deduced cure rate and censoring proportion (among susceptibles), for the simulated datasets used in our numerical experimentation.

\begin{table}[ht]
\centering
\begin{tabular}{cccccccccc} 
\toprule
\multicolumn{1}{c}{Scenario}&$\gamma$&$\lambda$&$\alpha_1$&$\alpha_2$&$\beta_0$&$\beta_1$&$\beta_2$&cure rate& cen. prop.\\\midrule
\multicolumn{1}{c}{A1}&1&1.5&0.8&0.8&1.5&1.5&-0.8& $5\%$& $10\%$\\
\multicolumn{1}{c}{A2}&1&1.5&0.8&0.8&1.5&1.5&-0.8& $5\%$& $20\%$\\\midrule
\multicolumn{1}{c}{B1}&1&1&0.5&0.5&-0.8&1.5&1.5&$25\%$& $10\%$\\
\multicolumn{1}{c}{B2}&1&1&0.5&0.5&-0.8&1.5&1.5&$25\%$& $20\%$\\\midrule
\multicolumn{1}{c}{C1}&1&1&1&1&-4&1&1&$60\%$& $10\%$\\
\multicolumn{1}{c}{C2}&1&1&1&1&-4&1&1&$60\%$& $20\%$\\\midrule
\multicolumn{1}{c}{D1}&-0.05&1&0.8&1&2&-1&1&$40\%$& $10\%$\\
\multicolumn{1}{c}{D2}&-0.05&1&0.8&1&2&-1&1&$40\%$& $20\%$\\\midrule
\multicolumn{1}{c}{E1}&-0.5&1&0.8&1&2&-0.7&1&$25\%$& $10\%$\\
\multicolumn{1}{c}{E2}&-0.5&1&0.8&1&2&-0.7&1&$25\%$& $20\%$\\\midrule
\multicolumn{1}{c}{F1}&-1&0.5&0.5&0.5&1&0&0&$0\%$& $10\%$\\
\multicolumn{1}{c}{F2}&-1&0.5&0.5&0.5&1&0&0&$0\%$& $20\%$\\
\multicolumn{1}{c}{F3}&-1&1&0.5&0.5&1&0&0&$0\%$& $30\%$\\
\multicolumn{1}{c}{F4}&-1&1&0.5&0.5&1&0&0&$0\%$& $40\%$\\
\bottomrule
\multicolumn{10}{l}{{\scriptsize $X_1\sim Uniform\{0,1\}$, for A1-B2, whereas $X_1\sim Uniform\{0,1,\ldots, 5\}$, for the rest cases.}}\\
\end{tabular}
\caption{The parameter values, the deduced cure rate and censoring proportion (among susceptibles), for the simulated datasets. For each scenario and sample size level $n\in\{500, 2000, 5000\}$, a total of $N_\mathrm{sim} = 20$ synthetic datasets were generated. }\label{t4b} 
\end{table}

\begin{figure}[p]
\begin{tabular}{cc}
     \includegraphics[scale = 0.36]{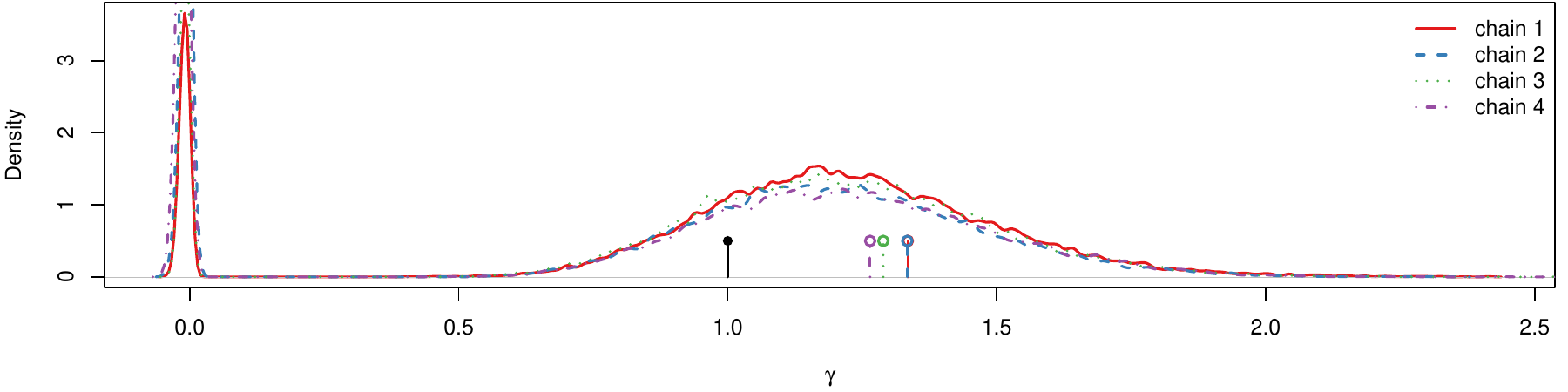}  \\
      \includegraphics[scale = 0.36]{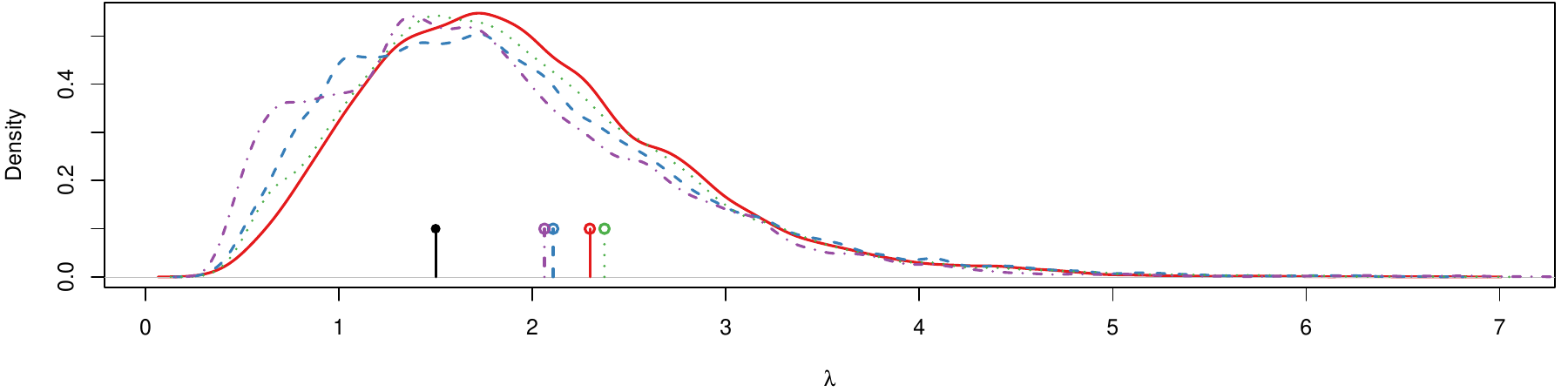}\\
     \includegraphics[scale = 0.36]{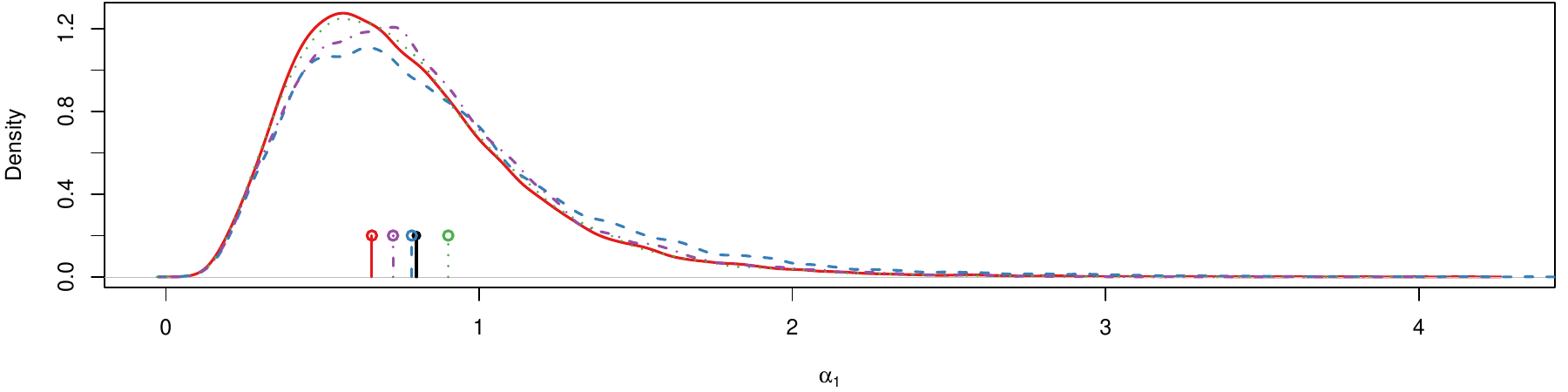}  \\
      \includegraphics[scale = 0.36]{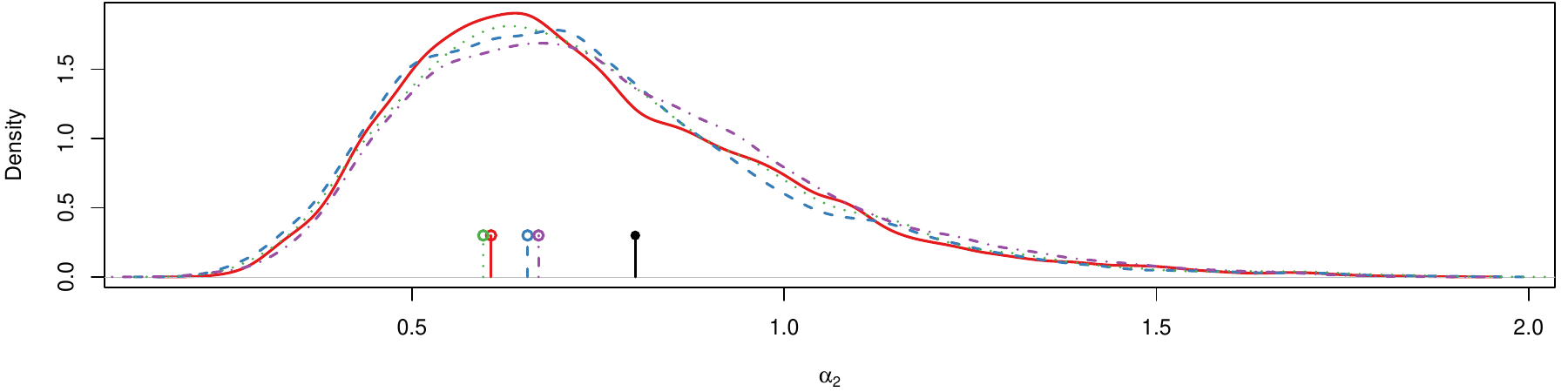}\\
     \includegraphics[scale = 0.36]{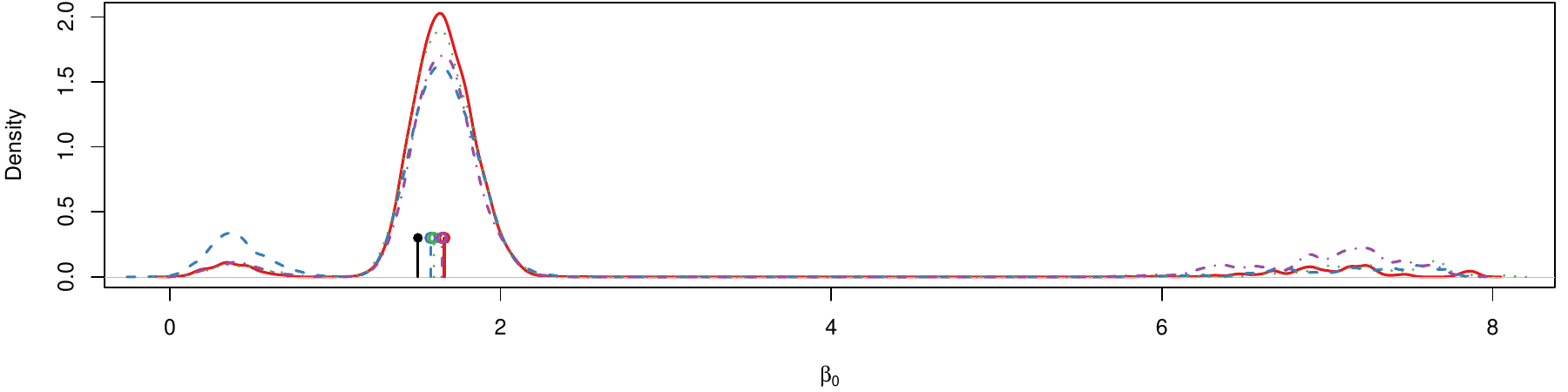}  \\
      \includegraphics[scale = 0.36]{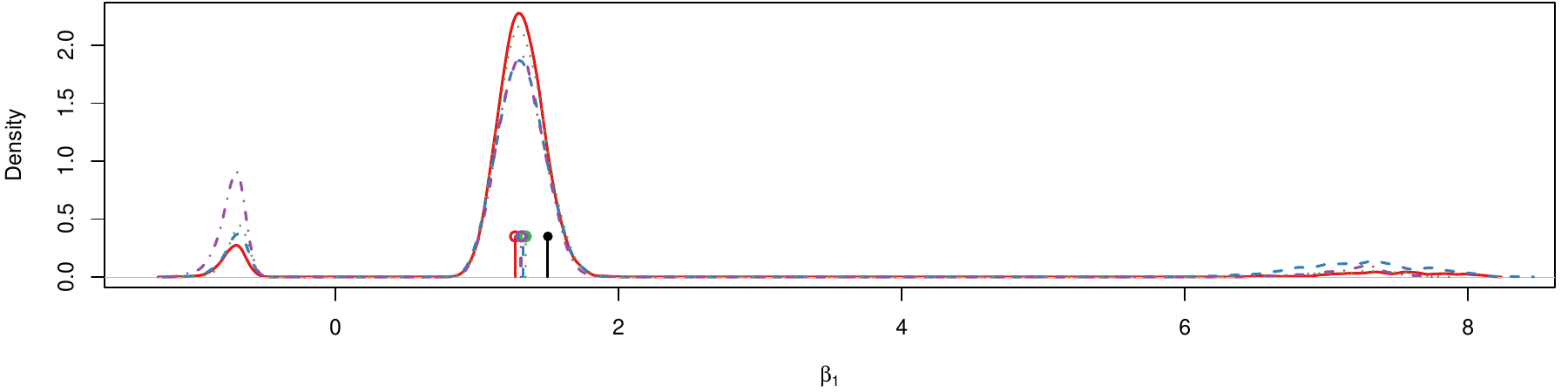}\\
     \includegraphics[scale = 0.36]{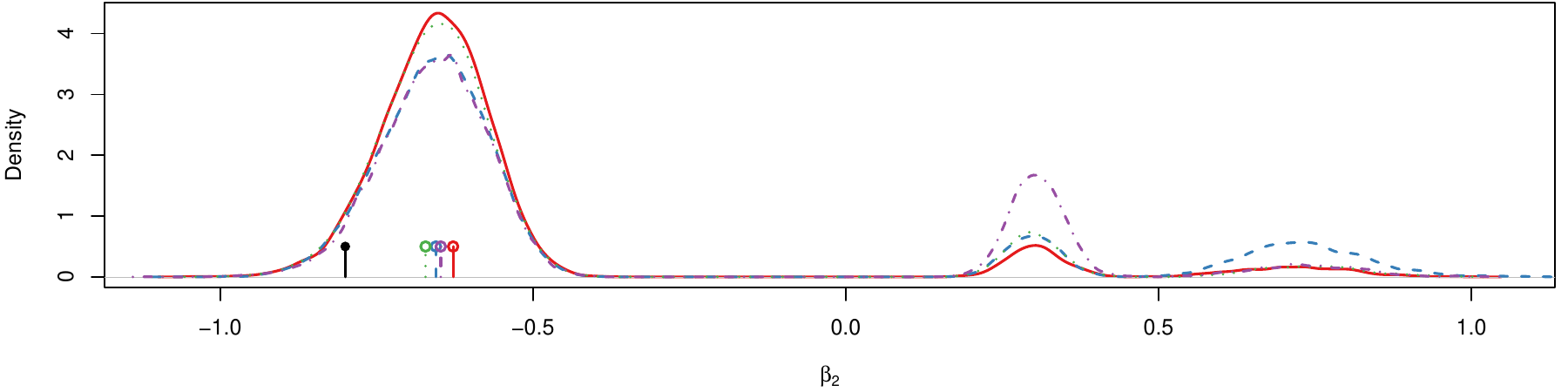}   \\
\end{tabular}
    \caption{Kernel density estimates of the marginal posterior distribution for each parameter, for 4 MCMC chains generated by Algorithm \ref{alg:mc3}. The black vertical line segment indicates the value used to generate the synthetic dataset (see Table \ref{t4b}, Scenario 1). The coloured vertical line segments indicate the MAP estimate per chain.}
    \label{fig:4chains_density}
\end{figure}

\begin{figure}[t]
    \centering
    \includegraphics[scale = 0.4]{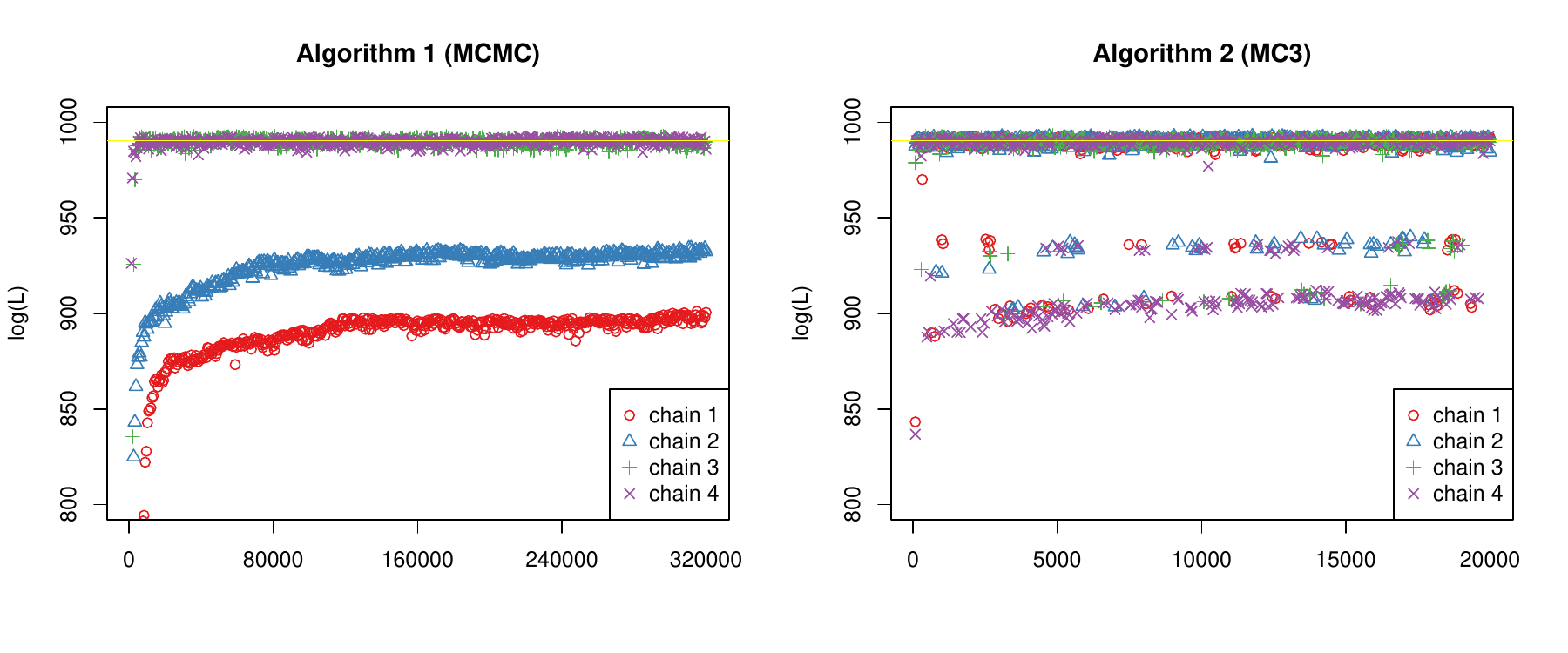}
    \caption{Observed log-likelihood values for four randomly initialized chains according to Algorithm \ref{alg:mala} and \ref{alg:mc3}. The horizontal yellow line indicates the log-likelihood value computed at the values used to generate the dataset.}
    \label{fig:logLvalues}
\end{figure}

 At first we focus on the results for one typical simulated dataset of $n=500$ observations generated from Scenario A1 (see Table \ref{t4b}). Figure \ref{fig:4chains_density} shows the resulting estimates of the marginal posterior distribution for 4 different starting values of Algorithm \ref{alg:mc3}, with $C=16$ tempered chains. The multi-modality in the marginal posterior distributions of $\gamma$, $\beta_0$, $\beta_1$ and $\beta_2$ is apparent. For this reason we will be using the Maximum A Posteriori (MAP) estimate as the point estimate of each parameter, i.e., the sampled parameter values $\boldsymbol\theta = (\gamma, \lambda, \alpha_1, \alpha_2, \beta_0, \beta_1, \beta_2)$ which correspond to the iteration with the highest (joint) posterior density
 \[
\hat{\boldsymbol\theta} = \mathrm{argmax}\{\pi(\boldsymbol{\theta}^{(t)}|\bs y, \bs \Delta, \bs x), t = 1,\ldots,N\}. 
 \]
 These correspond to the coloured vertical lines in Figure \ref{fig:4chains_density}. We can see that  the main mode of the posterior distribution (which corresponds to the one on top of the MAP estimate) contains the true parameter value.  For interval estimation we will be using Highest (posterior) Density Intervals (HDI) (which could be discontinuous intervals in general). Observe also that all four runs of our MC$^3$ sampler essentially lead to the same results, thus we can conclude that Algorithm \ref{alg:mc3} can successfully deal with this multimodal posterior distribution and produce a MCMC sample that converges in relatively small number of iterations in the posterior distribution. This is not the case for Algorithm \ref{alg:mala}, as shown in Figure \ref{fig:logLvalues}, where  the values of the observed log-likelihood are displayed for four randomly initialized chains. Observe that 2 among 4 chains (in particular chain 1 and chain 2) generated by Algorithm \ref{alg:mala} are trapped in a minor mode (see also Section \ref{sec:single_sim} of supplementary material).

Next, we summarize our findings across all $N_\mathrm{sim} = 20$ synthetic datasets and simulation scenarios. Our method is compared against the EM algorithm (the EM algorithm was used in  \cite{milienos2022reparameterization}), based on a small-EM  initialization procedure \citep{biernacki2003choosing} with 45 random starts (the starting values of the EM algorithm were obtained under the same random initialization scheme as in the MCMC sampler; see Section \ref{sec:mcmc_details} of supplementary material). Figure \ref{fig:simStudy_overall_regularized} compares point estimation accuracy for each parameter in terms of the Mean Absolute Error (MAE) between the true value and the point estimate according to each method. The MAE between the true parameter value $\theta_j\in\boldsymbol\theta = (\gamma,\lambda,\alpha_1,\alpha_2,\beta_0,\beta_1,\beta_2)$ and the corresponding estimate for each specific synthetic dataset $\hat\theta_j$ is estimated as 
\[
\mathrm{MAE}(\theta_j, \hat\theta_j) = \frac{1}{N_\mathrm{sim}}\sum_{i = 1}^{N_\mathrm{sim}}|\theta_j - \hat\theta_j^{(i)}|,\quad j = 1,\ldots,7
\]
where $\hat\theta_j^{(i)}$ denotes the estimate of $\theta_j$ arising from synthetic dataset $i$, $i = 1,\ldots,N_{\mathrm{sim}}$.

The bars represent the cumulative $\mathrm{MAE}=\sum_{j=1}^{7}\mathrm{MAE}(\theta_j, \hat\theta_j)$ (note that the contribution of each parameter is displayed using distinct colors). It is clear that our MCMC implementation is quite robust for both prior assumptions across all simulation scenarios. More specifically, the MAEs are almost equal for larger sample sizes ($n = 2000$ or $n=5000$). In the case of smaller sample size ($n=500$), the regularized prior distribution tends to produce smaller MAEs when compared to the vague prior. Both MCMC implementations outperform the EM algorithm, despite the fact that we have used a rather large number (45) of small runs to initialize the final algorithm. Further summary results per scenario are shown in  Figures \ref{fig:hdisA}, \ref{fig:hdisB}, \ref{fig:hdisC}, \ref{fig:hdisD}, \ref{fig:hdisE}, \ref{fig:hdisF} and \ref{fig:hdisF2} in Section \ref{ap1} of supplementary material. In almost all cases, the estimated marginal posterior distributions are multimodal and this mainly affects $\gamma$, $\beta_0$, $\beta_1$ and $\beta_2$. However, it is evident that the MAP estimate of each parameter is close to the ``true'' value. 

\begin{figure}[p]
\centering
\begin{tabular}{cccc}
\includegraphics[scale=0.3]{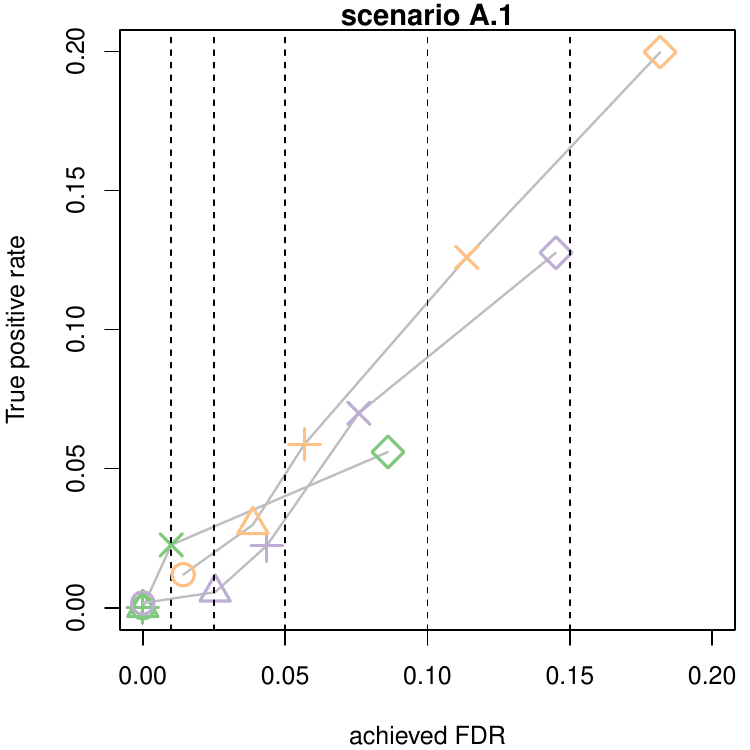} &
\includegraphics[scale=0.3]{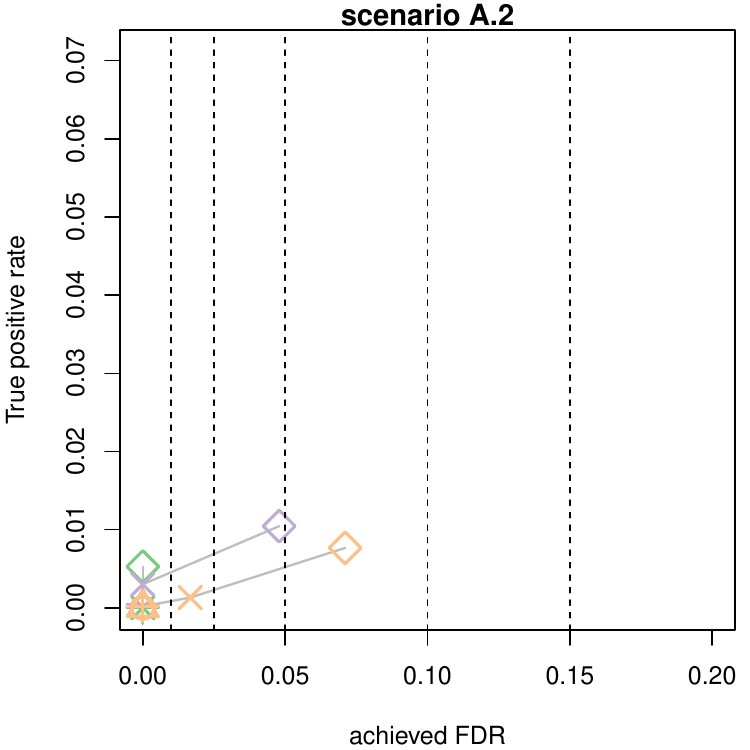} &
\includegraphics[scale=0.3]{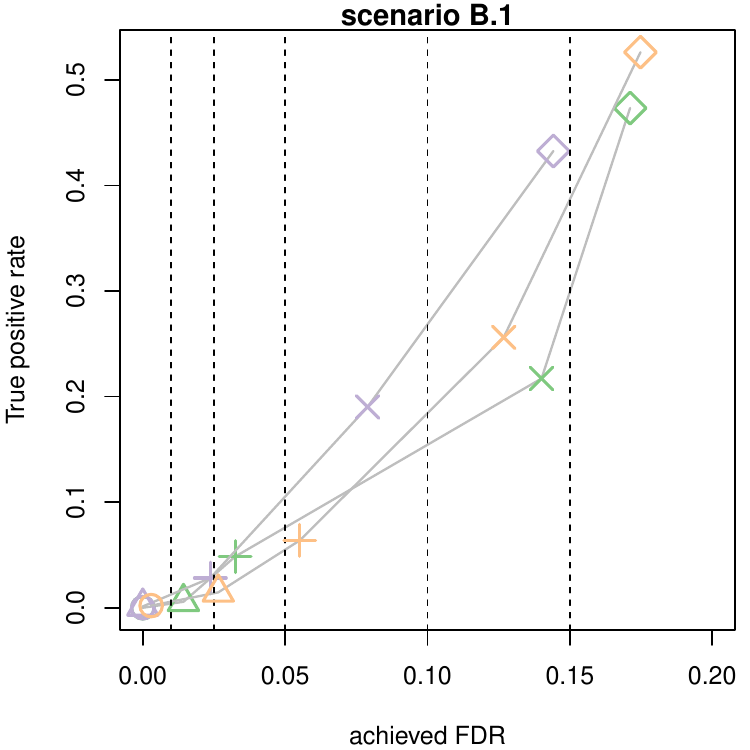} \\
\includegraphics[scale=0.3]{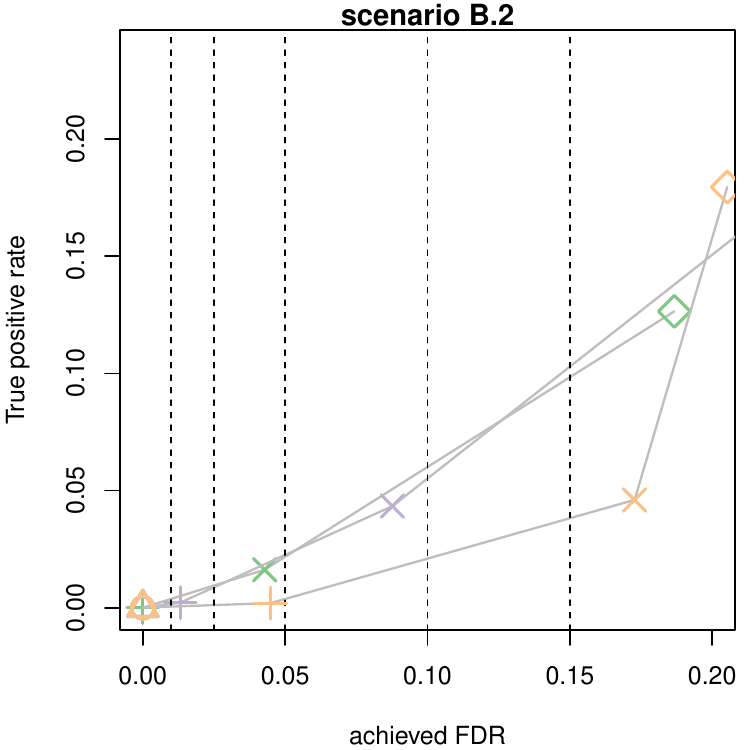} &
\includegraphics[scale=0.3]{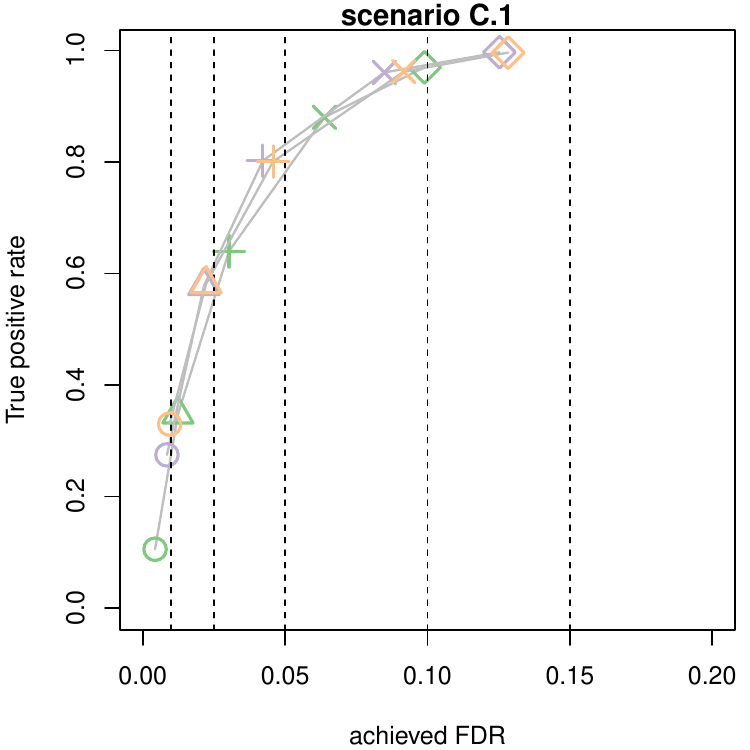} &
\includegraphics[scale=0.3]{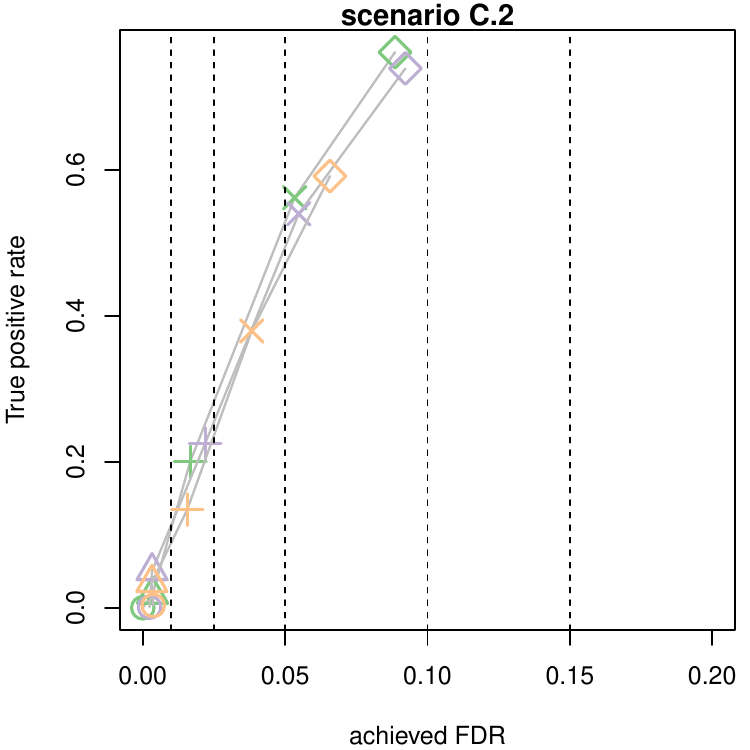} \\
\includegraphics[scale=0.3]{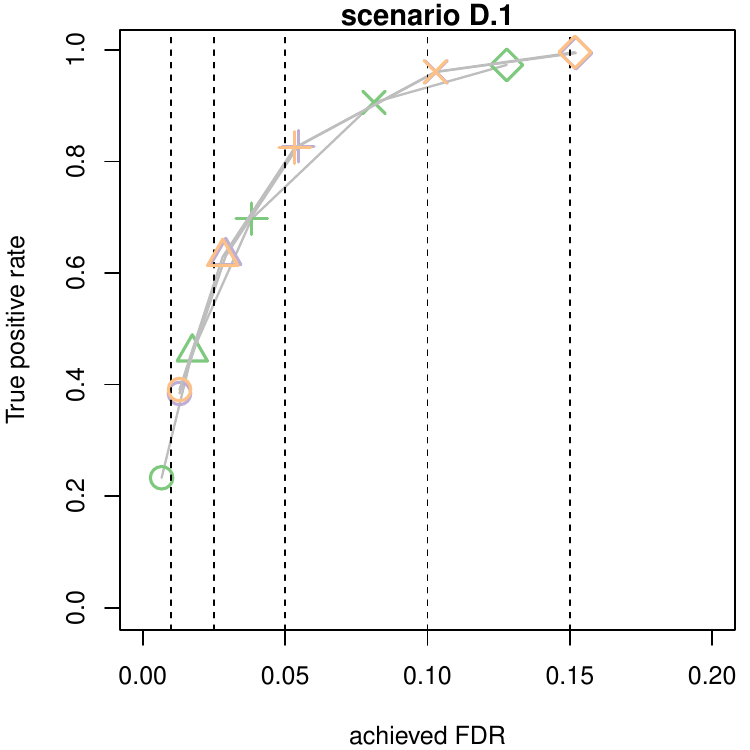} &
\includegraphics[scale=0.3]{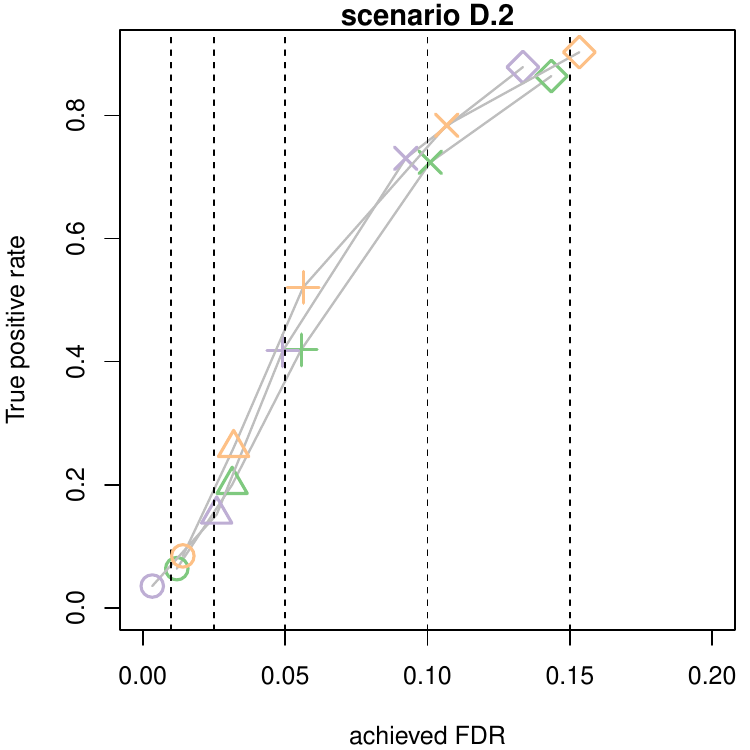} &
\includegraphics[scale=0.3]{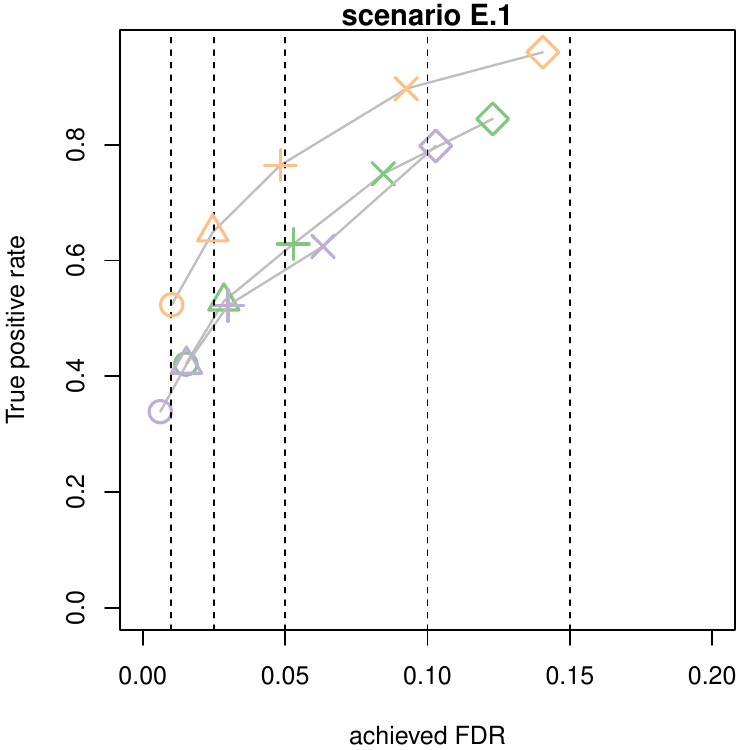} \\
\includegraphics[scale=0.3]{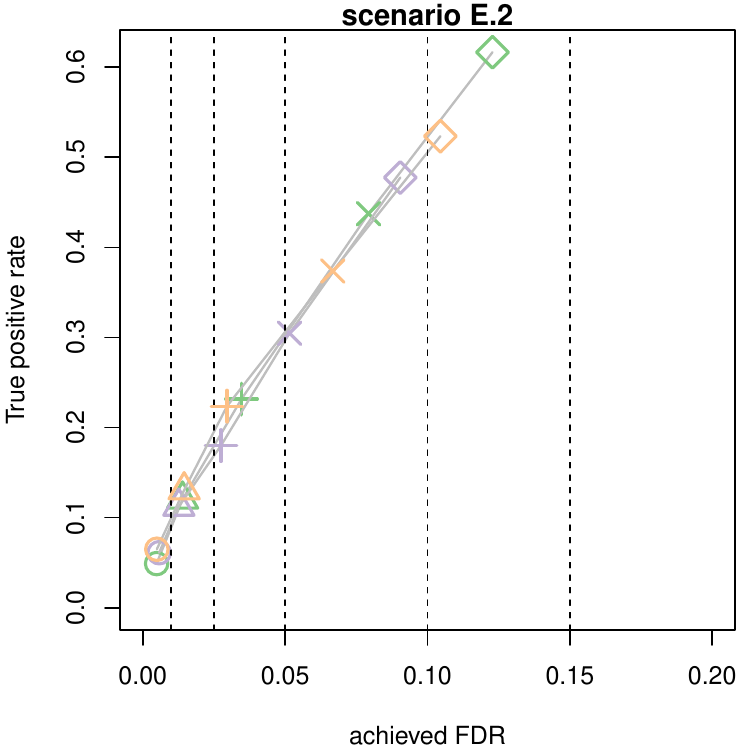} &
\includegraphics[scale=0.3]{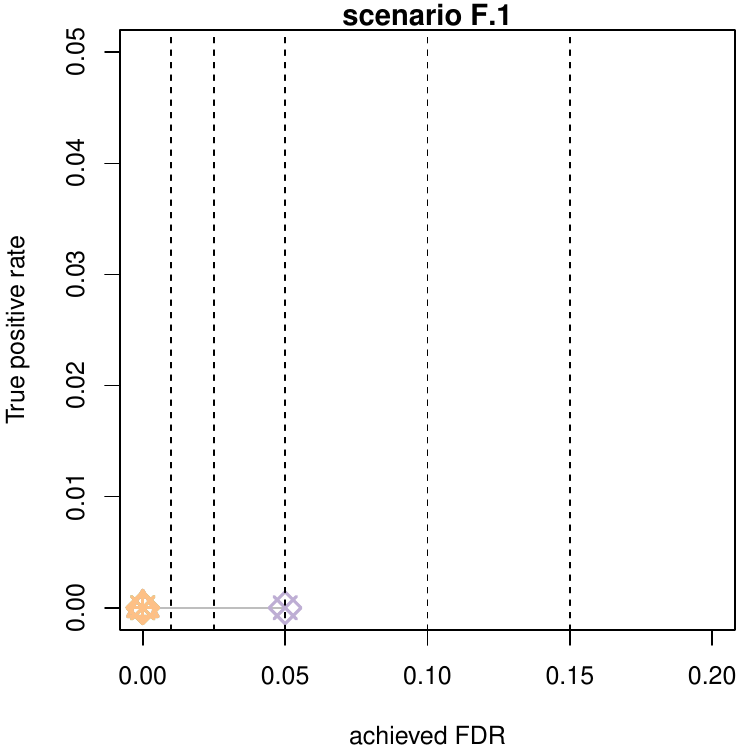} &
\includegraphics[scale=0.3]{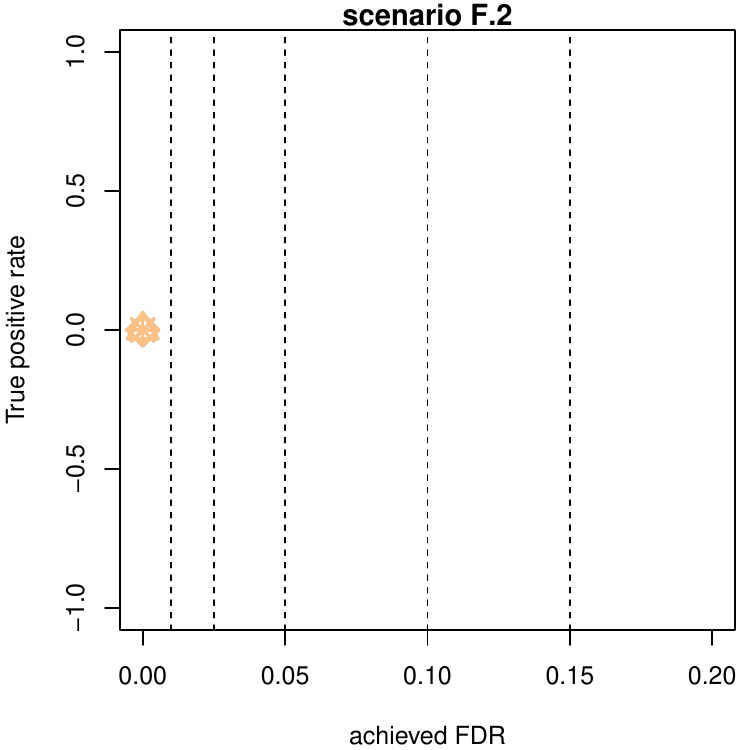} \\
\includegraphics[scale=0.3]{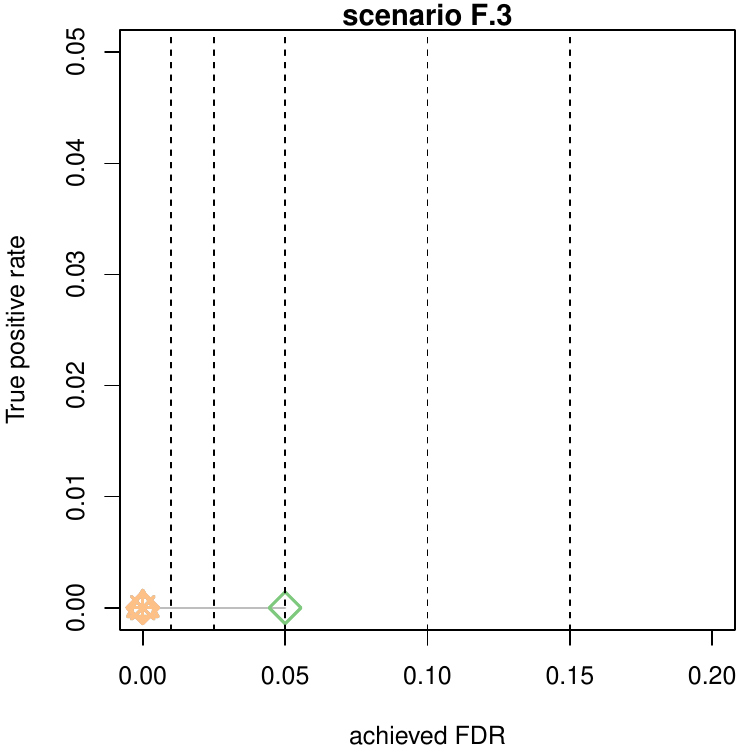} &
\includegraphics[scale=0.3]{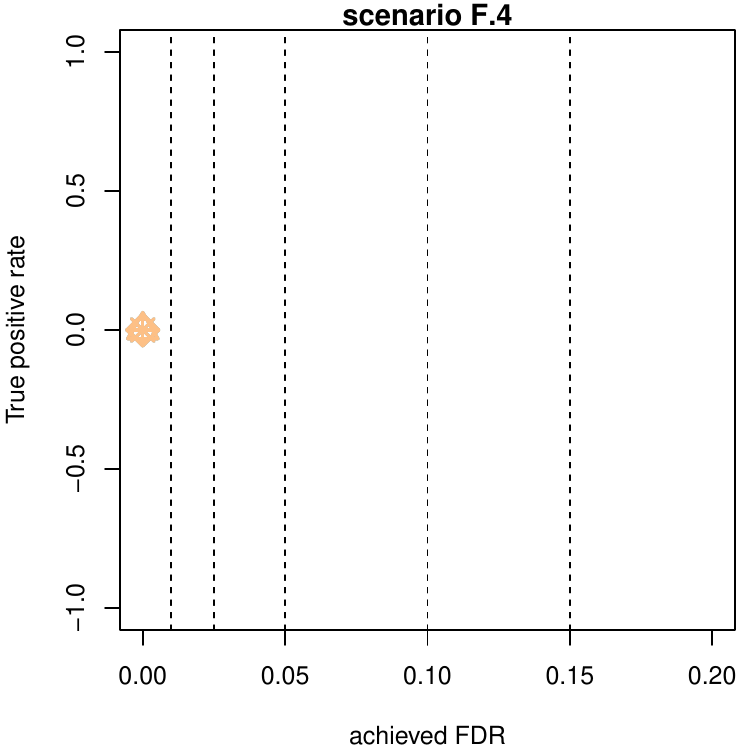} 
\end{tabular}
\caption{Achieved False Discovery Rates versus detection power (True Positive Rate) for simulation scenarios 1-16 and various sample sizes, for the regularized prior distribution: 500 (\textcolor{green}{---}), 2000 (\textcolor{violet}{---}), 5000 (\textcolor{orange}{---}).  Distinct point symbols are used to denote the levels of the target FDR $(\alpha)$ for each case: $0.01 = \circ$, $0.025 = \triangle, 0.05 = +, 0.1 = \times, 0.15 = \diamond$. }
\label{fig:fdr}
\end{figure}

Figure \ref{fig:fdr} illustrates the achieved FDR of cured subjects, for different levels of target FDR ($\alpha = 0.01, 0.025, 0.05, 0.1, 0.15$), when applying the decision rule in Equation \eqref{eq:fdr}. The $y$ axis corresponds to the detection power (or True Positive Rate) for each value of $\alpha$. Note that in almost every single case, the achieved FDR is smaller than the target FDR (each point is on the left of the corresponding vertical dotted line). This means that our method controls the FDR in the desired level. Most points in the last four scenarios are located to $(0,0)$, meaning that zero cured items are discovered (which is the desired behaviour, since there are no cured items in the sample in these cases). Note that these graphs correspond to the regularized prior distribution. Almost identical results are obtained under the vague prior distribution (results not shown).

\begin{figure}[p]
\centering
\begin{tabular}{c}
\hspace{-20ex}\includegraphics[scale=0.45]{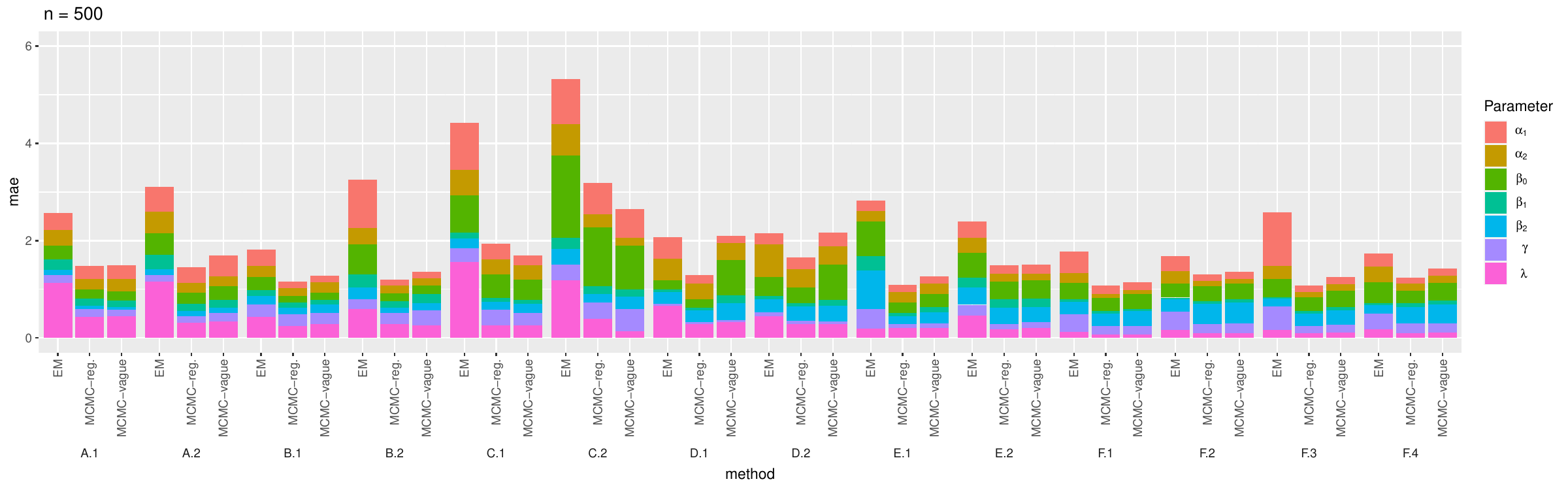}\\
\hspace{-20ex}\includegraphics[scale=0.45]{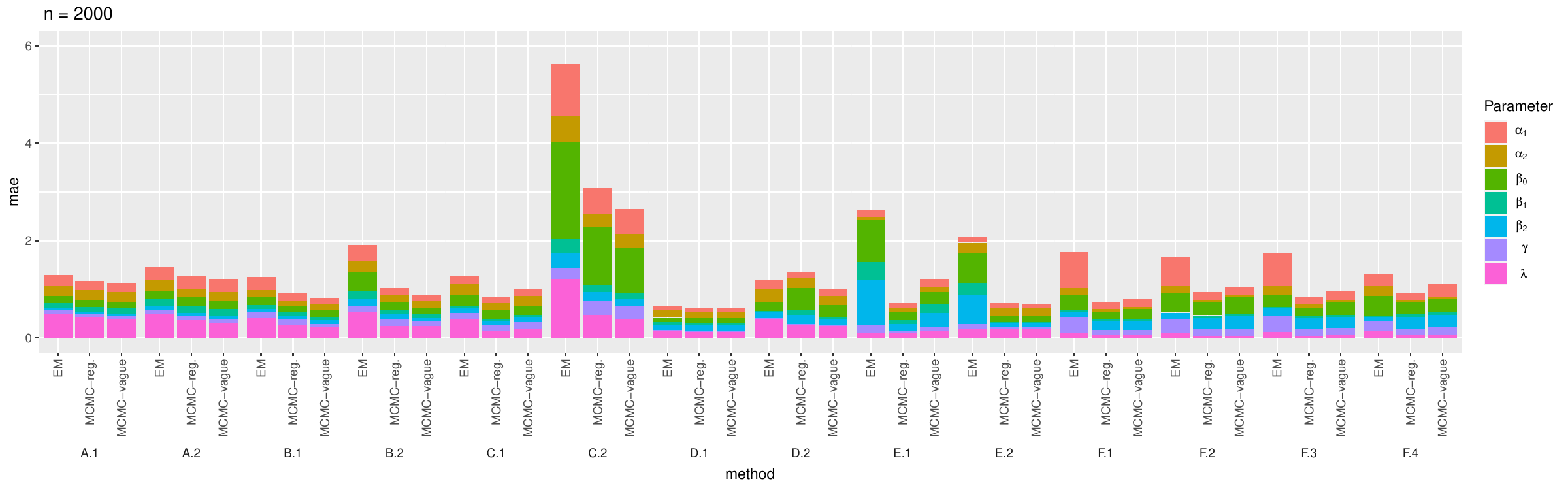}\\
\hspace{-20ex}\includegraphics[scale=0.45]{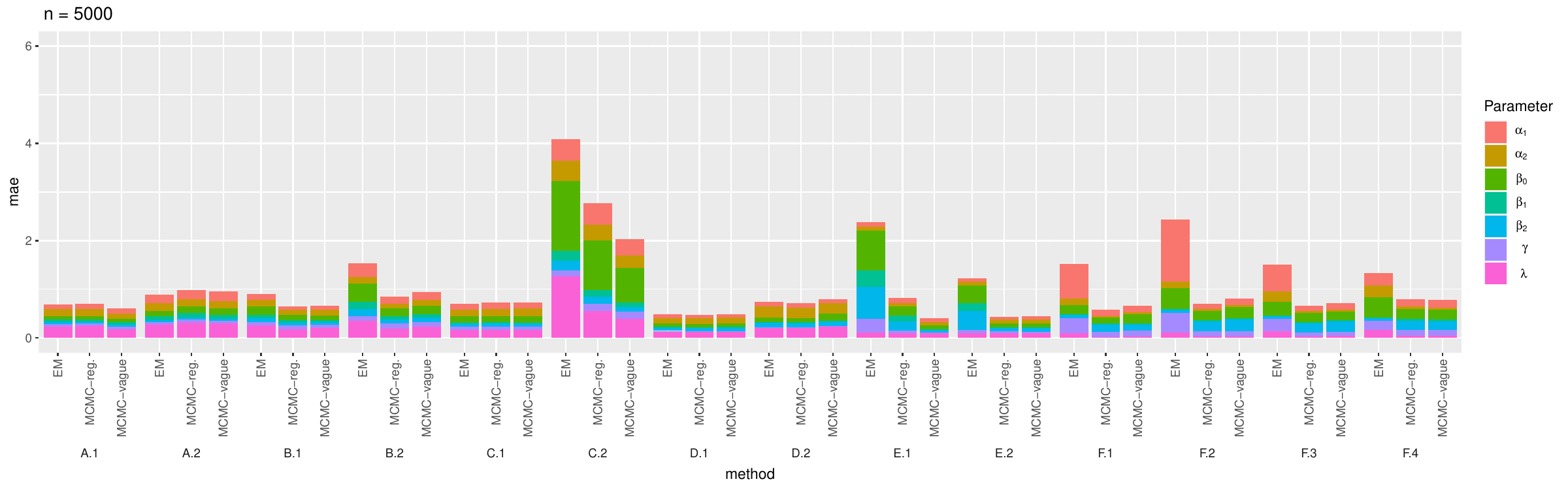}
\end{tabular}
\caption{The bars display cumulative Mean Absolute Errors (MAE) of parameter estimates for both prior schemes of the proposes MCMC sampler as well as the EM algorithm implementation (according to a small EM initialization scheme based on 45 random starts). MCMC-reg.~and MCMC-vague denote the regularized  and vague prior distributions, respectively. }
\label{fig:simStudy_overall_regularized}
\end{figure}

\subsection{Illustrative example: recidivism}\label{sec:recidivism}
In this section, the proposed methodology is illustrated using a  dataset which refers to recidivism for offenders released from prison. The dataset is provided by Iowa Department of Corrections (available for public use in \url{https://data.iowa.gov/}; dataset last updated in January 4, 2023 and retrieved in July 6, 2023); it is a set of offenders who left prison (parole/special sentence, work release, or discharge) and reentering the community, after a period of probation supervision. The event of interest is whether the offender was re-incarcerated after probation; every offender was followed for three years (i.e., the censoring time is degenerate at three years). 

\begin{figure}[p]
    \centering
    \includegraphics[scale = 0.4]{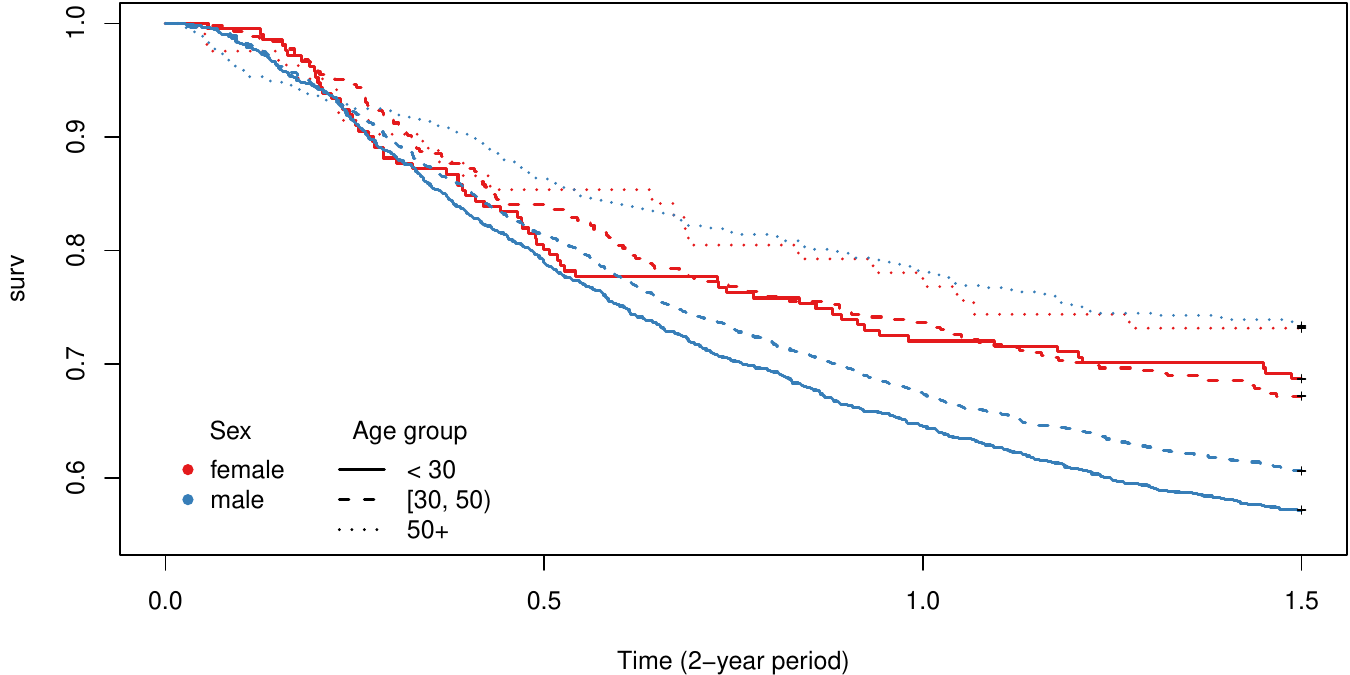}
    \caption{Kaplan-Meier survival function for the recidivism dataset, after grouping age into three categories. }
    \label{fig:recidivism_scatter}
\end{figure}

We considered a randomly selected subset of $n=5000$ observations. The following two covariates were taken into account: sex (0: female, 1: male) and age.  The recorded times were transformed into two-year periods, as in \cite{milienos2022reparameterization} wherein a previous version of the above dataset was analysed. The confusion matrix of censored items versus sex is shown in Table \ref{tab:recidivism}. In order to provide a rough visual summary of the main characteristics of the dataset, the Kaplan-Meier survival function is shown in Figure \ref{fig:recidivism_scatter} after grouping age into three (arbitrary) categories: $[18-30)$, $[30,50)$ and larger than $50$. However, in the subsequent analysis, the aforementioned grouping is not taken into account and age is treated as a continuous covariate.

\begin{table}[ht]
\centering
\begin{tabular}{rrr}
  \toprule
 & Female & Male \\ 
  \midrule
Censored  & 504 & 2601 \\ 
  Time-to-event & 234 & 1661 \\ 
   \bottomrule
\end{tabular}
\caption{Recidivism dataset: censored status for females and males.}
\label{tab:recidivism}
\end{table}

Before implementing our method we have transformed age so the sample mean and variance equal 0 and 1, respectively. Our results are based on 4 separate runs of Algorithm \ref{alg:mc3} with different random starting values. For each run we considered $C = 16$ tempered chains and a total of $N = 100,000$ MCMC cycles.  Our reported results are based on discarding the first $30,000$ cycles as burn-in period, and then thinning the chains by keeping every 10th sampled value.

\begin{table}[h]
\centering
\begin{tabular}{rrrrrrrr}
  \toprule
 & $\gamma$ & $\lambda$ & $\alpha_1$ & $\alpha_2$ & $\beta_0$ & $\beta_1$ (age) & $\beta_2$ (sex)\\ 
  \midrule
MAP estimate & -1.50 & 3.38 & 3.81 & 0.69 & 1.61 & 0.09 & -0.15 \\
$2.5\%$ quantile &-3.58 & 2.11 & 1.91 & 0.47 & -1.01 & -0.21 & -1.82 \\
$25\%$ quantile & -1.68 & 3.04 & 2.96 & 0.58 & -0.86 & -0.18 & 0.14 \\
$50\%$ quantile  & -0.95 & 3.75 & 3.99 & 0.65 & -0.77 & -0.16 & 0.24 \\
$75\%$ quantile & -0.38 & 4.64 & 5.55 & 0.73 & -0.62 & -0.10 & 0.31 \\ 
  $97.5\%$ quantile & 0.43 & 7.05 & 11.10 & 0.90 & 2.00 & 0.09 & 3.80 \\ 
 Gelman's diagnostic  & 1.08 & 1.01 & 1.03 & 1.01 & 1.06 & 1.02 & 1.07 \\ 
   \bottomrule
\end{tabular}
\caption{Posterior distribution summary for the parameters of the recidivism dataset.  }
\label{tab:recidivism_summary}
\end{table}

\begin{figure}[p]
\centering
\includegraphics[scale=0.4]{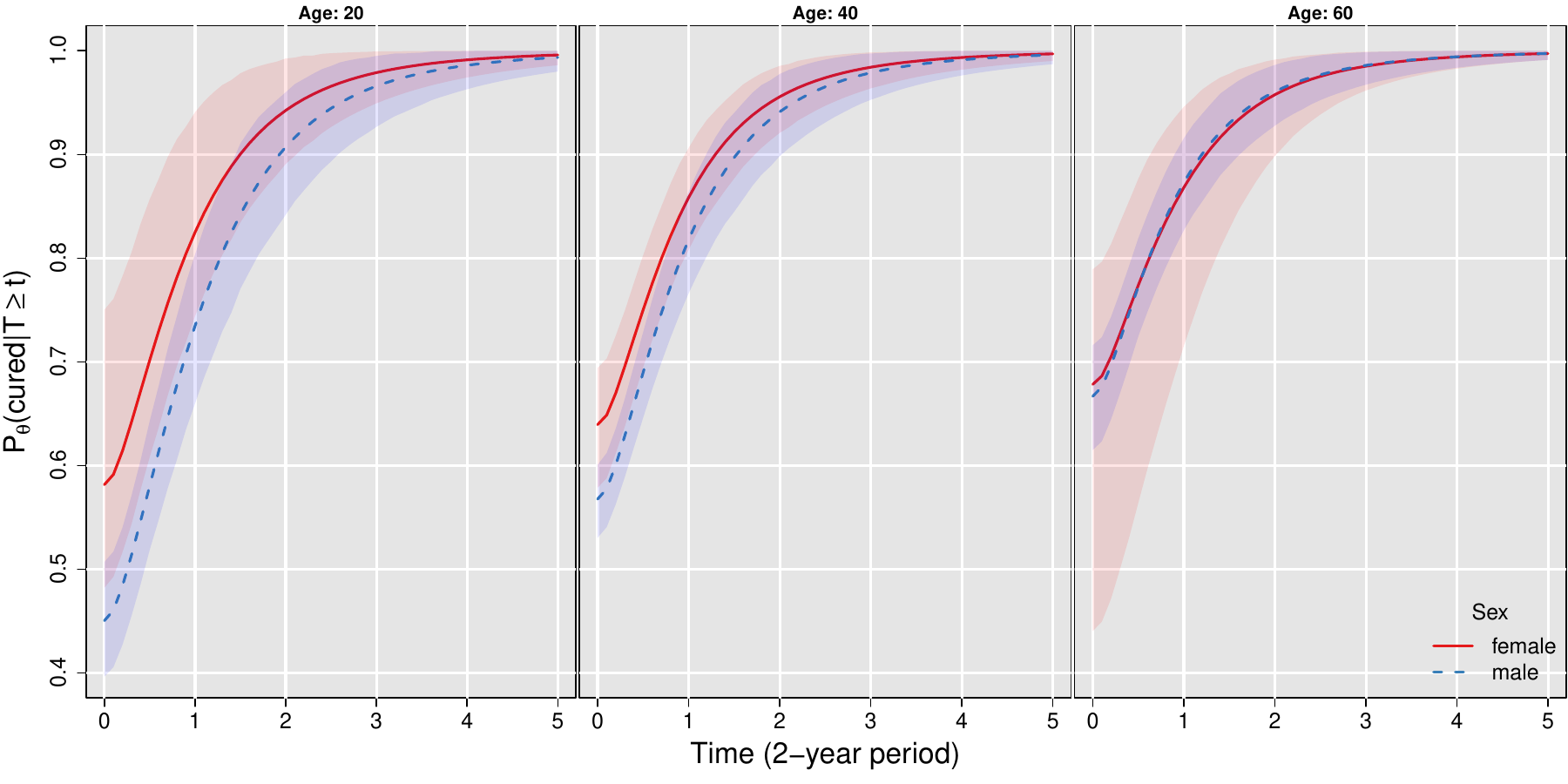}
\caption{Recidivism dataset: estimate of the posterior mean of cured probability for three values of age: 20, 40 and 60. The shaded regions correspond to (pointwise) $95\%$ HDIs.}
\label{fig:p_cured}
\end{figure}

Figure \ref{fig:p_cured} displays the estimated posterior mean of the probability of being cured, conditional on the event that $T \geqslant t$, for females and males, together with the corresponding $95\%$ HDIs. Three different values of the first covariate (age) are displayed, that is, 20, 40 and 60 years old. Clearly, there is a positive effect of age in the mean of cured probability: older offenders are less likely to commit another crime when compared to younger offenders. Moreover,  the posterior mean of cure probability is larger for women in smaller values of age. However, as the offender grows older (60 years) the probability of being cured is similar between men and women.

\begin{figure}[p]
\centering
\begin{tabular}{cc}
\includegraphics[scale=0.4]{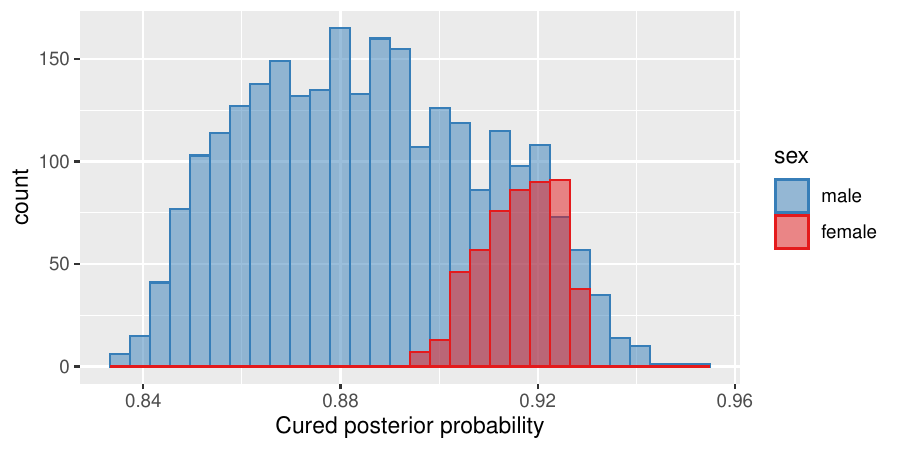}&
\includegraphics[scale=0.4]{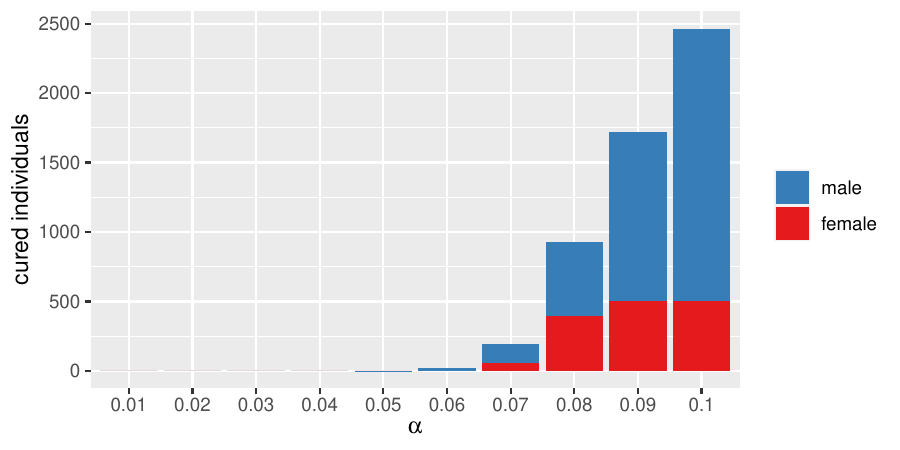}\\
(a) & (b)
\end{tabular}
\caption{Recidivism dataset. (a) histogram of cured posterior probabilities $\widehat{\mathrm{P}}(I_j = 0|\bs\Delta,\bs y, \bs x)$ for $j \in\bs\Delta_0$ and  (b) inferred number of cured individuals when controlling the FDR at level $\alpha$, for the regularized prior distribution.}
\label{fig:discoveries_vs_alpha}
\end{figure}

The estimated cured posterior probabilities among the 3105 censored offenders is shown in Figure \ref{fig:discoveries_vs_alpha}.(a). The corresponding number of cured subjects in the sample is shown in Figure \ref{fig:discoveries_vs_alpha}.(b), when controlling the FDR at various levels. We see that zero cured subjects are inferred when the target FDR $(\alpha)$ is lower than $0.05$. When $\alpha \in\{0.05,0.06,0.07,0.08,0.09,0.10\}$ the corresponding number of inferred cured items is equal to $2,   23,  191,  932, 1722, 2465$, respectively.  

\begin{figure}[p]
    \centering
\begin{tabular}{c}
     \includegraphics[scale = 0.36]{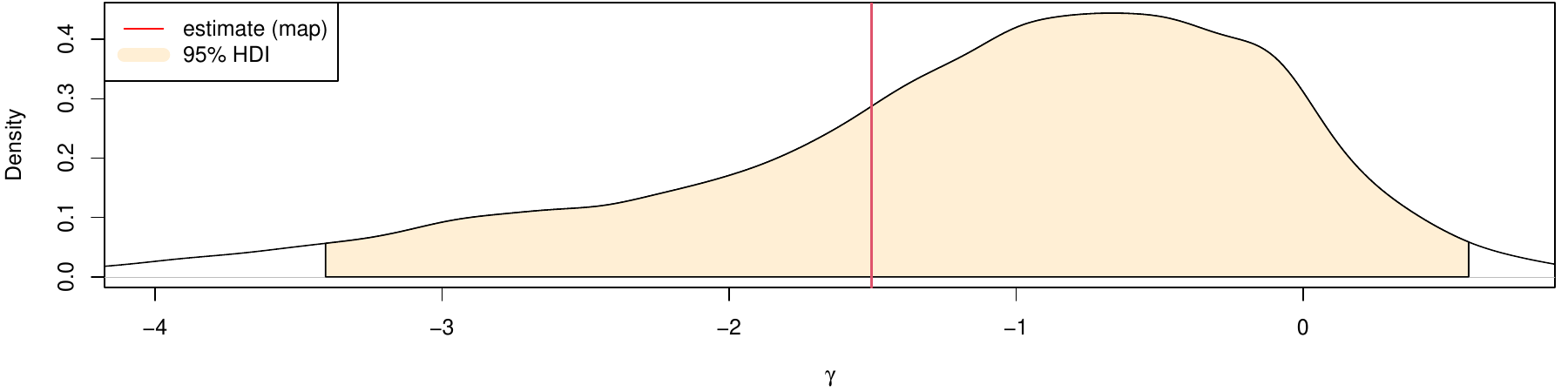}\\ 
     \includegraphics[scale = 0.36]{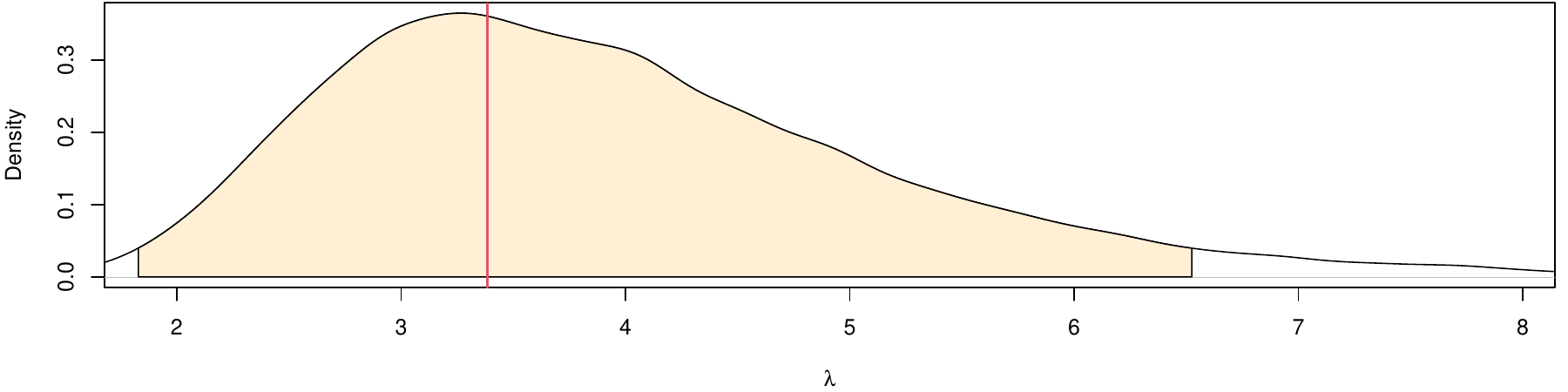}\\
     \includegraphics[scale = 0.36]{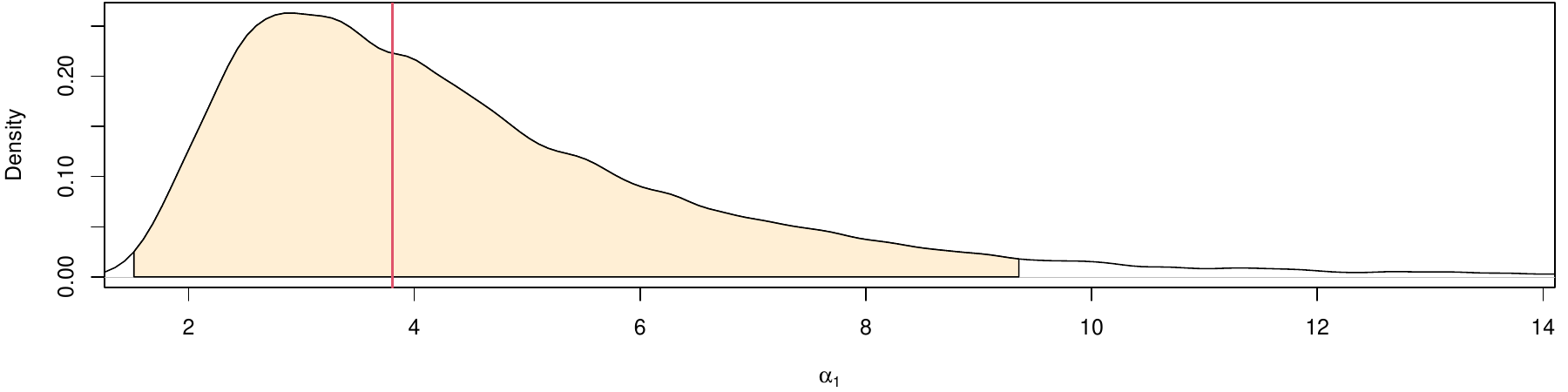}\\ 
     \includegraphics[scale = 0.36]{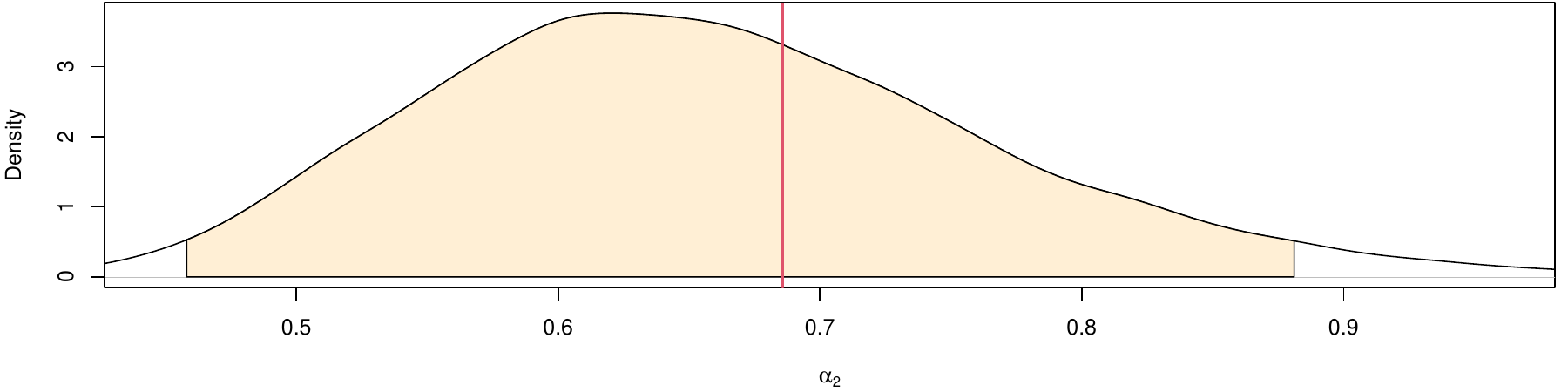}\\
     \includegraphics[scale = 0.36]{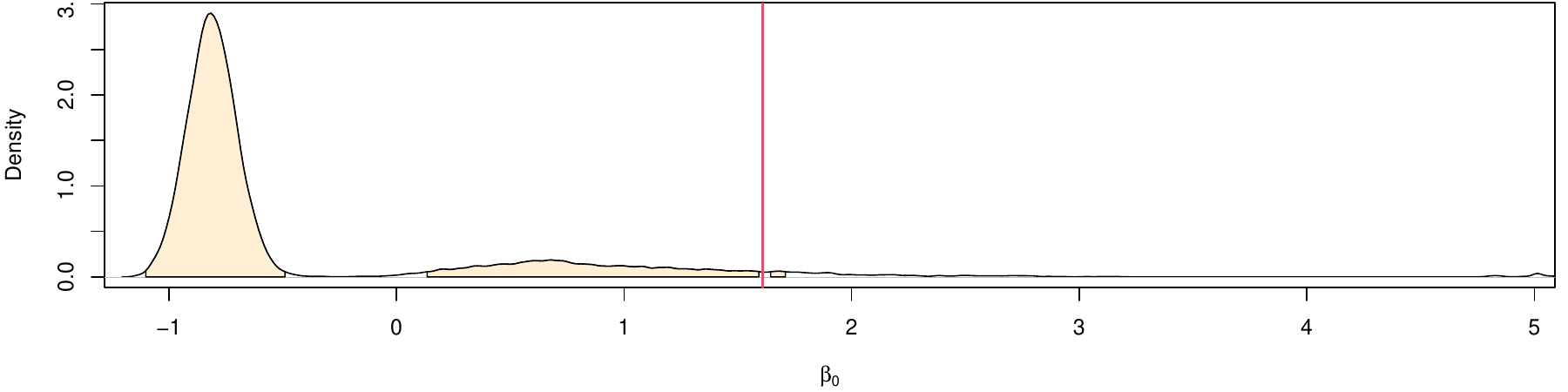}\\ 
     \includegraphics[scale = 0.36]{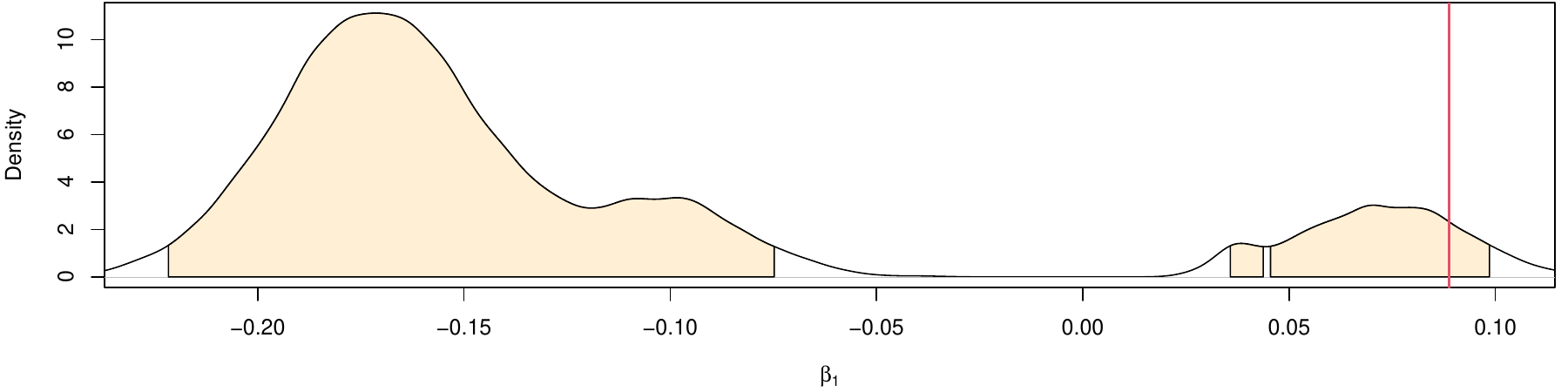}\\
     \includegraphics[scale = 0.36]{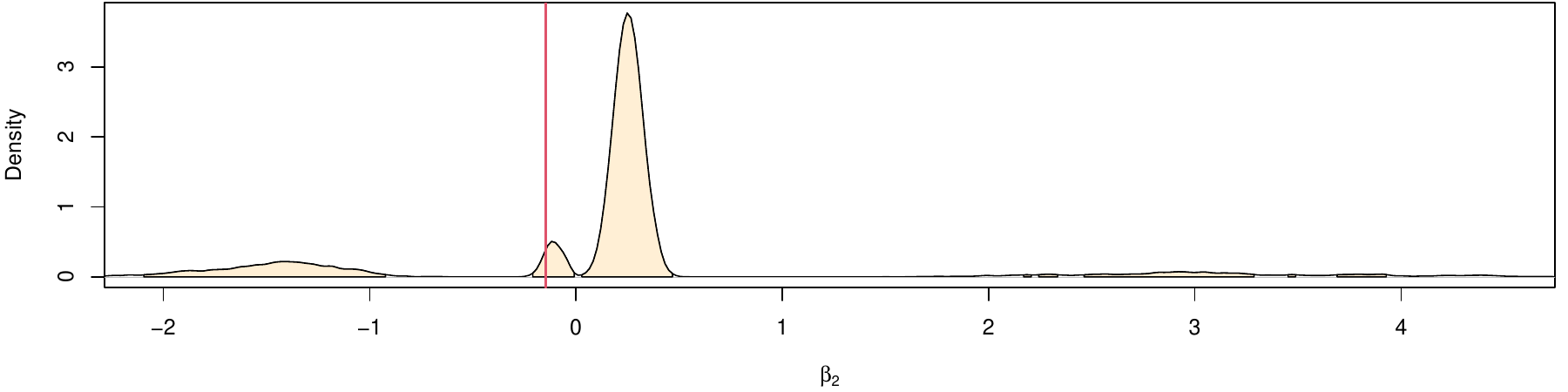}\\
\end{tabular}
    \caption{Recidivism dataset: kernel density estimates of marginal posterior distributions for each parameter. The shaded region corresponds to the $95\%$ HDI and the  vertical line denotes the MAP estimate. }
    \label{fig:recidivism_kernel}
\end{figure}

The MAP estimate of each parameter is reported in Table \ref{tab:recidivism_summary} along with the estimated quantiles. The potential scale reduction factor of the Gelman and Rubin's convergence diagnostic \citep{gelman1992inference} is smaller than 1.1 for all parameters, thus, it does not spot any convergence issues. See also Figure \ref{fig:recidivism_kernel} for the corresponding HDIs. Note that 0 is not included in the HDIs for $\beta_1$ and $\beta_2$, albeit both parameters exhibit multi-modality.


\section{Concluding remarks}

The family of cure models introduced by \cite{milienos2022reparameterization} incorporates many well studied cure models; for example, the mixture cure rate, the promotion time, the negative binomial and the binomial cure models are all special cases for certain values of the parameters. Typically, the likelihood surface as well as the posterior distribution are multimodal, in order to accommodate these alternative models in various parts of the parameter space. As a consequence, inference is rather challenging both from a likelihood as well as a Bayesian perspective. The proposed Bayesian method introduces a state-of-the-art MCMC sampler and deals with the aforementioned issues in an efficient manner. 

Our model is based on augmenting the data $\bs y = (y_1,\ldots,y_n)$ with the discrete latent cured status $\bs I = (I_1,\ldots,I_n)$ for each subject. Therefore, Hamiltonian  MCMC methods as the No-U-Turn Sampler \citep{hoffman2014no} implemented in the popular {\tt STAN} software \citep{carpenter2017stan}  are not directly applicable in this setting: only continuous parameters are allowed.  Of course,  one could integrate the discrete parameters ($\bs I$) out from the model and target the marginal posterior distribution (that is, $\pi(\bs\theta|\bs\Delta,\bs y,\bs x) = \sum_{\bs I}\pi(\bs\theta,\bs I|\bs\Delta,\bs y,\bs x)$), but there are two drawbacks with this approach. At first, inference for the latent cured status would not be straightforward. Second, the multimodality of the posterior distribution would make the Hamiltonian Monte Carlo sampler getting trapped into minor modes and the proper exploration of the posterior surface would be questionable. In a similar manner, even if we integrate the discrete parameters out, approximate Bayesian techniques such as the Integrated Nested Laplace Approximation \citep{gomez2020bayesian} are not expected to work well since the multimodal target posterior distribution would not justify to consider them at the first place.  

It is also worth noting that, although we assume the time-to-event is subject to non-informative random right censoring, left-censored and/or interval-censored data can be seamlessly incorporated into our methodology. This is because the cured status would remain unknown only within the right-censored group, ensuring that crucial aspects of our approach stay consistent. It goes without saying that the detailed performance of the estimation algorithm and its numerical aspects, in the presence of such data, need to be evaluated. We plan to address this issue in future research.

Moreover, a natural extension of the current study may be to explore issues regarding, for example, variable selection; within the Bayesian paradigm, this could be achieved by applying  stochastic search variable selection techniques \cite{george1995stochastic, dellaportas2002bayesian} or incorporating shrinkage prior distributions \citep{10.1093/acprof:oso/9780199694587.003.0017}. Model adequacy may be another issue that should be checked; this could be addressed under a predictive viewpoint using cross validation \citep{gelfand1992model}, posterior predictive $p$-values \citep{10.1214/aos/1176325622, 10.1214/13-EJS854} or Bayesian goodness of fit-testing procedures \citep{gelman2003bayesian}. Among authors' future research directions is also the development  of an {\tt R} package, which would implement the proposed methodology.

\backmatter


\section*{Declarations}

\begin{itemize}
\item Funding: Panagiotis Papastamoulis received funding from the reseach center of Athens University of Economics and Business.
\item Code and data availability: The (R/C++) code as well as the real dataset used in this paper is available online at \url{https://github.com/mqbssppe/Bayesian_cure_rate_model}. 
\item Conflict of interest: none.
\item We thank the Associate Editor and anonymous reviewers for their careful reading of our manuscript and their many
insightful comments and suggestions. 
\end{itemize}

\bibliography{sn-bibliography}


\begin{thebibliography}{71}
\ifx \bisbn   \undefined \def \bisbn  #1{ISBN #1}\fi
\ifx \binits  \undefined \def \binits#1{#1}\fi
\ifx \bauthor  \undefined \def \bauthor#1{#1}\fi
\ifx \batitle  \undefined \def \batitle#1{#1}\fi
\ifx \bjtitle  \undefined \def \bjtitle#1{#1}\fi
\ifx \bvolume  \undefined \def \bvolume#1{\textbf{#1}}\fi
\ifx \byear  \undefined \def \byear#1{#1}\fi
\ifx \bissue  \undefined \def \bissue#1{#1}\fi
\ifx \bfpage  \undefined \def \bfpage#1{#1}\fi
\ifx \blpage  \undefined \def \blpage #1{#1}\fi
\ifx \burl  \undefined \def \burl#1{\textsf{#1}}\fi
\ifx \doiurl  \undefined \def \doiurl#1{\url{https://doi.org/#1}}\fi
\ifx \betal  \undefined \def \betal{\textit{et al.}}\fi
\ifx \binstitute  \undefined \def \binstitute#1{#1}\fi
\ifx \binstitutionaled  \undefined \def \binstitutionaled#1{#1}\fi
\ifx \bctitle  \undefined \def \bctitle#1{#1}\fi
\ifx \beditor  \undefined \def \beditor#1{#1}\fi
\ifx \bpublisher  \undefined \def \bpublisher#1{#1}\fi
\ifx \bbtitle  \undefined \def \bbtitle#1{#1}\fi
\ifx \bedition  \undefined \def \bedition#1{#1}\fi
\ifx \bseriesno  \undefined \def \bseriesno#1{#1}\fi
\ifx \blocation  \undefined \def \blocation#1{#1}\fi
\ifx \bsertitle  \undefined \def \bsertitle#1{#1}\fi
\ifx \bsnm \undefined \def \bsnm#1{#1}\fi
\ifx \bsuffix \undefined \def \bsuffix#1{#1}\fi
\ifx \bparticle \undefined \def \bparticle#1{#1}\fi
\ifx \barticle \undefined \def \barticle#1{#1}\fi
\bibcommenthead
\ifx \bconfdate \undefined \def \bconfdate #1{#1}\fi
\ifx \botherref \undefined \def \botherref #1{#1}\fi
\ifx \url \undefined \def \url#1{\textsf{#1}}\fi
\ifx \bchapter \undefined \def \bchapter#1{#1}\fi
\ifx \bbook \undefined \def \bbook#1{#1}\fi
\ifx \bcomment \undefined \def \bcomment#1{#1}\fi
\ifx \oauthor \undefined \def \oauthor#1{#1}\fi
\ifx \citeauthoryear \undefined \def \citeauthoryear#1{#1}\fi
\ifx \endbibitem  \undefined \def \endbibitem {}\fi
\ifx \bconflocation  \undefined \def \bconflocation#1{#1}\fi
\ifx \arxivurl  \undefined \def \arxivurl#1{\textsf{#1}}\fi
\csname PreBibitemsHook\endcsname

\bibitem[\protect\citeauthoryear{Maller and Zhou}{1996}]{Mall96}
\begin{bbook}
\bauthor{\bsnm{Maller}, \binits{R.A.}},
\bauthor{\bsnm{Zhou}, \binits{X.}}:
\bbtitle{Survival {A}nalysis with {L}ong-term {S}urvivors}.
\bpublisher{John Wiley \& Sons},
\blocation{New York}
(\byear{1996})
\end{bbook}
\endbibitem

\bibitem[\protect\citeauthoryear{Peng and Yu}{2021}]{peng2021cure}
\begin{bbook}
\bauthor{\bsnm{Peng}, \binits{Y.}},
\bauthor{\bsnm{Yu}, \binits{B.}}:
\bbtitle{Cure Models: Methods, Applications, and Implementation}.
\bpublisher{Chapman and Hall/CRC},
\blocation{New York}
(\byear{2021})
\end{bbook}
\endbibitem

\bibitem[\protect\citeauthoryear{Peng and Taylor}{2014}]{pengtaylor14}
\begin{bchapter}
\bauthor{\bsnm{Peng}, \binits{Y.}},
\bauthor{\bsnm{Taylor}, \binits{J.M.}}:
\bctitle{Cure models}.
In: \beditor{\bsnm{Klein}, \binits{J.}},
\beditor{\bsnm{Houwelingen}, \binits{H.}},
\beditor{\bsnm{Ibrahim}, \binits{J.G.}},
\beditor{\bsnm{Scheike}, \binits{T.H.}} (eds.)
\bbtitle{Handbook of Survival Analysis},
pp. \bfpage{113}--\blpage{134}.
\bpublisher{Chapman \& Hall},
\blocation{~Boca Raton}
(\byear{2014}).
\bcomment{Chap. 6}
\end{bchapter}
\endbibitem

\bibitem[\protect\citeauthoryear{Amico and Van~Keilegom}{2018}]{amico2018cure}
\begin{barticle}
\bauthor{\bsnm{Amico}, \binits{M.}},
\bauthor{\bsnm{Van~Keilegom}, \binits{I.}}:
\batitle{Cure models in survival analysis}.
\bjtitle{Annual Review of Statistics and its Application}
\bvolume{5}(\bissue{1}),
\bfpage{311}--\blpage{342}
(\byear{2018})
\end{barticle}
\endbibitem

\bibitem[\protect\citeauthoryear{Rocha et~al.}{2017}]{rocha2017new}
\begin{barticle}
\bauthor{\bsnm{Rocha}, \binits{R.}},
\bauthor{\bsnm{Nadarajah}, \binits{S.}},
\bauthor{\bsnm{Tomazella}, \binits{V.}},
\bauthor{\bsnm{Louzada}, \binits{F.}}:
\batitle{A new class of defective models based on the {M}arshall--{O}lkin
  family of distributions for cure rate modeling}.
\bjtitle{Computational Statistics \& Data Analysis}
\bvolume{107},
\bfpage{48}--\blpage{63}
(\byear{2017})
\end{barticle}
\endbibitem

\bibitem[\protect\citeauthoryear{Schmidt and Witte}{1988}]{schmidt1988}
\begin{bbook}
\bauthor{\bsnm{Schmidt}, \binits{P.}},
\bauthor{\bsnm{Witte}, \binits{A.D.}}:
\bbtitle{Predicting Recidivism Using Survival Models}.
\bpublisher{Springer},
\blocation{New York}
(\byear{1988})
\end{bbook}
\endbibitem

\bibitem[\protect\citeauthoryear{Pel{\'a}ez
  et~al.}{2022}]{pelaez2022probability}
\begin{botherref}
\oauthor{\bsnm{Pel{\'a}ez}, \binits{R.}},
\oauthor{\bsnm{Van~Keilegom}, \binits{I.}},
\oauthor{\bsnm{Cao}, \binits{R.}},
\oauthor{\bsnm{Vilar}, \binits{J.}}:
Probability of default estimation in credit risk using mixture cure models.
Technical report,
Universidade da Coru{\~n}a
(2022)
\end{botherref}
\endbibitem

\bibitem[\protect\citeauthoryear{Yamaguchi}{1992}]{yamaguchi1992accelerated}
\begin{barticle}
\bauthor{\bsnm{Yamaguchi}, \binits{K.}}:
\batitle{Accelerated failure-time regression models with a regression model of
  surviving fraction: an application to the analysis of permanent employment in
  {J}apan}.
\bjtitle{Journal of the American Statistical Association}
\bvolume{87}(\bissue{418}),
\bfpage{284}--\blpage{292}
(\byear{1992})
\end{barticle}
\endbibitem

\bibitem[\protect\citeauthoryear{Yamaguchi}{2003}]{yamaguchi2003accelerated}
\begin{barticle}
\bauthor{\bsnm{Yamaguchi}, \binits{K.}}:
\batitle{Accelerated failure--time mover--stayer regression models for the
  analysis of last--episode data}.
\bjtitle{Sociological Methodology}
\bvolume{33}(\bissue{1}),
\bfpage{81}--\blpage{110}
(\byear{2003})
\end{barticle}
\endbibitem

\bibitem[\protect\citeauthoryear{Kalamatianou and
  McClean}{2003}]{kalamatianou2003perpetual}
\begin{barticle}
\bauthor{\bsnm{Kalamatianou}, \binits{A.G.}},
\bauthor{\bsnm{McClean}, \binits{S.}}:
\batitle{The perpetual student: modeling duration of undergraduate studies
  based on lifetime-type educational data}.
\bjtitle{Lifetime Data Analysis}
\bvolume{9}(\bissue{4}),
\bfpage{311}--\blpage{330}
(\byear{2003})
\end{barticle}
\endbibitem

\bibitem[\protect\citeauthoryear{Boag}{1948a}]{boag1948presentationI}
\begin{barticle}
\bauthor{\bsnm{Boag}, \binits{J.W.}}:
\batitle{The presentation and analysis of the results of radiotherapy. {Part
  I}}.
\bjtitle{The British Journal of Radiology}
\bvolume{21}(\bissue{243}),
\bfpage{128}--\blpage{138}
(\byear{1948})
\end{barticle}
\endbibitem

\bibitem[\protect\citeauthoryear{Boag}{1948b}]{boag1948presentationII}
\begin{barticle}
\bauthor{\bsnm{Boag}, \binits{J.W.}}:
\batitle{The presentation and analysis of the results of radiotherapy. {Part
  II}}.
\bjtitle{The British Journal of Radiology}
\bvolume{21}(\bissue{244}),
\bfpage{189}--\blpage{203}
(\byear{1948})
\end{barticle}
\endbibitem

\bibitem[\protect\citeauthoryear{Boag}{1949}]{boag49}
\begin{barticle}
\bauthor{\bsnm{Boag}, \binits{J.W.}}:
\batitle{Maximum likelihood estimates of the proportion of patients cured by
  cancer therapy}.
\bjtitle{Journal of the Royal Statistical Society: Series B}
\bvolume{11}(\bissue{1}),
\bfpage{15}--\blpage{53}
(\byear{1949})
\end{barticle}
\endbibitem

\bibitem[\protect\citeauthoryear{Yakovlev}{1994}]{yako1994}
\begin{barticle}
\bauthor{\bsnm{Yakovlev}, \binits{A.Y.}}:
\batitle{Letters to the editor: Parametric versus non-parametric methods for
  estimating cure rates based on censored survival data}.
\bjtitle{Statistics in Medicine}
\bvolume{13}(\bissue{9}),
\bfpage{983}--\blpage{986}
(\byear{1994})
\end{barticle}
\endbibitem

\bibitem[\protect\citeauthoryear{Yakovlev et~al.}{1993}]{yakovlev1993simple}
\begin{barticle}
\bauthor{\bsnm{Yakovlev}, \binits{A.Y.}},
\bauthor{\bsnm{Asselain}, \binits{B.}},
\bauthor{\bsnm{Bardou}, \binits{V.}},
\bauthor{\bsnm{Fourquet}, \binits{A.}},
\bauthor{\bsnm{Hoang}, \binits{T.}},
\bauthor{\bsnm{Rochefediere}, \binits{A.}},
\bauthor{\bsnm{Tsodikov}, \binits{A.}}:
\batitle{A simple stochastic model of tumor recurrence and its application to
  data on premenopausal breast cancer}.
\bjtitle{Biometrie et Analyse de Donnees Spatio-temporelles}
\bvolume{12},
\bfpage{66}--\blpage{82}
(\byear{1993})
\end{barticle}
\endbibitem

\bibitem[\protect\citeauthoryear{Hoang et~al.}{1996}]{hoang1996parametric}
\begin{barticle}
\bauthor{\bsnm{Hoang}, \binits{T.}},
\bauthor{\bsnm{Tsodikov}, \binits{A.}},
\bauthor{\bsnm{Asselain}, \binits{B.}},
\bauthor{\bsnm{Yakolev}, \binits{A.}}:
\batitle{A parametric analysis of tumor recurrence data}.
\bjtitle{Journal of Biological Systems}
\bvolume{4}(\bissue{03}),
\bfpage{391}--\blpage{403}
(\byear{1996})
\end{barticle}
\endbibitem

\bibitem[\protect\citeauthoryear{Greenhouse and
  Wolfe}{1984}]{greenhouse1984competing}
\begin{barticle}
\bauthor{\bsnm{Greenhouse}, \binits{J.B.}},
\bauthor{\bsnm{Wolfe}, \binits{R.A.}}:
\batitle{A competing risks derivation of a mixture model for the analysis of
  survival data}.
\bjtitle{Communications in Statistics-Theory and Methods}
\bvolume{13}(\bissue{25}),
\bfpage{3133}--\blpage{3154}
(\byear{1984})
\end{barticle}
\endbibitem

\bibitem[\protect\citeauthoryear{Larson and Dinse}{1985}]{larson1985mixture}
\begin{barticle}
\bauthor{\bsnm{Larson}, \binits{M.G.}},
\bauthor{\bsnm{Dinse}, \binits{G.E.}}:
\batitle{A mixture model for the regression analysis of competing risks data}.
\bjtitle{Journal of the Royal Statistical Society: Series C (Applied
  Statistics)}
\bvolume{34}(\bissue{3}),
\bfpage{201}--\blpage{211}
(\byear{1985})
\end{barticle}
\endbibitem

\bibitem[\protect\citeauthoryear{Meeker}{1987}]{meeker1987limited}
\begin{barticle}
\bauthor{\bsnm{Meeker}, \binits{W.Q.}}:
\batitle{Limited failure population life tests: application to integrated
  circuit reliability}.
\bjtitle{Technometrics}
\bvolume{29}(\bissue{1}),
\bfpage{51}--\blpage{65}
(\byear{1987})
\end{barticle}
\endbibitem

\bibitem[\protect\citeauthoryear{Meeker and
  LuValle}{1995}]{meeker1995accelerated}
\begin{barticle}
\bauthor{\bsnm{Meeker}, \binits{W.Q.}},
\bauthor{\bsnm{LuValle}, \binits{M.J.}}:
\batitle{An accelerated life test model based on reliability kinetics}.
\bjtitle{Technometrics}
\bvolume{37}(\bissue{2}),
\bfpage{133}--\blpage{146}
(\byear{1995})
\end{barticle}
\endbibitem

\bibitem[\protect\citeauthoryear{Tsodikov et~al.}{2003}]{Tsodikov03}
\begin{barticle}
\bauthor{\bsnm{Tsodikov}, \binits{A.D.}},
\bauthor{\bsnm{Ibrahim}, \binits{J.}},
\bauthor{\bsnm{Yakovlev}, \binits{A.}}:
\batitle{Estimating cure rates from survival data: an alternative to
  two-component mixture models}.
\bjtitle{Journal of the American Statistical Association}
\bvolume{98}(\bissue{464}),
\bfpage{1063}--\blpage{1078}
(\byear{2003})
\end{barticle}
\endbibitem

\bibitem[\protect\citeauthoryear{Cooner et~al.}{2007}]{cooner2007flexible}
\begin{barticle}
\bauthor{\bsnm{Cooner}, \binits{F.W.}},
\bauthor{\bsnm{Banerjee}, \binits{S.}},
\bauthor{\bsnm{Carlin}, \binits{B.P.}},
\bauthor{\bsnm{Sinha}, \binits{D.}}:
\batitle{Flexible cure rate modeling under latent activation schemes}.
\bjtitle{Journal of the American Statistical Association}
\bvolume{102}(\bissue{478}),
\bfpage{560}--\blpage{572}
(\byear{2007})
\end{barticle}
\endbibitem

\bibitem[\protect\citeauthoryear{Tsodikov}{2002}]{tsod02}
\begin{barticle}
\bauthor{\bsnm{Tsodikov}, \binits{A.D.}}:
\batitle{Semi-parametric models of long- and short-term survival: an
  application to the analysis of breast cancer survival in {U}tah by age and
  stage}.
\bjtitle{Statistics in Medicine}
\bvolume{21}(\bissue{6}),
\bfpage{895}--\blpage{920}
(\byear{2002})
\end{barticle}
\endbibitem

\bibitem[\protect\citeauthoryear{Tsodikov}{2003}]{tsodikov2003semiparametric}
\begin{barticle}
\bauthor{\bsnm{Tsodikov}, \binits{A.D.}}:
\batitle{Semiparametric models: a generalized self-consistency approach}.
\bjtitle{Journal of the Royal Statistical Society: Series B}
\bvolume{65}(\bissue{3}),
\bfpage{759}--\blpage{774}
(\byear{2003})
\end{barticle}
\endbibitem

\bibitem[\protect\citeauthoryear{Balakrishnan and
  Milienos}{2020}]{balakrishnan2020class}
\begin{barticle}
\bauthor{\bsnm{Balakrishnan}, \binits{N.}},
\bauthor{\bsnm{Milienos}, \binits{F.S.}}:
\batitle{On a class of non-linear transformation cure rate models}.
\bjtitle{Biometrical Journal}
\bvolume{62}(\bissue{5}),
\bfpage{1208}--\blpage{1222}
(\byear{2020})
\end{barticle}
\endbibitem

\bibitem[\protect\citeauthoryear{Yin and Ibrahim}{2005}]{yin05b}
\begin{barticle}
\bauthor{\bsnm{Yin}, \binits{G.}},
\bauthor{\bsnm{Ibrahim}, \binits{J.G.}}:
\batitle{Cure rate models: a unified approach}.
\bjtitle{The Canadian Journal of Statistics}
\bvolume{33}(\bissue{4}),
\bfpage{559}--\blpage{570}
(\byear{2005})
\end{barticle}
\endbibitem

\bibitem[\protect\citeauthoryear{Taylor and Liu}{2007}]{taylor2007statistical}
\begin{bchapter}
\bauthor{\bsnm{Taylor}, \binits{J.M.}},
\bauthor{\bsnm{Liu}, \binits{N.}}:
\bctitle{Statistical issues involved with extending standard models}.
In: \beditor{\bsnm{Nair}, \binits{V.}} (ed.)
\bbtitle{Advances in Statistical Modeling and Inference: Essays in Honor of
  Kjell A Doksum},
pp. \bfpage{299}--\blpage{311}.
\bpublisher{World Scientific}, \blocation{???}
(\byear{2007})
\end{bchapter}
\endbibitem

\bibitem[\protect\citeauthoryear{Peng and Xu}{2012}]{peng2012extended}
\begin{barticle}
\bauthor{\bsnm{Peng}, \binits{Y.}},
\bauthor{\bsnm{Xu}, \binits{J.}}:
\batitle{An extended cure model and model selection}.
\bjtitle{Lifetime Data Analysis}
\bvolume{18}(\bissue{2}),
\bfpage{215}--\blpage{233}
(\byear{2012})
\end{barticle}
\endbibitem

\bibitem[\protect\citeauthoryear{Diao and Yin}{2012}]{diao2012general}
\begin{barticle}
\bauthor{\bsnm{Diao}, \binits{G.}},
\bauthor{\bsnm{Yin}, \binits{G.}}:
\batitle{A general transformation class of semiparametric cure rate frailty
  models}.
\bjtitle{Annals of the Institute of Statistical Mathematics}
\bvolume{64}(\bissue{5}),
\bfpage{959}--\blpage{989}
(\byear{2012})
\end{barticle}
\endbibitem

\bibitem[\protect\citeauthoryear{Pal and
  Balakrishnan}{2017}]{pal2017expectation}
\begin{barticle}
\bauthor{\bsnm{Pal}, \binits{S.}},
\bauthor{\bsnm{Balakrishnan}, \binits{N.}}:
\batitle{Expectation maximization algorithm for {B}ox--{C}ox transformation
  cure rate model and assessment of model misspecification under {W}eibull
  lifetimes}.
\bjtitle{IEEE Journal of Biomedical and Health Informatics}
\bvolume{22}(\bissue{3}),
\bfpage{926}--\blpage{934}
(\byear{2017})
\end{barticle}
\endbibitem

\bibitem[\protect\citeauthoryear{Wang and Pal}{2022}]{wang2022two}
\begin{barticle}
\bauthor{\bsnm{Wang}, \binits{P.}},
\bauthor{\bsnm{Pal}, \binits{S.}}:
\batitle{A two-way flexible generalized gamma transformation cure rate model}.
\bjtitle{Statistics in Medicine}
\bvolume{41}(\bissue{13}),
\bfpage{2427}--\blpage{2447}
(\byear{2022})
\end{barticle}
\endbibitem

\bibitem[\protect\citeauthoryear{Pal and Roy}{2023}]{pal2023parameter}
\begin{botherref}
\oauthor{\bsnm{Pal}, \binits{S.}},
\oauthor{\bsnm{Roy}, \binits{S.}}:
On the parameter estimation of {B}ox-{C}ox transformation cure model.
Statistics in Medicine
\textbf{available online}
(2023)
\end{botherref}
\endbibitem

\bibitem[\protect\citeauthoryear{Gilks et~al.}{2018}]{10.2307/2986138}
\begin{barticle}
\bauthor{\bsnm{Gilks}, \binits{W.R.}},
\bauthor{\bsnm{Best}, \binits{N.G.}},
\bauthor{\bsnm{Tan}, \binits{K.K.C.}}:
\batitle{{Adaptive rejection Metropolis sampling within Gibbs sampling}}.
\bjtitle{Journal of the Royal Statistical Society Series C: Applied Statistics}
\bvolume{44}(\bissue{4}),
\bfpage{455}--\blpage{472}
(\byear{2018})
\doiurl{10.2307/2986138}
{\href{https://arxiv.org/abs/https://academic.oup.com/jrsssc/article-pdf/44/4/455/48749639/jrsssc\_44\_4\_455.pdf}{{https://academic.oup.com/jrsssc/article-pdf/44/4/455/48749639/jrsssc\_44\_4\_455.pdf}}}
\end{barticle}
\endbibitem

\bibitem[\protect\citeauthoryear{Tournoud and
  Ecochard}{2008}]{tournoud2008promotion}
\begin{barticle}
\bauthor{\bsnm{Tournoud}, \binits{M.}},
\bauthor{\bsnm{Ecochard}, \binits{R.}}:
\batitle{Promotion time models with time-changing exposure and heterogeneity:
  application to infectious diseases}.
\bjtitle{Biometrical Journal}
\bvolume{50}(\bissue{3}),
\bfpage{395}--\blpage{407}
(\byear{2008})
\end{barticle}
\endbibitem

\bibitem[\protect\citeauthoryear{Castro et~al.}{2009}]{Castro09}
\begin{barticle}
\bauthor{\bsnm{Castro}, \binits{M.d.}},
\bauthor{\bsnm{Cancho}, \binits{V.G.}},
\bauthor{\bsnm{Rodrigues}, \binits{J.}}:
\batitle{A {B}ayesian long-term survival model parametrized in the cured
  fraction}.
\bjtitle{Biometrical Journal}
\bvolume{51}(\bissue{3}),
\bfpage{443}--\blpage{455}
(\byear{2009})
\end{barticle}
\endbibitem

\bibitem[\protect\citeauthoryear{Ortega et~al.}{2012}]{ortega2012negative}
\begin{barticle}
\bauthor{\bsnm{Ortega}, \binits{E.M.}},
\bauthor{\bsnm{Cordeiro}, \binits{G.M.}},
\bauthor{\bsnm{Kattan}, \binits{M.W.}}:
\batitle{The negative binomial--beta {W}eibull regression model to predict the
  cure of prostate cancer}.
\bjtitle{Journal of Applied Statistics}
\bvolume{39}(\bissue{6}),
\bfpage{1191}--\blpage{1210}
(\byear{2012})
\end{barticle}
\endbibitem

\bibitem[\protect\citeauthoryear{Cordeiro et~al.}{2016}]{cordeiro2016model}
\begin{barticle}
\bauthor{\bsnm{Cordeiro}, \binits{G.M.}},
\bauthor{\bsnm{Cancho}, \binits{V.G.}},
\bauthor{\bsnm{Ortega}, \binits{E.M.}},
\bauthor{\bsnm{Barriga}, \binits{G.D.}}:
\batitle{A model with long-term survivors: negative binomial
  {B}irnbaum-{S}aunders}.
\bjtitle{Communications in Statistics-Theory and Methods}
\bvolume{45}(\bissue{5}),
\bfpage{1370}--\blpage{1387}
(\byear{2016})
\end{barticle}
\endbibitem

\bibitem[\protect\citeauthoryear{Rodrigues
  et~al.}{2018}]{rodrigues2018modeling}
\begin{botherref}
\oauthor{\bsnm{Rodrigues}, \binits{A.S.}},
\oauthor{\bsnm{Calsavara}, \binits{V.F.}},
\oauthor{\bsnm{Tomazella}, \binits{V.L.D.}}:
Modeling cure fraction with frailty term in latent risk: a {B}ayesian approach.
arXiv preprint arXiv:1803.08128
(2018)
\end{botherref}
\endbibitem

\bibitem[\protect\citeauthoryear{D'Andrea et~al.}{2018}]{d2018negative}
\begin{barticle}
\bauthor{\bsnm{D'Andrea}, \binits{A.}},
\bauthor{\bsnm{Rocha}, \binits{R.}},
\bauthor{\bsnm{Tomazella}, \binits{V.}},
\bauthor{\bsnm{Louzada}, \binits{F.}}:
\batitle{Negative binomial {K}umaraswamy-{G} cure rate regression model}.
\bjtitle{Journal of Risk and Financial Management}
\bvolume{11}(\bissue{6}),
\bfpage{1}--\blpage{14}
(\byear{2018})
\end{barticle}
\endbibitem

\bibitem[\protect\citeauthoryear{Le{\~a}o et~al.}{2021}]{leao2018negative}
\begin{barticle}
\bauthor{\bsnm{Le{\~a}o}, \binits{J.}},
\bauthor{\bsnm{Bourguignon}, \binits{M.}},
\bauthor{\bsnm{Saulo}, \binits{H.}},
\bauthor{\bsnm{Santos-Neto}, \binits{M.}},
\bauthor{\bsnm{Calsavara}, \binits{V.}}:
\batitle{The negative binomial beta prime regression model with cure rate:
  application with a melanoma dataset}.
\bjtitle{Journal of Statistical Theory and Practice}
\bvolume{15}(\bissue{3}),
\bfpage{1}--\blpage{21}
(\byear{2021})
\end{barticle}
\endbibitem

\bibitem[\protect\citeauthoryear{Pal}{2021}]{pal2021simplified}
\begin{barticle}
\bauthor{\bsnm{Pal}, \binits{S.}}:
\batitle{A simplified stochastic {EM} algorithm for cure rate model with
  negative binomial competing risks: an application to breast cancer data}.
\bjtitle{Statistics in Medicine}
\bvolume{40}(\bissue{28}),
\bfpage{6387}--\blpage{6409}
(\byear{2021})
\end{barticle}
\endbibitem

\bibitem[\protect\citeauthoryear{Zeng et~al.}{2006}]{zeng06}
\begin{barticle}
\bauthor{\bsnm{Zeng}, \binits{D.}},
\bauthor{\bsnm{Yin}, \binits{G.}},
\bauthor{\bsnm{Ibrahim}, \binits{J.G.}}:
\batitle{Semiparametric transformation models for survival data with a cure
  fraction}.
\bjtitle{Journal of the American Statistical Association}
\bvolume{101}(\bissue{474}),
\bfpage{670}--\blpage{684}
(\byear{2006})
\end{barticle}
\endbibitem

\bibitem[\protect\citeauthoryear{Koutras and
  Milienos}{2017}]{koutras2017flexible}
\begin{barticle}
\bauthor{\bsnm{Koutras}, \binits{M.V.}},
\bauthor{\bsnm{Milienos}, \binits{F.S.}}:
\batitle{A flexible family of transformation cure rate models}.
\bjtitle{Statistics in Medicine}
\bvolume{36}(\bissue{16}),
\bfpage{2559}--\blpage{2575}
(\byear{2017})
\end{barticle}
\endbibitem

\bibitem[\protect\citeauthoryear{Milienos}{2022}]{milienos2022reparameterization}
\begin{barticle}
\bauthor{\bsnm{Milienos}, \binits{F.S.}}:
\batitle{On a reparameterization of a flexible family of cure models}.
\bjtitle{Statistics in Medicine}
\bvolume{41}(\bissue{21}),
\bfpage{4091}--\blpage{4111}
(\byear{2022})
\end{barticle}
\endbibitem

\bibitem[\protect\citeauthoryear{Geman and Geman}{1984}]{geman1984stochastic}
\begin{botherref}
\oauthor{\bsnm{Geman}, \binits{S.}},
\oauthor{\bsnm{Geman}, \binits{D.}}:
Stochastic relaxation, gibbs distributions, and the bayesian restoration of
  images.
IEEE Transactions on Pattern Analysis and Machine Intelligence
(6),
721--741
(1984)
\end{botherref}
\endbibitem

\bibitem[\protect\citeauthoryear{Roberts and Tweedie}{1996}]{10.2307/3318418}
\begin{barticle}
\bauthor{\bsnm{Roberts}, \binits{G.O.}},
\bauthor{\bsnm{Tweedie}, \binits{R.L.}}:
\batitle{Exponential convergence of {L}angevin distributions and their discrete
  approximations}.
\bjtitle{Bernoulli}
\bvolume{2}(\bissue{4}),
\bfpage{341}--\blpage{363}
(\byear{1996})
\end{barticle}
\endbibitem

\bibitem[\protect\citeauthoryear{Altekar et~al.}{2004}]{altekar2004parallel}
\begin{barticle}
\bauthor{\bsnm{Altekar}, \binits{G.}},
\bauthor{\bsnm{Dwarkadas}, \binits{S.}},
\bauthor{\bsnm{Huelsenbeck}, \binits{J.P.}},
\bauthor{\bsnm{Ronquist}, \binits{F.}}:
\batitle{Parallel metropolis coupled markov chain monte carlo for bayesian
  phylogenetic inference}.
\bjtitle{Bioinformatics}
\bvolume{20}(\bissue{3}),
\bfpage{407}--\blpage{415}
(\byear{2004})
\end{barticle}
\endbibitem

\bibitem[\protect\citeauthoryear{Papastamoulis and
  Rattray}{2018}]{papastamoulis2018bayesian}
\begin{barticle}
\bauthor{\bsnm{Papastamoulis}, \binits{P.}},
\bauthor{\bsnm{Rattray}, \binits{M.}}:
\batitle{A {B}ayesian model selection approach for identifying differentially
  expressed transcripts from {R}{N}{A} sequencing data}.
\bjtitle{Journal of the Royal Statistical Society: Series C (Applied
  Statistics)}
\bvolume{67}(\bissue{1}),
\bfpage{3}--\blpage{23}
(\byear{2018})
\end{barticle}
\endbibitem

\bibitem[\protect\citeauthoryear{Bernhardt}{2016}]{bernhardt2016flexible}
\begin{barticle}
\bauthor{\bsnm{Bernhardt}, \binits{P.W.}}:
\batitle{A flexible cure rate model with dependent censoring and a known cure
  threshold}.
\bjtitle{Statistics in Medicine}
\bvolume{35}(\bissue{25}),
\bfpage{4607}--\blpage{4623}
(\byear{2016})
\end{barticle}
\endbibitem

\bibitem[\protect\citeauthoryear{Laska and
  Meisner}{1992}]{laska1992nonparametric}
\begin{barticle}
\bauthor{\bsnm{Laska}, \binits{E.M.}},
\bauthor{\bsnm{Meisner}, \binits{M.J.}}:
\batitle{Nonparametric estimation and testing in a cure model}.
\bjtitle{Biometrics}
\bvolume{48}(\bissue{4}),
\bfpage{1223}--\blpage{1234}
(\byear{1992})
\end{barticle}
\endbibitem

\bibitem[\protect\citeauthoryear{Safari et~al.}{2021}]{safari2021product}
\begin{barticle}
\bauthor{\bsnm{Safari}, \binits{W.C.}},
\bauthor{\bsnm{L{\'o}pez-de-Ullibarri}, \binits{I.}},
\bauthor{\bsnm{J{\'a}come}, \binits{M.A.}}:
\batitle{A product-limit estimator of the conditional survival function when
  cure status is partially known}.
\bjtitle{Biometrical Journal}
\bvolume{63}(\bissue{5}),
\bfpage{984}--\blpage{1005}
(\byear{2021})
\end{barticle}
\endbibitem

\bibitem[\protect\citeauthoryear{Safari et~al.}{2022}]{safari2022nonparametric}
\begin{barticle}
\bauthor{\bsnm{Safari}, \binits{W.C.}},
\bauthor{\bsnm{L{\'o}pez-de-Ullibarri}, \binits{I.}},
\bauthor{\bsnm{J{\'a}come}, \binits{M.A.}}:
\batitle{Nonparametric kernel estimation of the probability of cure in a
  mixture cure model when the cure status is partially observed}.
\bjtitle{Statistical Methods in Medical Research}
\bvolume{31}(\bissue{11}),
\bfpage{2164}--\blpage{2188}
(\byear{2022})
\end{barticle}
\endbibitem

\bibitem[\protect\citeauthoryear{Safari et~al.}{2023}]{safari2023latency}
\begin{botherref}
\oauthor{\bsnm{Safari}, \binits{W.C.}},
\oauthor{\bsnm{L{\'o}pez-de-Ullibarri}, \binits{I.}},
\oauthor{\bsnm{J{\'a}come}, \binits{M.A.}}:
Latency function estimation under the mixture cure model when the cure status
  is available.
Lifetime Data Analysis,
1--20
(2023)
\end{botherref}
\endbibitem

\bibitem[\protect\citeauthoryear{Wu et~al.}{2014}]{wu2014extension}
\begin{barticle}
\bauthor{\bsnm{Wu}, \binits{Y.}},
\bauthor{\bsnm{Lin}, \binits{Y.}},
\bauthor{\bsnm{Lu}, \binits{S.-E.}},
\bauthor{\bsnm{Li}, \binits{C.-S.}},
\bauthor{\bsnm{Shih}, \binits{W.J.}}:
\batitle{Extension of a {C}ox proportional hazards cure model when cure
  information is partially known}.
\bjtitle{Biostatistics}
\bvolume{15}(\bissue{3}),
\bfpage{540}--\blpage{554}
(\byear{2014})
\end{barticle}
\endbibitem

\bibitem[\protect\citeauthoryear{Roberts and
  Rosenthal}{1998}]{https://doi.org/10.1111/1467-9868.00123}
\begin{barticle}
\bauthor{\bsnm{Roberts}, \binits{G.O.}},
\bauthor{\bsnm{Rosenthal}, \binits{J.S.}}:
\batitle{Optimal scaling of discrete approximations to langevin diffusions}.
\bjtitle{Journal of the Royal Statistical Society: Series B (Statistical
  Methodology)}
\bvolume{60}(\bissue{1}),
\bfpage{255}--\blpage{268}
(\byear{1998})
\doiurl{10.1111/1467-9868.00123}
{\href{https://arxiv.org/abs/https://rss.onlinelibrary.wiley.com/doi/pdf/10.1111/1467-9868.00123}{{https://rss.onlinelibrary.wiley.com/doi/pdf/10.1111/1467-9868.00123}}}
\end{barticle}
\endbibitem

\bibitem[\protect\citeauthoryear{Girolami and
  Calderhead}{2011}]{girolami2011riemann}
\begin{barticle}
\bauthor{\bsnm{Girolami}, \binits{M.}},
\bauthor{\bsnm{Calderhead}, \binits{B.}}:
\batitle{Riemann manifold {L}angevin and {H}amiltonian {M}onte {C}arlo
  methods}.
\bjtitle{Journal of the Royal Statistical Society: Series B (Statistical
  Methodology)}
\bvolume{73}(\bissue{2}),
\bfpage{123}--\blpage{214}
(\byear{2011})
\end{barticle}
\endbibitem

\bibitem[\protect\citeauthoryear{Geyer}{1991}]{geyer1991markov}
\begin{bchapter}
\bauthor{\bsnm{Geyer}, \binits{C.J.}}:
\bctitle{Markov chain monte carlo maximum likelihood}.
In: \bbtitle{Computing Science and Statistics: Proceedings on the 23rd
  Symposium on the Interface},
pp. \bfpage{156}--\blpage{163}.
\bpublisher{Interface Foundation of North America}, \blocation{???}
(\byear{1991})
\end{bchapter}
\endbibitem

\bibitem[\protect\citeauthoryear{Geyer and Thompson}{1995}]{geyer1995annealing}
\begin{barticle}
\bauthor{\bsnm{Geyer}, \binits{C.J.}},
\bauthor{\bsnm{Thompson}, \binits{E.A.}}:
\batitle{Annealing markov chain monte carlo with applications to ancestral
  inference}.
\bjtitle{Journal of the American Statistical Association}
\bvolume{90}(\bissue{431}),
\bfpage{909}--\blpage{920}
(\byear{1995})
\end{barticle}
\endbibitem

\bibitem[\protect\citeauthoryear{Benjamini and
  Hochberg}{1995}]{benjamini1995controlling}
\begin{barticle}
\bauthor{\bsnm{Benjamini}, \binits{Y.}},
\bauthor{\bsnm{Hochberg}, \binits{Y.}}:
\batitle{Controlling the false discovery rate: a practical and powerful
  approach to multiple testing}.
\bjtitle{Journal of the Royal statistical society: series B (Methodological)}
\bvolume{57}(\bissue{1}),
\bfpage{289}--\blpage{300}
(\byear{1995})
\end{barticle}
\endbibitem

\bibitem[\protect\citeauthoryear{Storey}{2003}]{storey2003positive}
\begin{barticle}
\bauthor{\bsnm{Storey}, \binits{J.D.}}:
\batitle{The positive false discovery rate: a {B}ayesian interpretation and the
  q-value}.
\bjtitle{The Annals of Statistics}
\bvolume{31}(\bissue{6}),
\bfpage{2013}--\blpage{2035}
(\byear{2003})
\end{barticle}
\endbibitem

\bibitem[\protect\citeauthoryear{M{\"u}ller et~al.}{2004}]{muller2004optimal}
\begin{barticle}
\bauthor{\bsnm{M{\"u}ller}, \binits{P.}},
\bauthor{\bsnm{Parmigiani}, \binits{G.}},
\bauthor{\bsnm{Robert}, \binits{C.}},
\bauthor{\bsnm{Rousseau}, \binits{J.}}:
\batitle{Optimal sample size for multiple testing: the case of gene expression
  microarrays}.
\bjtitle{Journal of the American Statistical Association}
\bvolume{99}(\bissue{468}),
\bfpage{990}--\blpage{1001}
(\byear{2004})
\end{barticle}
\endbibitem

\bibitem[\protect\citeauthoryear{M{\"u}ller et~al.}{2006}]{muller2006fdr}
\begin{botherref}
\oauthor{\bsnm{M{\"u}ller}, \binits{P.}},
\oauthor{\bsnm{Parmigiani}, \binits{G.}},
\oauthor{\bsnm{Rice}, \binits{K.}}:
Fdr and bayesian multiple comparisons rules.
Johns Hopkins University, Dept. of Biostatistics Working Papers
\textbf{Working Paper 115}
(2006)
\end{botherref}
\endbibitem

\bibitem[\protect\citeauthoryear{Biernacki
  et~al.}{2003}]{biernacki2003choosing}
\begin{barticle}
\bauthor{\bsnm{Biernacki}, \binits{C.}},
\bauthor{\bsnm{Celeux}, \binits{G.}},
\bauthor{\bsnm{Govaert}, \binits{G.}}:
\batitle{Choosing starting values for the em algorithm for getting the highest
  likelihood in multivariate gaussian mixture models}.
\bjtitle{Computational Statistics \& Data Analysis}
\bvolume{41}(\bissue{3-4}),
\bfpage{561}--\blpage{575}
(\byear{2003})
\end{barticle}
\endbibitem

\bibitem[\protect\citeauthoryear{Gelman and Rubin}{1992}]{gelman1992inference}
\begin{barticle}
\bauthor{\bsnm{Gelman}, \binits{A.}},
\bauthor{\bsnm{Rubin}, \binits{D.B.}}:
\batitle{Inference from iterative simulation using multiple sequences}.
\bjtitle{Statistical Science}
\bvolume{7}(\bissue{4}),
\bfpage{457}--\blpage{472}
(\byear{1992})
\end{barticle}
\endbibitem

\bibitem[\protect\citeauthoryear{George and
  McCulloch}{1995}]{george1995stochastic}
\begin{barticle}
\bauthor{\bsnm{George}, \binits{E.I.}},
\bauthor{\bsnm{McCulloch}, \binits{R.E.}}:
\batitle{Stochastic search variable selection}.
\bjtitle{Markov chain Monte Carlo in Practice}
\bvolume{68}(\bissue{1}),
\bfpage{203}--\blpage{214}
(\byear{1995})
\end{barticle}
\endbibitem

\bibitem[\protect\citeauthoryear{Dellaportas
  et~al.}{2002}]{dellaportas2002bayesian}
\begin{barticle}
\bauthor{\bsnm{Dellaportas}, \binits{P.}},
\bauthor{\bsnm{Forster}, \binits{J.J.}},
\bauthor{\bsnm{Ntzoufras}, \binits{I.}}:
\batitle{On bayesian model and variable selection using mcmc}.
\bjtitle{Statistics and computing}
\bvolume{12}(\bissue{1}),
\bfpage{27}--\blpage{36}
(\byear{2002})
\end{barticle}
\endbibitem

\bibitem[\protect\citeauthoryear{Polson and
  Scott}{2011}]{10.1093/acprof:oso/9780199694587.003.0017}
\begin{bchapter}
\bauthor{\bsnm{Polson}, \binits{N.G.}},
\bauthor{\bsnm{Scott}, \binits{J.G.}}:
\bctitle{{Shrink Globally, Act Locally: Sparse Bayesian Regularization and
  Prediction}}.
In: \bbtitle{{Bayesian Statistics 9}}.
\bpublisher{Oxford University Press}, \blocation{???}
(\byear{2011}).
\doiurl{10.1093/acprof:oso/9780199694587.003.0017} .
\burl{https://doi.org/10.1093/acprof:oso/9780199694587.003.0017}
\end{bchapter}
\endbibitem

\bibitem[\protect\citeauthoryear{Gelfand et~al.}{1992}]{gelfand1992model}
\begin{barticle}
\bauthor{\bsnm{Gelfand}, \binits{A.E.}},
\bauthor{\bsnm{Dey}, \binits{D.K.}},
\bauthor{\bsnm{Chang}, \binits{H.}}:
\batitle{Model determination using predictive distributions with implementation
  via sampling-based methods}.
\bjtitle{Bayesian Statistics}
\bvolume{4},
\bfpage{147}--\blpage{167}
(\byear{1992})
\end{barticle}
\endbibitem

\bibitem[\protect\citeauthoryear{Meng}{1994}]{10.1214/aos/1176325622}
\begin{barticle}
\bauthor{\bsnm{Meng}, \binits{X.-L.}}:
\batitle{Posterior predictive $p$-values}.
\bjtitle{The Annals of Statistics}
\bvolume{22}(\bissue{3}),
\bfpage{1142}--\blpage{1160}
(\byear{1994})
\doiurl{10.1214/aos/1176325622}
\end{barticle}
\endbibitem

\bibitem[\protect\citeauthoryear{Gelman}{2013}]{10.1214/13-EJS854}
\begin{barticle}
\bauthor{\bsnm{Gelman}, \binits{A.}}:
\batitle{{Two simple examples for understanding posterior p-values whose
  distributions are far from uniform}}.
\bjtitle{Electronic Journal of Statistics}
\bvolume{7}(\bissue{none}),
\bfpage{2595}--\blpage{2602}
(\byear{2013})
\doiurl{10.1214/13-EJS854}
\end{barticle}
\endbibitem

\bibitem[\protect\citeauthoryear{Gelman}{2003}]{gelman2003bayesian}
\begin{barticle}
\bauthor{\bsnm{Gelman}, \binits{A.}}:
\batitle{A bayesian formulation of exploratory data analysis and
  goodness-of-fit testing}.
\bjtitle{International Statistical Review}
\bvolume{71}(\bissue{2}),
\bfpage{369}--\blpage{382}
(\byear{2003})
\end{barticle}
\endbibitem

\end{thebibliography}

\appendix
\section*{Appendices}
\addcontentsline{toc}{section}{Appendices}
\renewcommand{\thesubsection}{\Alph{subsection}}
\renewcommand\thesubsection{\thesection.\arabic{subsection}}
\renewcommand{\theequation}{\thesection.\arabic{equation}}
\renewcommand{\thefigure}{\thesection.\arabic{figure}}





\section{Hyperparameters and MCMC sampler details}\label{sec:mcmc_details}

\begin{table}[t]
    \centering
\begin{tabular}{ccccccccccc}
    \toprule
          & $a_\gamma$ & $b_\gamma$ & $a_\lambda$ &$b_\lambda$ & $a_1$ & $b_1$ & $a_2$ & $b_2$ & $\mu$ & $\Sigma$\\ 
          \midrule
   vague  &  0.2 & 0.1 & 2.001 & 1 & 2.001 & 1 &  2.001 & 1 & $(0,\ldots,0)^\top$ & $100 \mathrm{I}_{p}$\\
   regularized & 1 & 1 & 2.1 & 1.1 & 2.1 & 1.1 & 2.1 & 1.1  & $(0,\ldots,0)^\top$ & $10 \mathrm{I}_{p}$\\
   \bottomrule
\end{tabular}
    \caption{Parameters of the prior distribution.}
    \label{tab:prior_pars}
\end{table}

We have used two sets of hyper-parameters in the prior distribution as seen in Table \ref{tab:prior_pars}. The ``vague'' prior distribution corresponds to a case where the prior variance of each parameter is large. On the other hand, the second choice heavily penalizes large (absolute) values of the parameters and can be seen as a regularization prior on the absolute values of each parameter. 

In order to select values of the scale parameters of our Metropolis-Hastings and MALA proposals with a reasonable acceptance rate betweeen proposed moves, the MCMC sampler runs for an initial warm-up period at Step 0 of Algorithm \ref{alg:mc3}. During this period, each parameter is adaptively tuned as the MCMC sampler progresses in order to achieve acceptance rates of the proposed updates between pre-specified limits. The final value of each parameter is then selected as the one that will be used in the subsequent main MCMC sampler.  More specifically, the scale parameter of the proposal distributions were tuned in order to achieve acceptance rates of the proposed moves between $15\%-30\%$. The scale parameter ($\tau$)  of the MALA proposal in Equation \ref{eq:proposal}  was tuned until an acceptance rate between $40\%-60\%$ is achieved (see also \citealp{https://doi.org/10.1111/1467-9868.00123}).

A total of $C = 16$ tempered chains was considered. The temperature of each chain was selected according to the following scheme:
\[
h_c = \frac{1}{(1+\varepsilon)^{c^{d} - 1}},\quad c=1,\ldots,C,
\]
where $\varepsilon > 0$, and $d > 0$. In our simulations the values  $\varepsilon = 0.001$ and $d = 2.5$ were used. The MCMC sampler was run for a total of $N=20000$ MCMC cycles, with each cycle consisting of $m_1 = 10$ MCMC iterations. 

Each chain uses  randomly selected initial values based on the following distributions (all independent)
\begin{align*}
\gamma&\sim\mathcal N(0, 4)\\
\lambda&\sim\mathcal E(1)\\
\alpha_1&\sim\mathcal E(1)\\
\alpha_2&\sim\mathcal E(1)\\
\beta_j&\sim\mathcal N(0,4),\quad j=0,1,\ldots,p.
\end{align*}

\section{Details of MH and MALA updates}\label{sec:mh_mala}
\subsection{Metropolis-Hastings step}\label{sec:mh-details}

Firstly, we propose to update $\gamma$ using the following normal proposal distribution
\begin{equation}
\label{eq:gamma_proposal}
\tilde\gamma\sim\mathcal N\left(\gamma,s^2_\gamma\right).
\end{equation}
All remaining parameters are held fixed (as well as the latent variables $I_1,\ldots,I_n$), and  hence this move proposes an update of the form $$\boldsymbol{\theta} = \left(\gamma,\lambda,\alpha_1,\alpha_2,\boldsymbol{\beta}\right)\rightarrow 
\tilde{\boldsymbol{\theta}} = \left(\tilde\gamma,\lambda,\alpha_1,\alpha_2,\boldsymbol{\beta}\right).
$$
Because the density of the proposal distribution in \eqref{eq:gamma_proposal} is symmetric to $\gamma^{(t-1)}$ and $\tilde \gamma$, the Metropolis-Hastings acceptance probability reduces to 
\begin{equation}
\label{eq:gamma_prob}
A_{1,h}\left(\boldsymbol{\theta},\tilde{\boldsymbol{\theta}}\right) = \min\left\{1,\frac{
\dot f_h(\boldsymbol y| \bs\Delta, \tilde{\boldsymbol{\theta}},\boldsymbol{I}, \boldsymbol x)\dot \pi_h(\tilde{\gamma})
}{\dot f_h(\boldsymbol y| \bs\Delta,\boldsymbol{\theta},\boldsymbol{I}, \boldsymbol x)\dot \pi_h(\gamma)}\right\}.
\end{equation}
If the move is accepted we set $\boldsymbol{\theta} = \tilde{\boldsymbol{\theta}}$, otherwise $\boldsymbol{\theta} = \boldsymbol{\theta}$.

Next, we propose to update $\lambda$ using the following log-normal proposal distribution 
\begin{equation}
\label{eq:lambda_proposal}
\tilde\lambda\sim\mathcal{LN}\left(\log\left(\lambda\right),s^2_\lambda\right).
\end{equation}
All remaining parameters are held fixed (as well as the latent variables $I_1,\ldots,I_n$), and therefore this move proposes an update of the form $$\boldsymbol{\theta} = \left(\gamma,\lambda,\alpha_1,\alpha_2,\boldsymbol{\beta}\right)\rightarrow 
\tilde{\boldsymbol{\theta}} = \left(\gamma,\tilde\lambda,\alpha_1,\alpha_2,\boldsymbol{\beta}\right).
$$
The ratio of the densities of the proposal distribution in \eqref{eq:lambda_proposal} is equal to $$\frac{q(\tilde\lambda\rightarrow\lambda)}{q(\lambda\rightarrow\tilde\lambda)}=\frac{\tilde\lambda}{\lambda};$$ hence, the Metropolis-Hastings acceptance probability reduces to 
\begin{equation}
\label{eq:lambda_prob}
A_{2,h}\left(\boldsymbol{\theta},\tilde{\boldsymbol{\theta}}\right) = \min\left\{1,\frac{
\dot f_h(\boldsymbol y|\bs\Delta, \tilde{\boldsymbol{\theta}},\boldsymbol{I}, \boldsymbol x)\dot \pi_h(\tilde{\lambda})\tilde\lambda
}{\dot f_h(\boldsymbol y|\bs\Delta, \boldsymbol{\theta},\boldsymbol{I}, \boldsymbol x)\dot \pi_h(\lambda)\lambda}\right\}.
\end{equation}
If the move is accepted, we update $\boldsymbol{\theta} = \tilde{\boldsymbol{\theta}}$, otherwise $\boldsymbol{\theta}$ is held fixed at its current value.

Next, we propose to update $\alpha_1$ using the following log-normal proposal distribution
\begin{equation}
\label{eq:alpha1_proposal}
\tilde\alpha_1\sim\mathcal{LN}\left(\log\left(\alpha_1\right),s^2_{\alpha_1}\right).
\end{equation}
All remaining parameters are held fixed (as well as the latent variables $I_1,\ldots,I_n$), and this move proposes an update of the form $$\boldsymbol{\theta} = \left(\gamma,\lambda,\alpha_1,\alpha_2,\boldsymbol{\beta}\right)\rightarrow 
\tilde{\boldsymbol{\theta}} = \left(\gamma,\lambda,\tilde\alpha_1,\alpha_2,\boldsymbol{\beta}\right).
$$
The Metropolis-Hastings acceptance probability reduces to 
\begin{equation}
\label{eq:a1_prob}
A_{3,h}\left(\boldsymbol{\theta},\tilde{\boldsymbol{\theta}}\right) = \min\left\{1,\frac{
\dot f_h(\boldsymbol y| \bs\Delta, \tilde{\boldsymbol{\theta}},\boldsymbol{I}, \boldsymbol x)\dot \pi_h(\tilde{\alpha}_1)\tilde\alpha_1
}{\dot f_h(\boldsymbol y|\bs\Delta, \boldsymbol{\theta},\boldsymbol{I}, \boldsymbol x)\dot \pi_h(\alpha_1)\alpha_1}\right\}.
\end{equation}
If the move is accepted, we update $\boldsymbol{\theta} = \tilde{\boldsymbol{\theta}}$, otherwise $\boldsymbol{\theta}$ is held fixed at its current value.

Next, we propose to update $\alpha_2$ using the following log-normal proposal distribution
\begin{equation}
\label{eq:alpha2_proposal}
\tilde\alpha_2\sim\mathcal{LN}\left(\log\left(\alpha_2\right),s^2_{\alpha_2}\right).
\end{equation}
Again, all remaining parameters are held fixed (as well as the latent variables $I_1,\ldots,I_n$), and this move proposes an update of the form $$\boldsymbol{\theta} = \left(\gamma,\lambda,\alpha_1,\alpha_2,\boldsymbol{\beta}\right)\rightarrow 
\tilde{\boldsymbol{\theta}} = \left(\gamma,\lambda,\alpha_1,\tilde\alpha_2,\boldsymbol{\beta}\right).
$$
The Metropolis-Hastings acceptance probability reduces to 
\begin{equation}
\label{eq:a2_prob}
A_{4,h}\left(\boldsymbol{\theta},\tilde{\boldsymbol{\theta}}\right) = \min\left\{1,\frac{
\dot f_h(\boldsymbol y|\bs\Delta, \tilde{\boldsymbol{\theta}},\boldsymbol{I}, \boldsymbol x)\dot \pi_h(\tilde{\alpha}_2)\tilde\alpha_2
}{\dot f_h(\boldsymbol y|\bs\Delta, \boldsymbol{\theta},\boldsymbol{I}, \boldsymbol x)\dot \pi_h(\alpha_2)\alpha_2}\right\}.
\end{equation}
If the move is accepted, we update $\boldsymbol{\theta} = \tilde{\boldsymbol{\theta}}$, otherwise $\boldsymbol{\theta}$ is held fixed at its current value.

The next step of this move is to propose to simultaneously update all regression coefficients $\boldsymbol\beta$, using independent normal proposal distributions, that is,
\begin{equation}
\label{eq:beta_proposal}
\tilde{\boldsymbol{\beta}}\sim\mathcal{N}_p\left(\boldsymbol{\beta},\boldsymbol{\nu} \mathcal I_p\right) 
\end{equation}
where $\boldsymbol{\nu} = (\tau_0,\nu_1,\ldots,\nu_k)$ denotes a vector of fixed positive numbers corresponding to the variance of the proposal distribution. Because the density of the proposal distribution in \eqref{eq:gamma_proposal} is symmetric to $\boldsymbol{\beta}$ and $\tilde{\boldsymbol{\beta}}$, the Metropolis-Hastings acceptance probability reduces to 
\begin{equation}
\label{eq:beta_prob}
A_{5,h}\left(\boldsymbol{\beta},\tilde{\boldsymbol{\beta}}\right) = \min\left\{1,\frac{
\dot f_h(\boldsymbol y|\bs\Delta, \tilde{\boldsymbol{\theta}},\boldsymbol{I}, \boldsymbol x)\dot \pi_h(\tilde{\boldsymbol{\beta}})
}{\dot f_h(\boldsymbol y|\bs\Delta, \boldsymbol{\theta},\boldsymbol{I}, \boldsymbol x)\dot \pi_h(\boldsymbol{\beta})}\right\}.
\end{equation}

In all previous acceptance probabilities of the proposed moves, the term $\dot f_h(\boldsymbol y|\bs\Delta, \boldsymbol{\theta},\boldsymbol{I}, \boldsymbol x)$ is proportional to $\dot f_h(\boldsymbol y,\bs I|\bs\Delta, \boldsymbol{\theta}, \boldsymbol x)$ defined in Equation \eqref{eq:f_h}. Thus, in all cases, the ratio of conditional densities is equal to
$
\frac{\dot f_h(\boldsymbol y|\bs\Delta, \tilde{\boldsymbol{\theta}},\boldsymbol{I}, \boldsymbol x)
}{\dot f_h(\boldsymbol y|\bs\Delta, \boldsymbol{\theta},\boldsymbol{I}, \boldsymbol x)} = 
\frac{\dot f_h(\boldsymbol y,\boldsymbol{I}|\bs\Delta, \tilde{\boldsymbol{\theta}}, \boldsymbol x)
}{\dot f_h(\boldsymbol y,\boldsymbol{I}|\bs\Delta, \boldsymbol{\theta}, \boldsymbol x)}.
$

\subsection{MALA step}\label{sec:mala-details}

The proposal in \eqref{eq:proposal} is accepted according to the usual Metropolis-Hastings probability, that is,
\begin{align}
\label{eq:map}
A_{6,h}(\bs\theta, \tilde{\bs\theta}) = \min\left\{1, 
\frac{\dot f_h(\bs y|\bs\Delta,\tilde{\bs\theta}, \bs x, \bs I)\dot \pi_h(\tilde{\bs\theta})}{\dot f_h(\bs y|\bs\Delta,\bs\theta,\bs x, \bs I)\dot \pi_h(\bs\theta)}
\frac{
\mathrm{P}\left(\tilde{\bs\theta}\rightarrow\bs\theta\right)
}{
\mathrm{P}\left(\bs\theta\rightarrow\tilde{\bs\theta}\right)
}
\right\},
\end{align}
where $\mathrm{P}\left(a\rightarrow b\right)$ denotes the probability density of proposing state $b$ while in $a$.  From \eqref{eq:proposal} we have that $\mathrm{P}\left(\bs\theta\rightarrow\tilde{\bs\theta}\right)$ is the density of the $$\mathcal N_{p+4}\left(\bs\theta + \tau\nabla\log \dot f_h(\bs\theta|\bs y, \bs x, \bs I), 2\tau \mathcal I_{p+4}\right)$$
distribution, evaluated at $\tilde{\bs\theta}$. The density of the reverse transition $\left(\tilde{\bs\theta}\rightarrow\bs\theta\right)$ is equal to the density of the distribution 
$$\mathcal N_{p+4}\left(\tilde{\bs\theta} + \tau\nabla\log \dot f_h(\tilde{\bs\theta}|\bs\Delta,\bs y, \bs x, \bs I), 2\tau \mathcal I_{p+4}\right)$$
evaluated at $\bs\beta^{(t)}$.
 
From the last Equation it is obvious that if a proposed value lies outside the parameter space, the acceptance probability is zero. In such a case the proposed move is immediately rejected. Certain details for computing the the gradient vector of the logarithm of the joint posterior of $\bs \theta$, conditional on $\bs I$ can be found in Appendix \ref{ap2}.

\section{Partial derivatives of the complete log-likelihood function}\label{ap2}
For computing the partial derivatives of the complete log-likelihood 
\begin{align*}
\log(L_c)=\sum_{i\in \Delta_1} \log f_P(y_i;{\bm \theta},{\bm x}_i)+ \sum_{i\in \Delta_0} \log p_0({\bm x}_i;{\bm \theta})+\sum_{i\in \Delta_0} I_i \log \left(\frac{S_P(y_i;{\bm \theta},{\bm x}_i)}{p_0({\bm x}_i;{\bm \theta})}-1\right),
\end{align*}
we need the partial derivatives of the functions $f_P(y_i;{\bm \theta},{\bm x}_i), p_0({\bm x}_i;{\bm \theta})$  and $S_P(y_i;{\bm \theta},{\bm x}_i)$. Hence, denoting by $f_{P0}(y_i;{\bm \theta},{\bm x}_i)=(1+ \gamma \vartheta({\bm x}_i)c^{\gamma \vartheta({\bm x}_i)} F(y_i;\alpha_1, \alpha_2)^\lambda)^{-\frac{1}{\gamma}-1}$, we have 
\begin{align*}
&\frac{\partial \log f_P(y_i;{\bm \theta},{\bm x}_i)}{\partial  \lambda }=\lambda^{-1}+\log F(y_i;\alpha_1, \alpha_2)+\frac{\partial \log f_{P0}(y_i;{\bm \theta},{\bm x}_i)}{\partial  \lambda}\\
&\frac{\partial \log f_P(y_i;{\bm \theta},{\bm x}_i)}{\partial  \gamma }=\theta e^{-1}+\frac{\partial \log f_{P0}(y_i;{\bm \theta},{\bm x}_i)}{\partial  \gamma}\\
&\frac{\partial \log f_P(y_i;{\bm \theta},{\bm x}_i)}{\partial  \alpha_1}=(\lambda-1) \frac{\partial \log F(y_i;\alpha_1, \alpha_2)}{\partial  \alpha_1 }+\frac{\partial \log f(y_i;\alpha_1, \alpha_2)}{\partial  \alpha_1}+\frac{\partial \log f_{P0}(y_i;{\bm \theta},{\bm x}_i)}{\partial  \alpha_1}\\
&\frac{\partial \log  f_P(y_i;{\bm \theta},{\bm x}_i)}{\partial  \alpha_2}=(\lambda-1) \frac{\partial \log F(y_i;\alpha_1, \alpha_2)}{\partial  \alpha_2 }+\frac{\partial \log f(y_i;\alpha_1, \alpha_2)}{\partial  \alpha_2}+\frac{\partial \log f_{P0}(y_i;{\bm \theta},{\bm x}_i)}{\partial  \alpha_2}\\
&\frac{\partial \log  f_P(y_i;{\bm \theta},{\bm x}_i)}{\partial  \beta_j }=\left(\frac{1}{\vartheta({\bm x}_i)}+\gamma e^{-1}+\frac{\partial \log f_{P0}(y_i;{\bm \theta},{\bm x}_i)}{\partial \vartheta({\bm x}_i) }\right)\exp(\beta_0+{\bm\beta} {\bm x}_i) x_j^{I[j>0]}
\end{align*}
with
\begin{align*}
&\frac{\partial F(y_i;\alpha_1, \alpha_2)}{\partial  \alpha_1 }=\alpha_2 y_i (\alpha_1 y_i)^{\alpha_2-1} \exp(-(\alpha_1 y_i)^{\alpha_2}),\\
&\frac{\partial F(y_i;\alpha_1, \alpha_2)}{\partial  \alpha_2 }=  \log(\alpha_1 y_i) (\alpha_1 y_i)^{\alpha_2} \exp(-(\alpha_1 y_i)^{\alpha_2}),\\
&\frac{\partial f(y_i;\alpha_1, \alpha_2)}{\partial  \alpha_1 }=(\alpha_1 y_i)^{\alpha_2-1}\alpha_2^2 \exp(-(\alpha_1 y_i)^{\alpha_2})\left(1 - (\alpha_1 y_i)^{\alpha_2} \right)\\
&\frac{\partial f(y_i;\alpha_1, \alpha_2)}{\partial  \alpha_2 }=\frac{1}{y_i}\exp(-(\alpha_1 y_i)^{\alpha_2}) (\alpha_1 y_i)^{\alpha_2} \left( 1 +\alpha_2 \log(\alpha_1 y_i)-\alpha_2  (\alpha_1 y_i)^{\alpha_2} \log(\alpha_1 y_i)\right)
\end{align*}
and
\begin{align*}
&\frac{\partial \log f_{P0}(y_i;{\bm \theta},{\bm x}_i)}{\partial  \lambda}=\left(-\frac{1}{\gamma}-1\right) \frac{\gamma \vartheta({\bm x}_i) c^{\gamma \vartheta({\bm x}_i) } F(y_i;\alpha_1, \alpha_2)^\lambda \log F(y_i;\alpha_1, \alpha_2)}{1+ \gamma \vartheta({\bm x}_i) c^{\gamma \vartheta({\bm x}_i) } F(y_i;\alpha_1, \alpha_2)^\lambda}\\
&\frac{\partial \log f_{P0}(y_i;{\bm \theta},{\bm x}_i)}{\partial  \gamma}=\frac{1}{\gamma^2}\log (1+ \gamma \vartheta({\bm x}_i) c^{\gamma \vartheta({\bm x}_i) } F(y_i;\alpha_1, \alpha_2)^\lambda)\\
&~~~~~~~~~~~~~~+\left(-\frac{1}{\gamma}-1\right) \vartheta({\bm x}_i) c^{\gamma \vartheta({\bm x}_i) } F(y_i;\alpha_1, \alpha_2)^\lambda \frac{1+\gamma \vartheta({\bm x}_i) e^{-1}}{1+ \gamma \vartheta({\bm x}_i) c^{\gamma \vartheta({\bm x}_i) } F(y_i;\alpha_1, \alpha_2)^\lambda}\\
&\frac{\partial \log f_{P0}(y_i;{\bm \theta},{\bm x}_i)}{\partial  \alpha_1}=\left(-\frac{1}{\gamma}-1\right) \frac{\lambda\gamma \vartheta({\bm x}_i) c^{\gamma \vartheta({\bm x}_i) } F(y_i;\alpha_1, \alpha_2)^{\lambda-1}\frac{\partial F(y_i;\alpha_1, \alpha_2)}{\partial  \alpha_1 }}{1+ \gamma \vartheta({\bm x}_i) c^{\gamma \vartheta({\bm x}_i) } F(y_i;\alpha_1, \alpha_2)^\lambda}\\
&\frac{\partial \log f_{P0}(y_i;{\bm \theta},{\bm x}_i)}{\partial  \alpha_2}=\left(-\frac{1}{\gamma}-1\right) \frac{\lambda\gamma \vartheta({\bm x}_i) c^{\gamma \vartheta({\bm x}_i) } F(y_i;\alpha_1, \alpha_2)^{\lambda-1}\frac{\partial F(y_i;\alpha_1, \alpha_2)}{\partial  \alpha_2 }}{1+ \gamma \vartheta({\bm x}_i) c^{\gamma \vartheta({\bm x}_i) } F(y_i;\alpha_1, \alpha_2)^\lambda}\\
&\frac{\partial \log f_{P0}(y_i;{\bm \theta},{\bm x}_i)}{\partial  \vartheta({\bm x}_i)}=\left(-\frac{1}{\gamma}-1\right) \gamma c^{\gamma \vartheta({\bm x}_i) } F(y_i;\alpha_1, \alpha_2)^\lambda \frac{1+\gamma \vartheta({\bm x}_i) e^{-1}}{1+ \gamma \vartheta({\bm x}_i) c^{\gamma \vartheta({\bm x}_i) } F(y_i;\alpha_1, \alpha_2)^\lambda}.
\end{align*}
Moreover, we have 
\begin{align*}
&\frac{\partial\log p_0({\bm x}_i;{\bm \theta})}{\partial  \gamma}=\frac{1}{\gamma} \left(\frac{\log(1+ \gamma \vartheta({\bm x}_i) c^{\gamma \vartheta({\bm x}_i)})}{\gamma}- \vartheta({\bm x}_i)c^{\gamma \vartheta({\bm x}_i)}\frac{1+\gamma e^{-1} \vartheta({\bm x}_i)}{1+ \gamma \vartheta({\bm x}_i)c^{\gamma \vartheta({\bm x}_i)}}\right)\\
&\frac{\partial \log p_0({\bm x}_i;{\bm \theta})}{\partial  \beta_j}=-c^{\gamma \vartheta({\bm x}_i)}\frac{1+\gamma\vartheta({\bm x}_i)e^{-1}}{1+\gamma\vartheta({\bm x}_i)c^{\gamma \vartheta({\bm x}_i)}}\exp(\beta_0+{\bm\beta} {\bm x}_i) x_j^{I[j>0]}
\end{align*}
and
\begin{align*}
&\frac{\partial S_P(y_i;{\bm \theta},{\bm x}_i)}{\partial  \gamma }=S_P(y_i;{\bm \theta},{\bm x}_i) \left(\frac{1}{\gamma^2}\log(1+ \gamma \vartheta({\bm x}_i) c^{\gamma\vartheta({\bm x}_i)} F(y_i;\alpha_1, \alpha_2)^\lambda)\right.\\
&\left.-\frac{F(y_i;\alpha_1, \alpha_2)^\lambda\theta({\bm x}_i) c^{\gamma\vartheta({\bm x}_i) }}{\gamma}\frac{1+\gamma \theta({\bm x}_i) e^{-1}}{1+ \gamma \vartheta({\bm x}_i) c^{\gamma\vartheta({\bm x}_i) } F(y_i;\alpha_1, \alpha_2)^\lambda}\right)\\
&\frac{\partial S_P(y_i;{\bm \theta},{\bm x}_i)}{\partial  \lambda }=-f_{P0}(y_i;{\bm \theta},{\bm x}_i) \vartheta({\bm x}_i) c^{\gamma\vartheta({\bm x}_i) } F(y_i;\alpha_1, \alpha_2)^{\lambda} \log(F(y_i;\alpha_1, \alpha_2))\\
&\frac{\partial S_P(y_i;{\bm \theta},{\bm x}_i)}{\partial  \alpha_1}=-f_{P0}(y_i;{\bm \theta},{\bm x}_i) \vartheta({\bm x}_i) c^{\gamma\vartheta({\bm x}_i) } \lambda F(y_i;\alpha_1, \alpha_2)^{\lambda-1} \frac{\partial F(y_i;\alpha_1, \alpha_2)}{\partial  \alpha_1}\\
&\frac{\partial S_P(y_i;{\bm \theta},{\bm x}_i)}{\partial  \alpha_2}=-f_{P0}(y_i;{\bm \theta},{\bm x}_i) \vartheta({\bm x}_i) c^{\gamma\vartheta({\bm x}_i) } \lambda F(y_i;\alpha_1, \alpha_2)^{\lambda-1} \frac{\partial F(y_i;\alpha_1, \alpha_2)}{\partial  \alpha_2}\\
&\frac{\partial S_P(y_i;{\bm \theta},{\bm x}_i)}{\partial  \beta_j}=-f_{P0}(y_i;{\bm \theta},{\bm x}_i)c^{\gamma\vartheta({\bm x}_i) }F(y_i;\alpha_1, \alpha_2)^{\lambda} (1+\gamma\vartheta({\bm x}_i)e^{-1})\exp(\beta_0+{\bm\beta} {\bm x}_i) x_j^{I[j>0]}.
\end{align*}

\section{Further simulation results}\label{sec:sim_more}

\subsection{Comparison of Algorithms \ref{alg:mala} and \ref{alg:mc3} on a single simulated dataset }\label{sec:single_sim}

In this section, we provide additional insights to the single synthetic dataset presented in Section \ref{sec:sims}. In order to highlight the benefits of the parallel tempering sampling scheme we compare long runs  of Algorithm \ref{alg:mala} with shorter runs of parallel tempered chains in Algorithm  \ref{alg:mc3}. Recall that the results presented in Figure \ref{fig:4chains_density} are based on $N = 20000$ cycles of Algorithm \ref{alg:mc3} with $C=16$ tempered chains. Hence, a total of $m = 16 \times 20000 = 320000$ MCMC iterations were considered to Algorithm \ref{alg:mala}. Each Algorithm ran 4 times from different random starting values.

\begin{figure}[p]
    \centering
    \begin{tabular}{c}
        \includegraphics[scale=0.55]{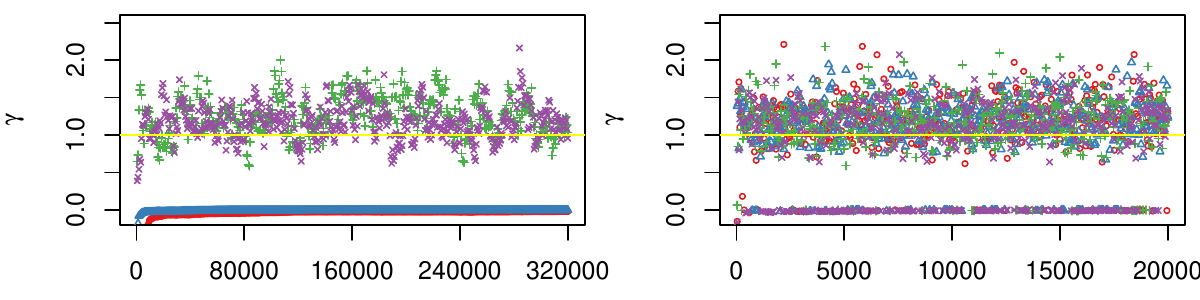}  \\
        \includegraphics[scale=0.55]{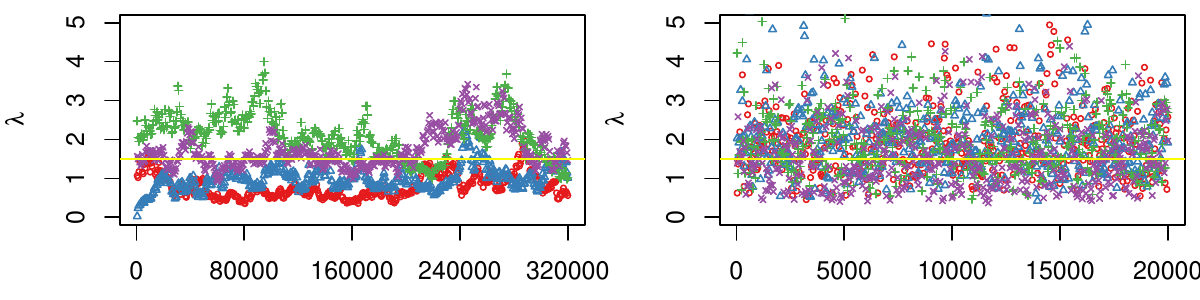}  \\          
        \includegraphics[scale=0.55]{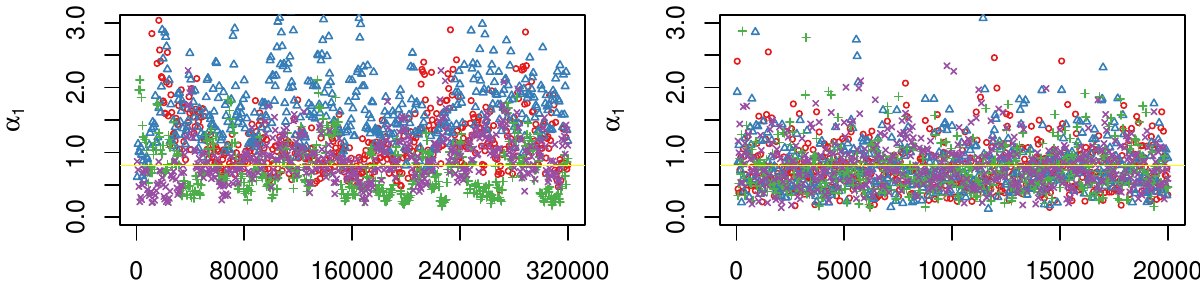}  \\                  
        \includegraphics[scale=0.55]{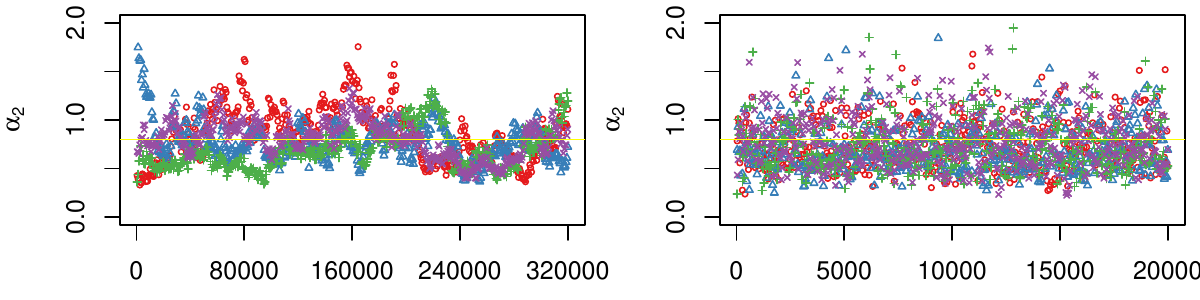}  \\                          
        \includegraphics[scale=0.55]{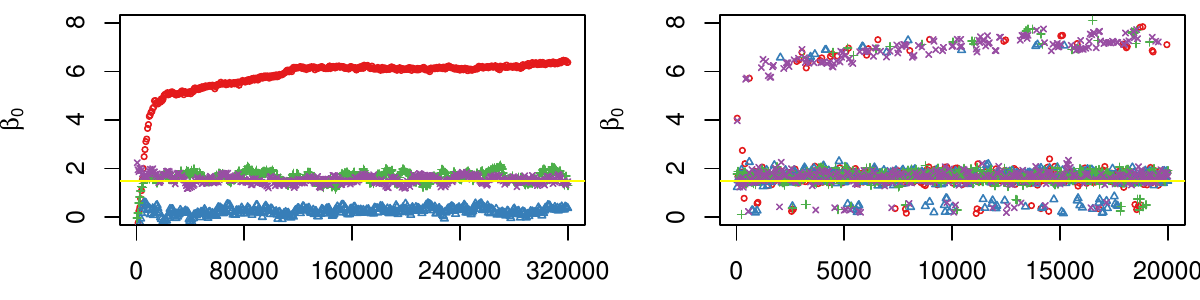}  \\                          
\includegraphics[scale=0.55]{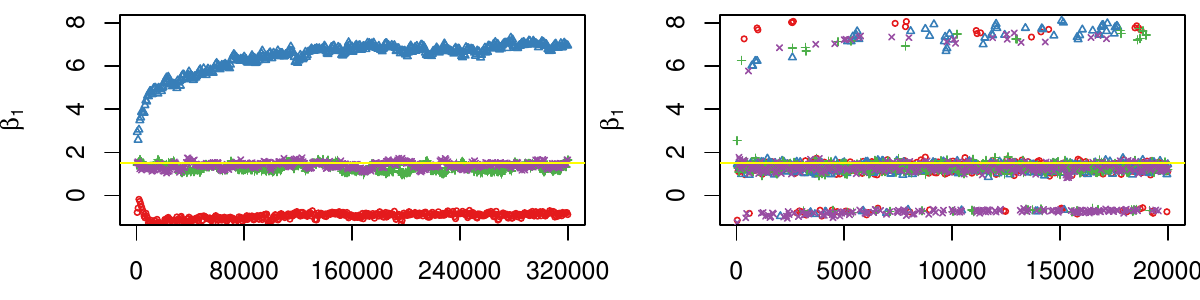}  \\                          
\includegraphics[scale=0.55]{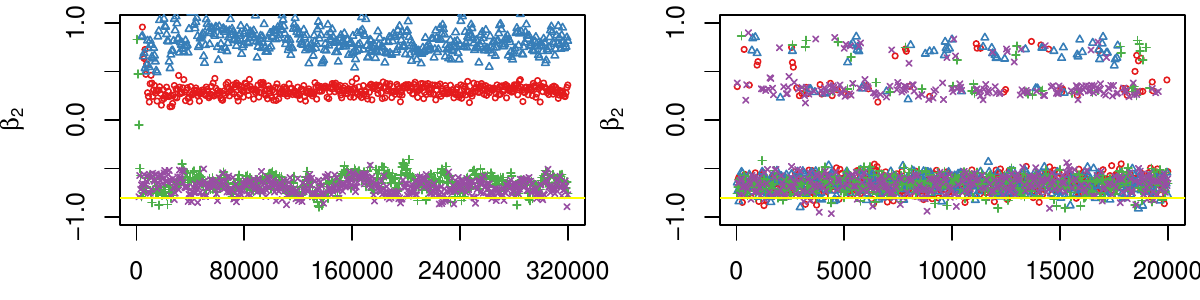}  
    \end{tabular}
    \caption{Thinned MCMC outputs for four randomly initialized chains according to  Algorithm \ref{alg:mala} (left) and Algorithm \ref{alg:mc3} (right) for a simulated from Scenario A1 with sample size equal to $n = 500$. The horizontal yellow line indicates the values used to generate the dataset.}
    \label{fig:multimodality}
\end{figure}

The corresponding traces of sampled parameter values are displayed  in Figure \ref{fig:multimodality} (see for example the generated values of $\gamma$ and $\beta_0$). In such cases it is quite challenging to properly explore the posterior surface, since the simulated MCMC sample may be trapped in a local mode. In Figure \ref{fig:multimodality} it is clearly seen that  the MC$^3$ implementation (using $C=16$ heated chains) in Algorithm \ref{alg:mc3} enables the generated parameter values consecutively alternating between the main and the minor ones (this behaviour is vividly illustrated particularly for $\gamma$ and $\beta_0$), therefore, the chains are able to freely explore the posterior distribution and avoid getting trapped to the minor mode(s).

\begin{figure}[ht]
    \centering
\includegraphics[scale = 0.55]{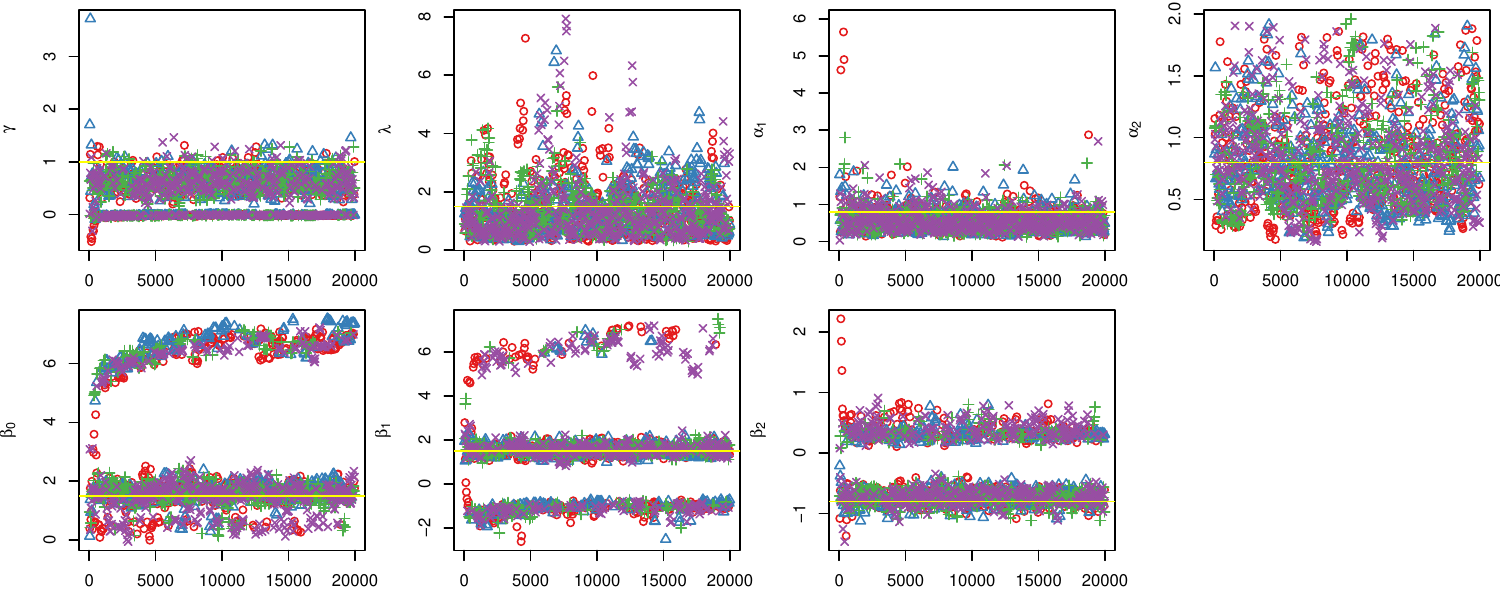}
    \caption{Thinned MCMC outputs for four randomly initialized chains according to  Algorithm \ref{alg:mc3}  for a simulated from Scenario A1 with sample size equal to $n = 300$. The horizontal yellow line indicates the values used to generate the dataset.}
    \label{fig:multimodality_smaller_n}
\end{figure}

We have also checked the validity of our findings in cases of smaller sample sizes. Therefore, we have repeated the present analysis when reducing the size of the simulated data set to $n = 300$ observations. We concluded that the MC$^{3}$ sampler leads to similar answers in four distinct runs under different starting values, as indicated in Figure \ref{fig:multimodality_smaller_n}. On the other hand, when using Algorithm \ref{alg:mala} the generated MCMC sample might get trapped into minor modes of the posterior distribution (results not displayed).

\subsection{Additional simulation study results}\label{ap1}

\begin{figure}[p]
\centering
\begin{tabular}{cc}
\hspace{-20ex}\includegraphics[scale=0.20]{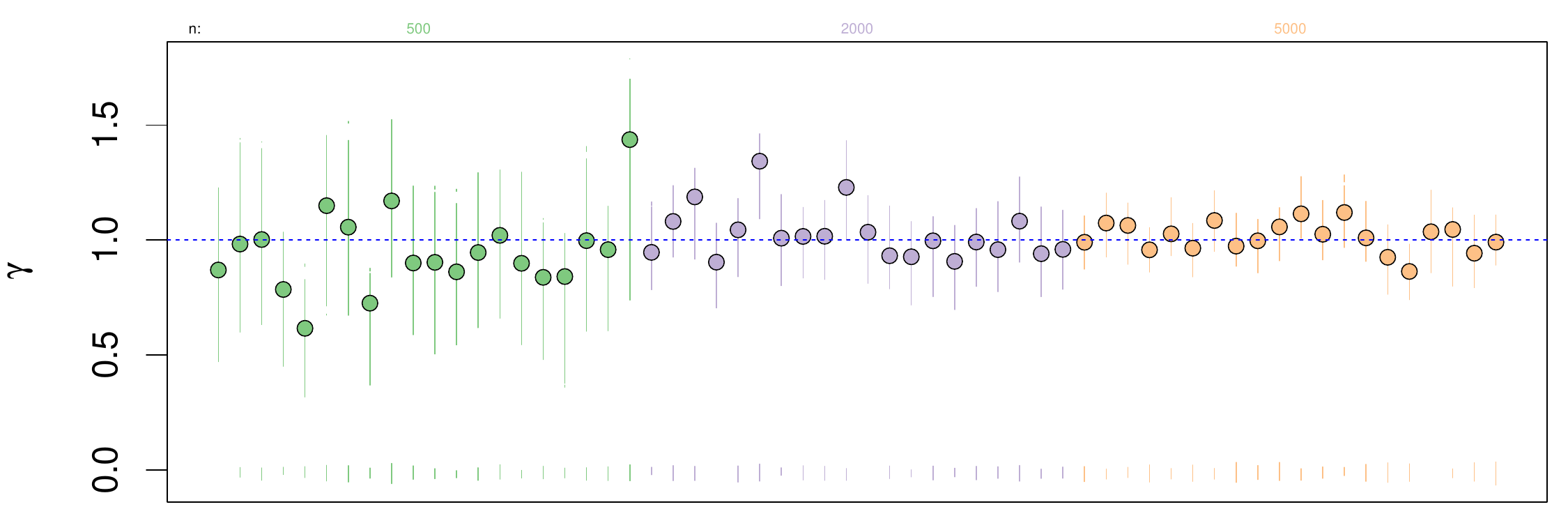}&
\includegraphics[scale=0.20]{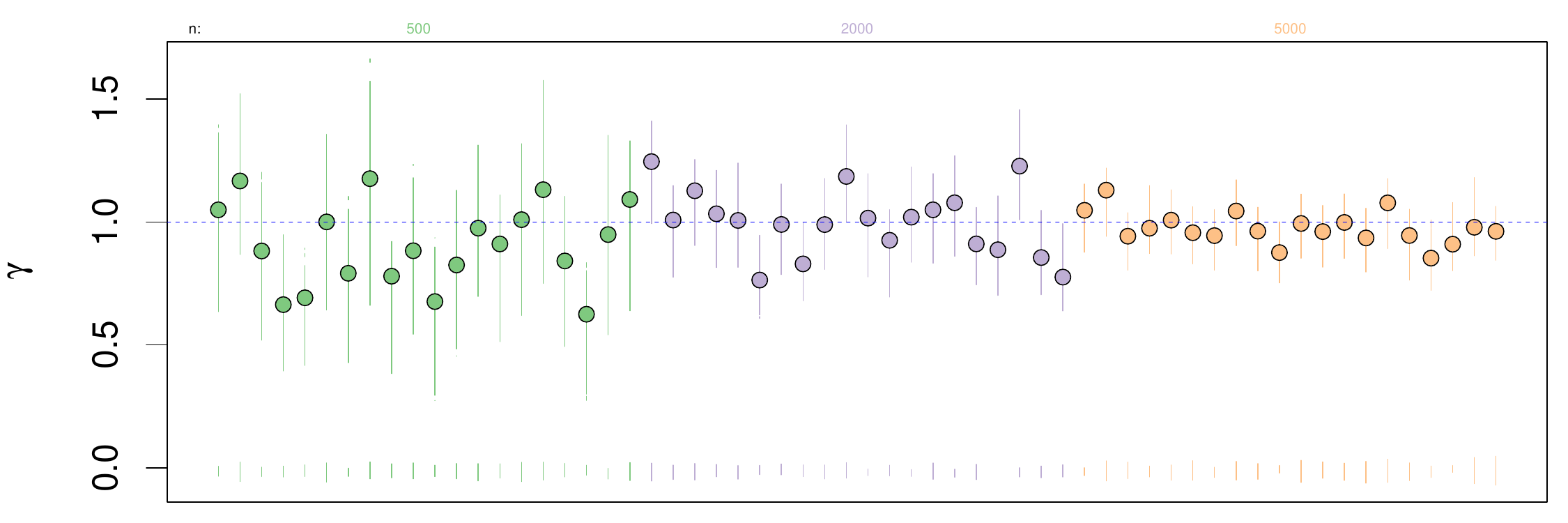}\\
\hspace{-20ex}\includegraphics[scale=0.20]{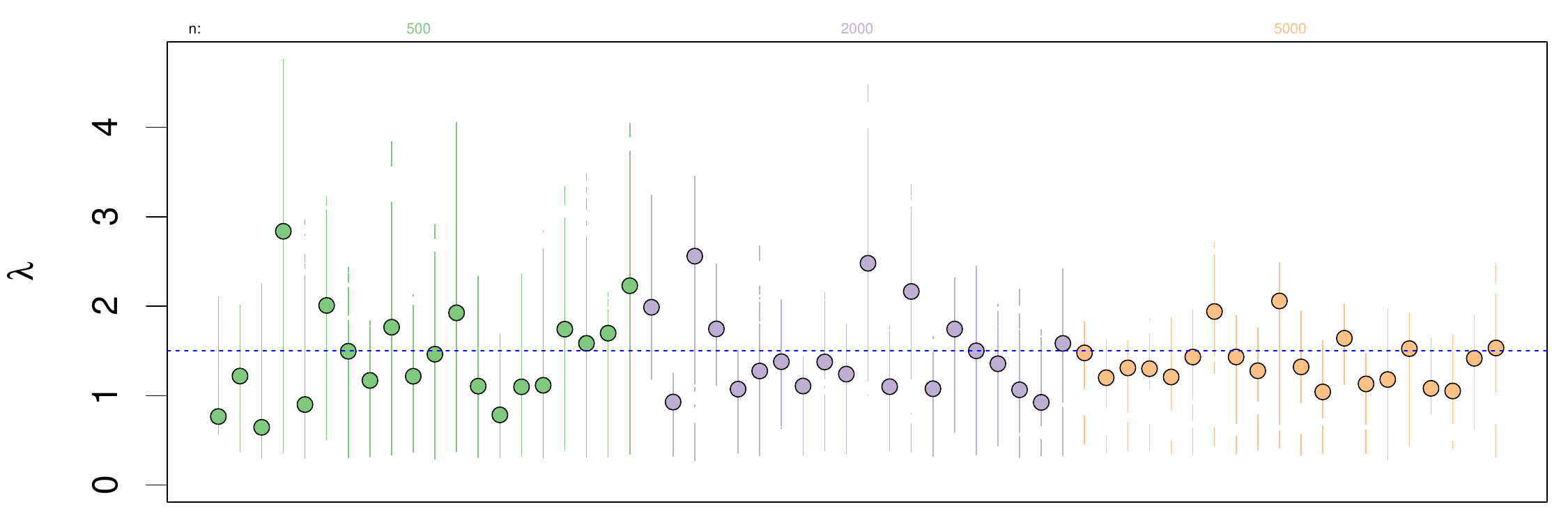}&
\includegraphics[scale=0.20]{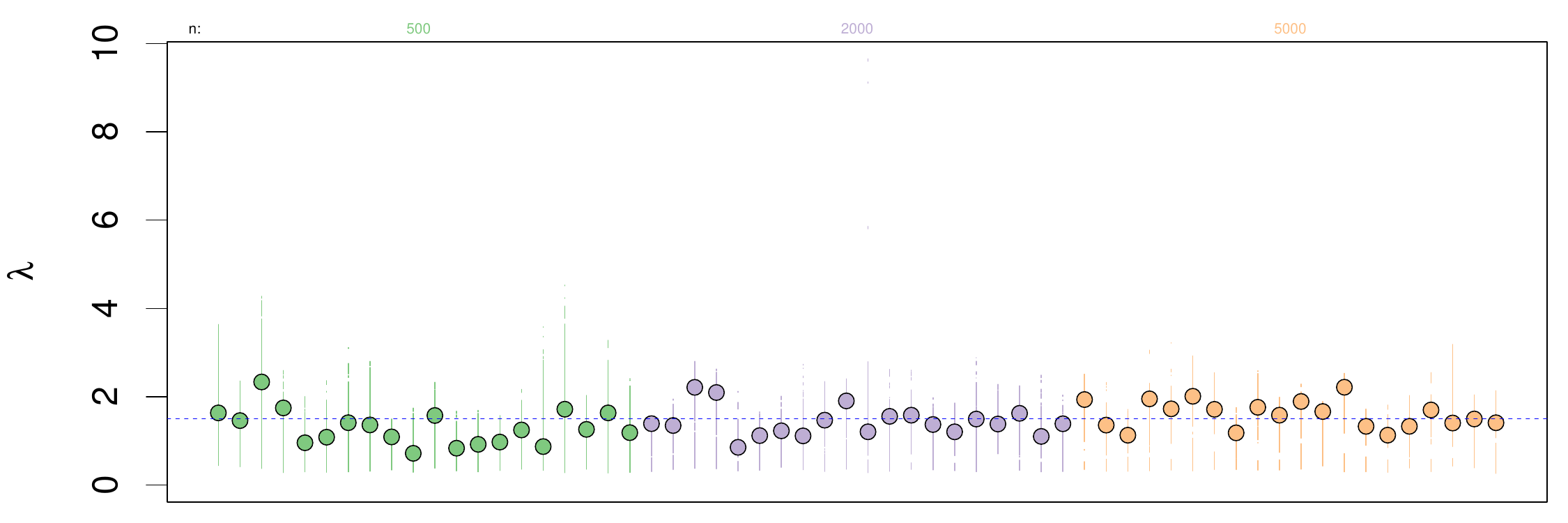}\\

\hspace{-20ex}\includegraphics[scale=0.20]{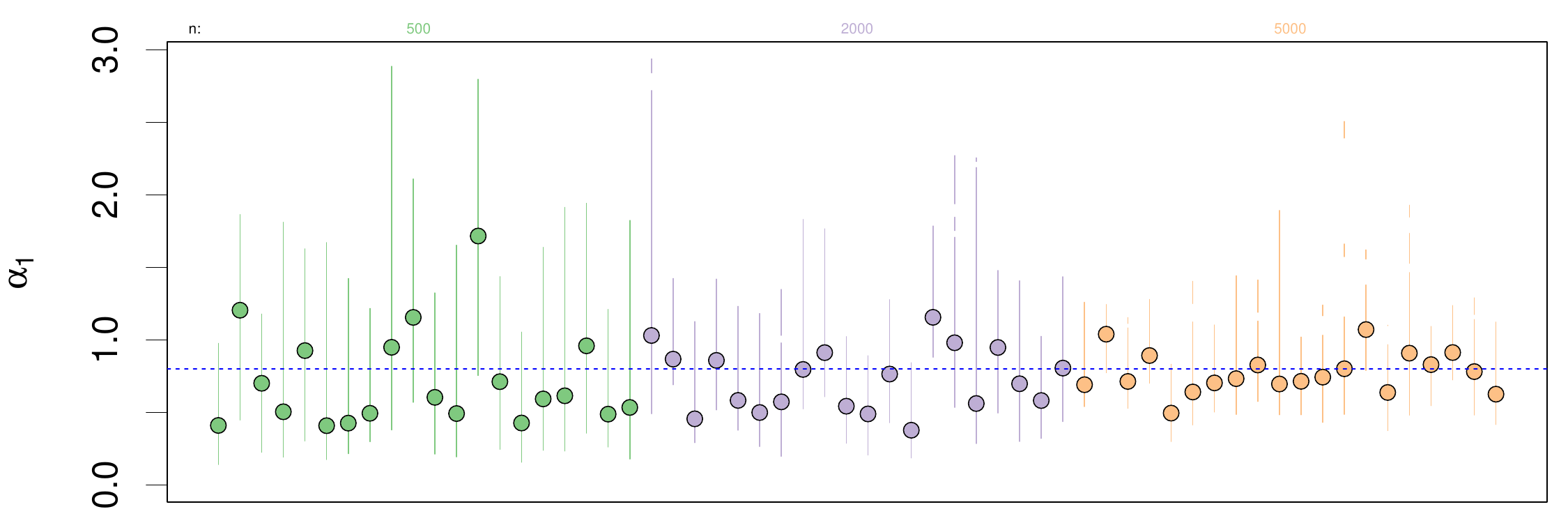}&
\includegraphics[scale=0.20]{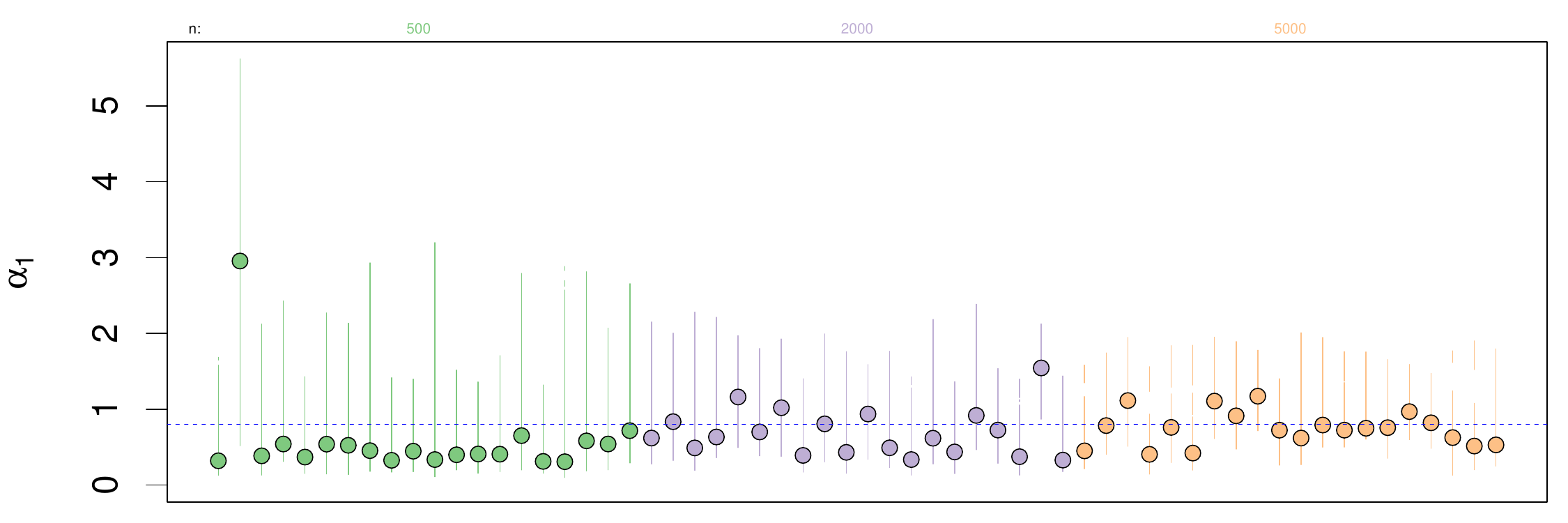}\\

\hspace{-20ex}\includegraphics[scale=0.20]{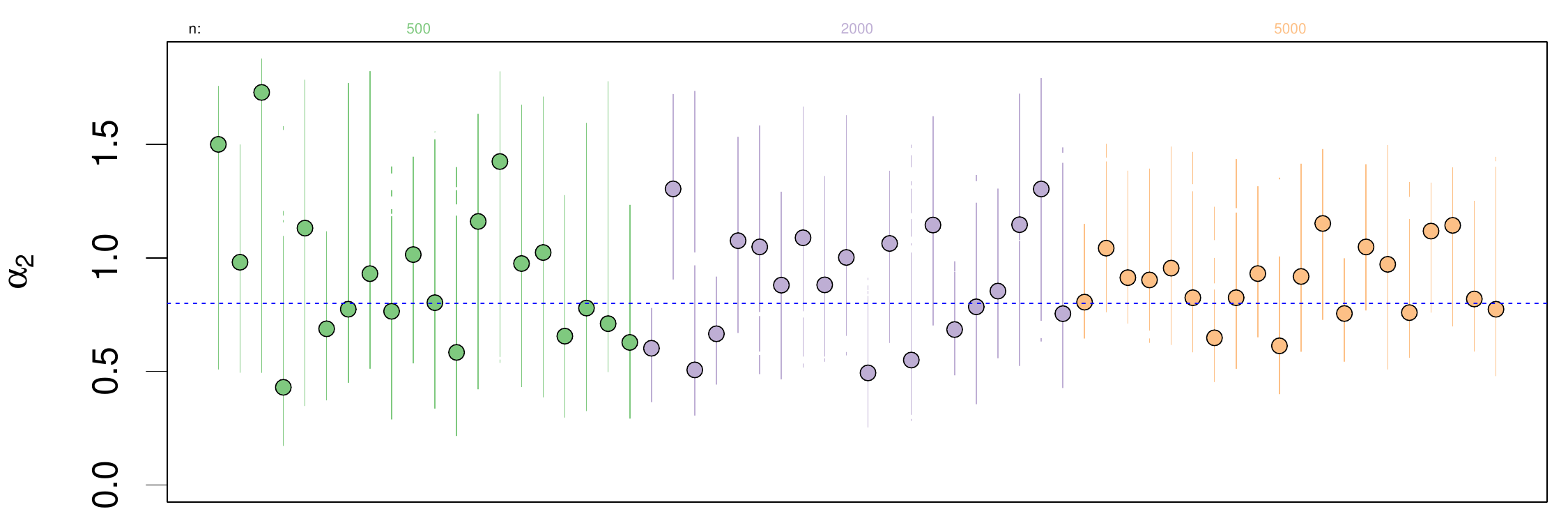}&
\includegraphics[scale=0.20]{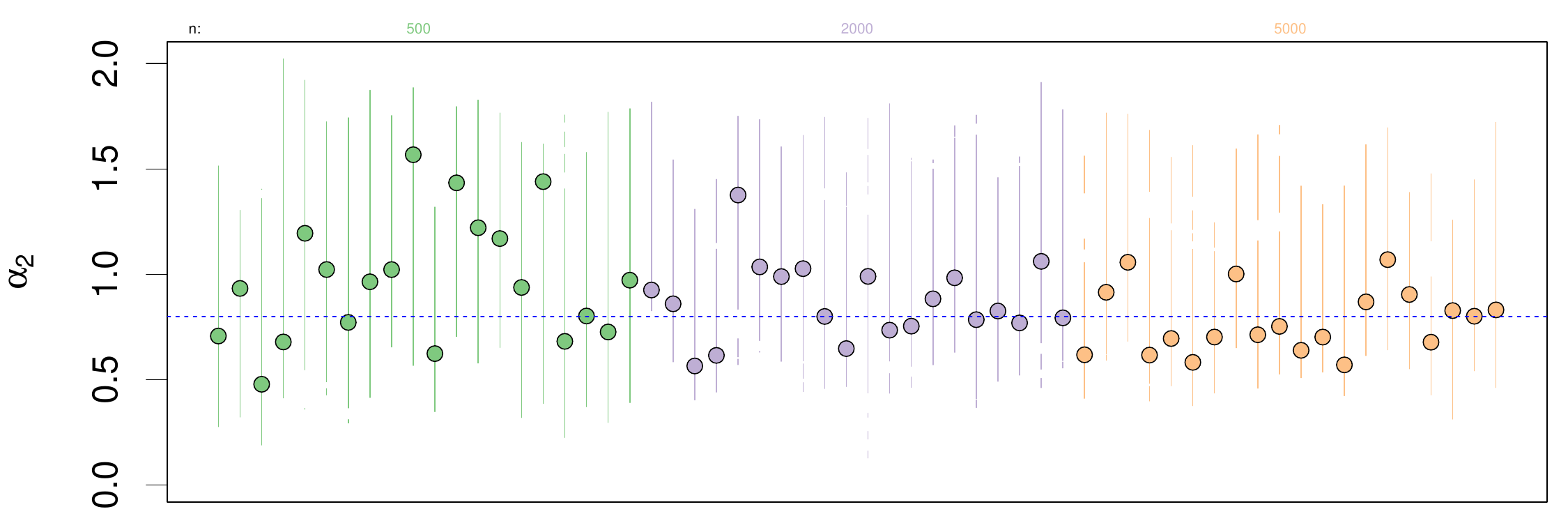}\\

\hspace{-20ex}\includegraphics[scale=0.20]{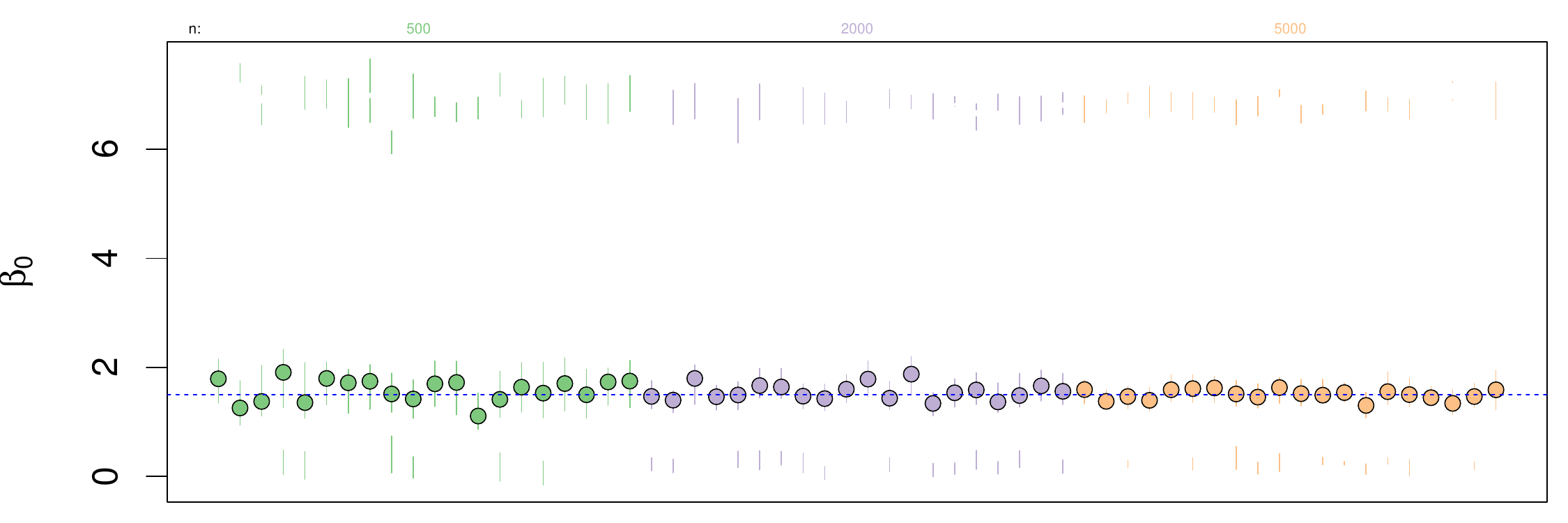}&
\includegraphics[scale=0.20]{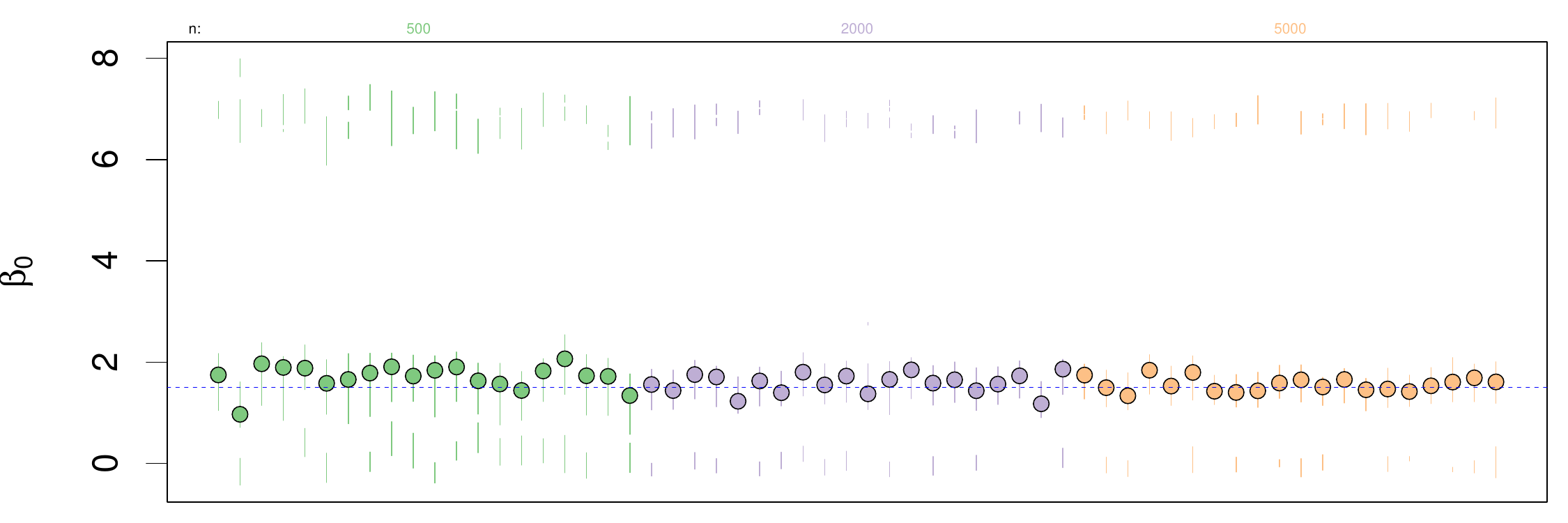}\\

\hspace{-20ex}\includegraphics[scale=0.20]{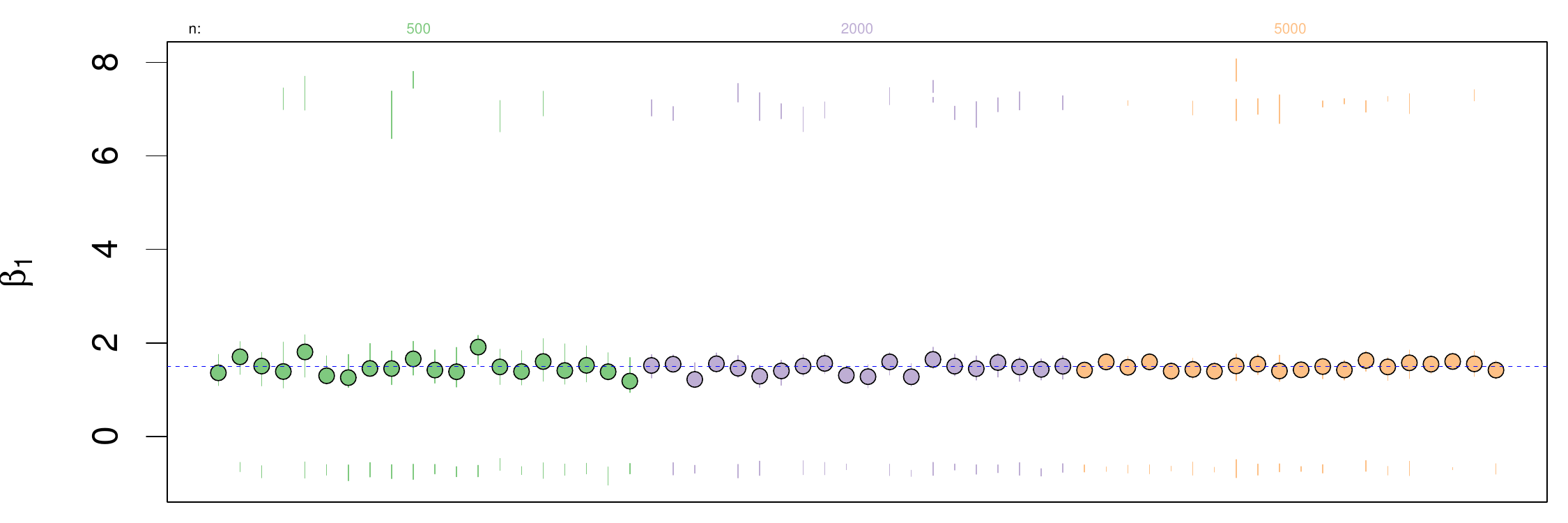}&
\includegraphics[scale=0.20]{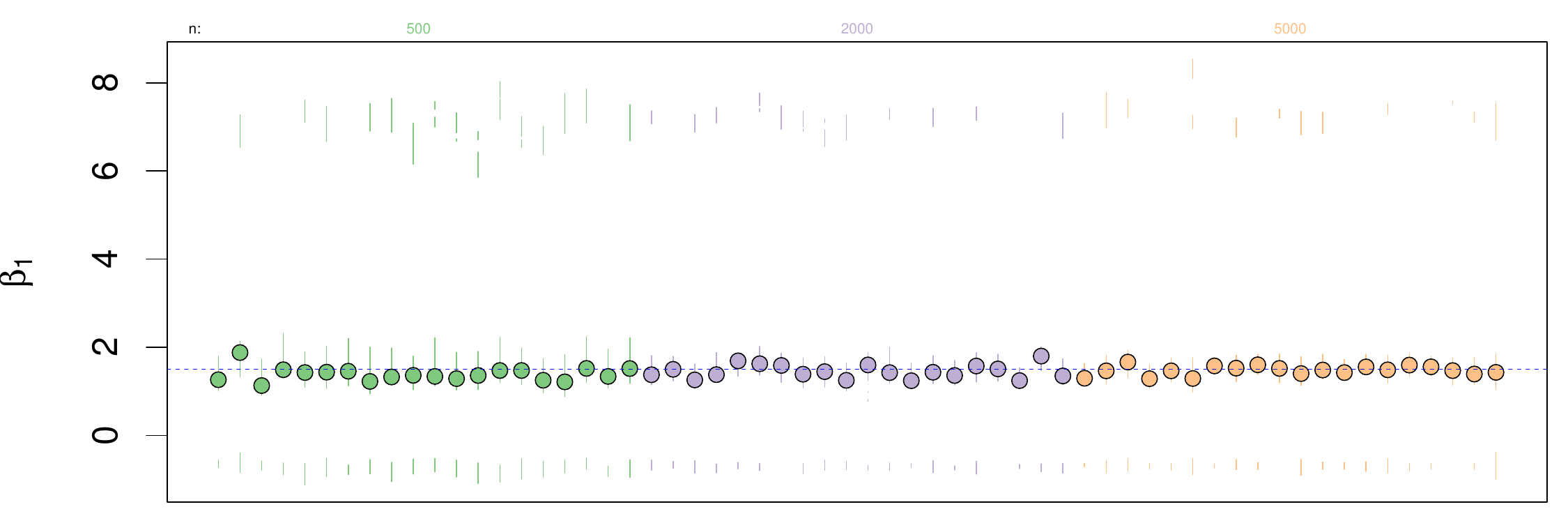}\\

\hspace{-20ex}\includegraphics[scale=0.20]{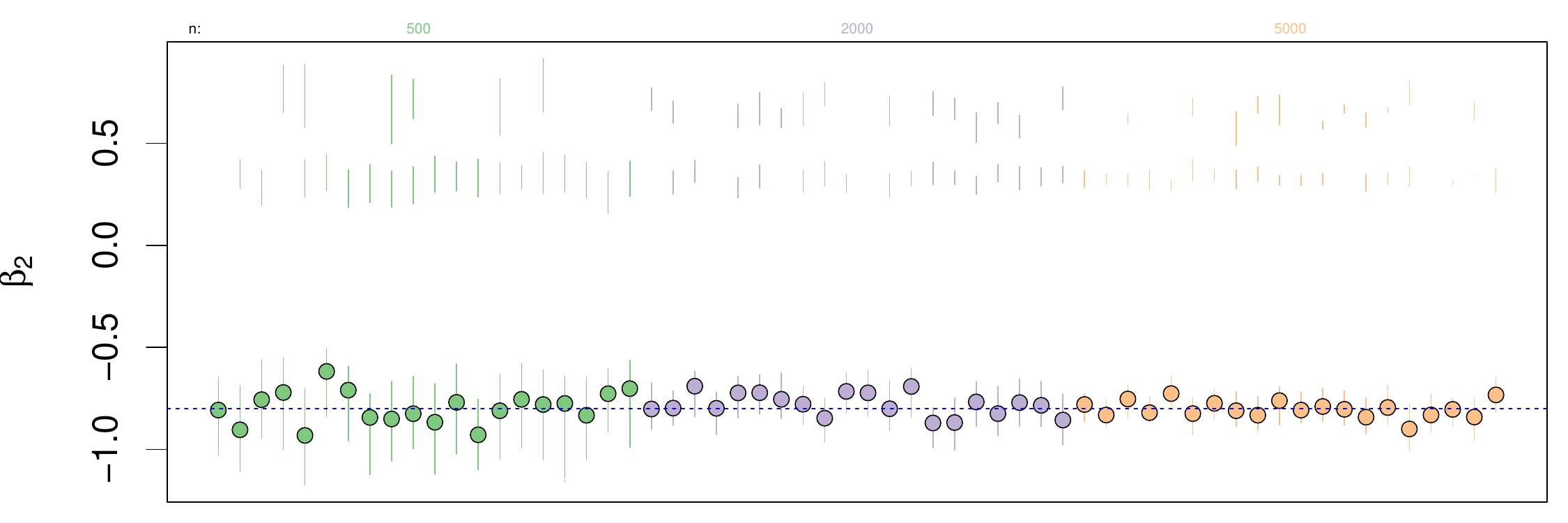}&
\includegraphics[scale=0.20]{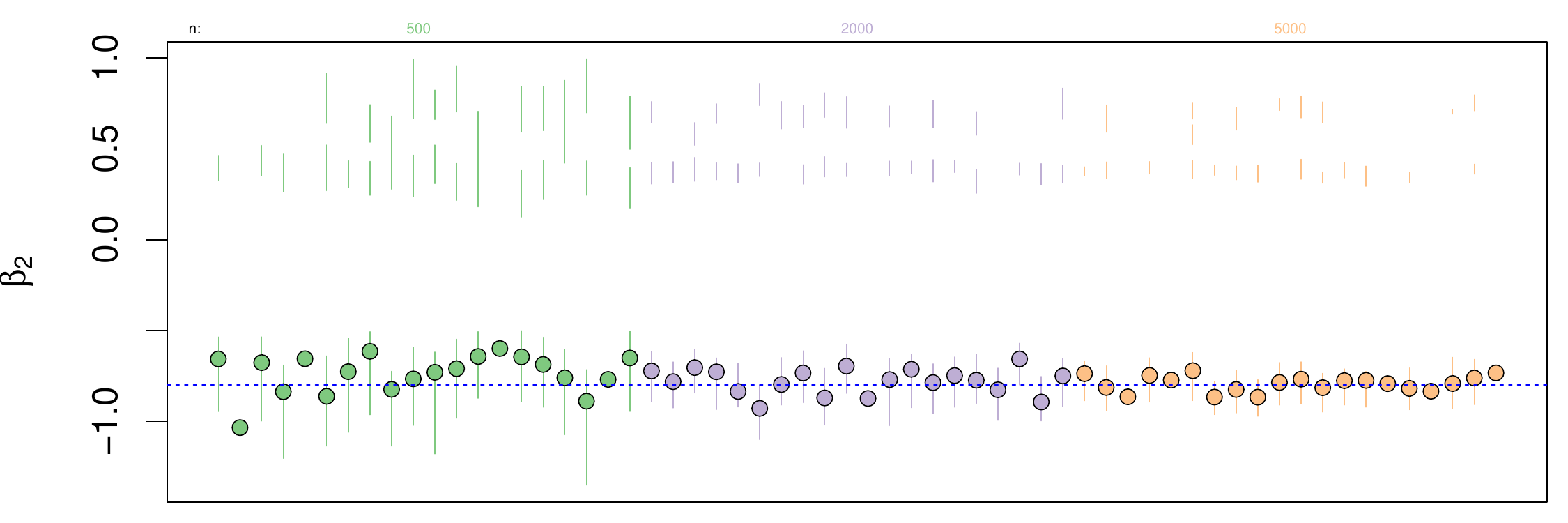}\\
scenario A.1 & scenario A.2
\end{tabular}
\caption{Point estimates (MAP) with $95\%$ Highest Density Intervals for our simulated datasets. Different colour indicate levels of sample size: 500 (\textcolor{green}{---}), 2000 (\textcolor{violet}{---}), 5000 (\textcolor{orange}{---}). The horizontal line indicates the true value (see Scenarios A1 and A2 in Table \ref{t4b}).} 
\label{fig:hdisA}
\end{figure}

\begin{figure}[p]
\centering
\begin{tabular}{cc}
\hspace{-20ex}\includegraphics[scale=0.20]{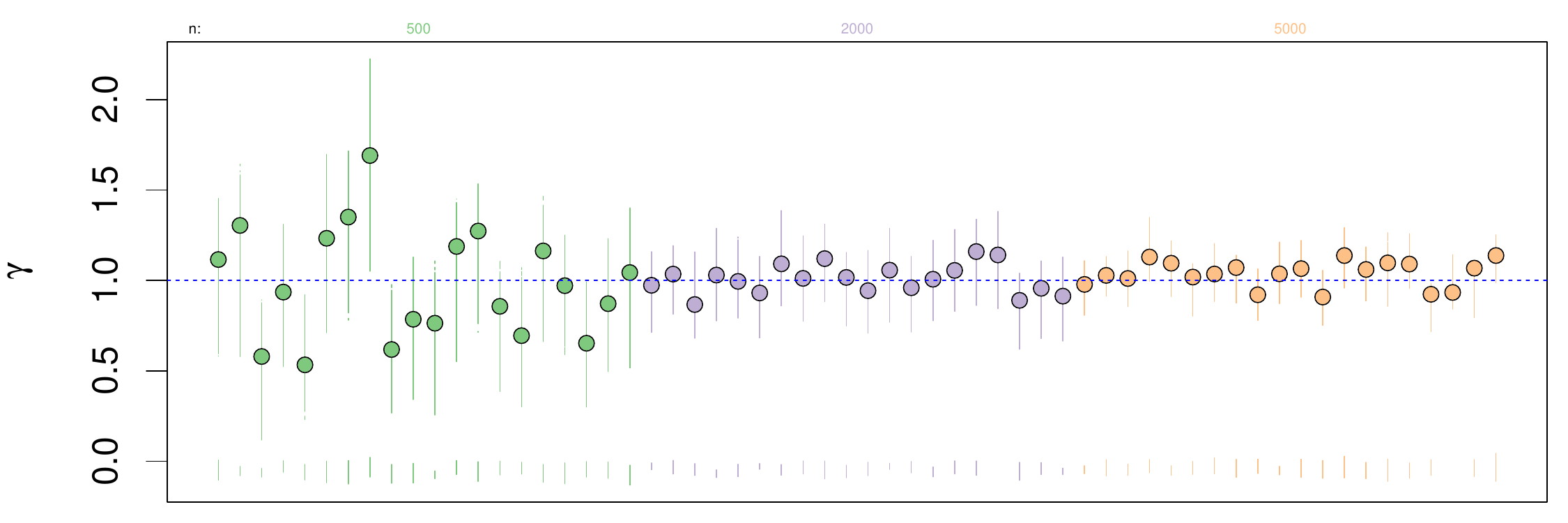}&
\includegraphics[scale=0.20]{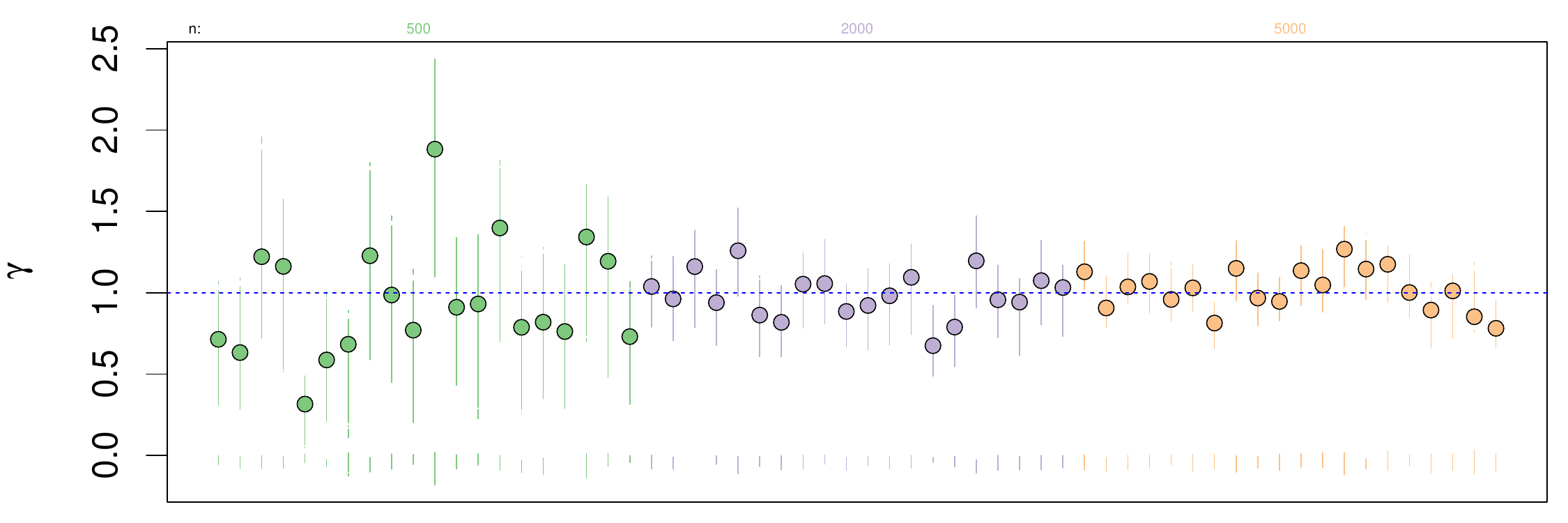}\\

\hspace{-20ex}\includegraphics[scale=0.20]{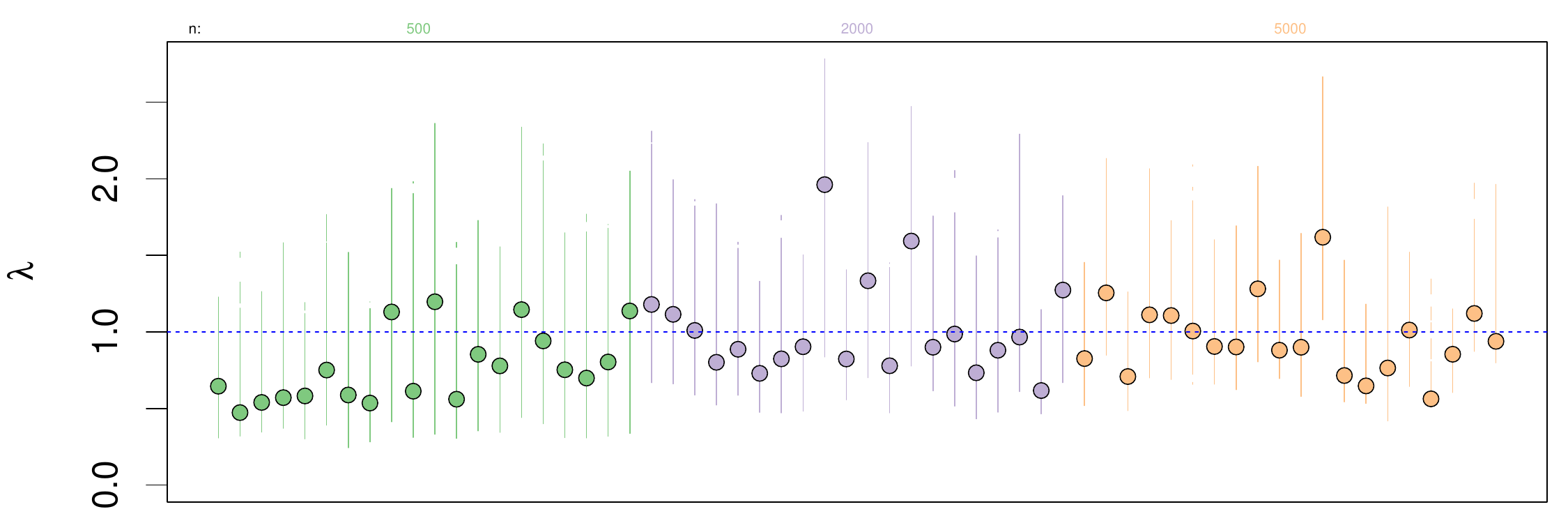}&
\includegraphics[scale=0.20]{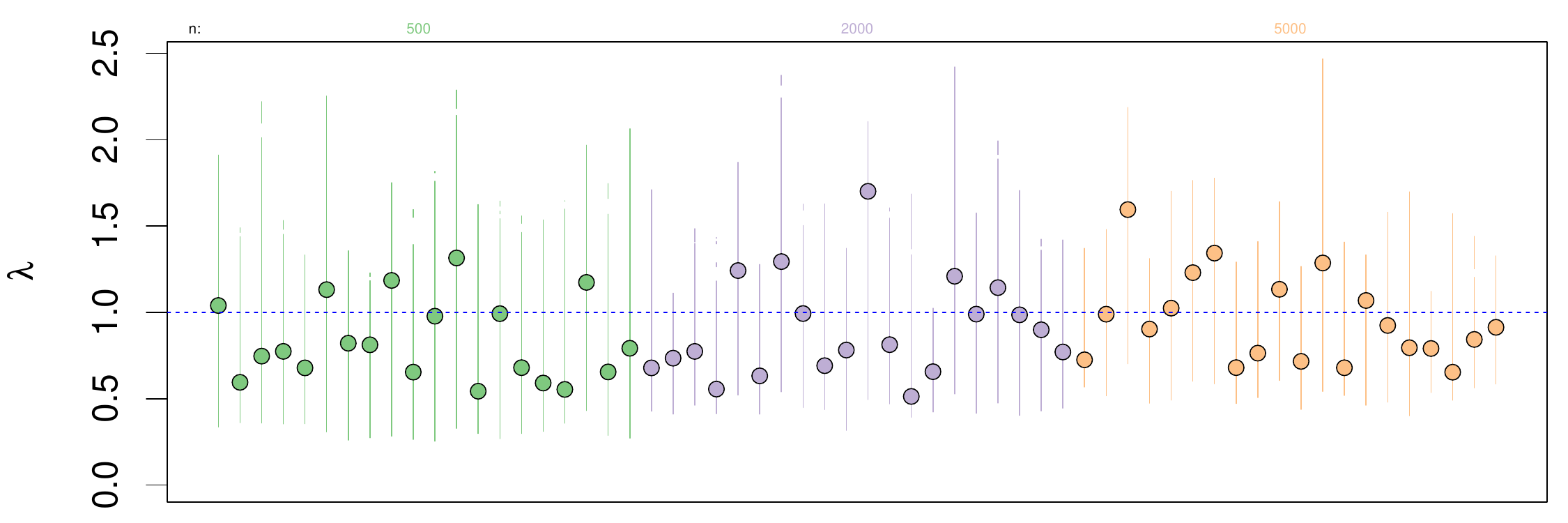}\\

\hspace{-20ex}\includegraphics[scale=0.20]{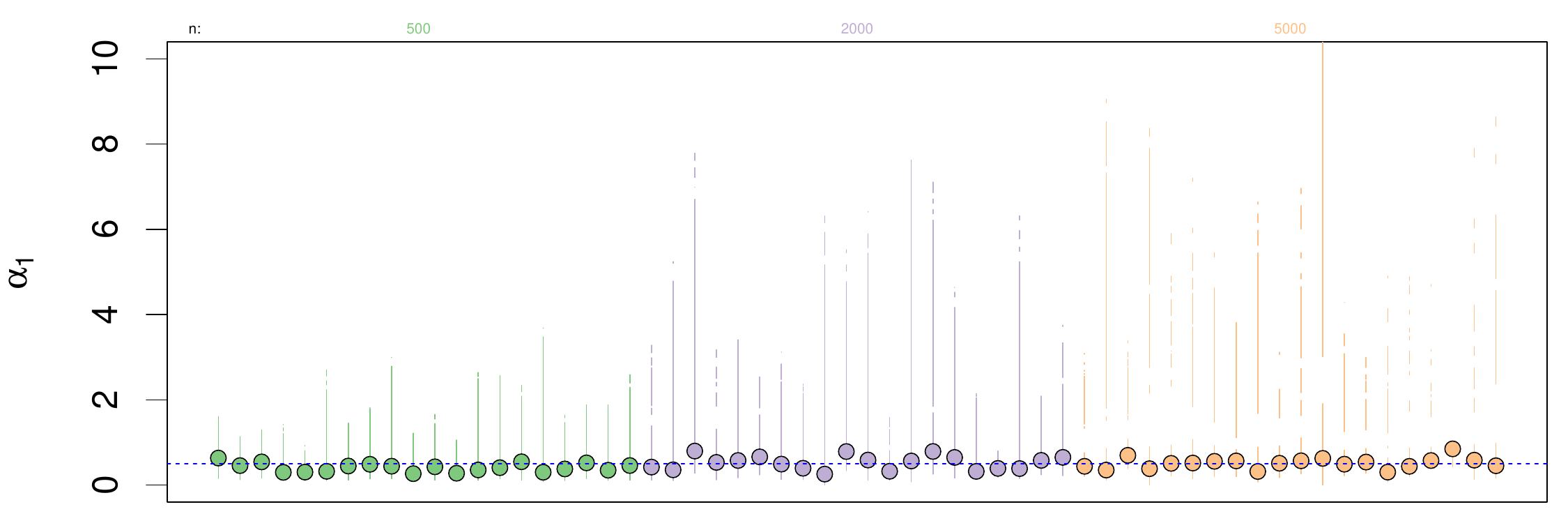}&
\includegraphics[scale=0.20]{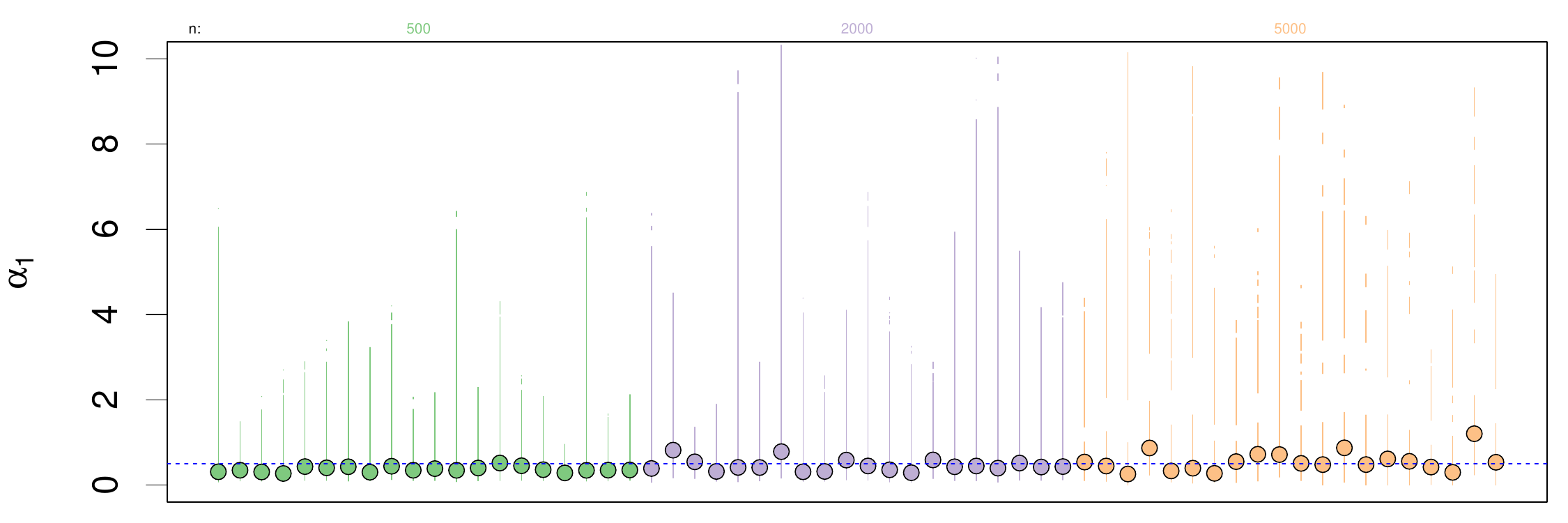}\\

\hspace{-20ex}\includegraphics[scale=0.20]{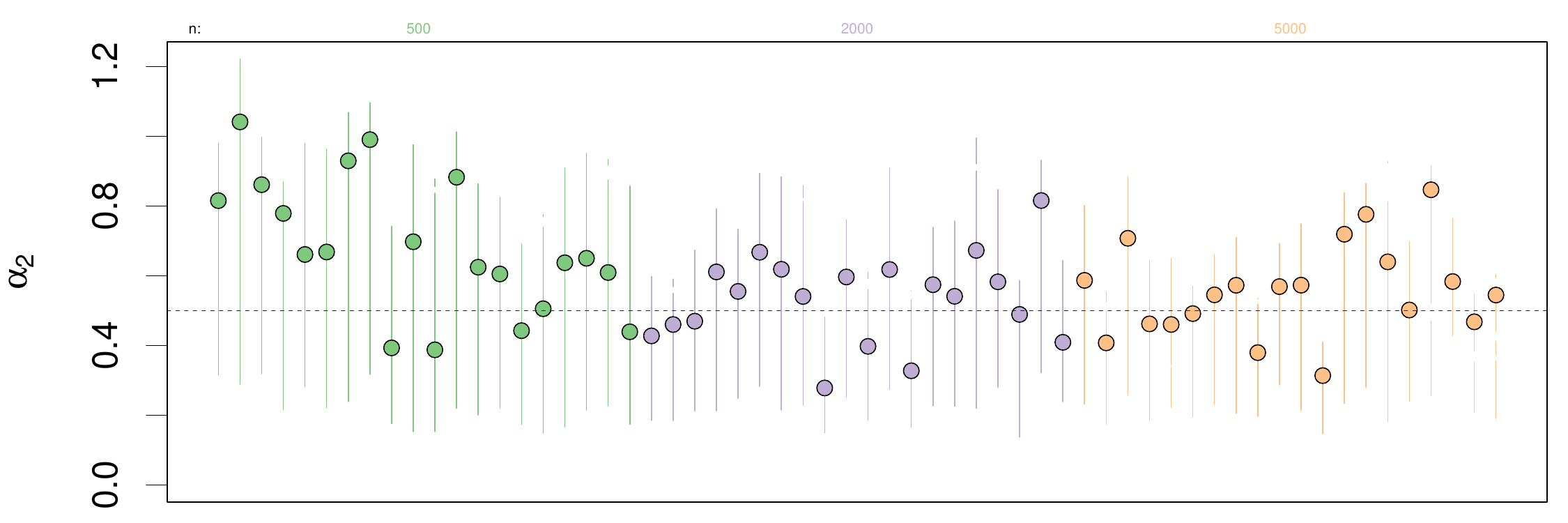}&
\includegraphics[scale=0.20]{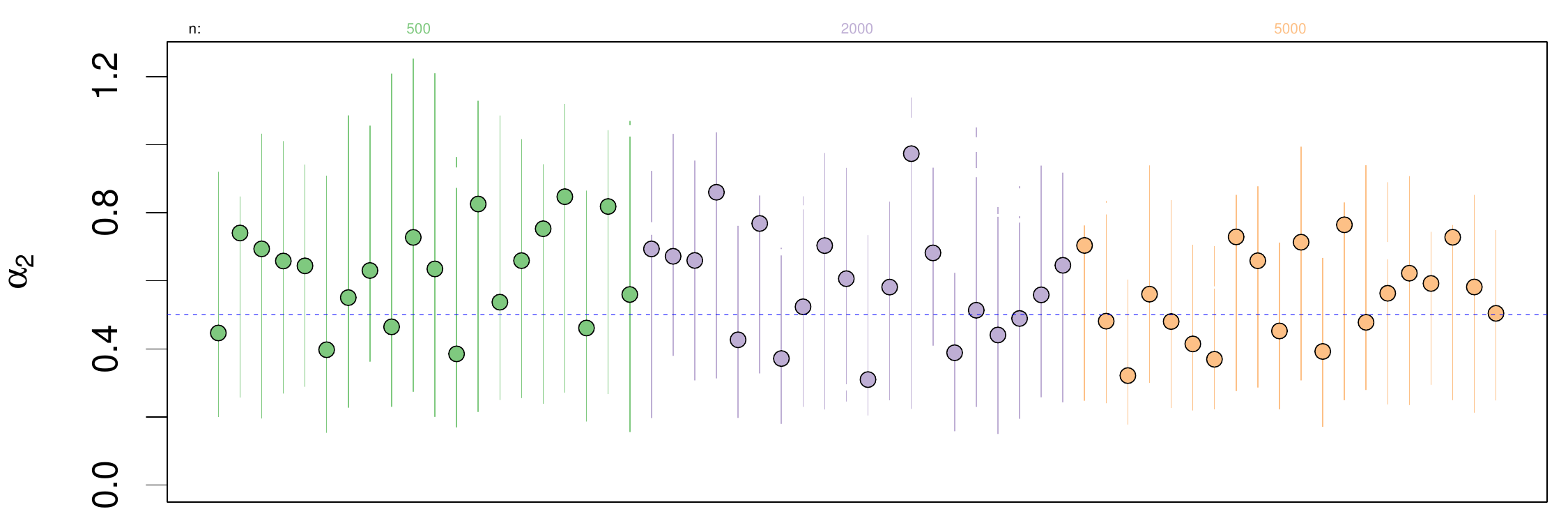}\\

\hspace{-20ex}\includegraphics[scale=0.20]{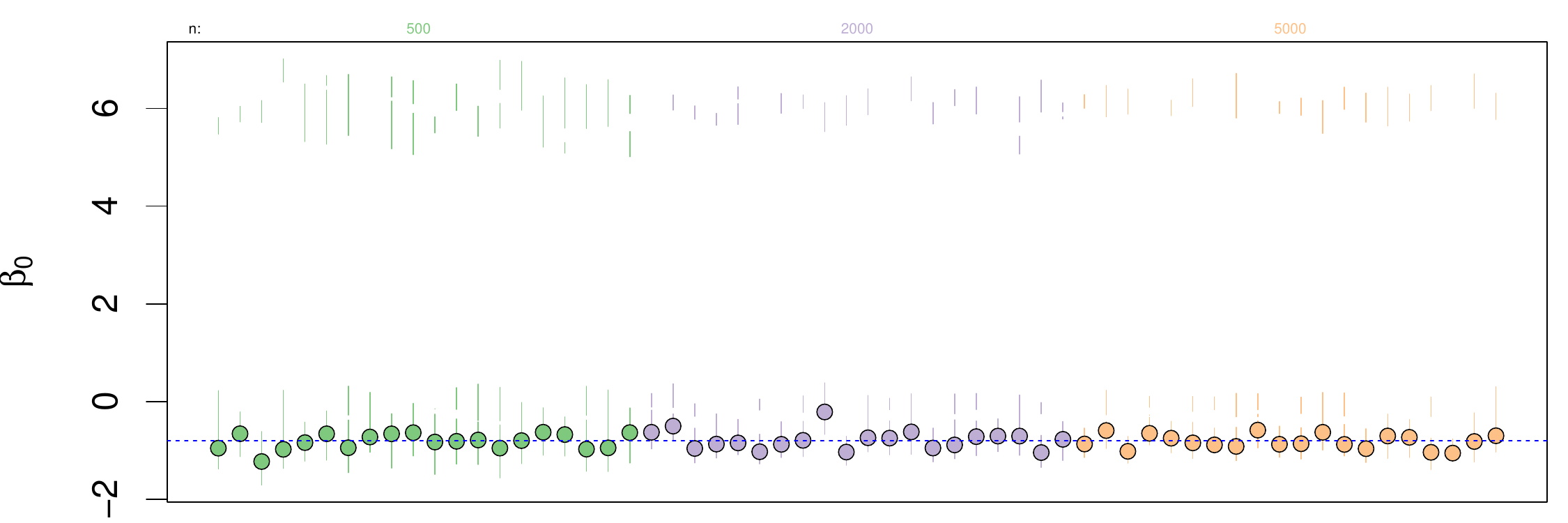}&
\includegraphics[scale=0.20]{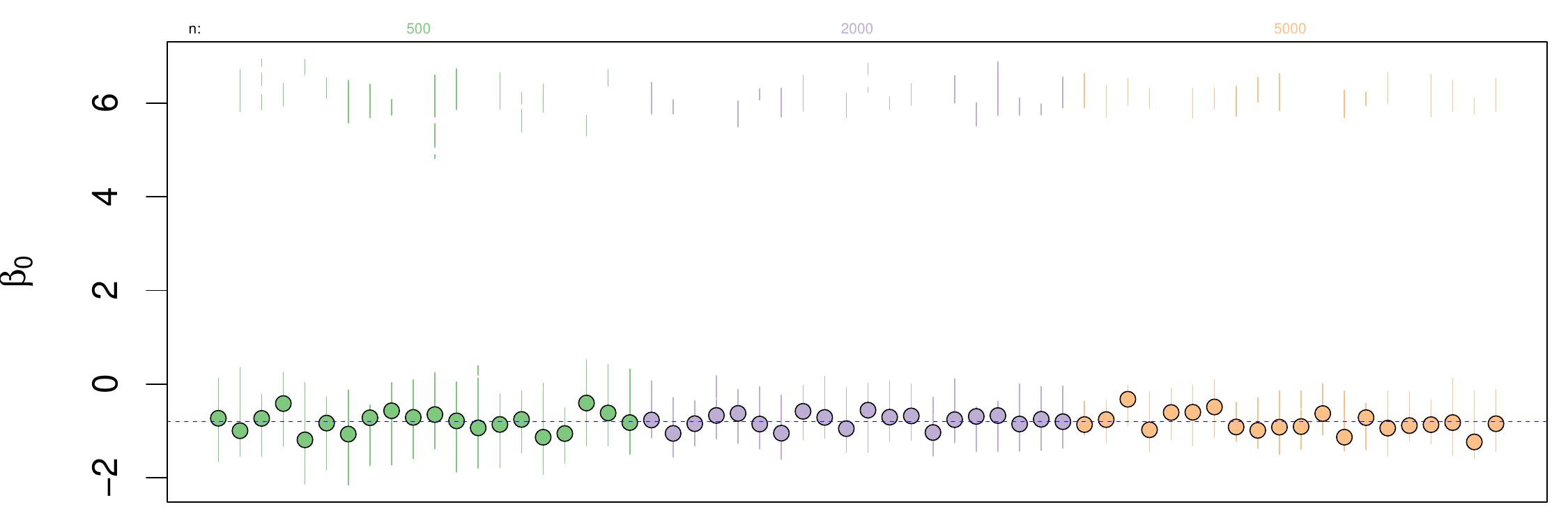}\\

\hspace{-20ex}\includegraphics[scale=0.20]{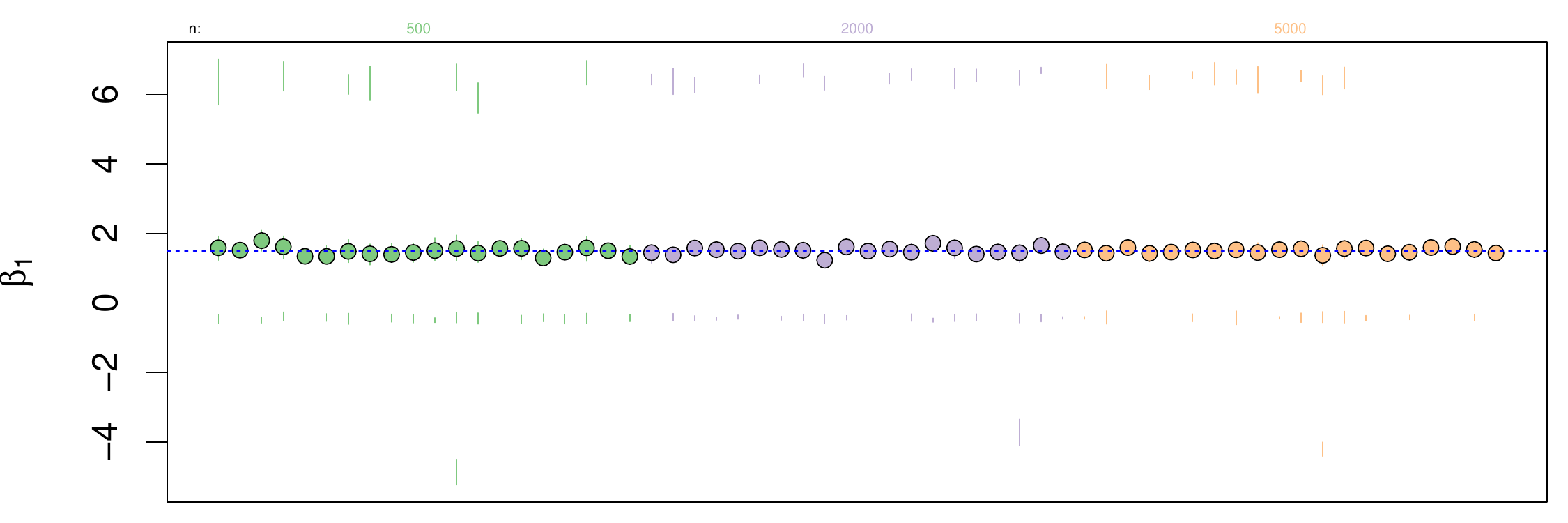}&
\includegraphics[scale=0.20]{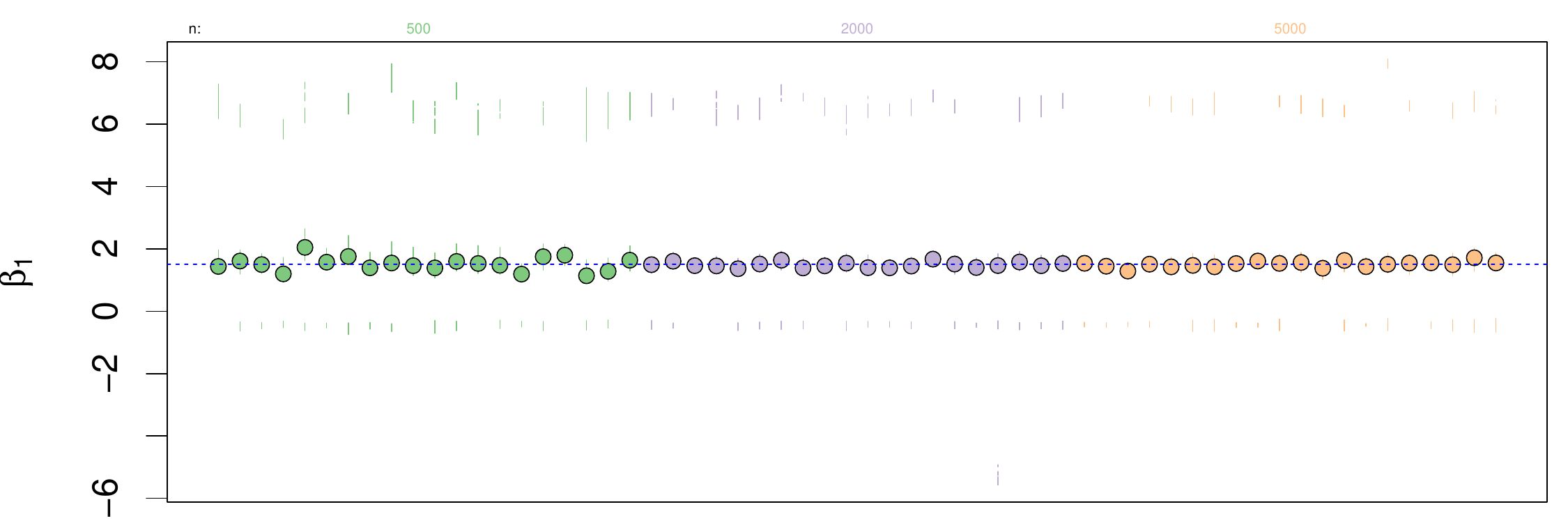}\\

\hspace{-20ex}\includegraphics[scale=0.20]{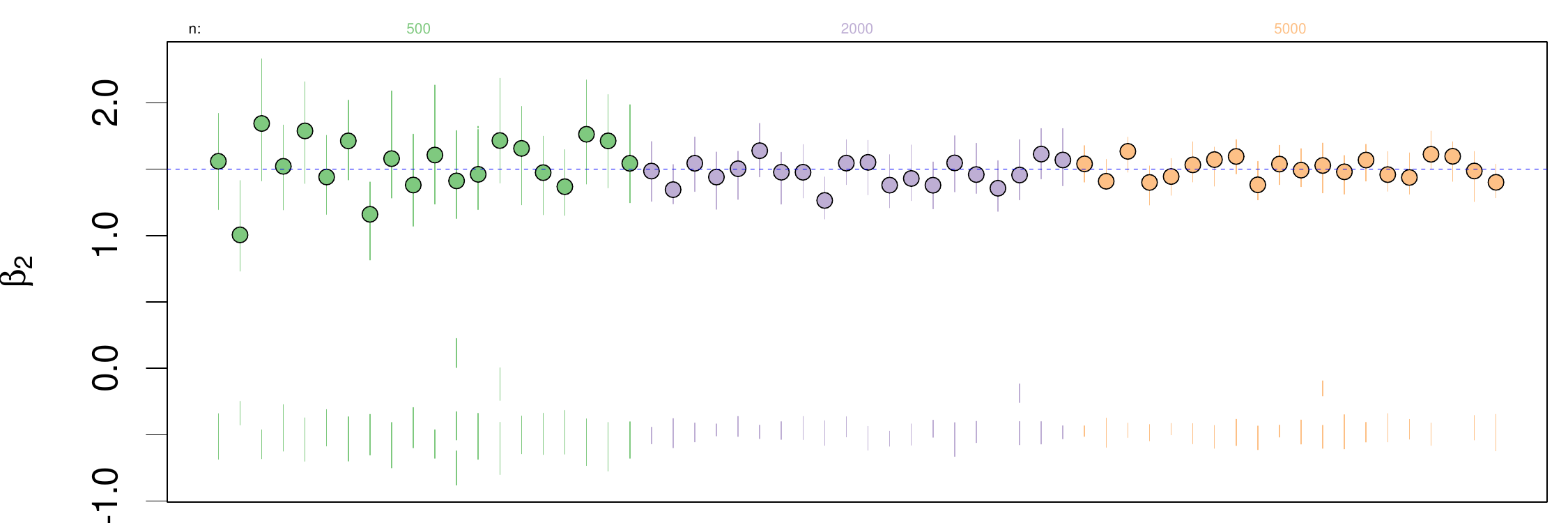}&
\includegraphics[scale=0.20]{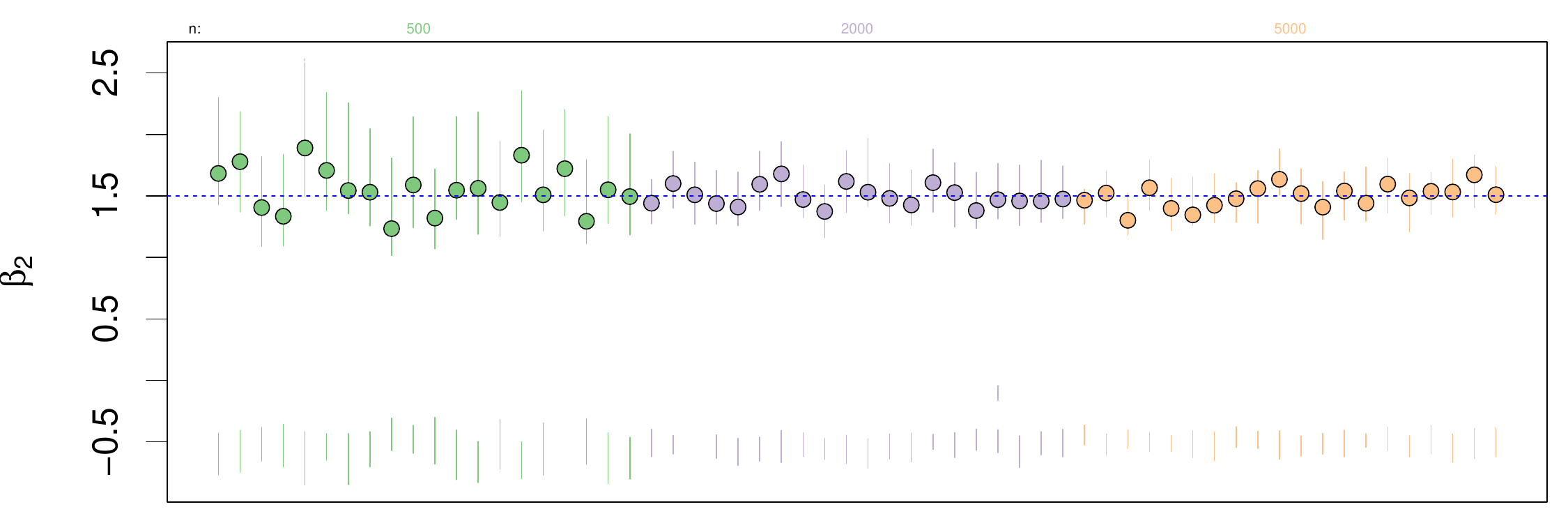}\\
scenario B.1 & scenario B.2
\end{tabular}
\caption{Point estimates (MAP) with $95\%$ Highest Density Intervals for our simulated datasets. Different colour indicate levels of sample size: 500 (\textcolor{green}{---}), 2000 (\textcolor{violet}{---}), 5000 (\textcolor{orange}{---}). The horizontal line indicates the true value (see Scenarios B1 and B2 in Table \ref{t4b}).} 
\label{fig:hdisB}
\end{figure}

\begin{figure}[p]
\centering
\begin{tabular}{cc}
\hspace{-20ex}\includegraphics[scale=0.20]{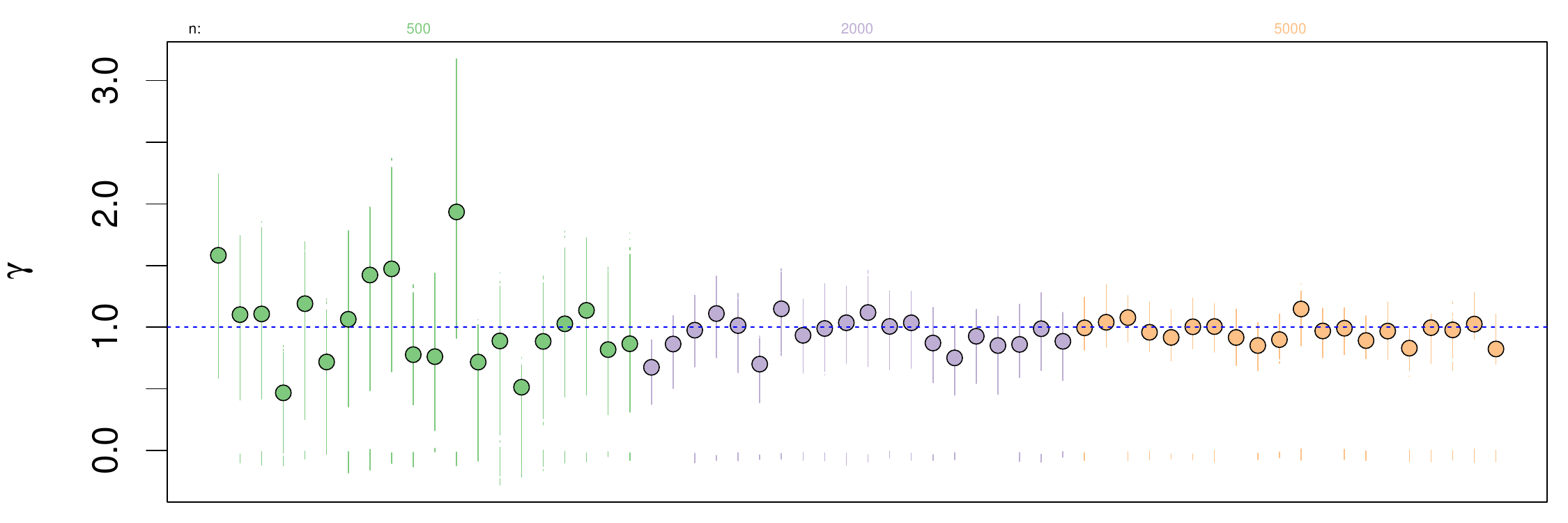}&
\includegraphics[scale=0.20]{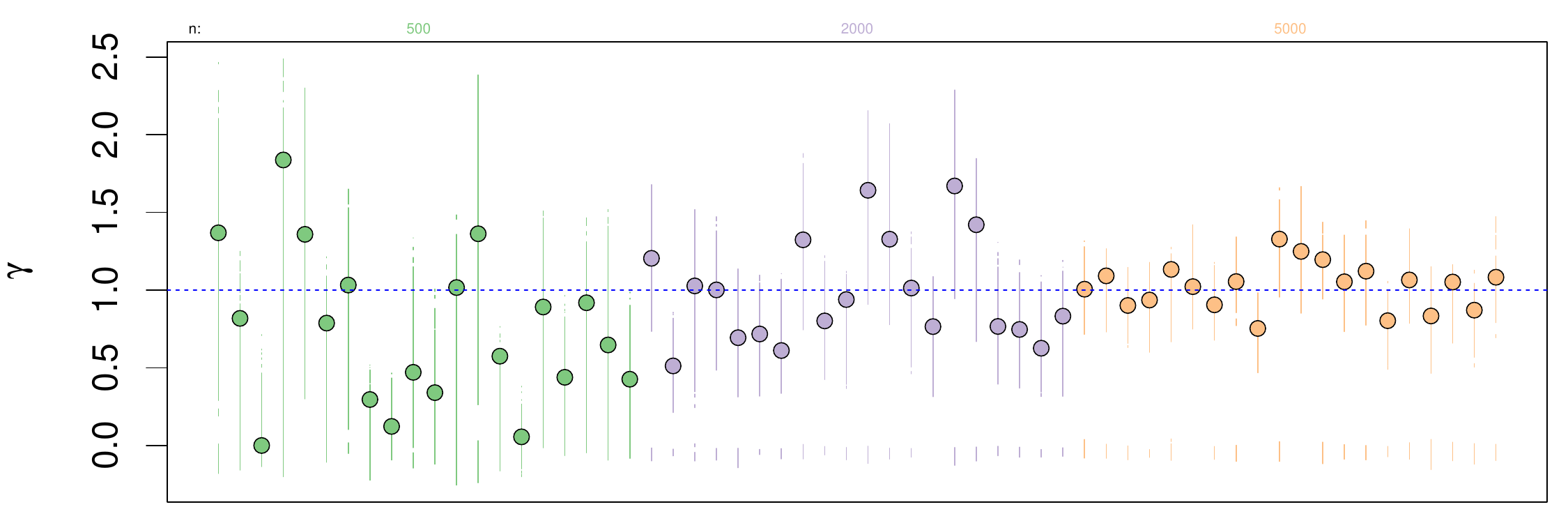}
\\

\hspace{-20ex}\includegraphics[scale=0.20]{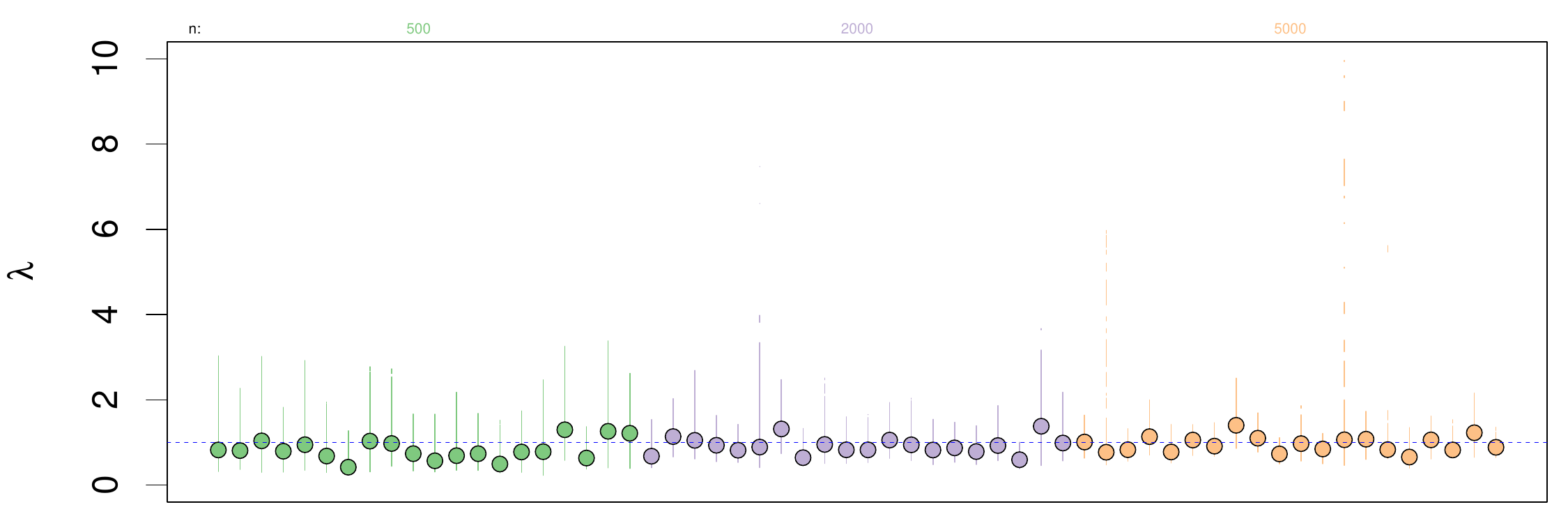}&
\includegraphics[scale=0.20]{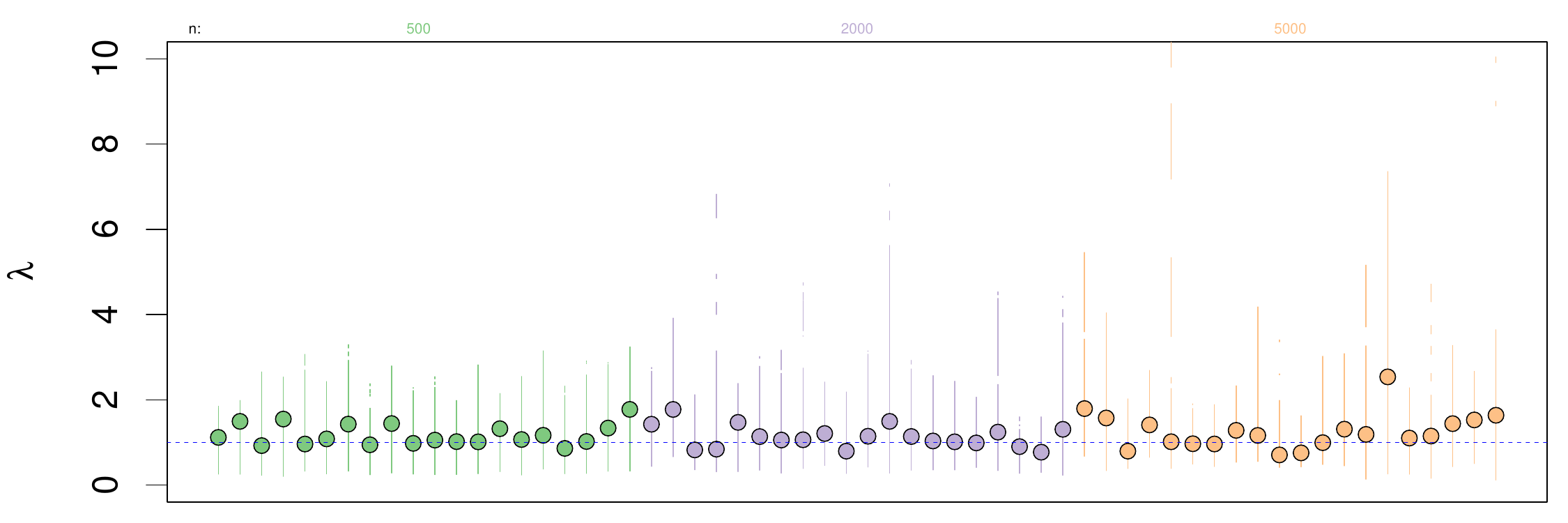}
\\

\hspace{-20ex}\includegraphics[scale=0.20]{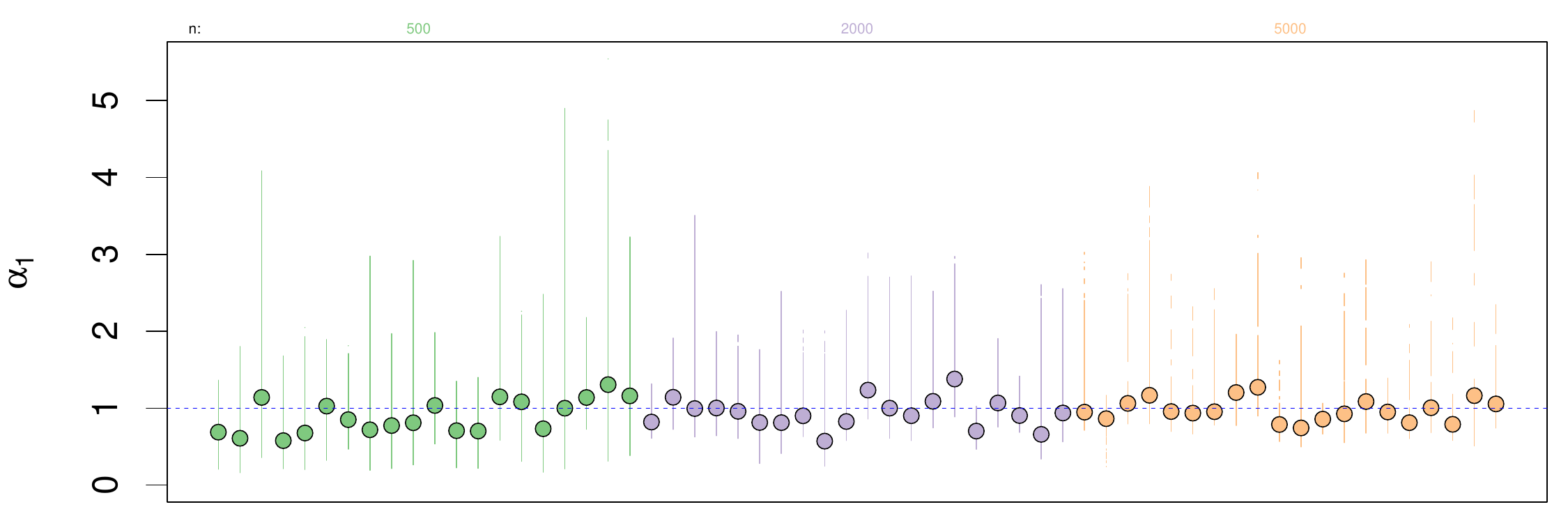}&
\includegraphics[scale=0.20]{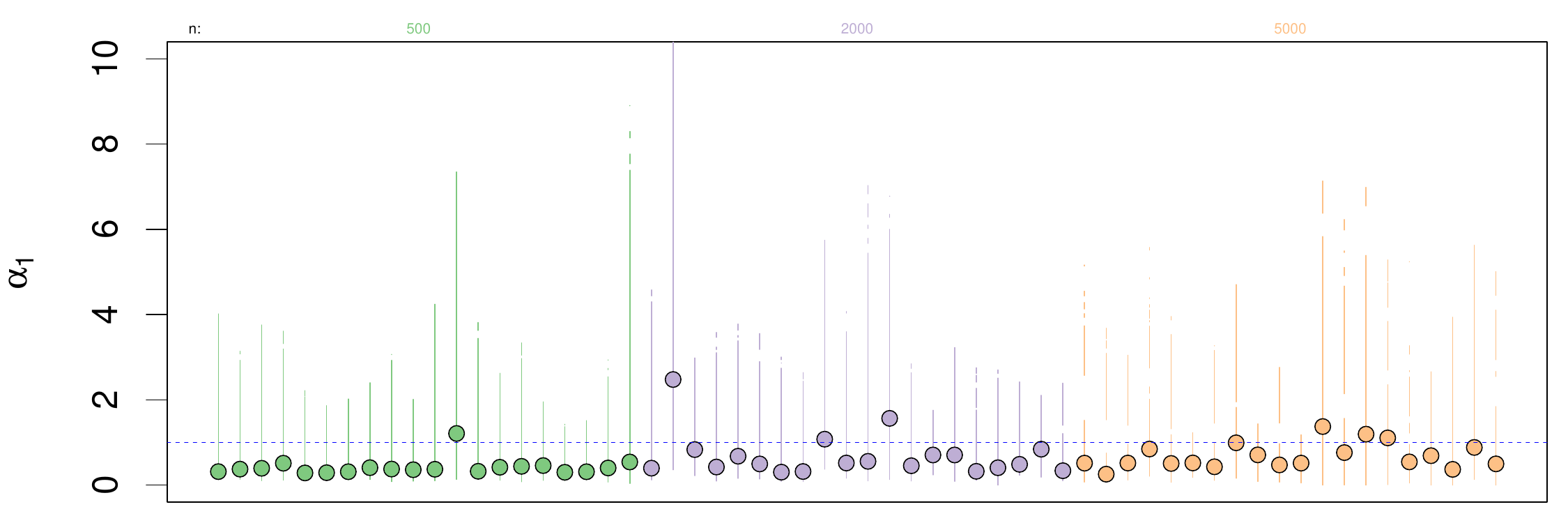}
\\

\hspace{-20ex}\includegraphics[scale=0.20]{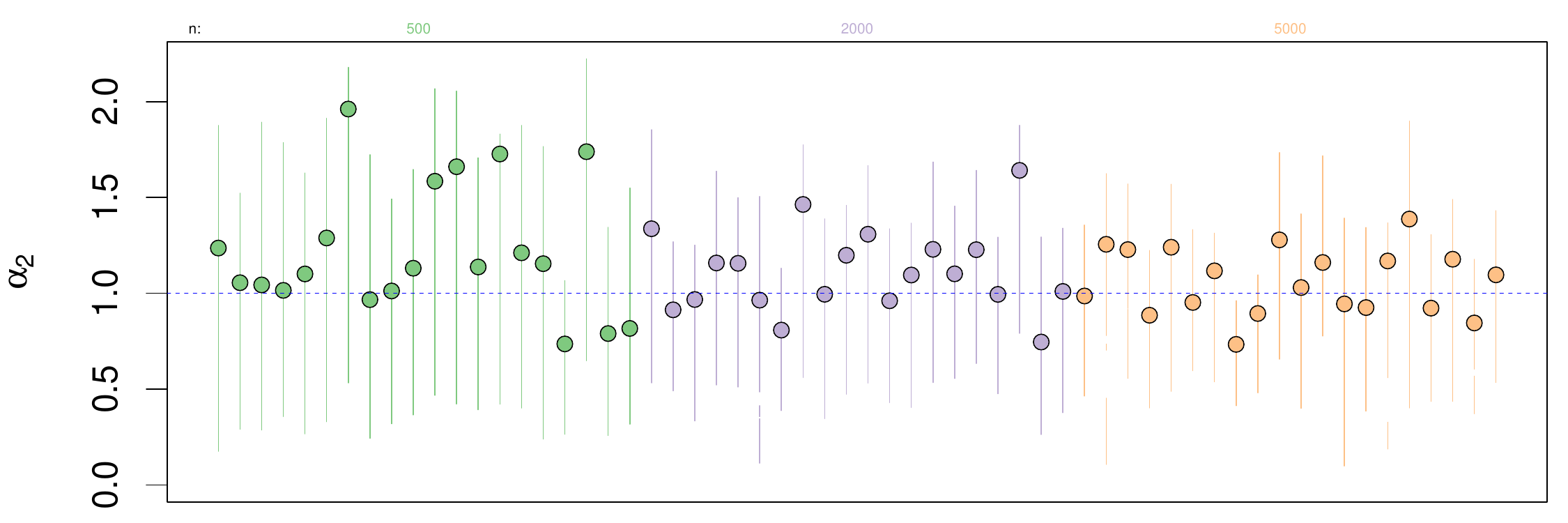}&
\includegraphics[scale=0.20]{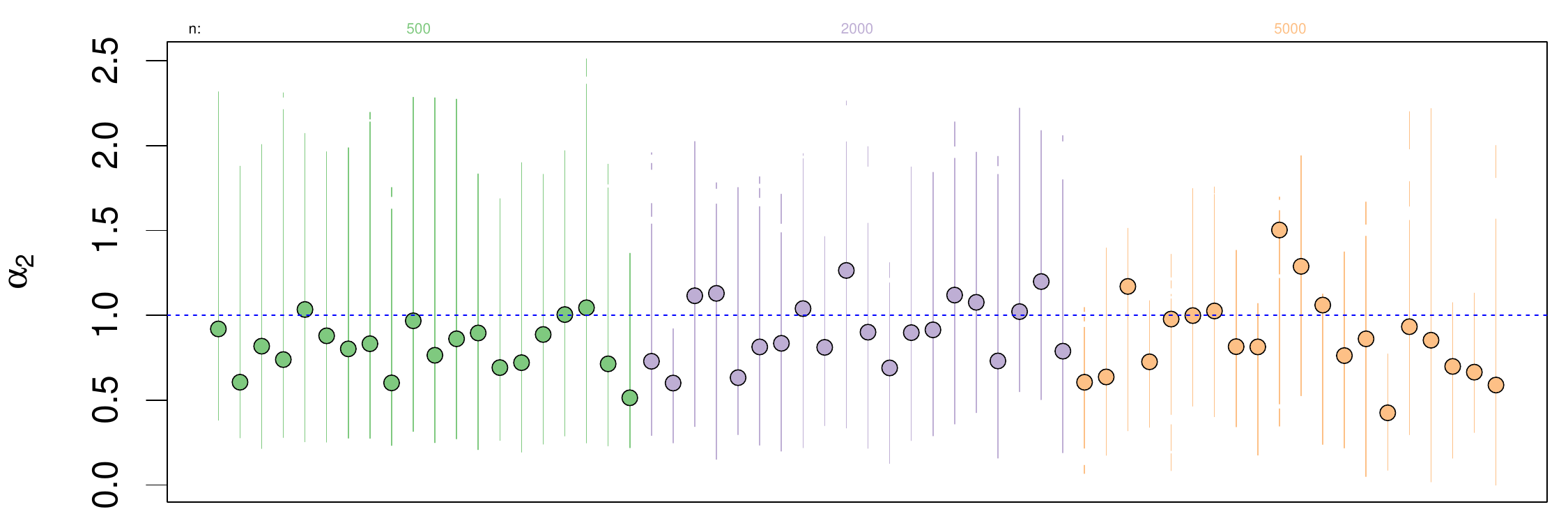}
\\

\hspace{-20ex}\includegraphics[scale=0.20]{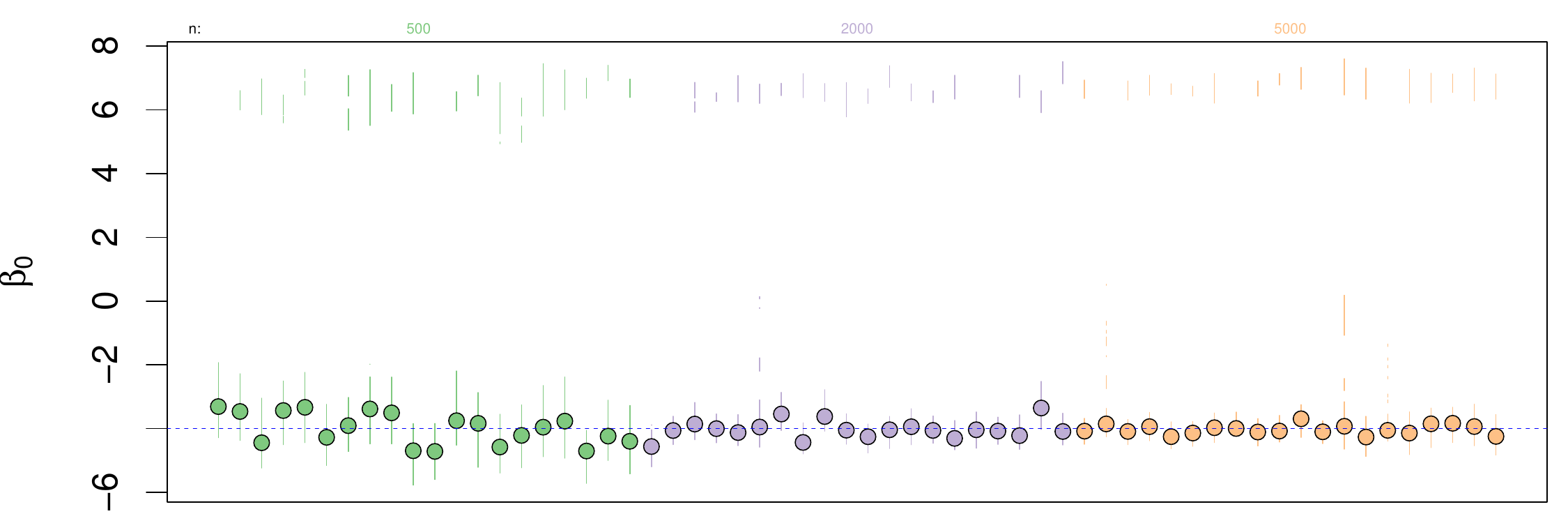}&
\includegraphics[scale=0.20]{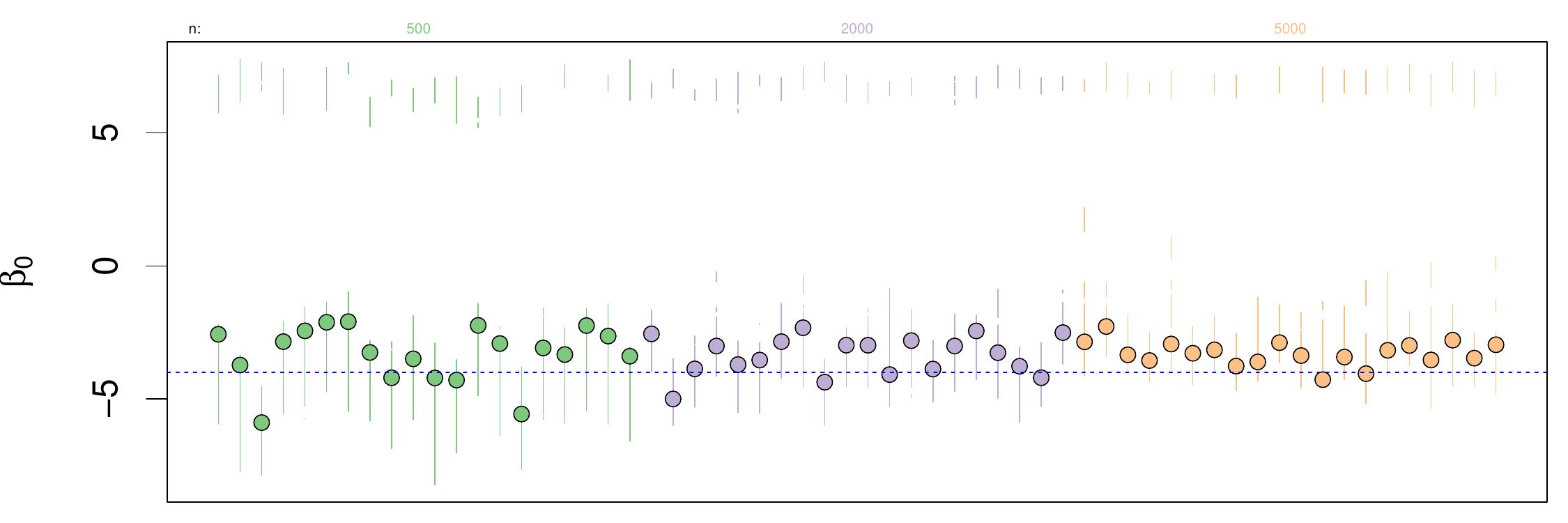}
\\

\hspace{-20ex}\includegraphics[scale=0.20]{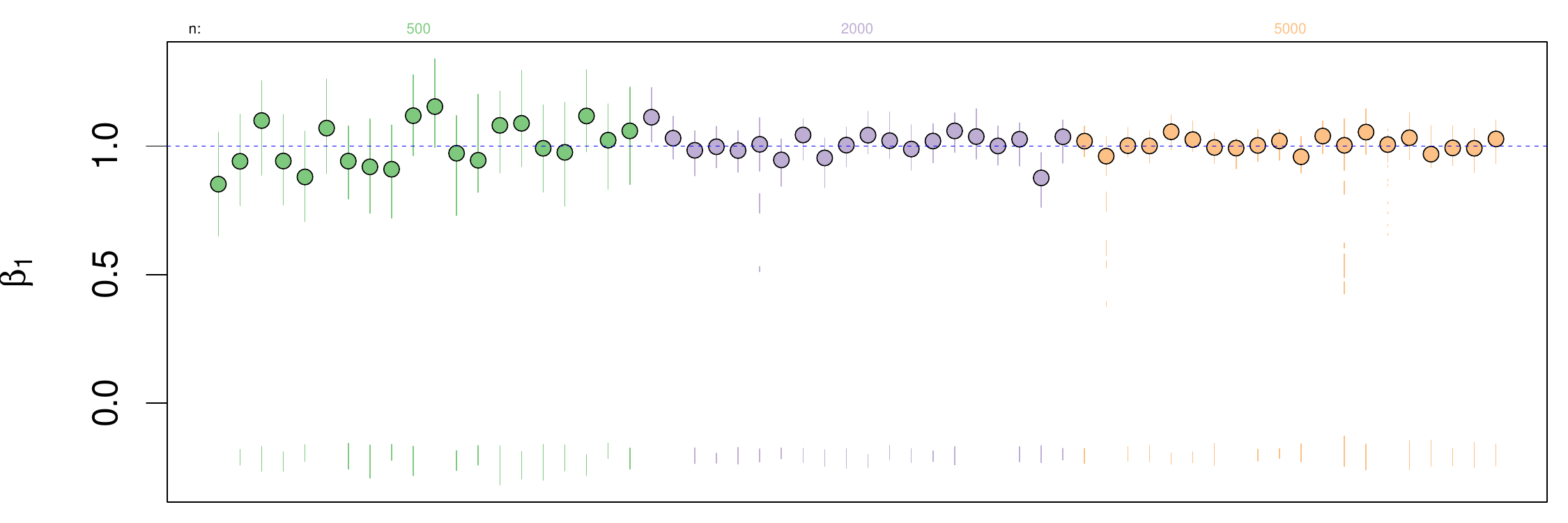}&
\includegraphics[scale=0.20]{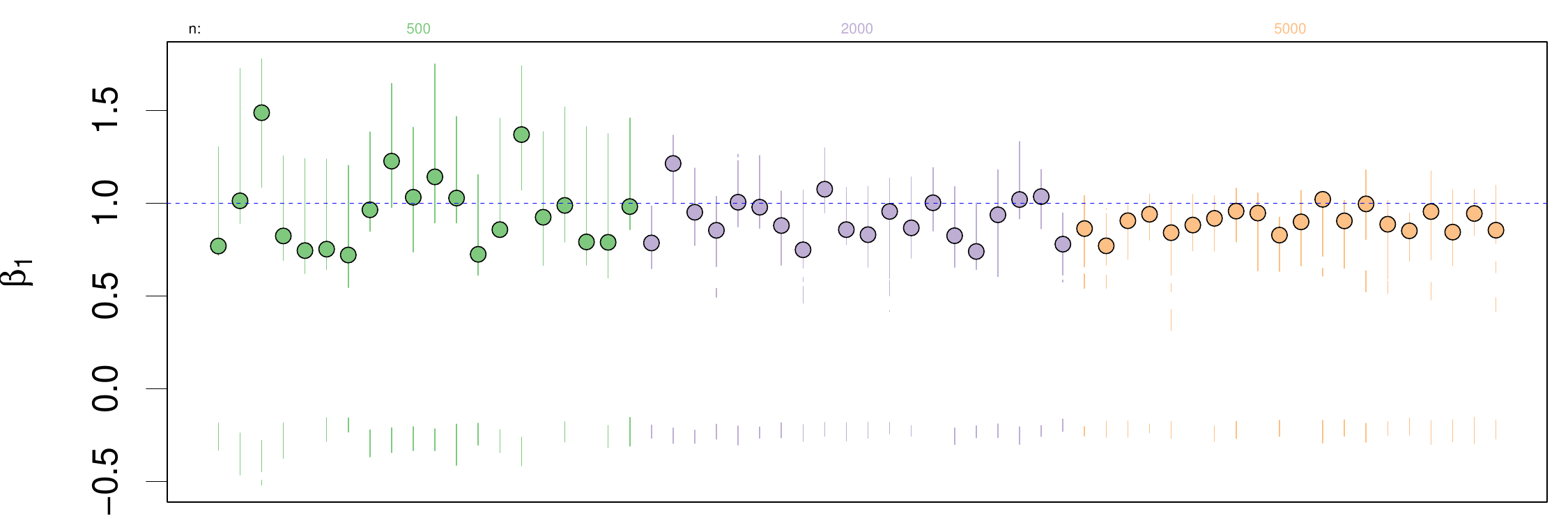}
\\

\hspace{-20ex}\includegraphics[scale=0.20]{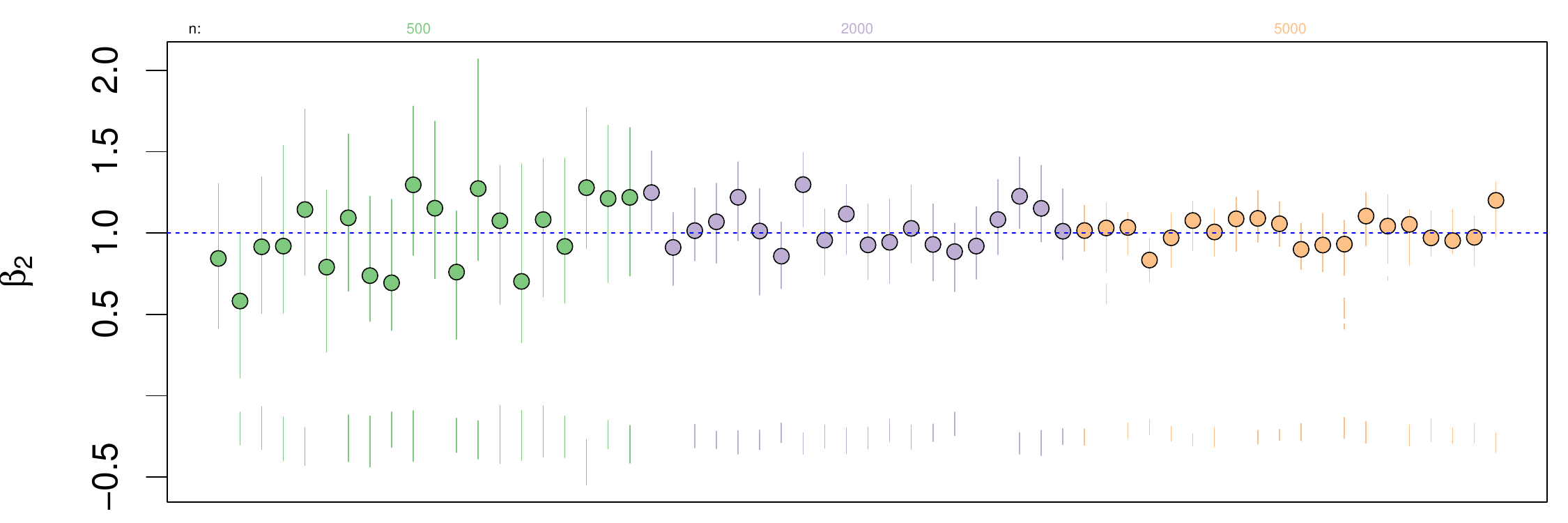}&
\includegraphics[scale=0.20]{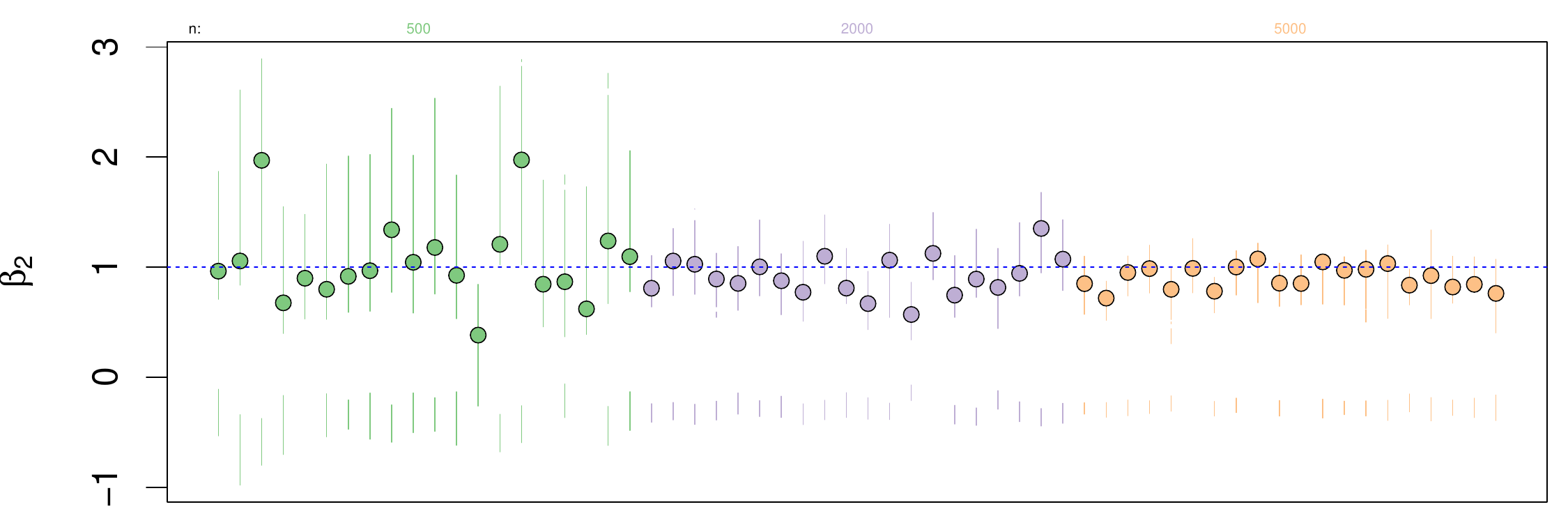}
\\
scenario C.1 & scenario C.2
\end{tabular}
\caption{Point estimates (MAP) with $95\%$ Highest Density Intervals for our simulated datasets. Different colour indicate levels of sample size: 500 (\textcolor{green}{---}), 2000 (\textcolor{violet}{---}), 5000 (\textcolor{orange}{---}). The horizontal line indicates the true value (see Scenarios C1 and C2 in Table \ref{t4b}).} 
\label{fig:hdisC}
\end{figure}

\begin{figure}[p]
\centering
\begin{tabular}{cc}
\hspace{-20ex}\includegraphics[scale=0.20]{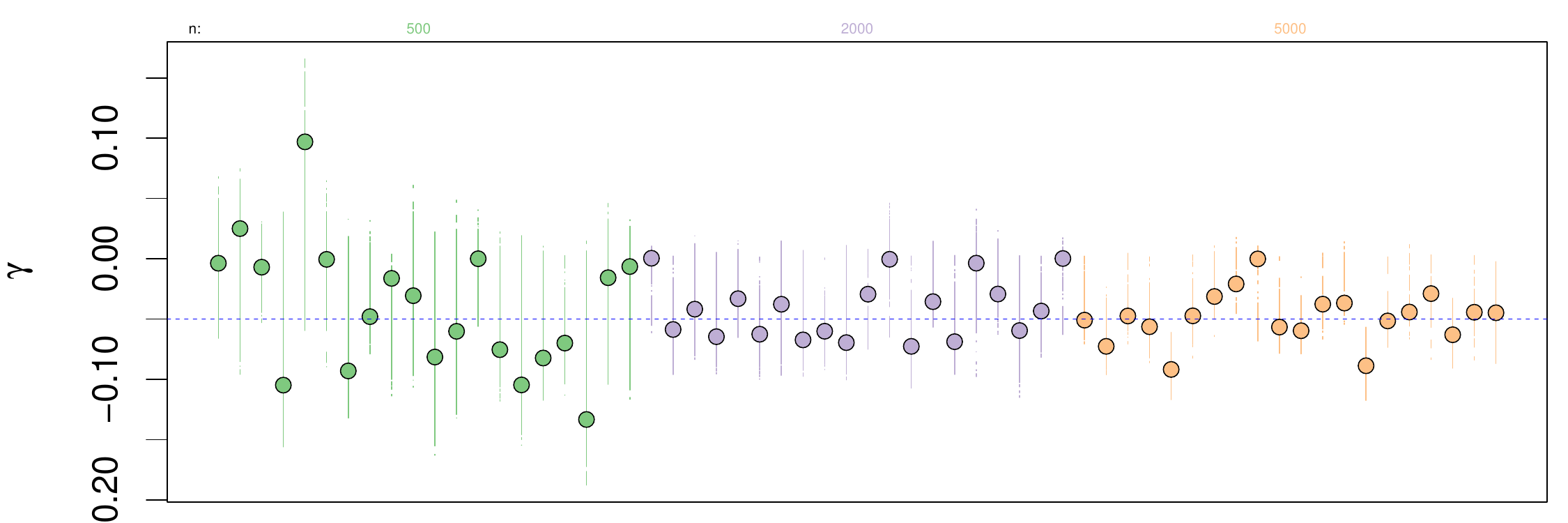}&
\includegraphics[scale=0.20]{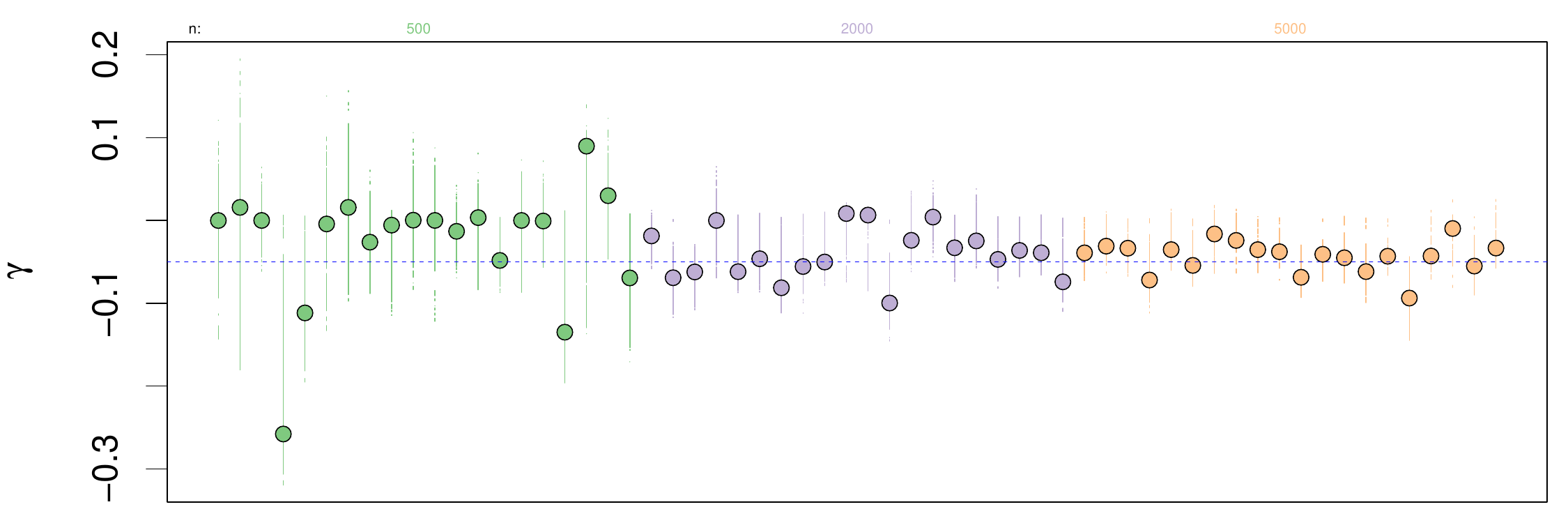}
\\

\hspace{-20ex}\includegraphics[scale=0.20]{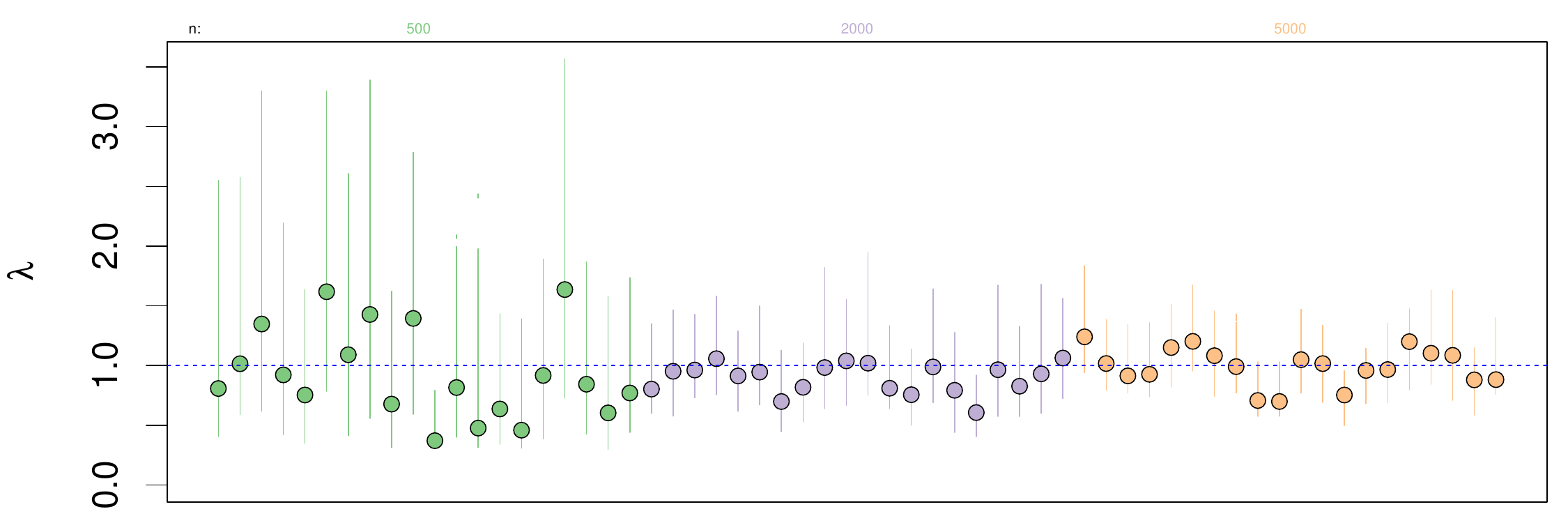}&
\includegraphics[scale=0.20]{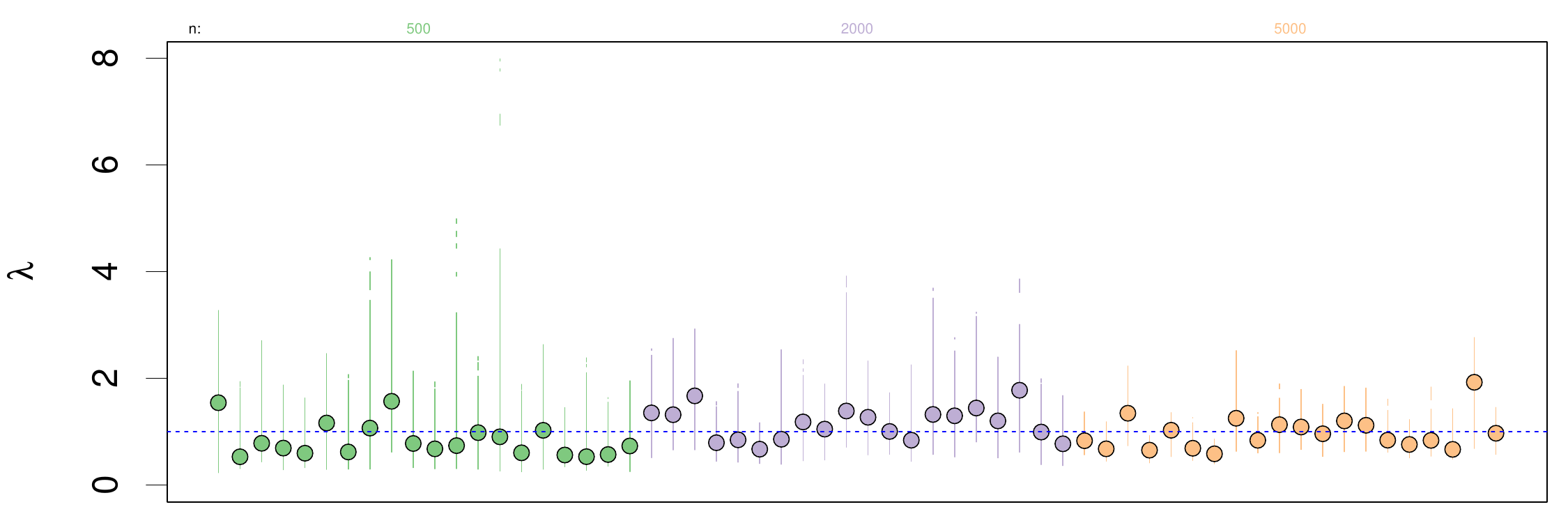}
\\

\hspace{-20ex}\includegraphics[scale=0.20]{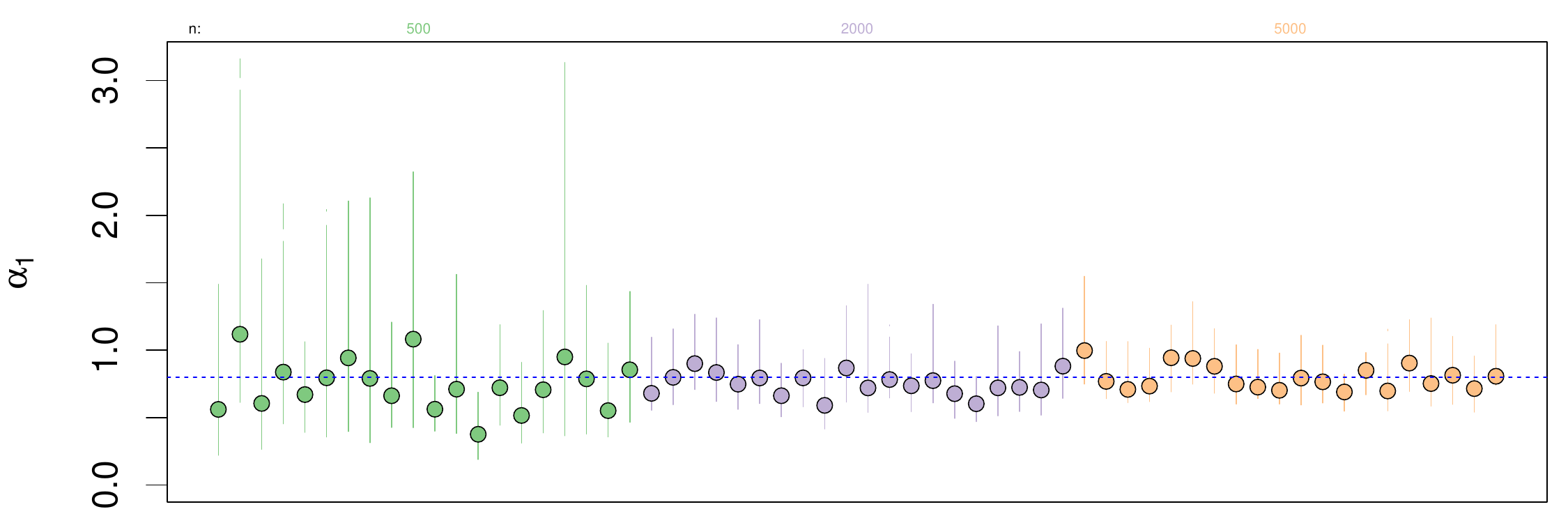}&
\includegraphics[scale=0.20]{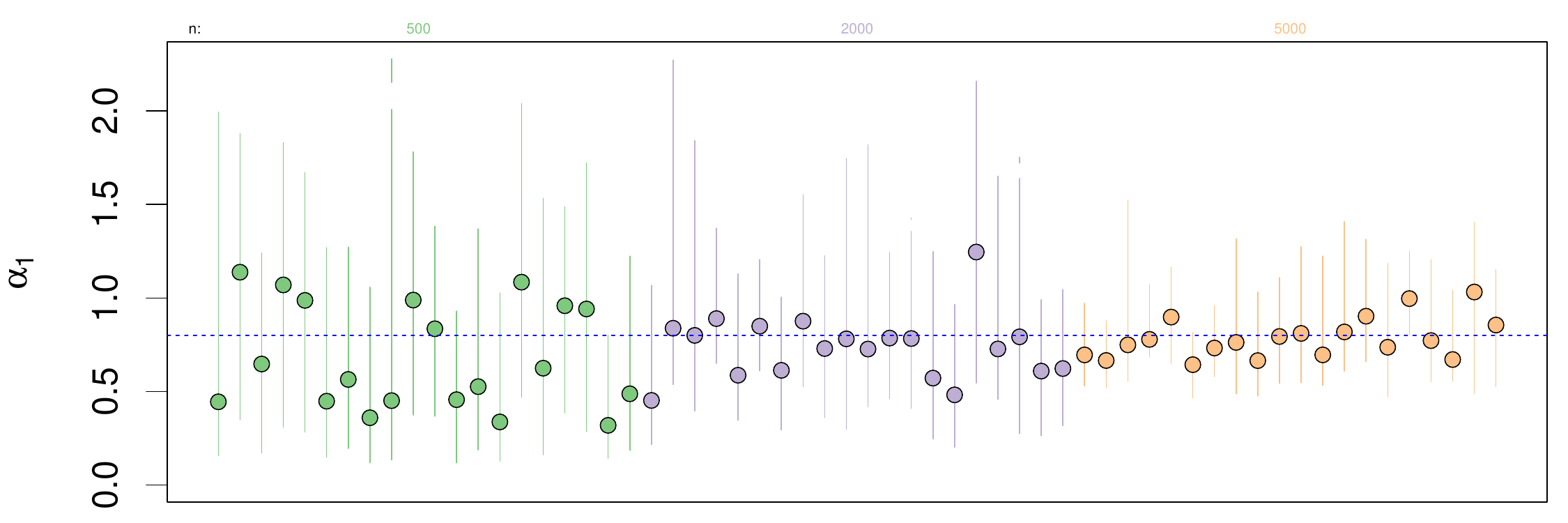}
\\

\hspace{-20ex}\includegraphics[scale=0.20]{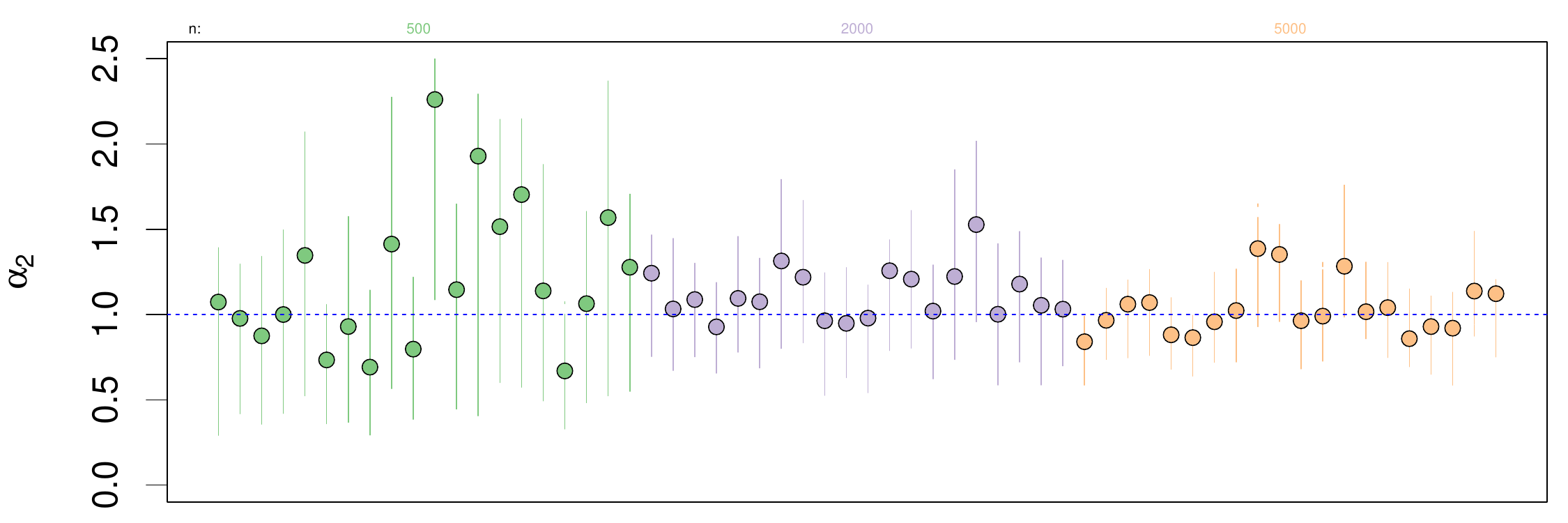}&
\includegraphics[scale=0.20]{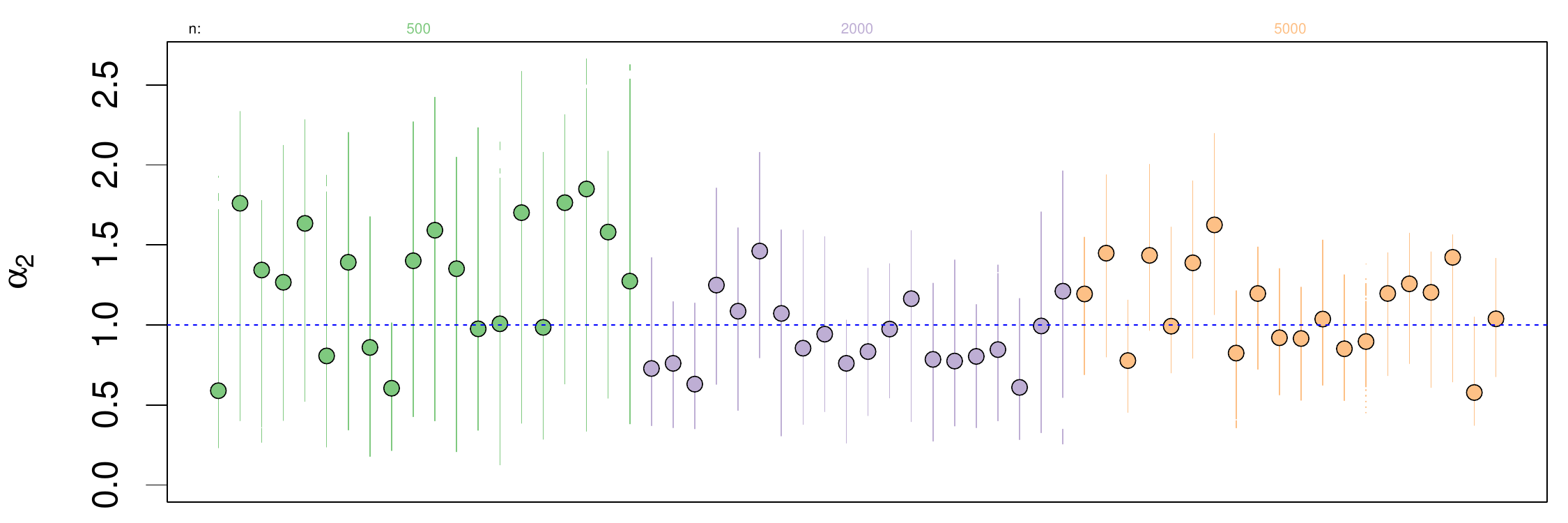}
\\

\hspace{-20ex}\includegraphics[scale=0.20]{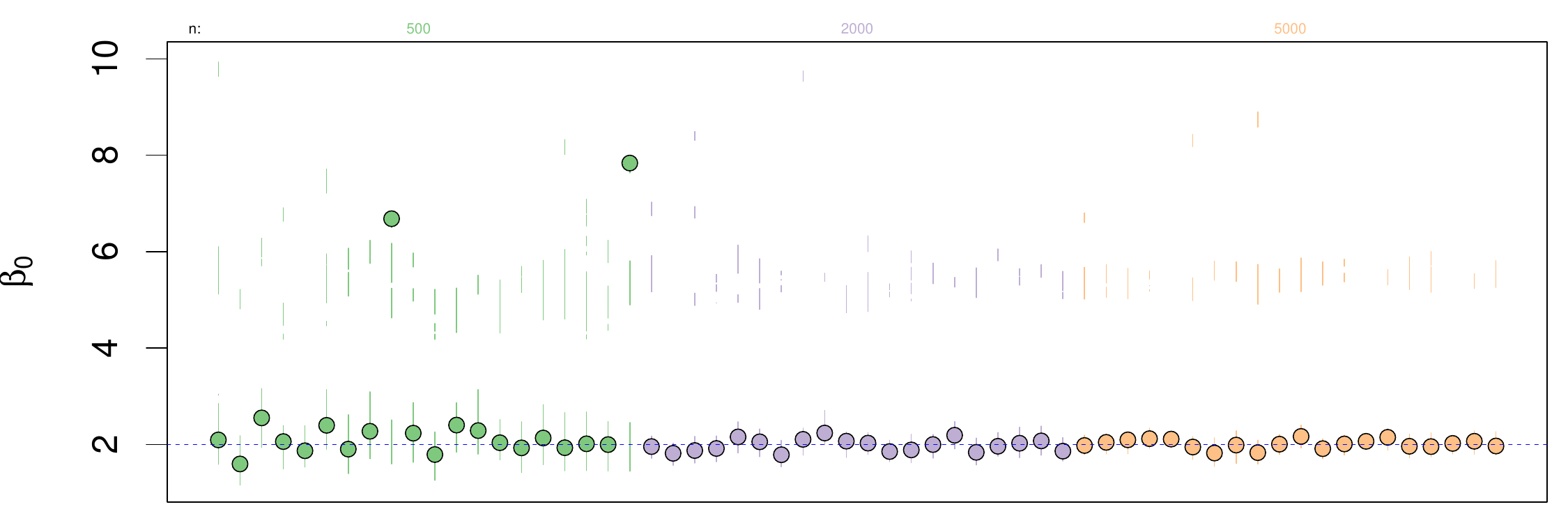}&
\includegraphics[scale=0.20]{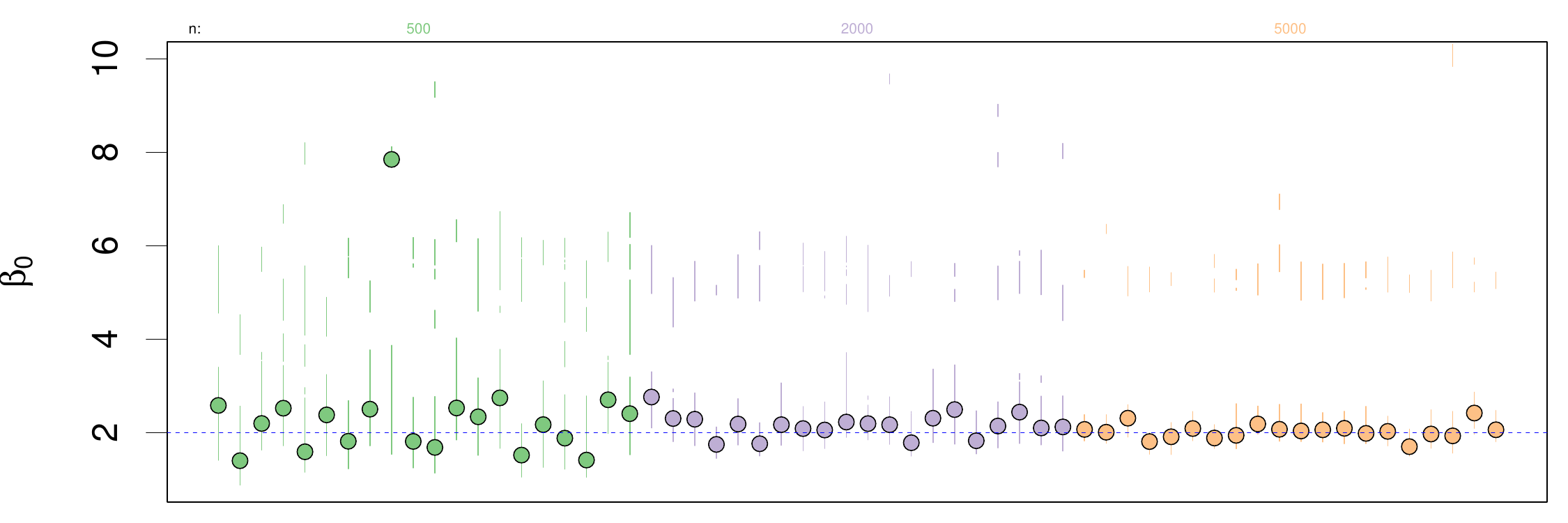}
\\

\hspace{-20ex}\includegraphics[scale=0.20]{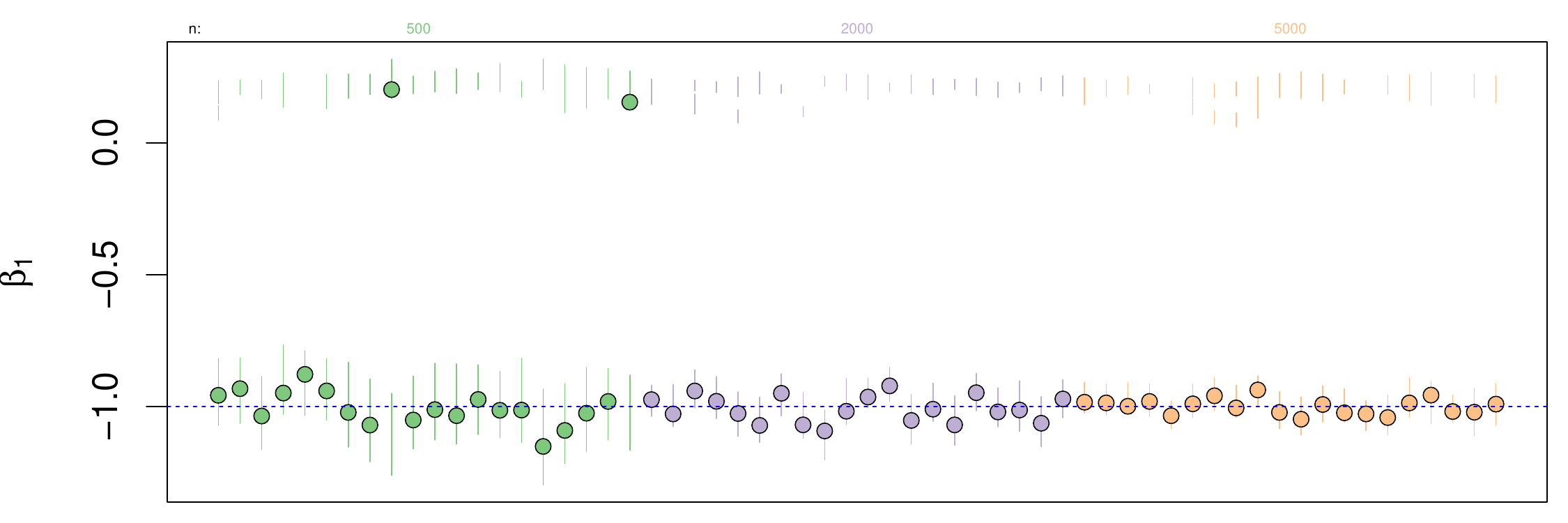}&
\includegraphics[scale=0.20]{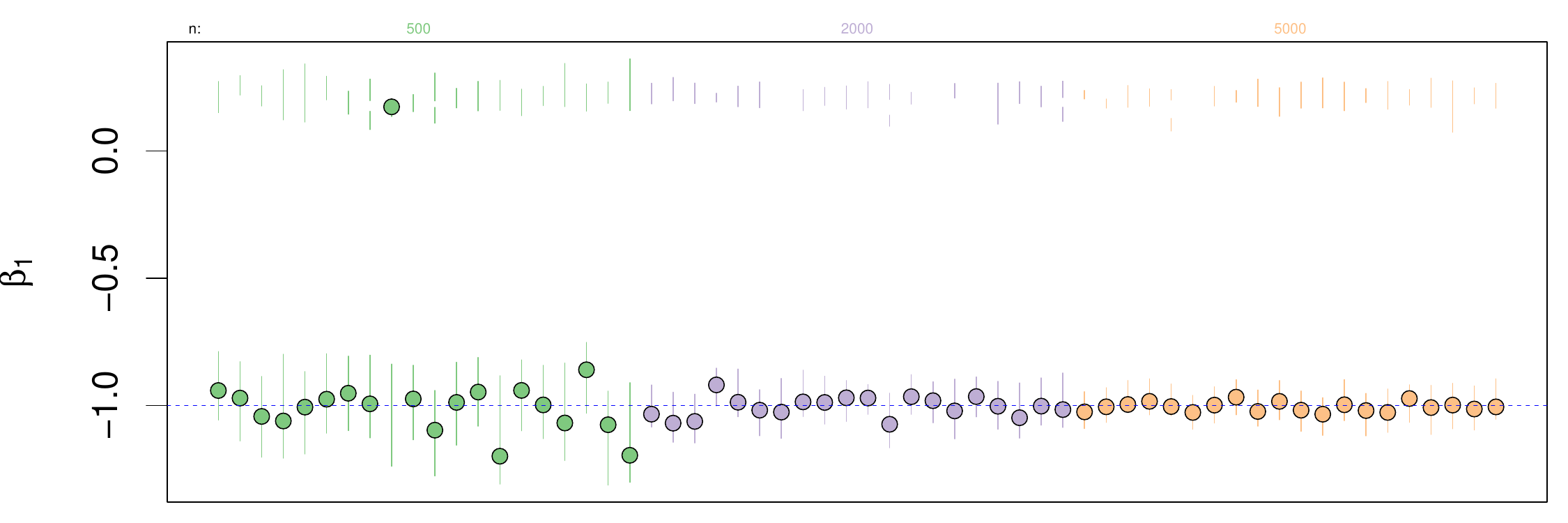}
\\

\hspace{-20ex}\includegraphics[scale=0.20]{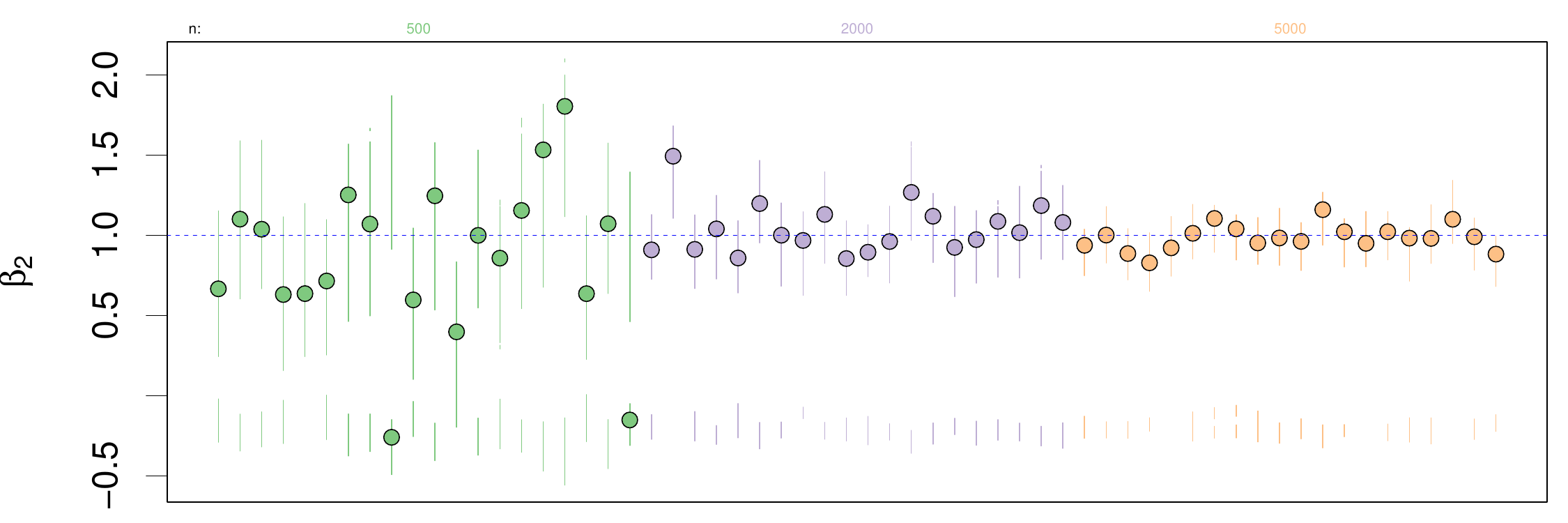}&
\includegraphics[scale=0.20]{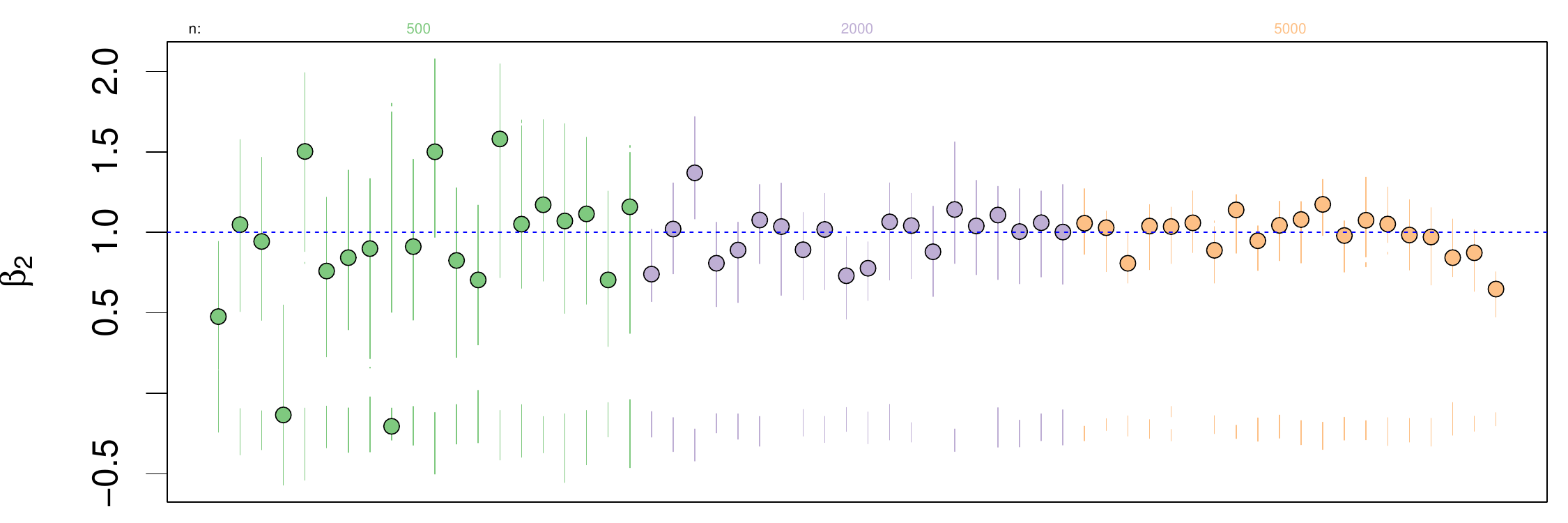}
\\
scenario D.1 & scenario D.2
\end{tabular}
\caption{Point estimates (MAP) with $95\%$ Highest Density Intervals for our simulated datasets. Different colour indicate levels of sample size: 500 (\textcolor{green}{---}), 2000 (\textcolor{violet}{---}), 5000 (\textcolor{orange}{---}). The horizontal line indicates the true value (see Scenarios D1 and D2 in Table \ref{t4b}).} 
\label{fig:hdisD}
\end{figure}

\begin{figure}[p]
\centering
\begin{tabular}{cc}
\hspace{-20ex}\includegraphics[scale=0.20]{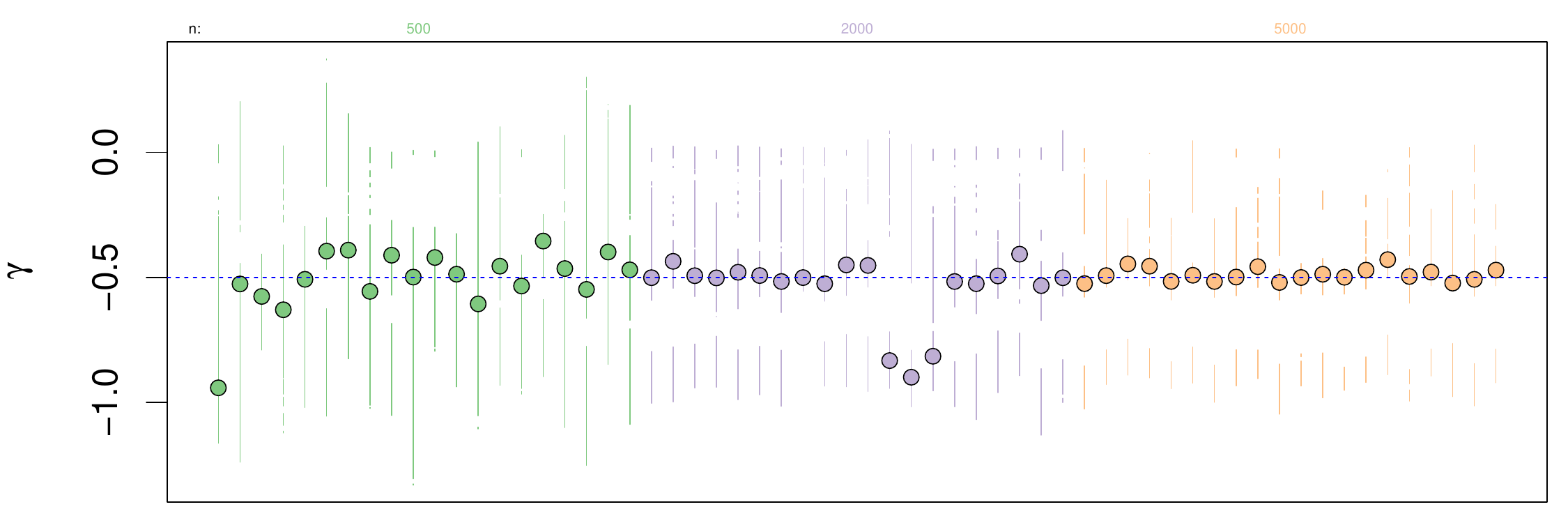}&
\includegraphics[scale=0.20]{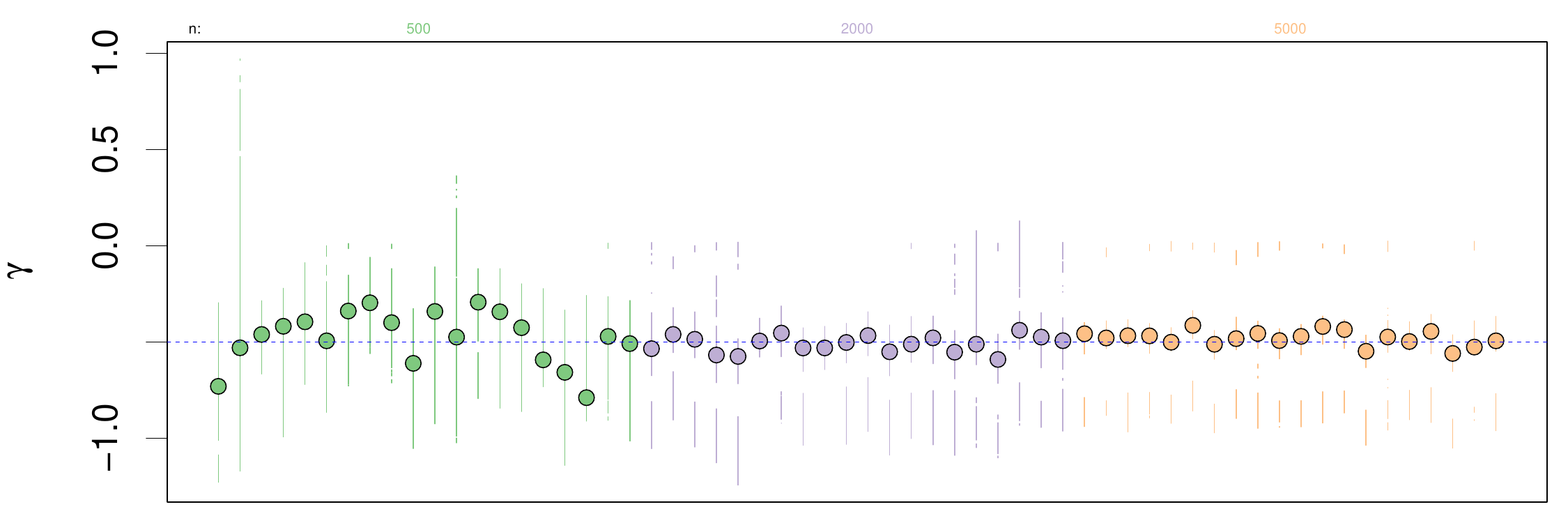}
\\

\hspace{-20ex}\includegraphics[scale=0.20]{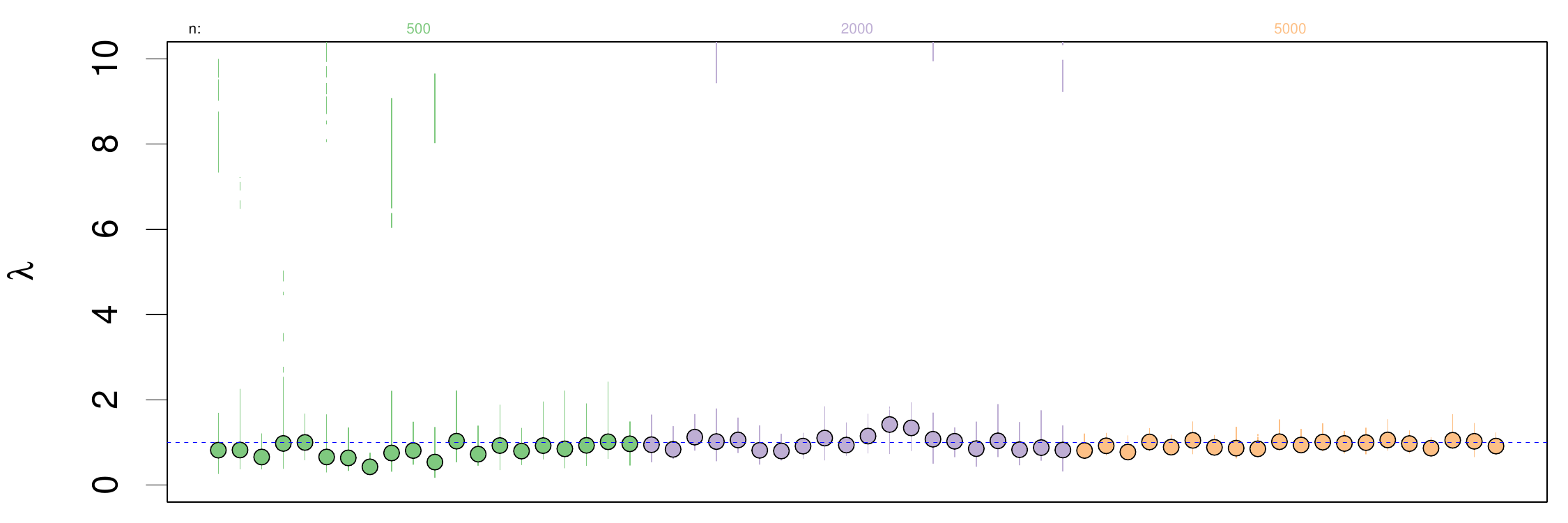}&
\includegraphics[scale=0.20]{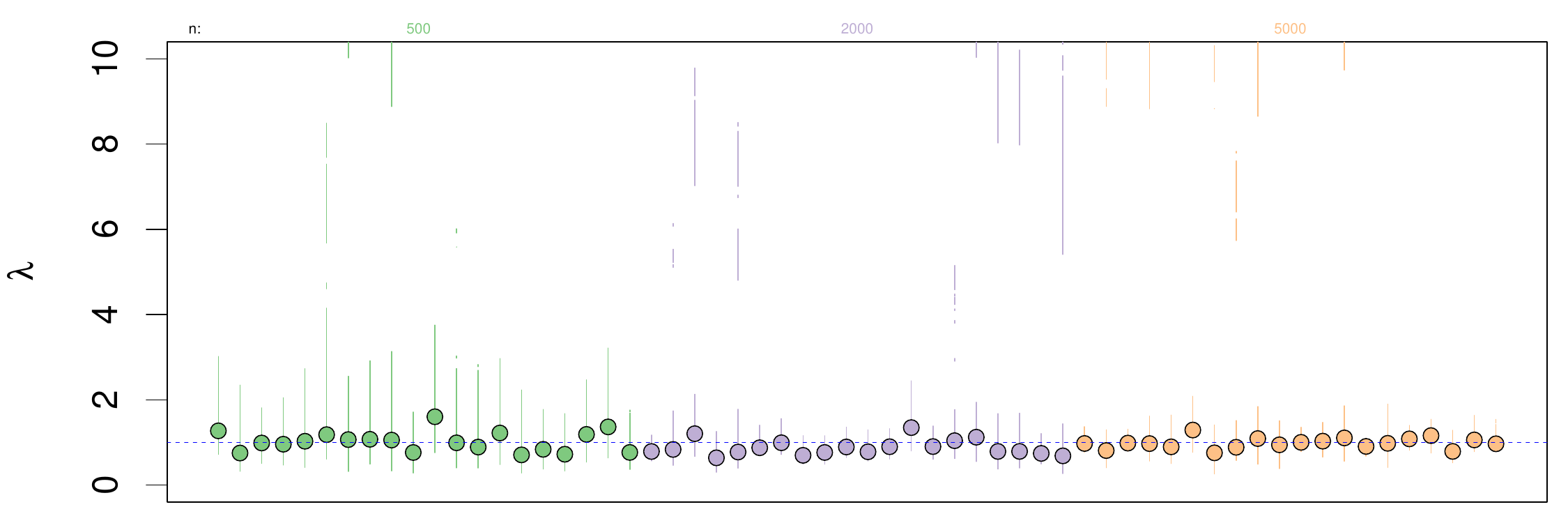}
\\

\hspace{-20ex}\includegraphics[scale=0.20]{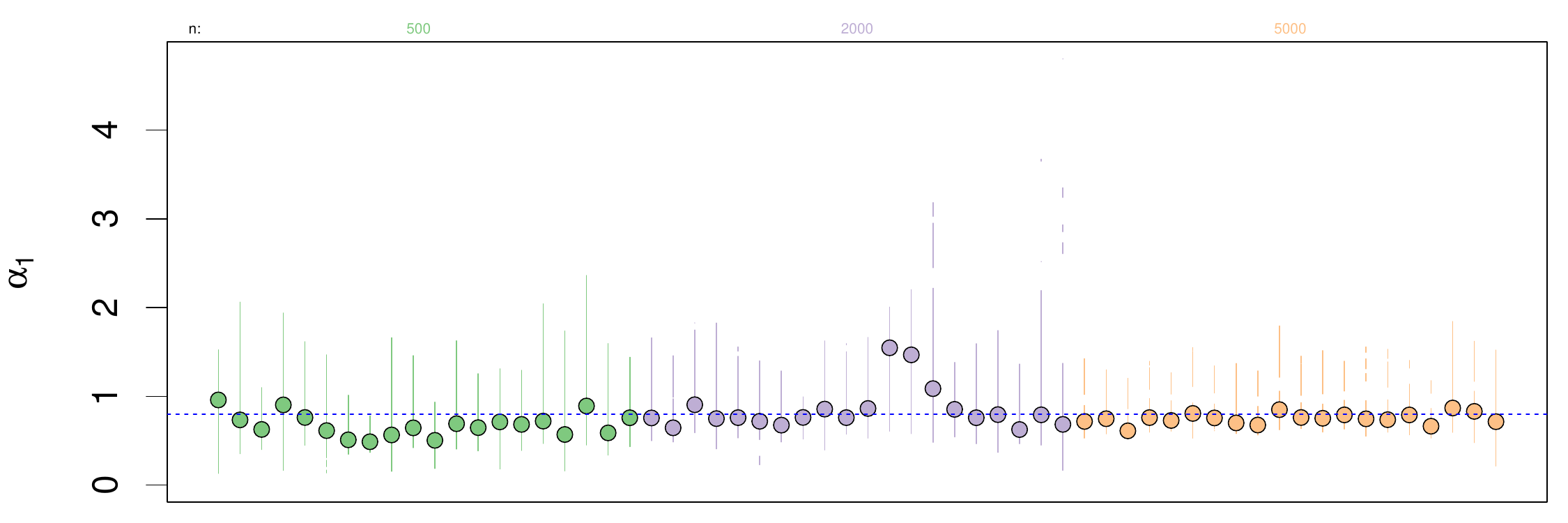}&
\includegraphics[scale=0.20]{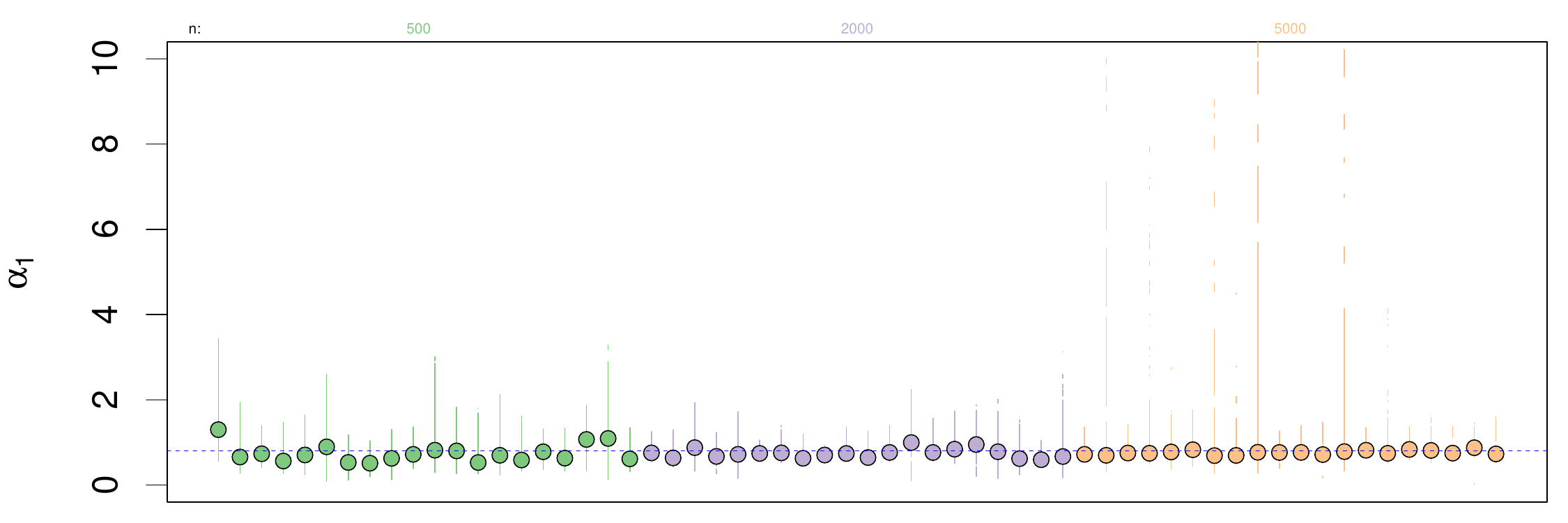}
\\

\hspace{-20ex}\includegraphics[scale=0.20]{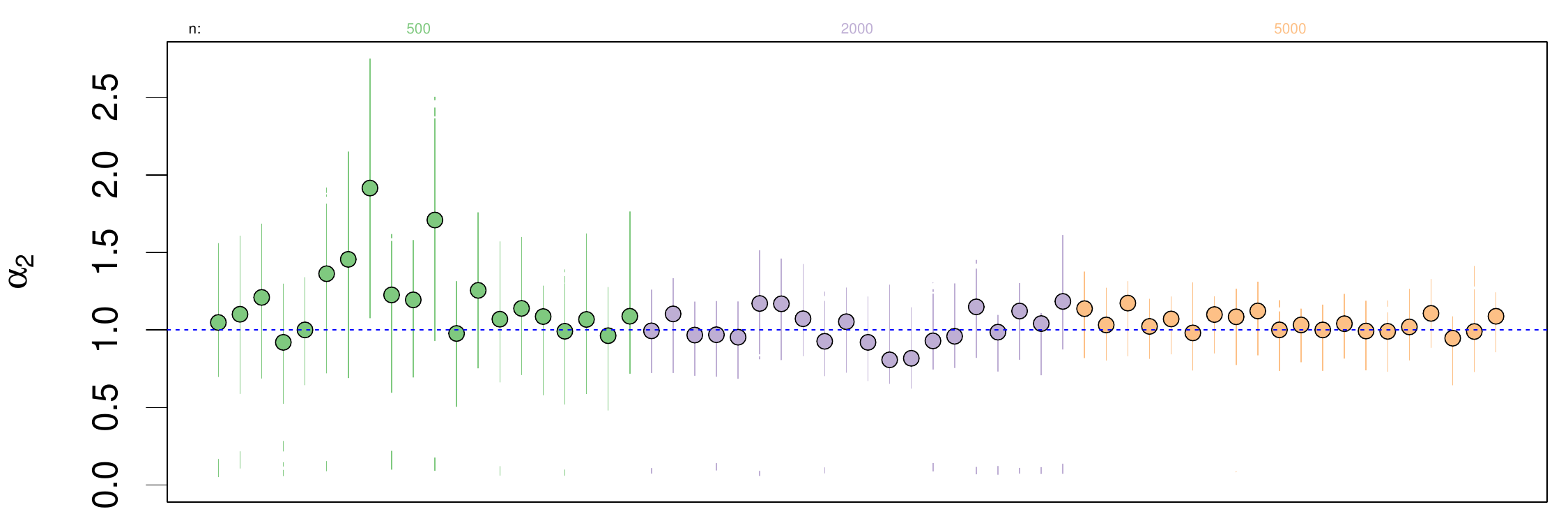}&
\includegraphics[scale=0.20]{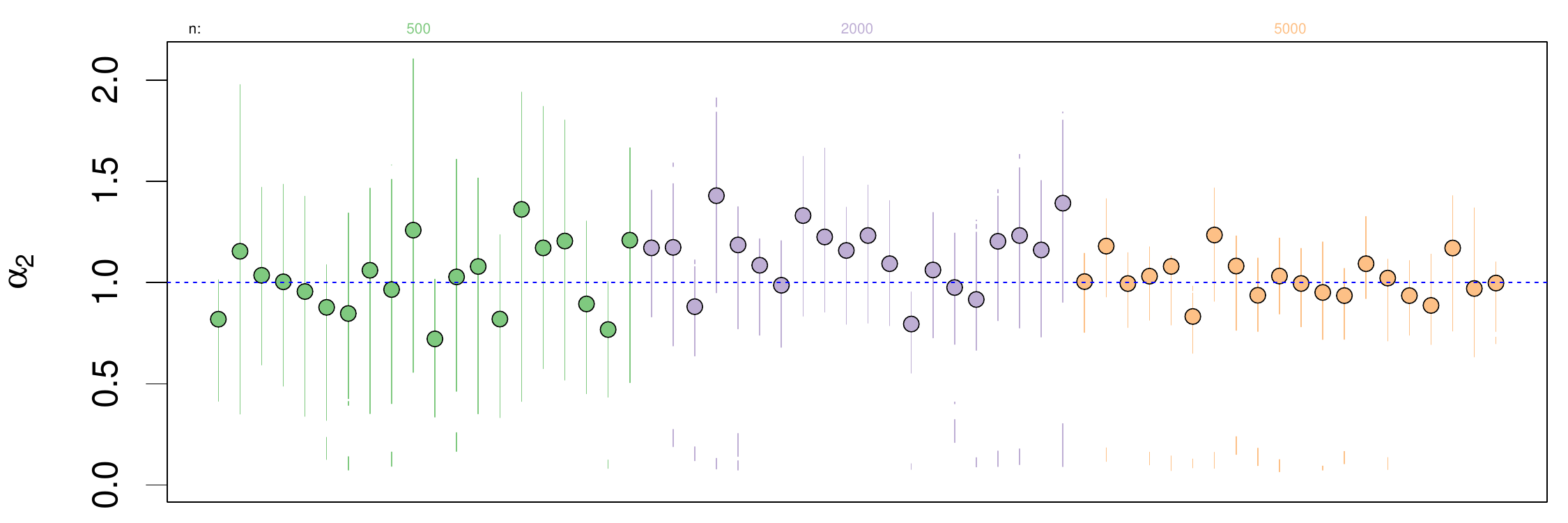}
\\

\hspace{-20ex}\includegraphics[scale=0.20]{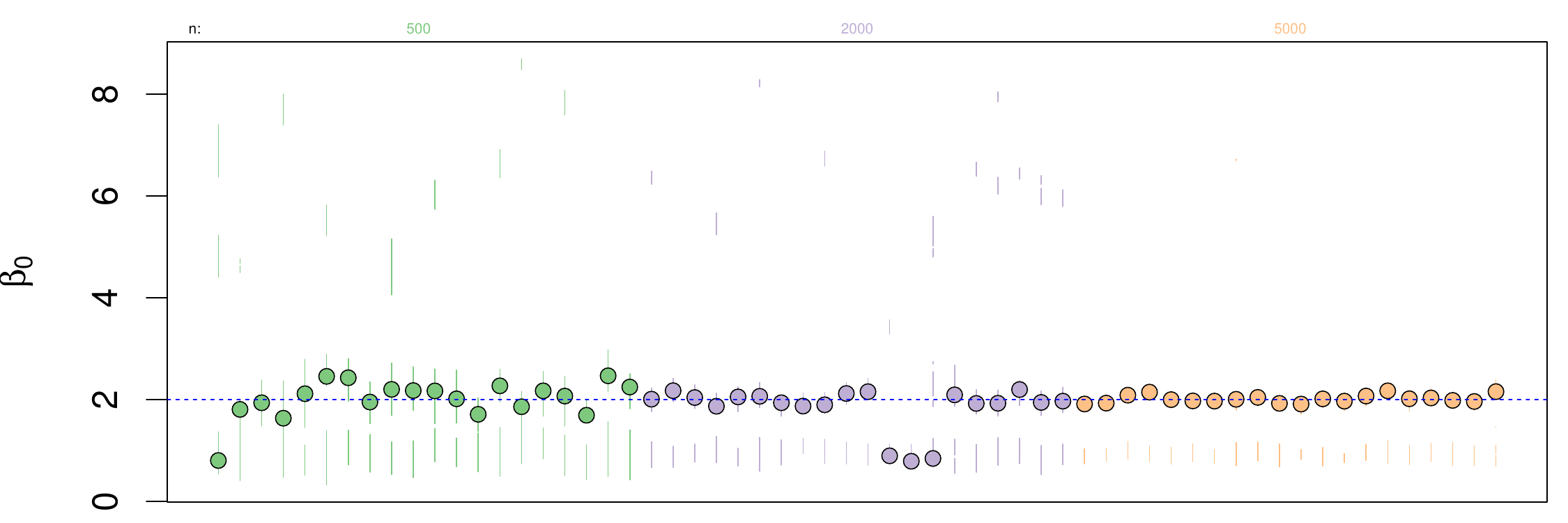}&
\includegraphics[scale=0.20]{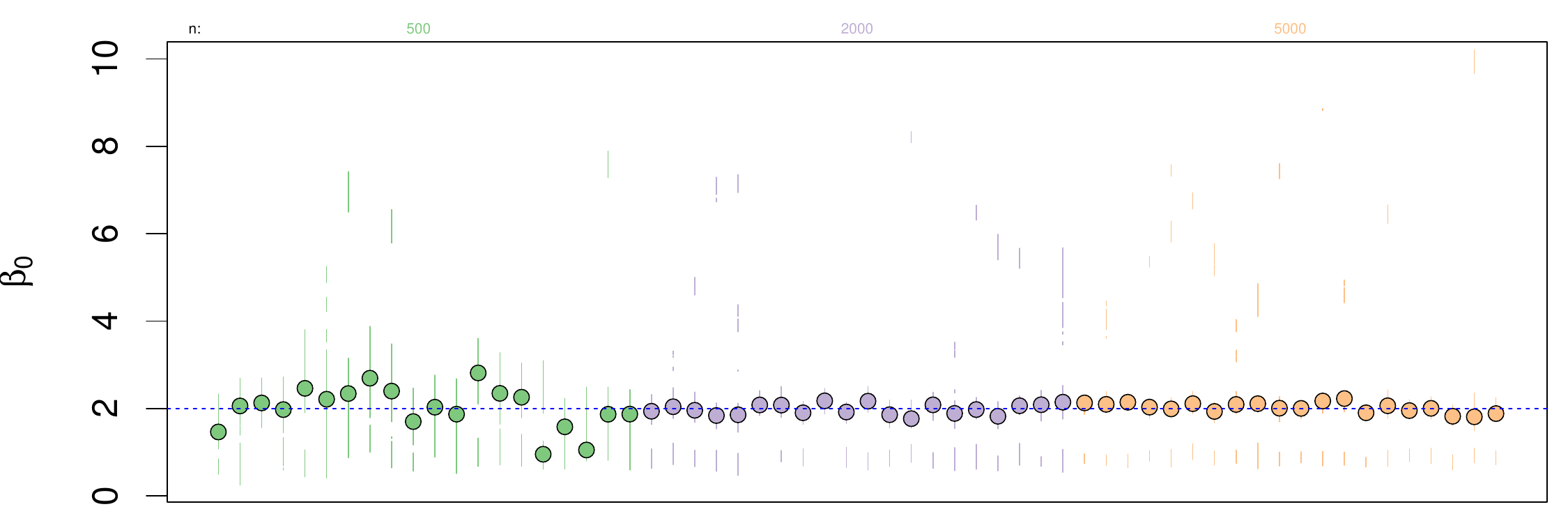}
\\

\hspace{-20ex}\includegraphics[scale=0.20]{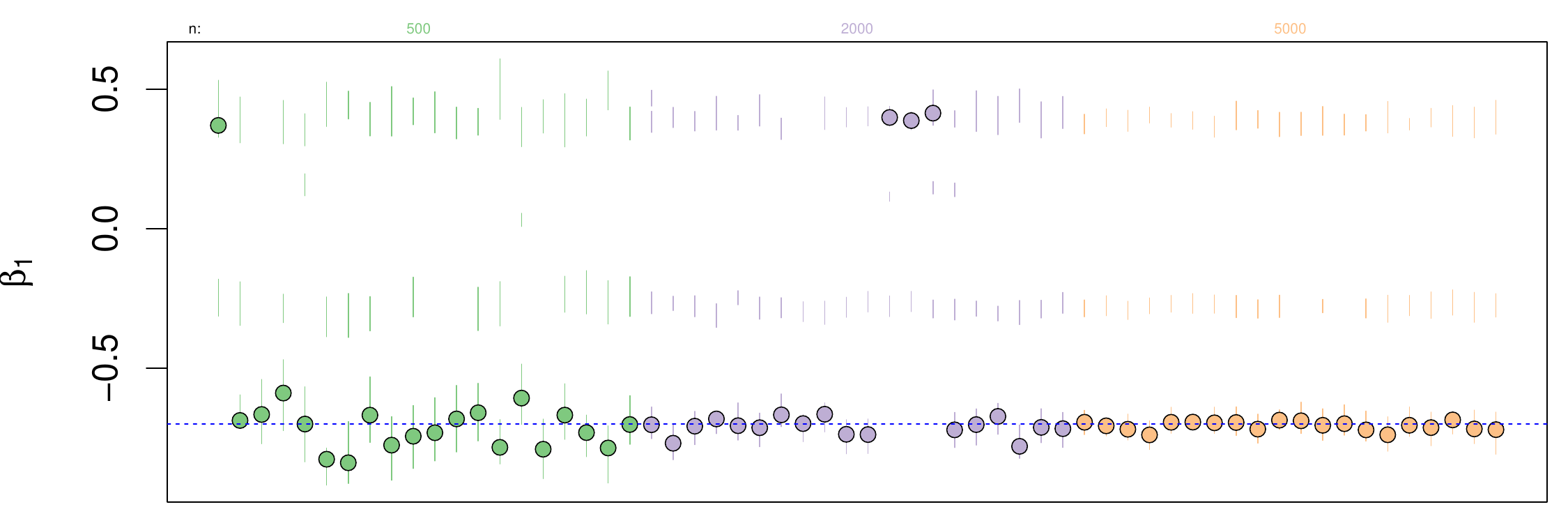}&
\includegraphics[scale=0.20]{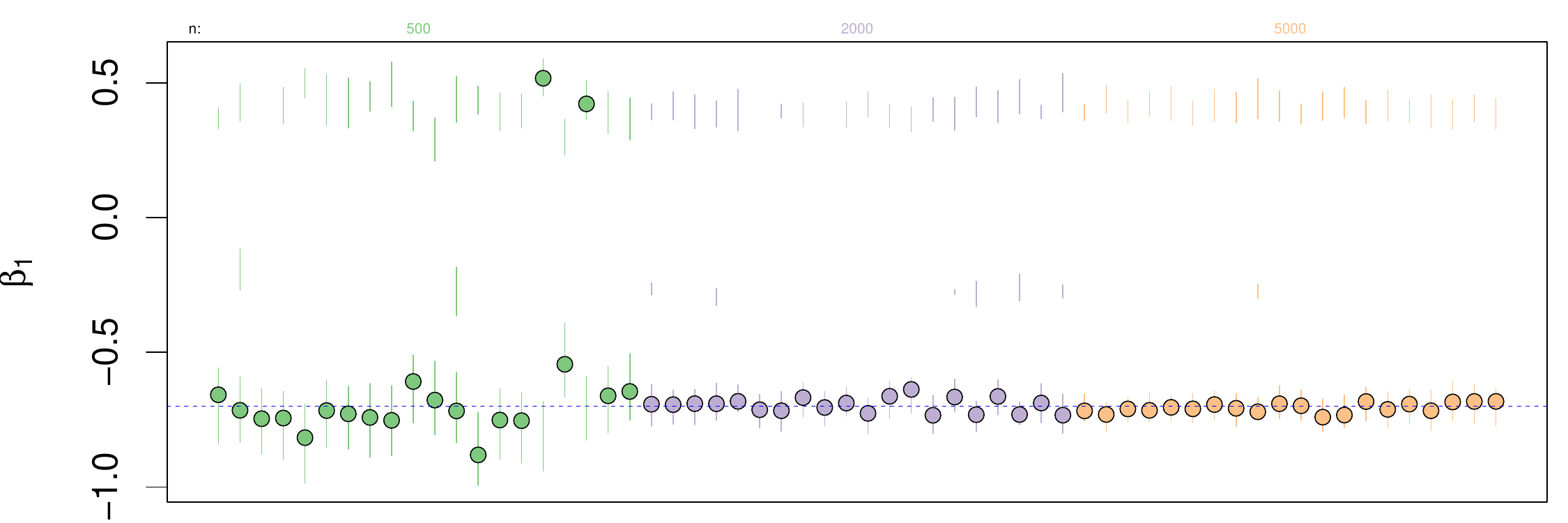}
\\

\hspace{-20ex}\includegraphics[scale=0.20]{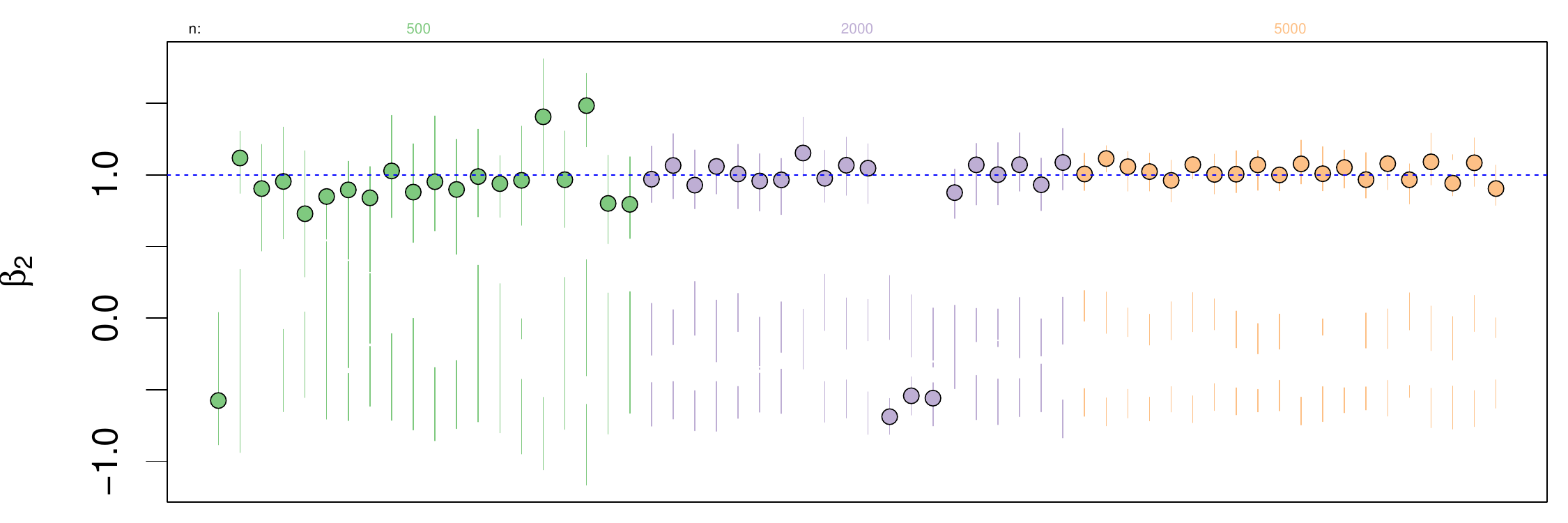}&
\includegraphics[scale=0.20]{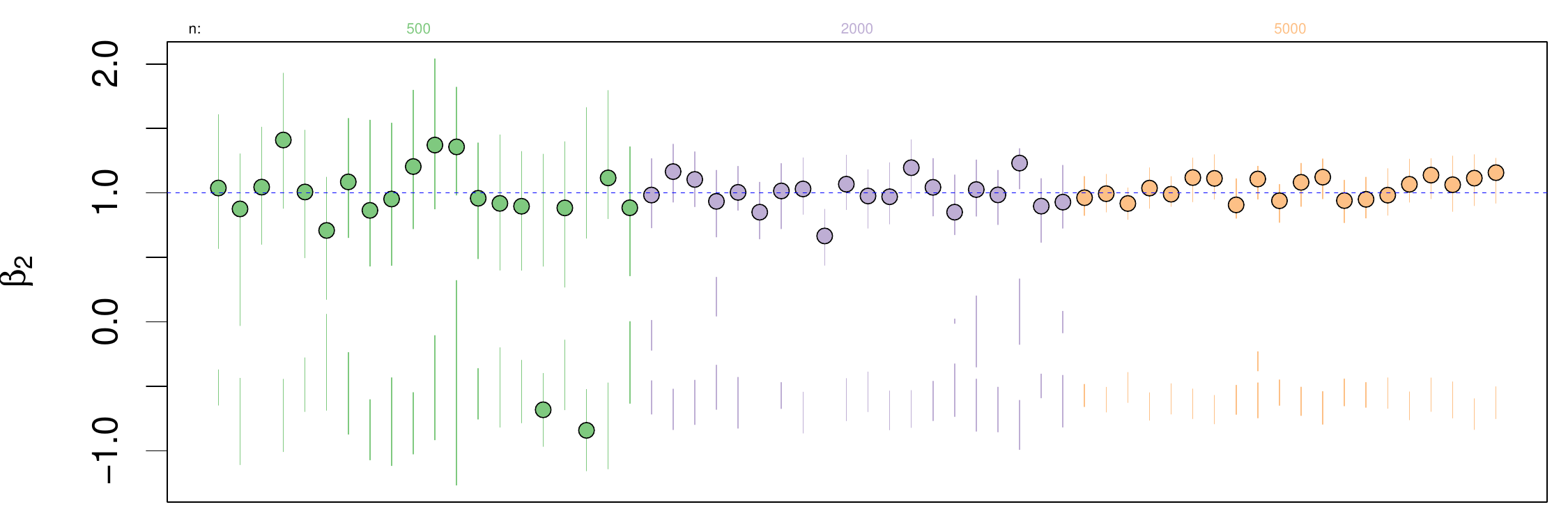}
\\
scenario E.1 & scenario E.2
\end{tabular}
\caption{Point estimates (MAP) with $99\%$ Highest Density Intervals for our simulated datasets. Different colour indicate levels of sample size: 500 (\textcolor{green}{---}), 2000 (\textcolor{violet}{---}), 5000 (\textcolor{orange}{---}). The horizontal line indicates the true value (see Scenarios E1 and E2 in Table \ref{t4b}).} 
\label{fig:hdisE}
\end{figure}

\begin{figure}[p]
\centering
\begin{tabular}{cc}
\hspace{-20ex}\includegraphics[scale=0.20]{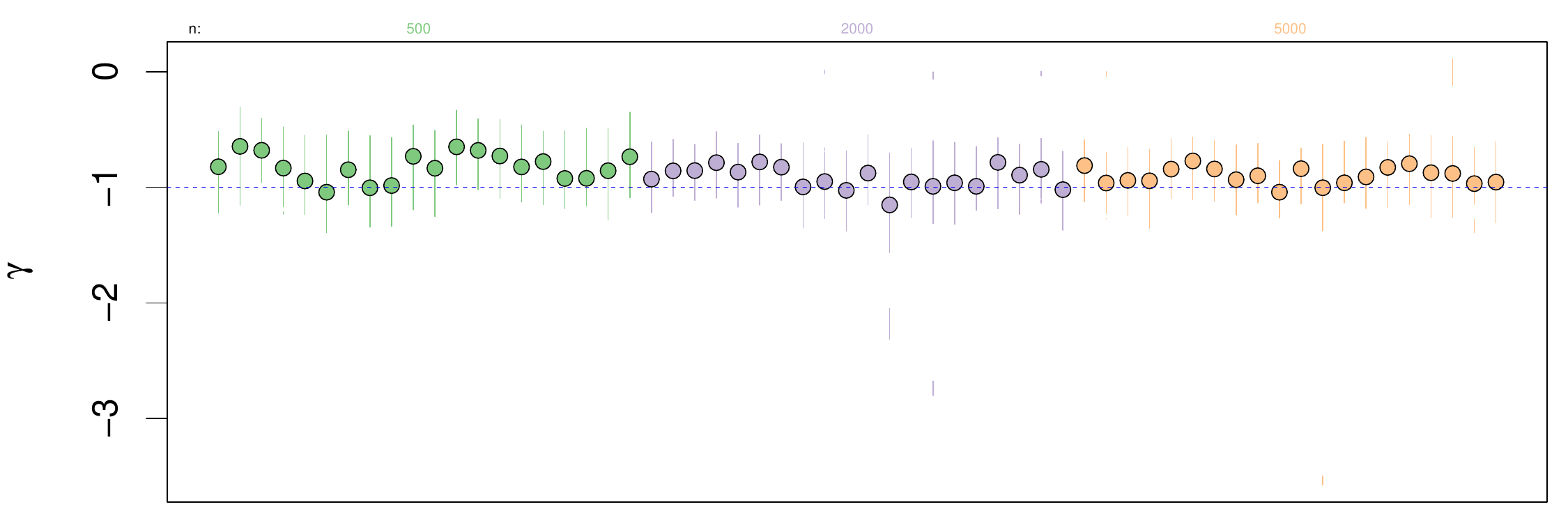}&
\includegraphics[scale=0.20]{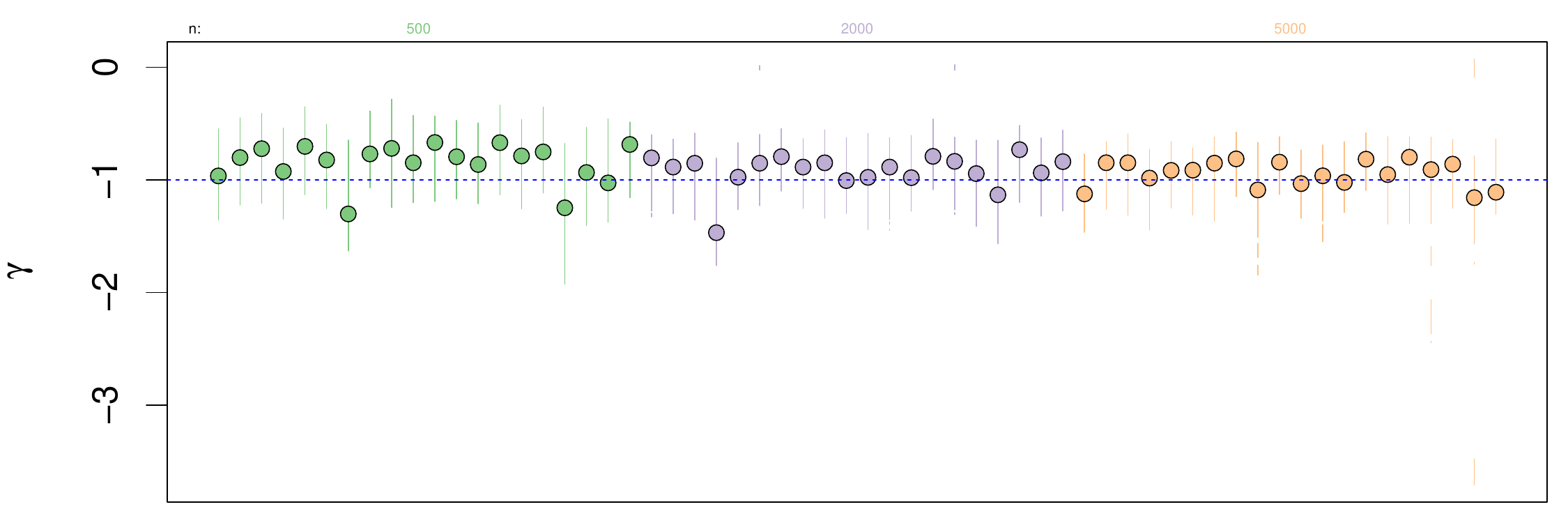}
\\

\hspace{-20ex}\includegraphics[scale=0.20]{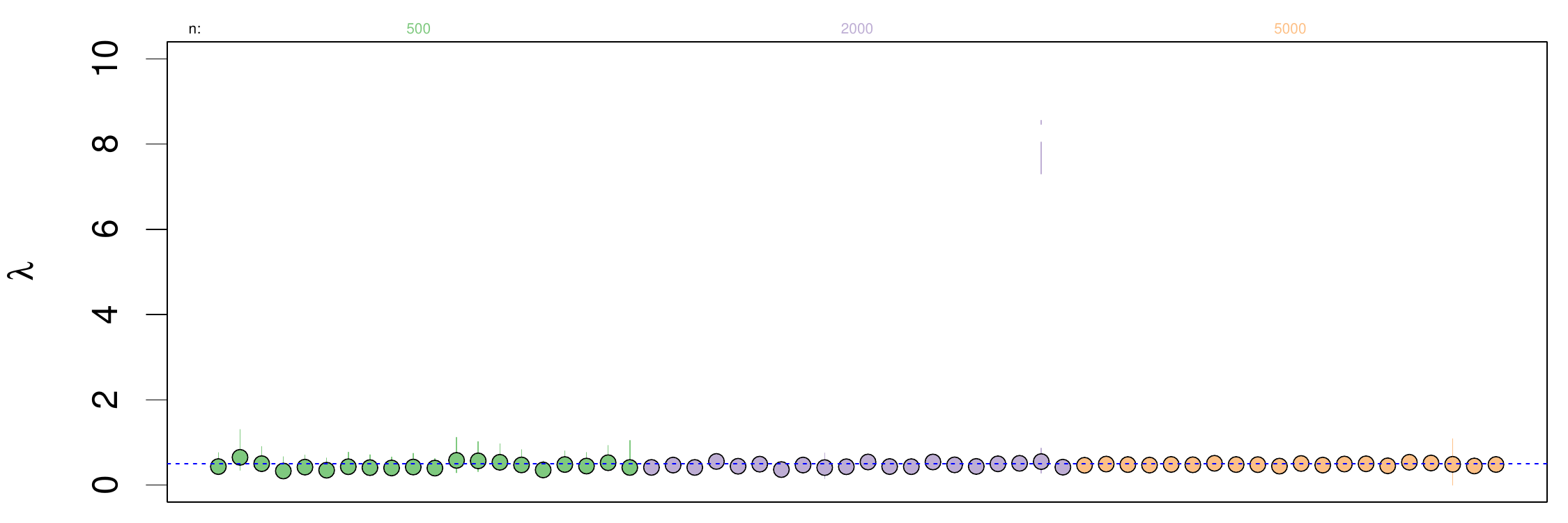}&
\includegraphics[scale=0.20]{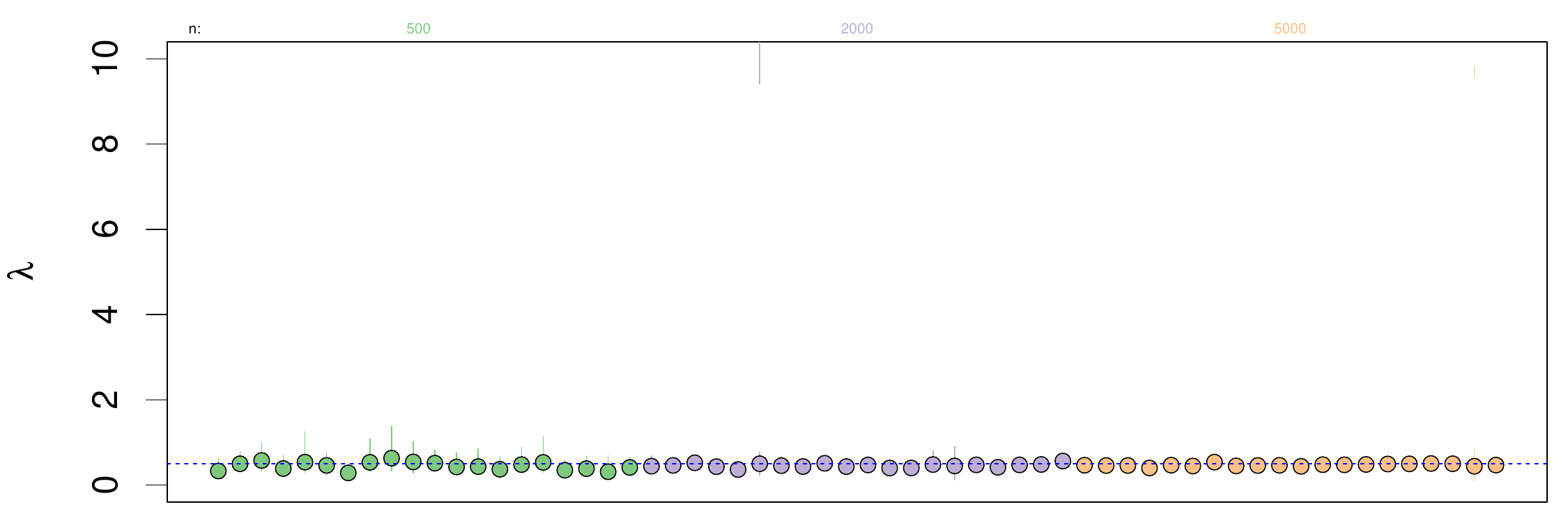}
\\

\hspace{-20ex}\includegraphics[scale=0.20]{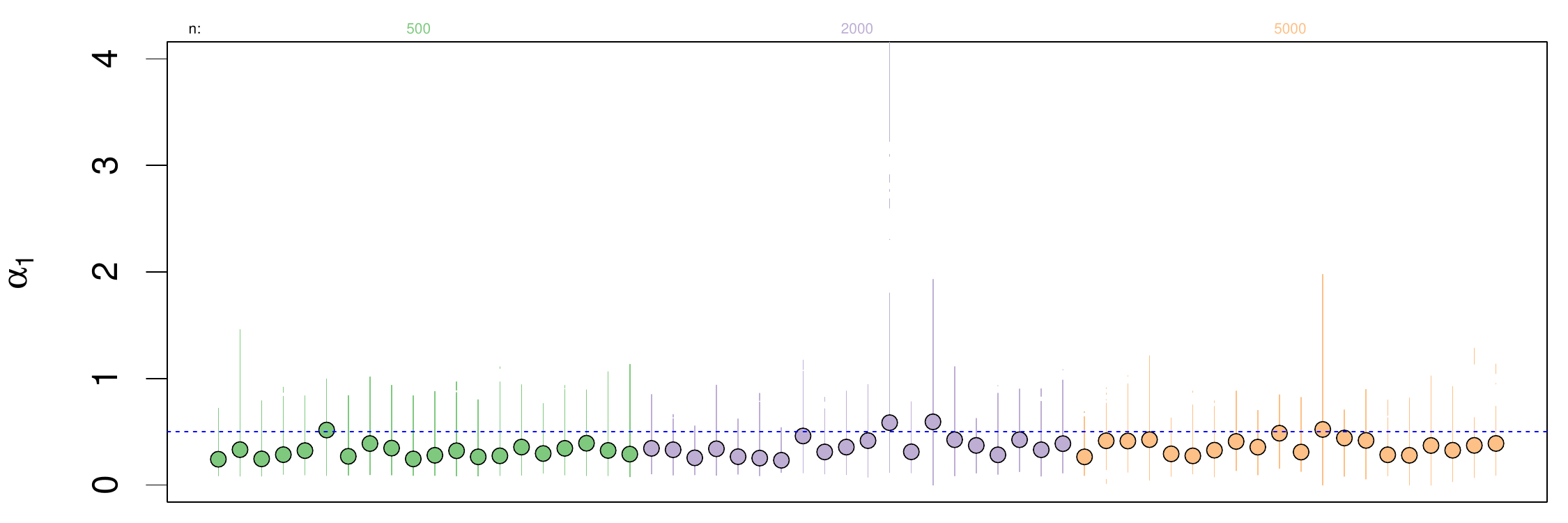}&
\includegraphics[scale=0.20]{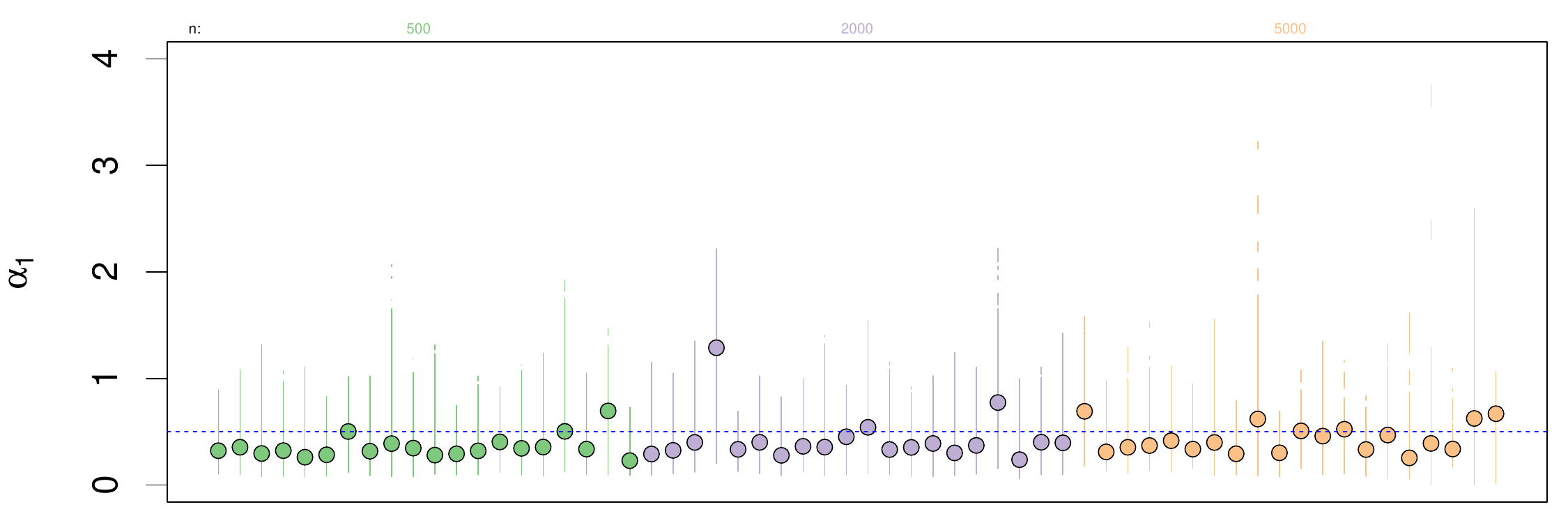}
\\

\hspace{-20ex}\includegraphics[scale=0.20]{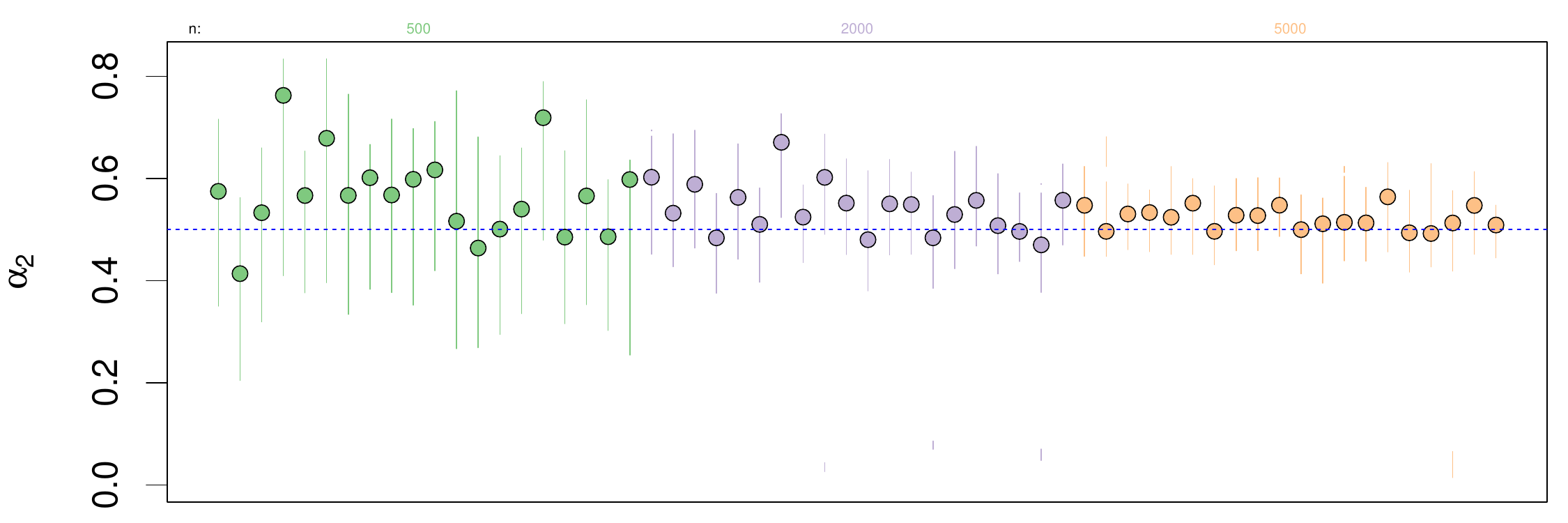}&
\includegraphics[scale=0.20]{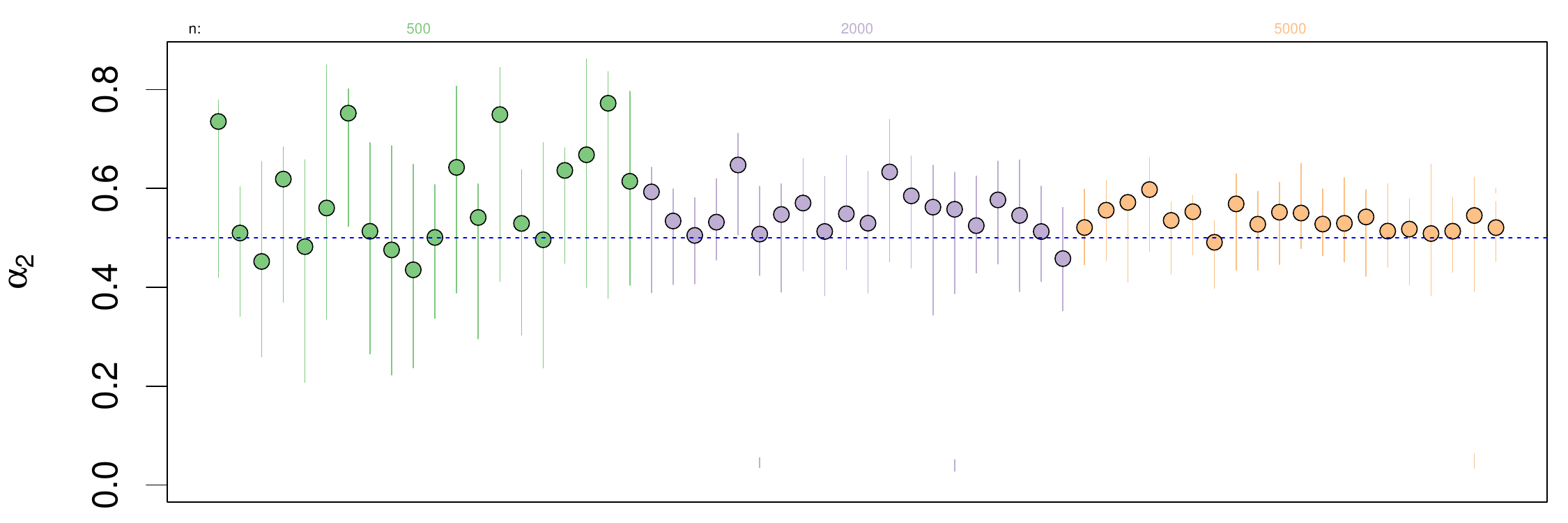}
\\

\hspace{-20ex}\includegraphics[scale=0.20]{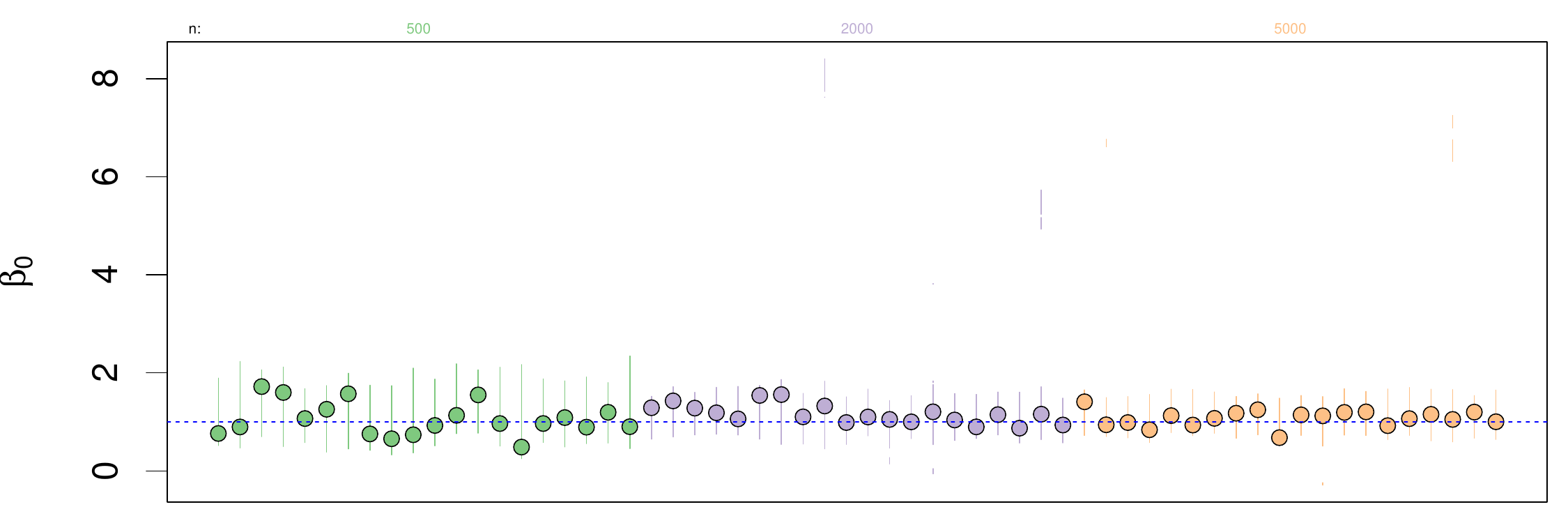}&
\includegraphics[scale=0.20]{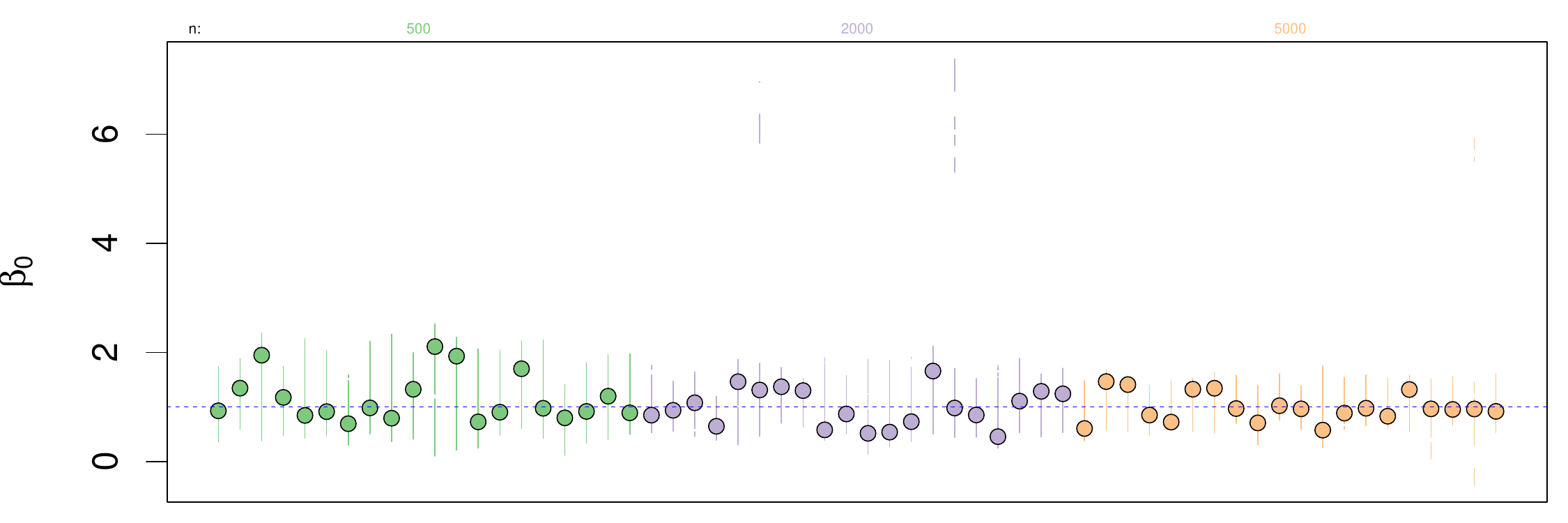}
\\

\hspace{-20ex}\includegraphics[scale=0.20]{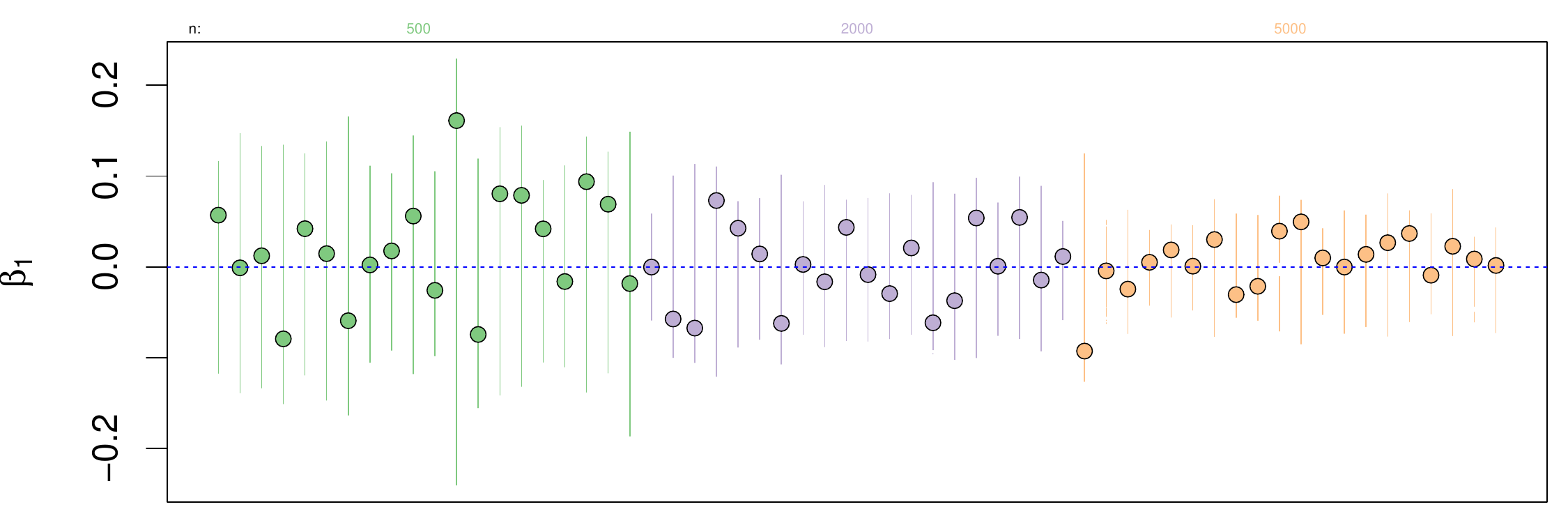}&
\includegraphics[scale=0.20]{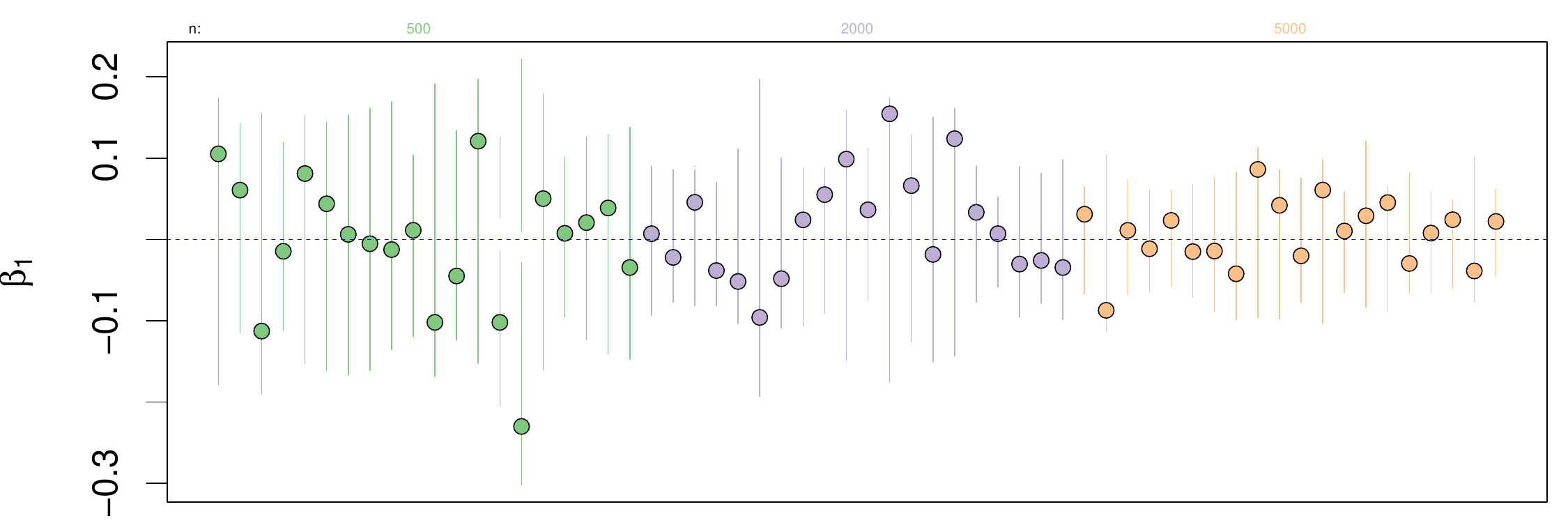}
\\

\hspace{-20ex}\includegraphics[scale=0.20]{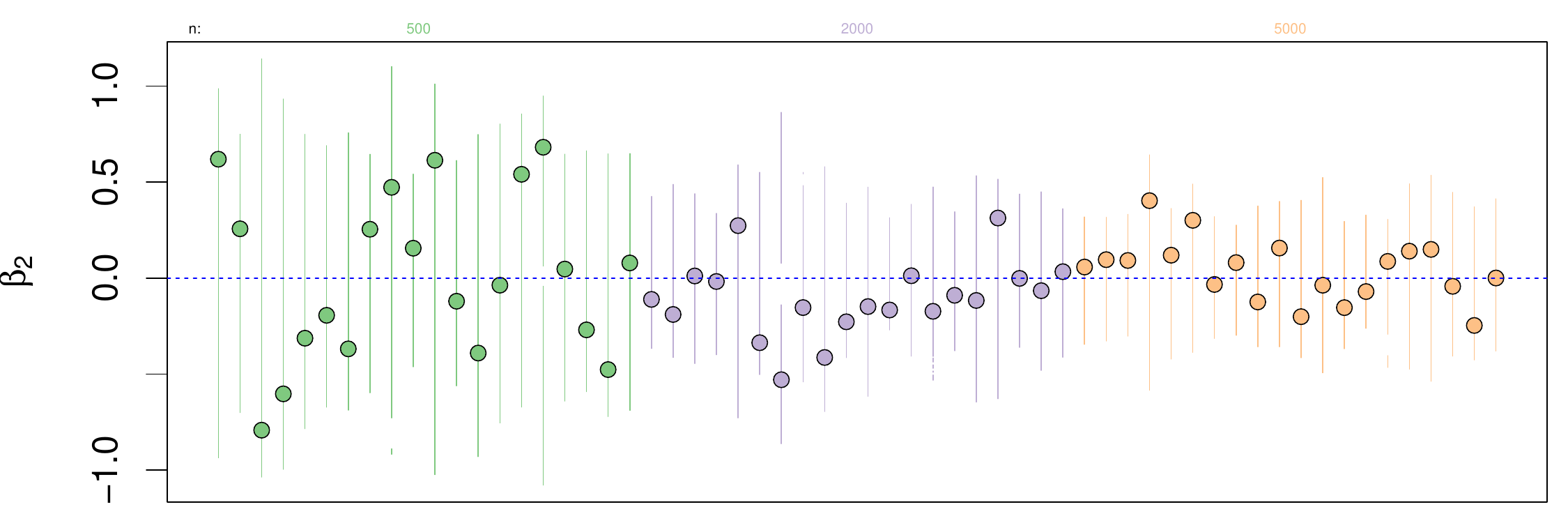}&
\includegraphics[scale=0.20]{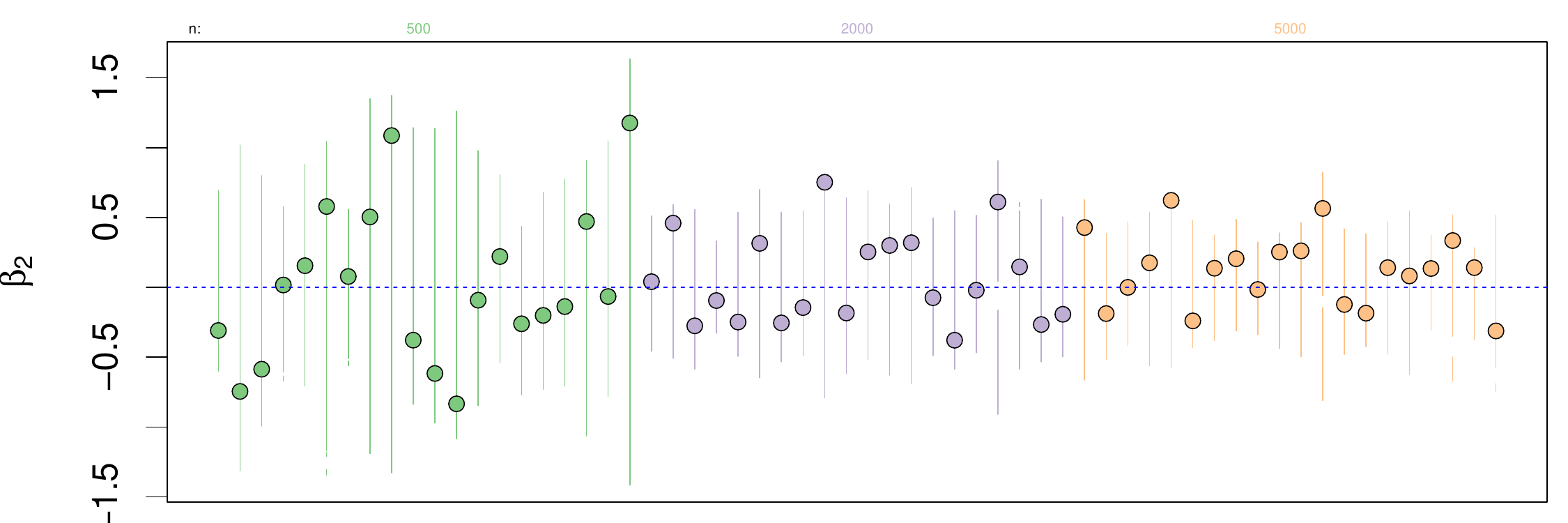}
\\
scenario F.1 & scenario F.2
\end{tabular}
\caption{Point estimates (MAP) with $95\%$ Highest Density Intervals for our simulated datasets. Different colour indicate levels of sample size: 500 (\textcolor{green}{---}), 2000 (\textcolor{violet}{---}), 5000 (\textcolor{orange}{---}). The horizontal line indicates the true value (see Scenarios F1 and F2 in Table \ref{t4b}).} 
\label{fig:hdisF}
\end{figure}

\begin{figure}[p]
\centering
\begin{tabular}{cc}
\hspace{-20ex}\includegraphics[scale=0.20]{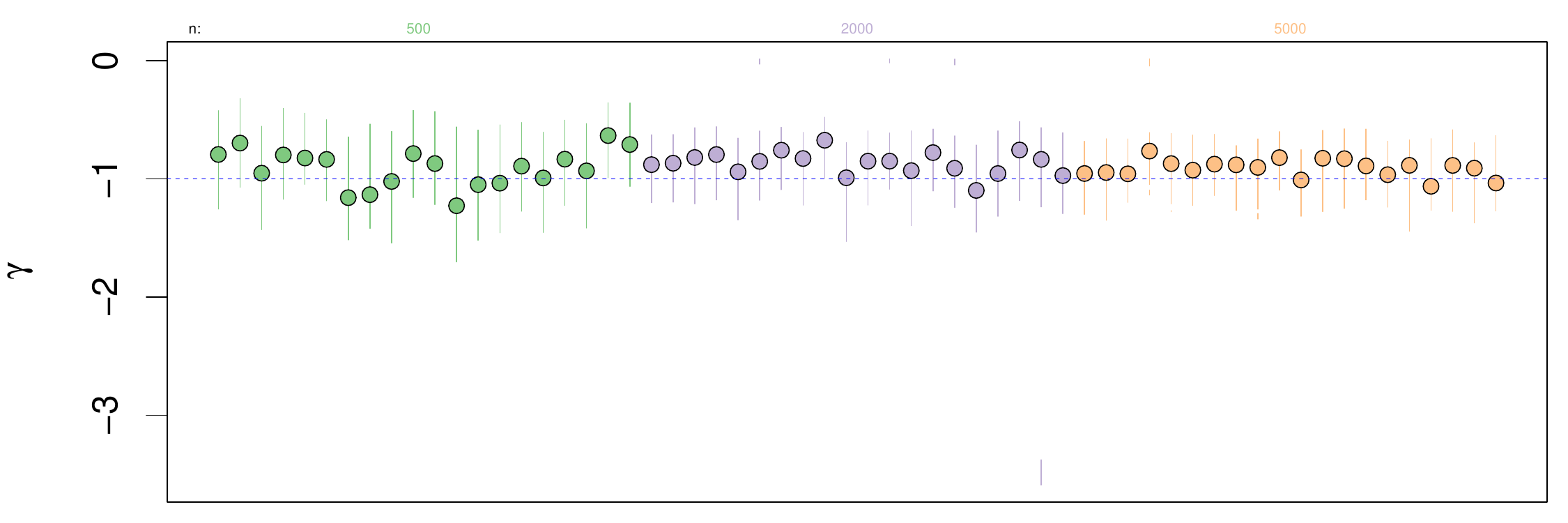}&
\includegraphics[scale=0.20]{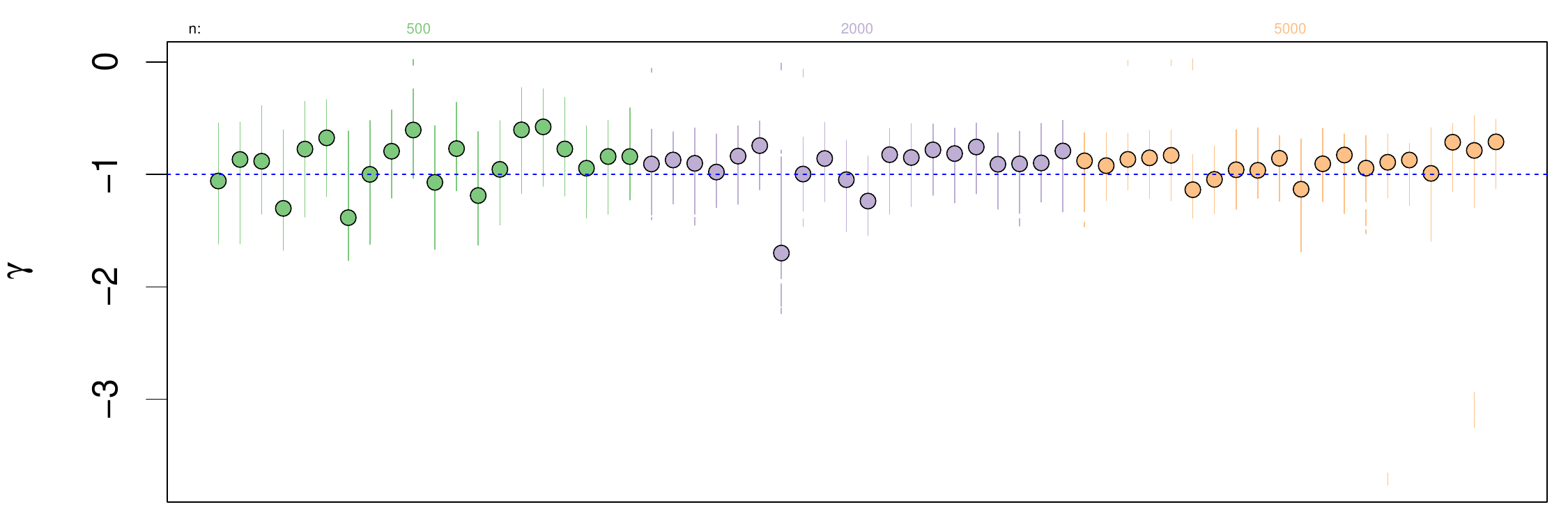}
\\

\hspace{-20ex}\includegraphics[scale=0.20]{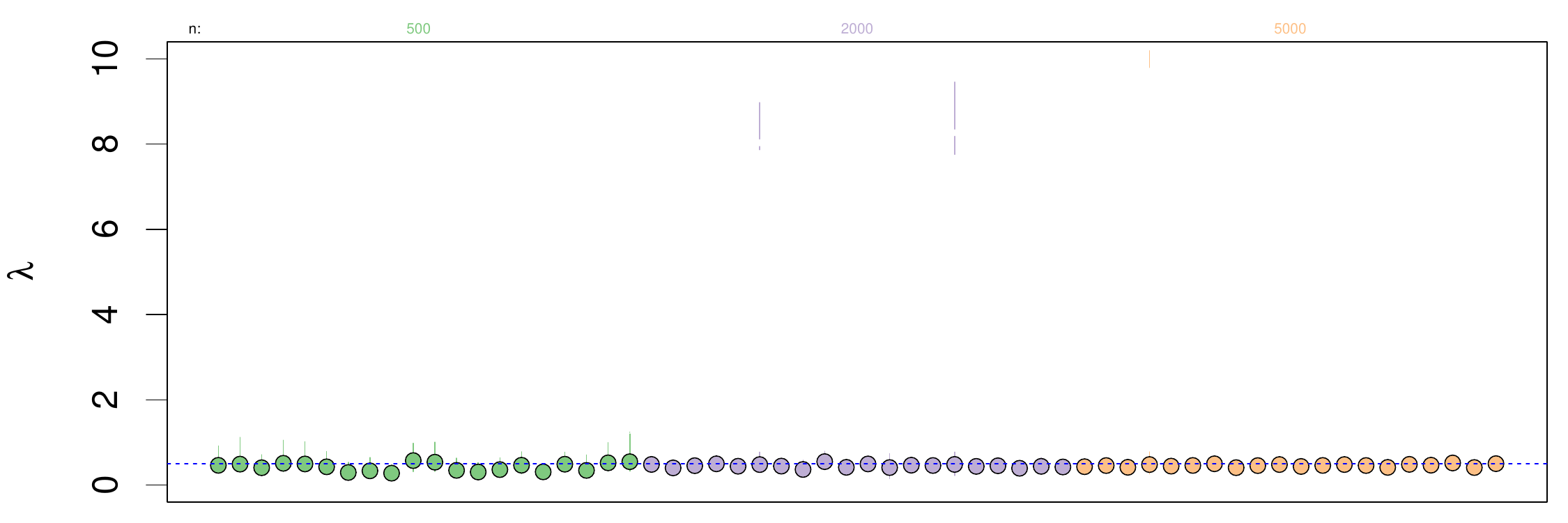}&
\includegraphics[scale=0.20]{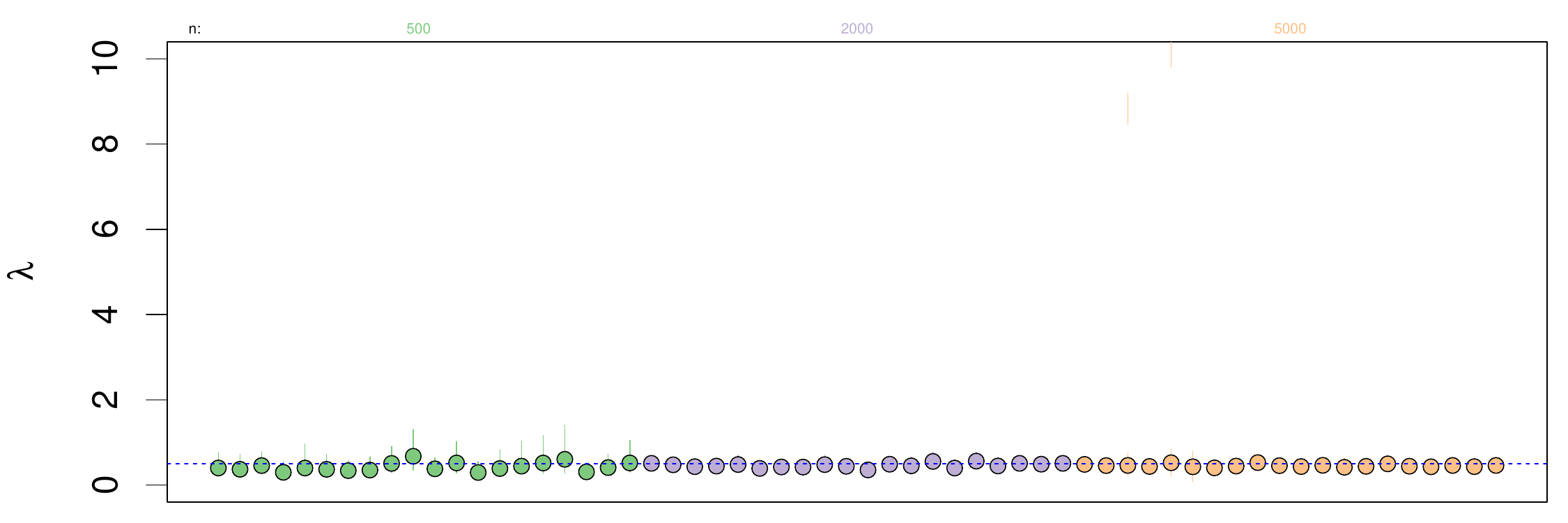}
\\

\hspace{-20ex}\includegraphics[scale=0.20]{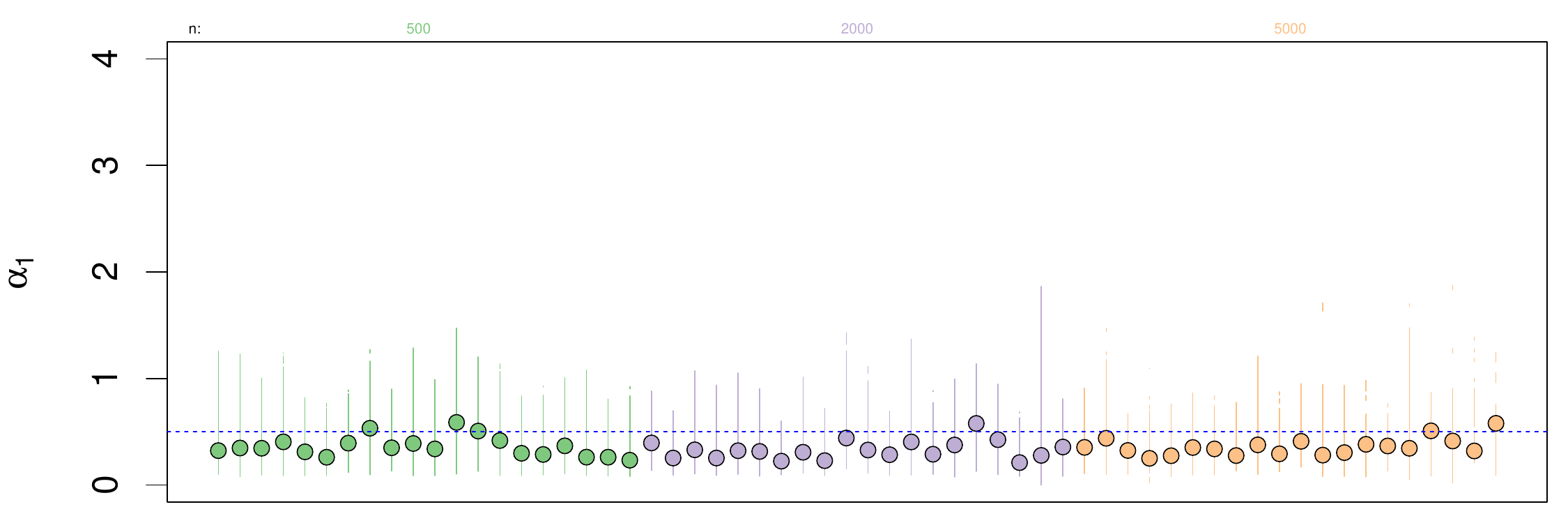}&
\includegraphics[scale=0.20]{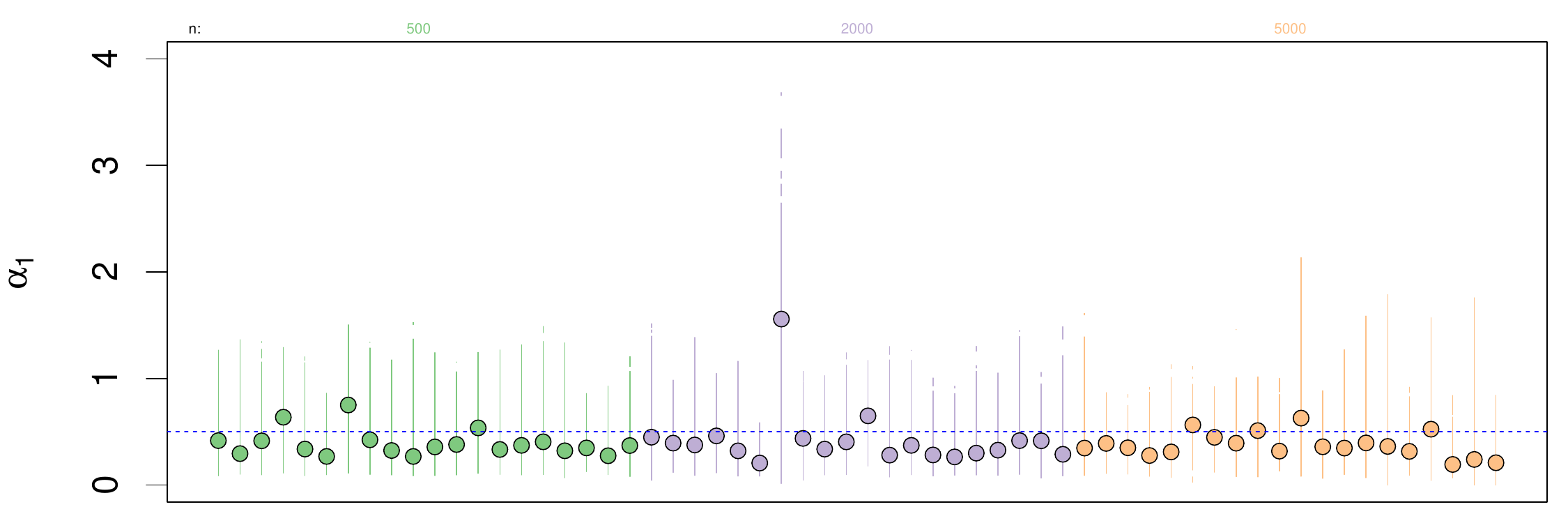}
\\

\hspace{-20ex}\includegraphics[scale=0.20]{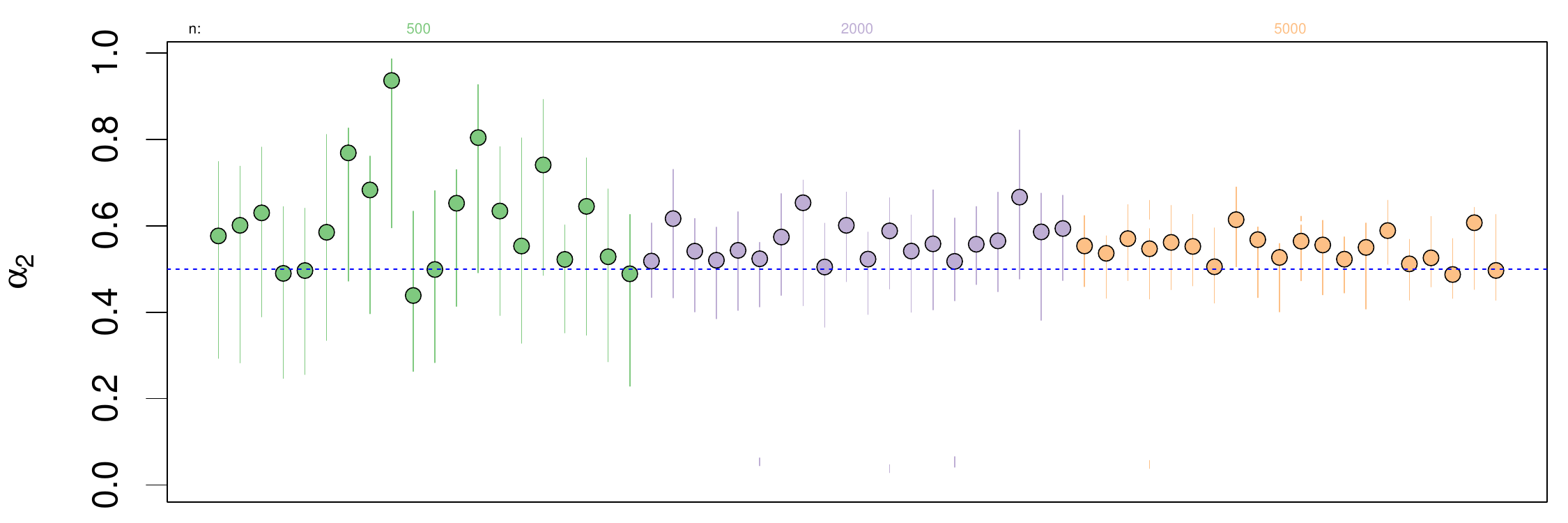}&
\includegraphics[scale=0.20]{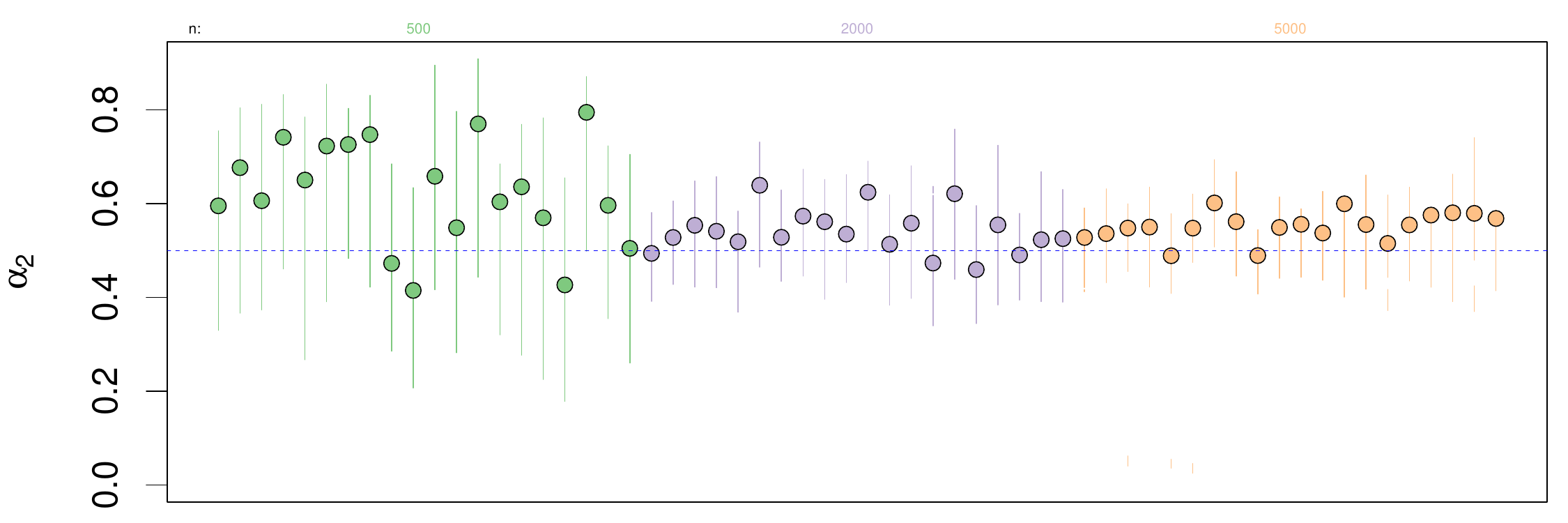}
\\

\hspace{-20ex}\includegraphics[scale=0.20]{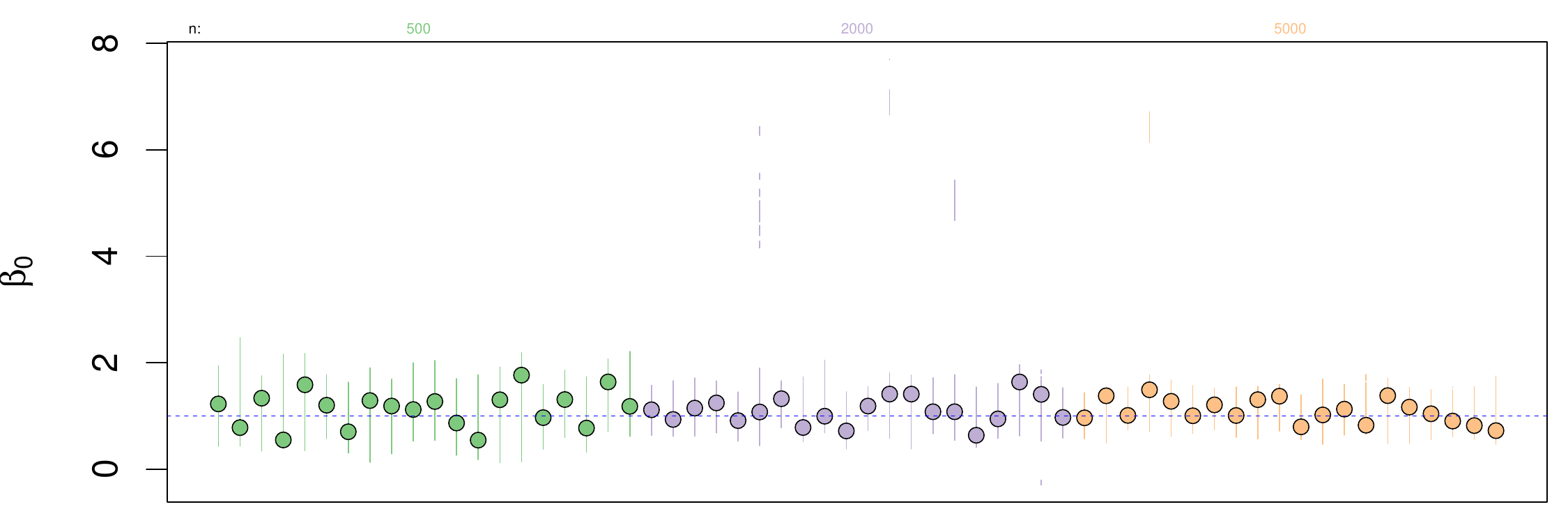}&
\includegraphics[scale=0.20]{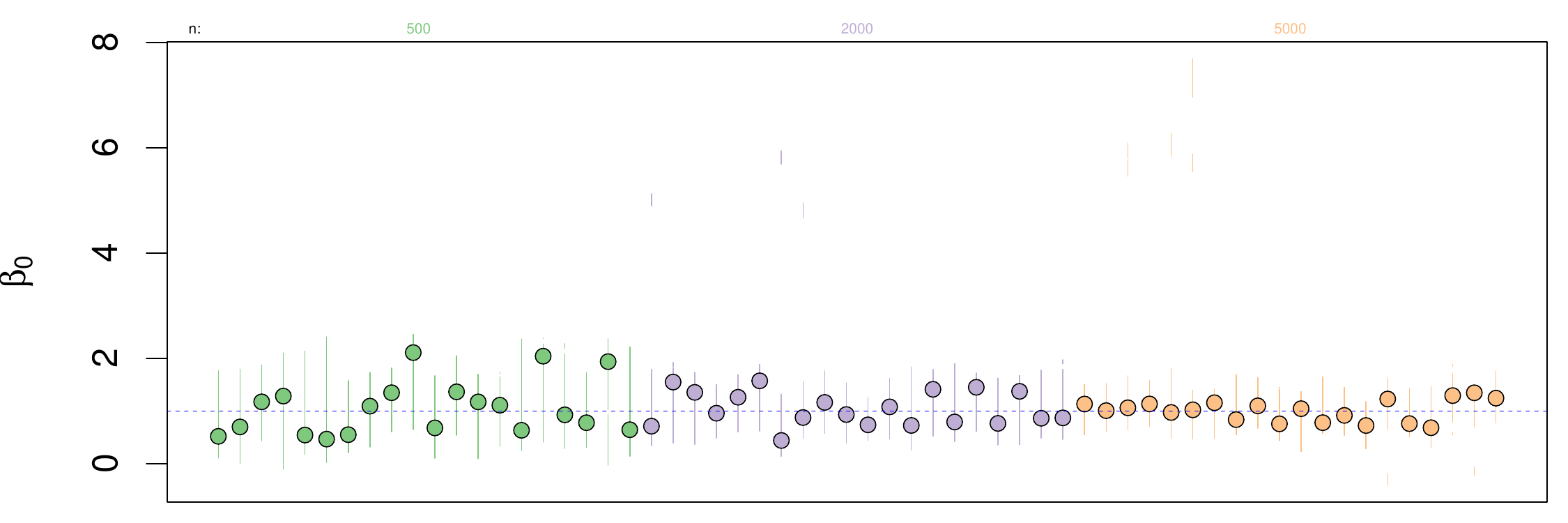}
\\

\hspace{-20ex}\includegraphics[scale=0.20]{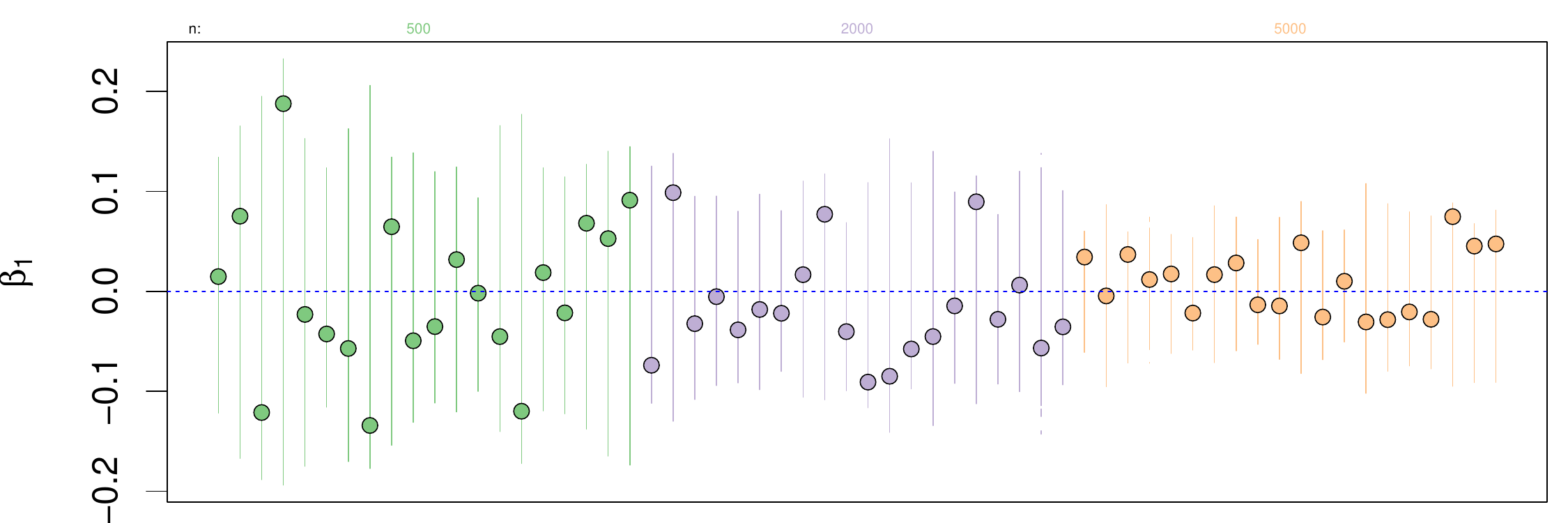}&
\includegraphics[scale=0.20]{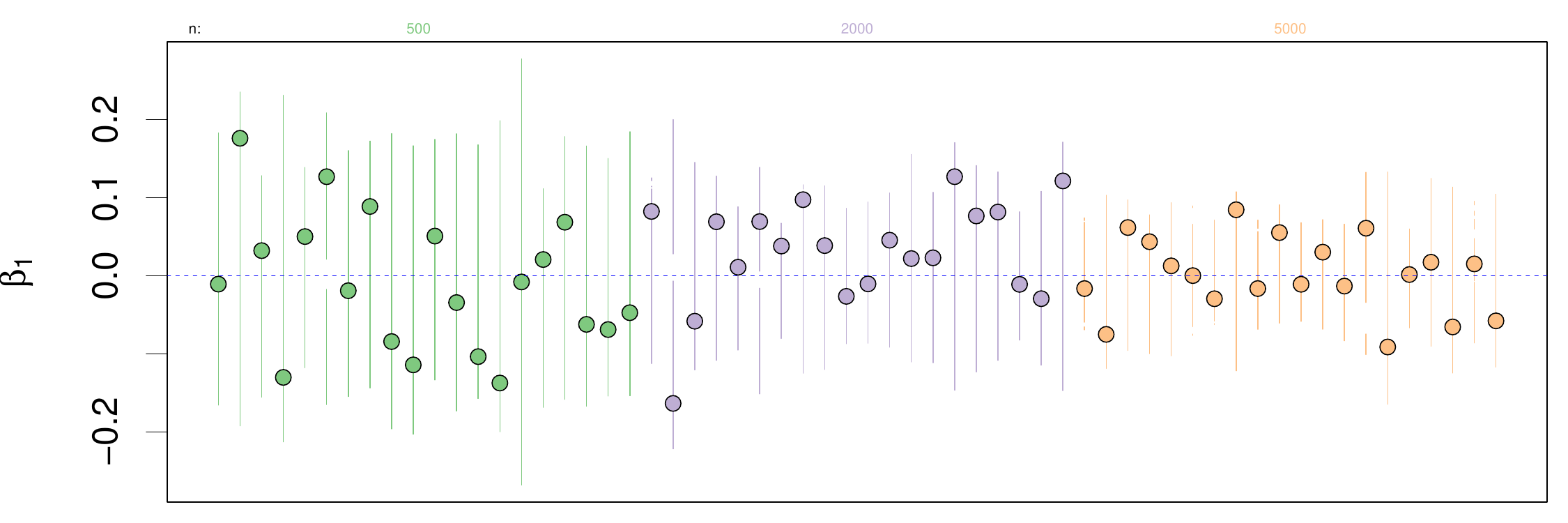}
\\

\hspace{-20ex}\includegraphics[scale=0.20]{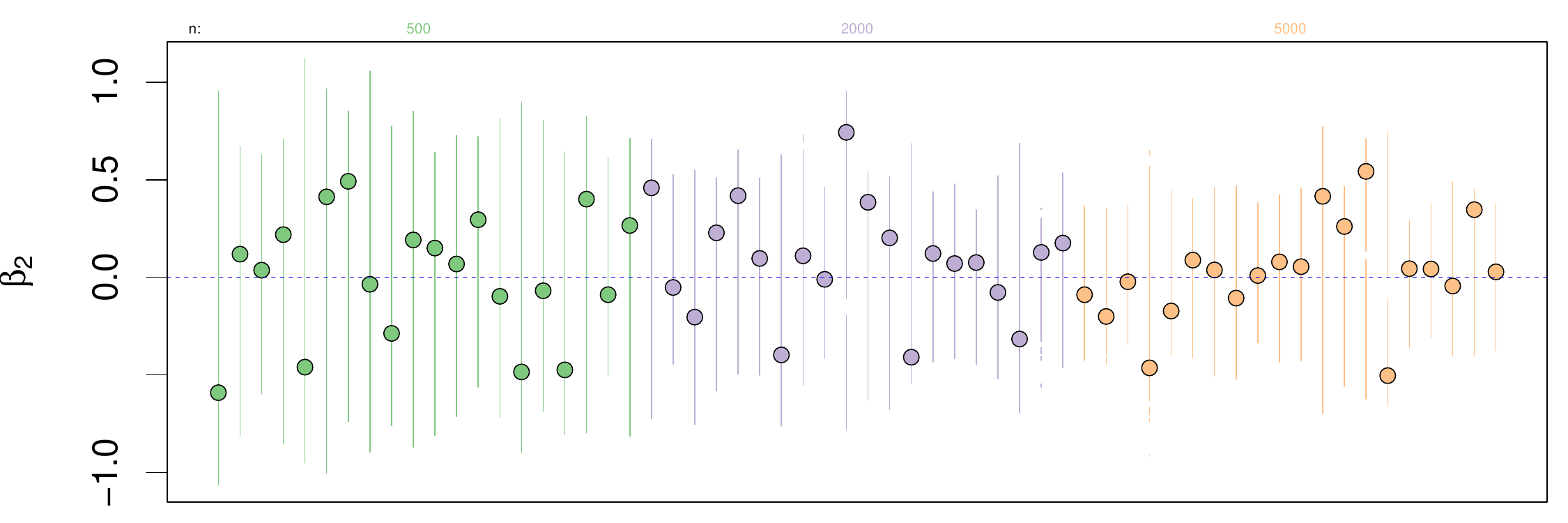}&
\includegraphics[scale=0.20]{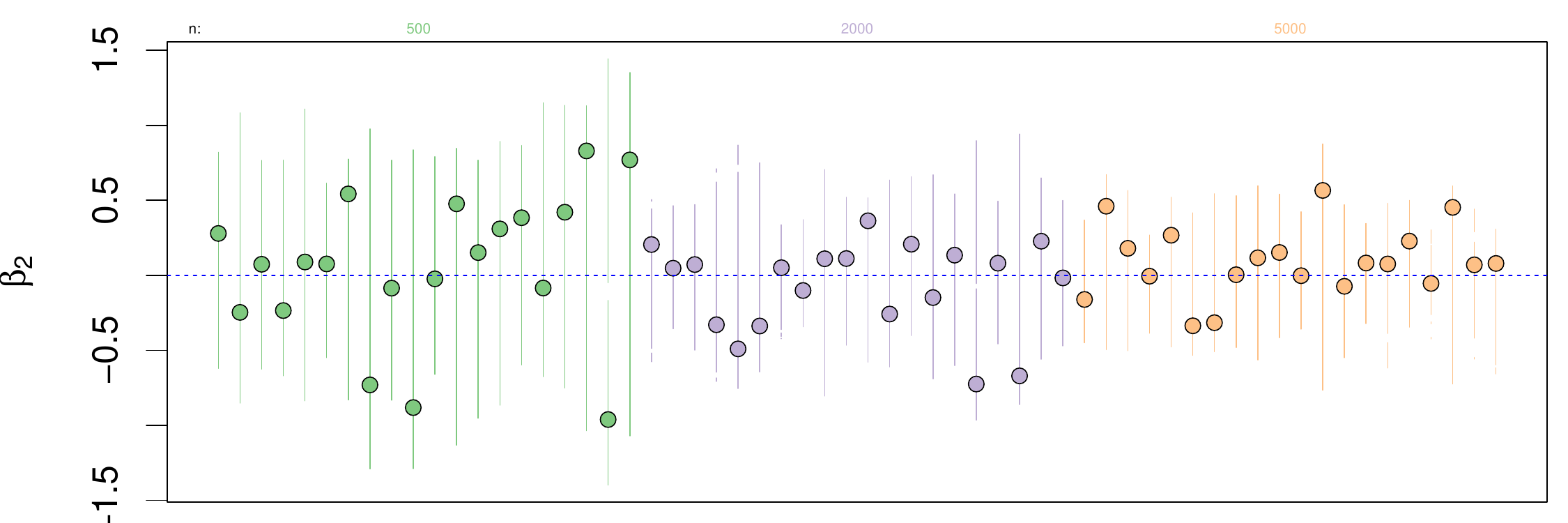}
\\
scenario F.3 & scenario F.4
\end{tabular}
\caption{Point estimates (MAP) with $95\%$ Highest Density Intervals for our simulated datasets. Different colour indicate levels of sample size: 500 (\textcolor{green}{---}), 2000 (\textcolor{violet}{---}), 5000 (\textcolor{orange}{---}). The horizontal line indicates the true value (see Scenarios F3 and F4 in Table \ref{t4b}).} 
\label{fig:hdisF2}
\end{figure}

This section contains supplementary Figures \ref{fig:hdisA}, \ref{fig:hdisB}, \ref{fig:hdisC}, \ref{fig:hdisD}, \ref{fig:hdisE}, \ref{fig:hdisF} and \ref{fig:hdisF2}, discussed in Section \ref{sec:sims}.

\subsection{Time comparison}\label{time}

\begin{figure}
    \centering
    \includegraphics[scale = 0.5]{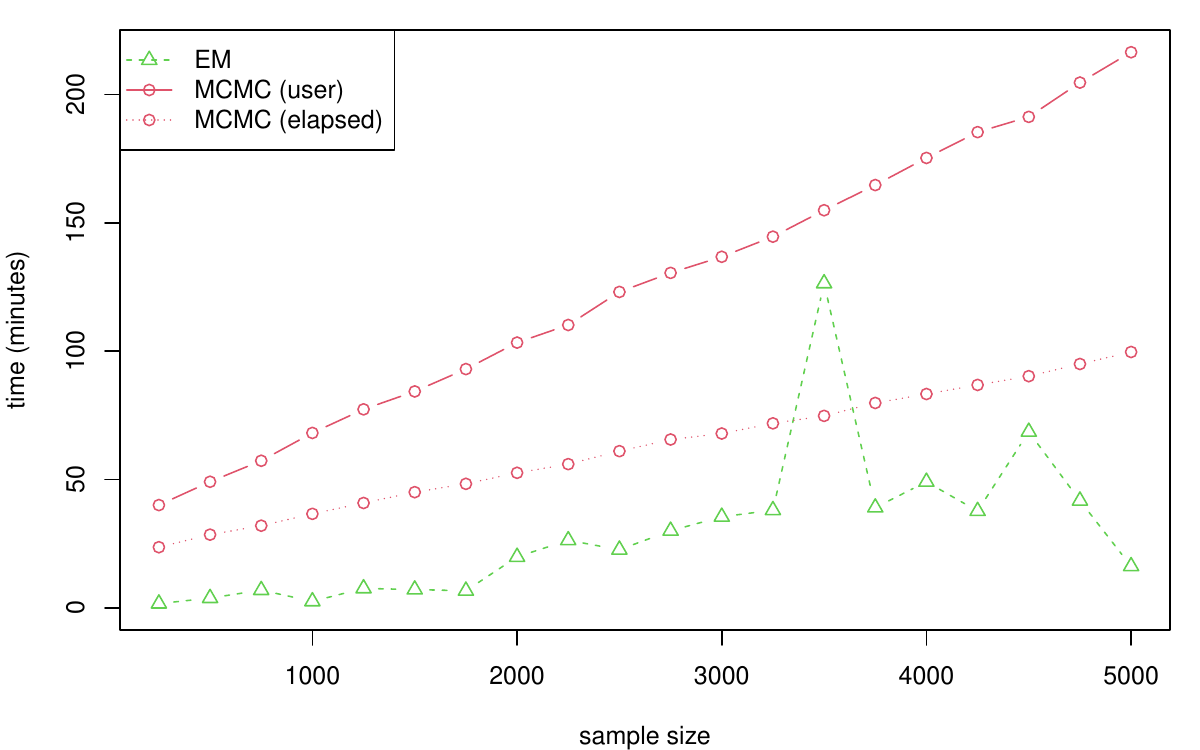}
    \caption{Time needed for applying the MCMC and EM algorithms. The elapsed time is smaller than user time for the MCMC implementation because each heated chain runs in parallel cores. Details of algorithms: the EM algorithm uses a random small-EM scheme based on 45 starts and convergence was assessed by iterating the Expectation and Maximization step until the maximum difference between parameter updates in successive steps becomes smaller than 0.005. The MC$^3$ scheme in our MCMC algorithm uses 16 heated chains in total, for a total of $M= 20000$ cycles. Each cycle consists of 10 usual MCMC iterations.}
    \label{fig:time}
\end{figure}

In this section we assess the computational complexity of our method by examining the CPU time needed to run our MC$^{3}$ algorithm for various levels of sample size. We are also reporting the corresponding time needed by the EM algorithm implementation. For this purpose, we generated 20 synthetic datasets as described in the main simulation study, considering that the sample size ($n$) ranges in the set $n\in \{250, 500, \ldots, 5000\}$. For each dataset we ran the the EM and MCMC algorithms as described in Section \ref{sec:sims} and measured CPU time. The results are displayed in Figure \ref{fig:time}. Note that for the MCMC implementation we also utilized parallel processing of the heated chains so we are reporting both user and elapsed time. All benchmarks were performed in a Linux workstation with the following specifications: Operating System: Ubuntu 22.04.4 LTS, 11th Gen Intel Core i9-11900 @ 2.50GHz $\times$ 16.

\clearpage
\newpage

\end{document}